\documentclass[final,1p,sort&compress]{elsarticle}
\usepackage{mathrsfs}
\usepackage{amsmath}
\usepackage{amssymb}

\usepackage{multirow}
\usepackage{array}

\usepackage{tikz}
\usepackage{circuitikz}
\usepackage{color}
\usepackage{graphicx}
\usepackage{epstopdf}

\journal{Physics Reports}

\usepackage{units}

\usepackage{enumerate}

\usepackage{url}
\usepackage[colorlinks]{hyperref}
\hypersetup{%
	plainpages=true,
	breaklinks=true,
	hypertexnames=false,
	pageanchor=true,
	colorlinks=true,
	linkcolor={blue},
	citecolor={red},
	urlcolor={blue},
	anchorcolor={black}
}

\newcommand{\rd}{\ensuremath{\mathrm{d}}}
\newcommand{\id}{\ensuremath{\,\rd}}

\newcommand{\bra}[1]{\left\langle{#1}\right|}
\newcommand{\ket}[1]{\left|{#1}\right\rangle}

\newcommand{\ketbra}[2]{\left|{#1}\rangle \langle{#2}\right|}
\newcommand{\brakket}[3]{\langle{#1}|{#2}|{#3}\rangle}
\newcommand{\expec}[1]{\left\langle{#1}\right\rangle}

\newcommand{\expecc}[1]{\text{E}\left[#1\right]}

\newcommand{\comm}[2]{\left[ #1, #2 \right]}
\newcommand{\lind}[1]{\mathcal{D}\left[#1\right]}
\newcommand{\meas}[1]{\mathcal{M}\left[#1\right]}

\newcommand{\measg}[1]{\mathcal{G}\left[#1\right]}

\newcommand{\sz}{\sigma_z}
\newcommand{\sx}{\sigma_x}

\newcommand{\sm}{\sigma_-}
\renewcommand{\sp}{\sigma_+}

\newcommand{\abs}[1]{\left|#1\right|}

\newcommand{\nn}{\nonumber}

\newcommand{\figref}[1]{\mbox{Fig.~\ref{#1}}}
\newcommand{\tabref}[1]{\mbox{Table~\ref{#1}}}
\newcommand{\secref}[1]{\mbox{Sec.~\ref{#1}}}

\newcommand{\appref}[1]{\mbox{Appendix~\ref{#1}}}
\renewcommand{\eqref}[1]{\mbox{Eq.~(\ref{#1})}}

\newcommand{\be}{\begin{equation}}
\newcommand{\ee}{\end{equation}}
\newcommand{\bea}{\begin{eqnarray}}
\newcommand{\eea}{\end{eqnarray}}

\def \cc{\text{c.c.}}

\begin{document}
	
\begin{frontmatter}

\title{Microwave photonics with superconducting quantum circuits}

\author[FirstAddress,SecondAddress]{Xiu Gu \fnref{myfootnote}}

\author[SecondAddress]{Anton Frisk Kockum \fnref{myfootnote}}

\fntext[myfootnote]{These authors contributed equally to this work.}

\author[SecondAddress,ThirdAddress]{Adam Miranowicz}
\author[FirstAddress,SecondAddress,FourthAddress]{Yu-xi Liu\corref{corresponding1}}
\cortext[corresponding1]{Corresponding author}
\ead{yuxiliu@mail.tsinghua.edu.cn}

\author[SecondAddress,FifthAddress]{Franco Nori\corref{corresponding}}
\cortext[corresponding]{Corresponding author}
\ead{fnori@riken.jp}

\address[FirstAddress]{Institute of Microelectronics, Tsinghua University, Beijing 100084, China}
\address[SecondAddress]{Center for Emergent Matter Science, RIKEN, Saitama 351-0198, Japan}
\address[ThirdAddress]{Faculty of Physics, Adam Mickiewicz University, 61-614 Pozna\'n, Poland}
\address[FourthAddress]{Tsinghua National Laboratory for Information Science and Technology (TNList), Beijing 100084, China}
\address[FifthAddress]{Physics Department, The University of Michigan, Ann Arbor, Michigan 48109-1040, USA}

\date{\today}
	
\begin{abstract}

In the past 20 years, impressive progress has been made both experimentally and theoretically in superconducting quantum circuits, which provide a platform for manipulating microwave photons. This emerging field of superconducting quantum microwave circuits has been driven by many new interesting phenomena in microwave photonics and quantum information processing. For instance, the interaction between superconducting quantum circuits and single microwave photons can reach the regimes of strong, ultra-strong, and even deep-strong coupling. Many higher-order effects, unusual and less familiar in traditional cavity quantum electrodynamics with natural atoms, have been experimentally observed, e.g., giant Kerr effects, multi-photon processes, and single-atom induced bistability of microwave photons. These developments may lead to improved understanding of the counterintuitive properties of quantum mechanics, and speed up applications ranging from microwave photonics to superconducting quantum information processing. In this article, we review experimental and theoretical progress in microwave photonics with superconducting quantum circuits. We hope that this global review can provide a useful roadmap for this rapidly developing field.

\end{abstract}

	
\end{frontmatter}


\newpage
\tableofcontents
\newpage

\section{Introduction}
\label{sec:Introduction}

The interaction between matter and electromagnetic fields is one of the most fundamental processes occurring in nature. This interaction is explored in \emph{cavity quantum electrodynamics} (cavity QED)~\cite{Haroche2006}, the field of QED that studies the interactions between atoms and quantized electromagnetic fields inside a high-$Q$ cavity. This research field is very important for both atomic physics and quantum optics. In the past, research in cavity QED was focused on phenomena in the optical domain~\cite{Vahala2004}.

Starting from experiments on macroscopic quantum coherence in devices where Cooper pairs tunnel between small superconducting islands~\cite{Nakamura1999}, the last two decades have seen a greatly increased interest in using \emph{superconducting quantum circuits} (SQCs), based on Josephson junctions, for implementing quantum bits -- the basic units of quantum information processing. It has also been demonstrated that SQCs possess discrete energy levels and behave like atoms. For this reason, SQCs are often referred to as \textit{superconducting artificial atoms}. The typical voltage applied to a Josephson junction is a few microvolts, so SQCs usually operate at frequencies in the microwave domain. The study of properties of SQCs and the interaction between SQCs and a classical microwave field is related to atomic physics. However, the study of the interaction between SQCs and a quantized microwave electromagnetic field, which is usually referred to as \emph{circuit QED}, requires more knowledge of quantum optics.

\begin{figure}
\centering
\includegraphics[width=\linewidth]{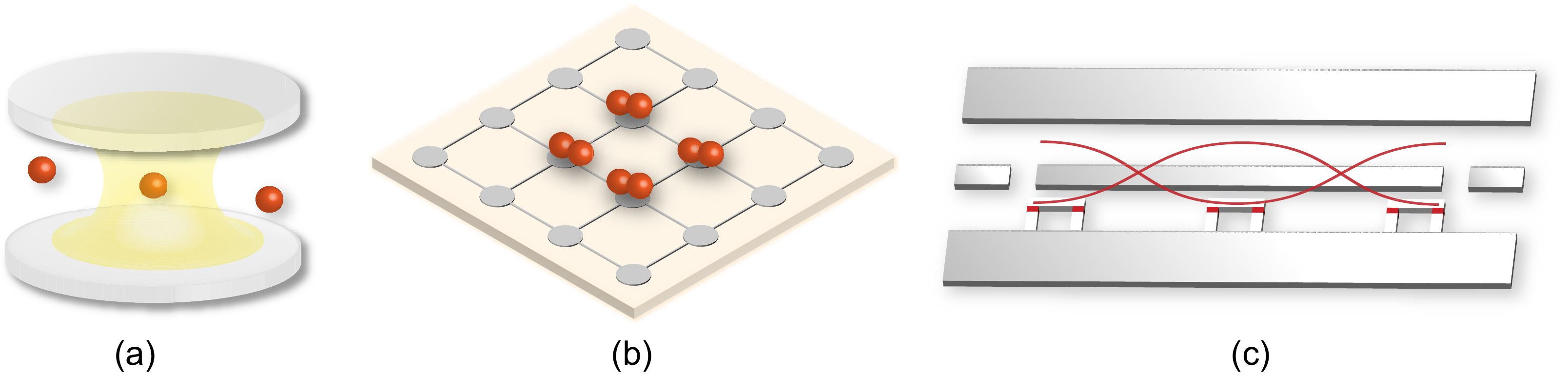}
\caption{Schematic illustrations of natural and artificial atoms interacting with quantized bosonic modes. (a) Cavity QED. Flying neutral atoms interact with photons confined in a high-$Q$ cavity. (b) Two-dimensional array of ion traps.
An ensemble of ions are trapped by electric and magnetic fields at each site. Lasers manipulate the collective vibration modes of the ions. These vibrational modes couple to the ions' internal degrees of freedom. (c) Circuit QED. Superconducting qubits with Josephson junctions (red blocks) fabricated on chip interact with the electric fields in a transmission-line resonator. Courtesy of Ze-Liang Xiang. 
\label{fig:Atoms}}
\end{figure}

Phenomena which occur in atomic physics and quantum optics, using natural atoms and optical photons, are expected to also manifest in setups using superconducting artificial atoms and microwave photons. For example, experiments where superconducting qubits are strongly coupled to a superconducting transmission-line resonator open many new possibilities for studying the strong interaction between microwave photons and various quantum devices, in analogy with the interaction between optical photons and atoms in quantum optics~\cite{You2003, Wallraff2004, Blais2004}. That is, the studies in cavity QED can be extended to quantized electrical circuits, constructed by Josephson junctions and various electronic components. This circuit QED has developed into a rapidly growing field of studying quantum optics in the microwave domain with SQCs. For a basic comparison between SQCs and natural atoms, see \figref{fig:Atoms} and \tabref{tab:comparison}.

Many well-known physical effects in quantum optics and atomic physics have been demonstrated with SQCs, including dressed states, Autler--Townes splitting, electromagnetically induced transparency, the Mollow triplet, and other phenomena. Also, much theoretical and experimental activity is presently concerned with further exploring atomic physics and quantum optics in the microwave domain with SQCs. To distinguish such studies from traditional quantum optics and atomic physics, we refer to this research field as \emph{microwave photonics with SQCs}.

\begin{table}
\centering
\renewcommand{\arraystretch}{1.2}
\renewcommand{\tabcolsep}{0.15cm}
\providecommand{\tabularnewline}{\\}
\small
	\begin{tabular}{|>{\centering}m{2cm}|>{\centering}m{3.3cm}|>{\centering}m{3.3cm}|>{\centering}m{3.6cm}|}
		\hline 
		\multirow{2}{3cm}{}  & \multicolumn{2}{c|}{\textbf{Natural atoms}} & \textbf{Artificial atoms} \tabularnewline
		\cline{2-4} 
		
		                    & Neutral atoms & Trapped ions & Superconducting qubits \tabularnewline
		\hline 
		\textbf{Qubits}     & Atoms & Ions & Josephson-junction devices\tabularnewline
		\hline 
		\textbf{Dimensions} & $\sim\unit[10^{-10}]{m}$ & $\sim\unit[10^{-10}]{m}$ & $\sim\unit[10^{-6}]{m}$ \tabularnewline
		\hline 
		\textbf{Energy gap}  & $\sim\unit[10^{14}]{Hz}$ (optical), GHz (hyperfine)  & $\sim\unit[10^{14}]{Hz}$ (optical), \\ GHz (hyperfine) & $\sim\unit[1\textup{--}10]{GHz}$ \tabularnewline
		\hline 
		\textbf{Quantized bosonic modes} & Photons & Collective vibration modes of ions \\(phonons) & $LC$ circuits (photons), surface acoustic waves (phonons)\tabularnewline 
		\hline 
		\textbf{Frequency range} & Microwave, optical & Microwave, optical  & Microwave \tabularnewline
		\hline 
		
		\textbf{Controls}   & Lasers & Lasers, electric/magnetic fields & Microwave pulses, voltages, currents \tabularnewline
		\hline
		
		\multirow{6}{3cm}{\textbf{Components}}  & Mirrors & Electrodes & Capacitors \tabularnewline
		\cline{2-4} 
	
		                             & Optical and microwave cavities & Optical cavities, vibration modes & $LC$ and transmission-line resonators, 3D cavities \tabularnewline
	    	\cline{2-4}                         
		                             & Optical fibers & Optical fibers & Transmission lines \tabularnewline
		\cline{2-4}                         
		                             & Beam-splitters & Beam-splitters  & Hybrid couplers, Josephson mixers \tabularnewline
		\hline 
		\textbf{Temperature} & nK--$\mu$K & $\mu$K--mK & $\sim\unit[10]{mK}$ \tabularnewline
		\hline 
		\textbf{Advantages} & Homogeneous (parameters set \\ by nature) & Long coherence times & Strong and controllable coupling, tunable in situ, fabricated on chip \tabularnewline
		\hline 
	\end{tabular}
\caption{Comparison between natural atoms and artificial atoms based on superconducting circuits and their interaction with quantized bosonic modes.
\label{tab:comparison}}
\end{table}

In contrast to natural atoms, the superconducting artificial atoms can be designed, fabricated, and controlled for various research purposes. Furthermore, the interaction between artificial atoms and electromagnetic fields can be artificially engineered. Hence, SQCs can possess some features, which are very different from those of natural atoms. That is, circuit QED can be used to demonstrate phenomena that cannot be realized or observed in atomic physics and quantum optics. For instance, single- and two-photon processes can coexist in SQCs, and the coupling between SQCs and microwave fields can become ultrastrong.

In the past twenty years, significant progress has been made in this emerging research field of circuit QED. Although there are several brief reviews on SQCs and circuit QED, a comprehensive and pedagogical introduction to the field is still needed. Thus, we believe that it is a good moment to summarize the results of the field and consider future perspectives.

In this review, we mainly focus on the recent progress of atomic physics and quantum optics with SQCs. Quantum computing using SQCs is only partially covered in the last Section on applications. We first review several different SQCs, which are considered as superconducting artificial atoms in different configurations, and describe various models of microwaves interacting with SQCs. Then we summarize the progress achieved so far in quantum optics and atomic physics in the microwave domain with SQCs. We also describe some research subjects and devices (e.g., circulators and mixers) which are not common in quantum optics and atomic physics. Throughout the review, we discuss the advantages and drawbacks of SQCs compared to natural atoms.

The sections of the review are organized as follows. In~\secref{sec:BasicConcepts}, we introduce basic components, concepts, and models of various superconducting artificial atoms, superconducting qubits, and resonators. In~\secref{sec:MicrowaveComponents}, we introduce other microwave components such as beam-splitters, circulators, switches, and mixers. In~\secref{sec:CircuitQED}, we discuss the interaction between a quantized microwave field and SQCs. Starting from a general theoretical description, we show how the strength of the interaction defines various parameter regimes of interest, including the regime of ultrastrong coupling. In~\secref{sec:WaveguideQED}, we provide an overview of the progress in waveguide QED, where SQCs interact with a continuum of modes in an open transmission line rather than a single mode in a resonator. In~\secref{sec:QuantumOptics}, we summarize a series of typical phenomena in microwave quantum optics and atomic physics that have been demonstrated with SQCs. These phenomena include electromagnetically induced transparency, Autler--Townes splitting, lasing with and without population inversion, sideband transitions, squeezed states, photon blockade, and quantum jumps. In particular, we pay extra attention to phenomena which cannot be demonstrated using traditional quantum optics and atom physics. In~\secref{sec:NonlinearProcess}, some new nonlinear phenomena in superconducting artificial atoms are summarized. In~\secref{sec:PhotonGeneration} and~\secref{sec:PhotonDetection}, we describe theoretical and experimental progress for generating and detecting microwave photons using SQCs. Compared to optical photons, microwave photons are easier to generate and manipulate, but harder to detect. In~\secref{sec:Applications}, we briefly review superconducting quantum information processing and quantum metamaterials using superconducting qubits. Finally, we provide a very brief summary and some perspectives on the field in~\secref{sec:Summary}. As a complement to the first few sections of the review, we present more details about circuit quantization in~\appref{sec:circuit-quantization} and about unitary transformations and the Jaynes--Cummings model in~\appref{sec:UnitaryTransfJC}. A list of acronyms can be found in~\appref{sec:Acronyms}.

\section{Basic concepts: qubits and resonators}
\label{sec:BasicConcepts}

In microwave photonics with superconducting quantum circuits (SQCs), the basic building blocks of cavity QED, \textit{natural atoms} and \textit{optical cavities}, are replaced with \textit{superconducting artificial atoms} and \textit{resonators}, respectively. In this section, we provide an overview of these fundamental parts of a circuit-QED setup and demonstrate the wide choice of designs possible for artificial atoms. We begin in~\secref{sec:SQCsAndJJs} by introducing the Josephson junction and explaining why it is essential to SQCs. We then describe three basic types of SQCs functioning as artificial atoms (\secref{sec:ThreeSQCTypes}) and show how these three designs have been extended to improve various properties (\secref{sec:SQCExtensions}). After discussing selection rules for (\secref{sec:SelectionRules}), multi-level versions of (\secref{sec:ThreeLevel}), and readout of (\secref{sec:Readout}) the artificial atoms, we end in~\secref{sec:Resonators} by reviewing different types of resonators that the SQCs can be coupled to. The interaction between resonators and SQCs is treated in~\secref{sec:CircuitQED}. While the current sections give Hamiltonians for many SQCs, the procedure for deriving such Hamiltonians from electric circuits is reviewed in~\appref{sec:circuit-quantization}.

\subsection{Superconducting quantum circuits and Josephson junctions}
\label{sec:SQCsAndJJs}

The prototypes of SQCs date back to the study of \textit{Josephson junctions} in the 1980s (e.g., Ref.~\cite{Martinis1985}). This research was related to the question whether macroscopic systems can behave quantum mechanically. Recently, the study of SQCs has been attracting extensive attention, and one can say that SQCs have opened a new research area with many potential applications in quantum-information processing. In addition to Josephson junctions, SQCs are constructed with inductors, capacitors, and other electronic elements. The Josephson junctions are equivalent to \textit{nonlinear inductors}. 

SQCs are macroscopic in scale, and operate at milli-Kelvin temperatures to maintain their superconducting states. Similar to natural atoms, SQCs have discrete energy levels. Therefore, SQCs are also referred to as \textit{superconducting artificial atoms}. However, in contrast to natural atoms, SQCs can be artificially designed and fabricated for different research purposes and their energy levels can be adjusted by external parameters, e.g., currents and voltages, or magnetic and electric fields.  Also, the coupling strength between SQCs and their electromagnetic environments can be tuned. A detailed description of these unique properties has been presented in a number of excellent reviews on superconducting artificial atoms~\cite{Makhlin2001, Devoret2004, You2005, Wendin2005, Wendin2007, Flatte2007, Zagoskin2007, Clarke2008, Schoelkopf2008, Girvin2009, Martinis2009, You2011, Buluta2011, Siddiqi2011, Zagoskin2011, Oliver2013, Langford2013, Xiang2013, Devoret2013, QuantumMachines2014}. Here, for the completeness of this article, we present a brief summary of these reviews.

A basic requirement for the SQCs to function as artificial atoms is the \textit{anharmonicity} of their energy-level spacing. Josephson junctions play an important role in SQCs because of the strong nonlinearity they provide, which is the key to making superconducting artificial atoms anharmonic. Another important property of the Josephson junctions is that they, like the other circuit elements in an SQC, have negligibly small energy dissipation.

A Josephson junction is made of two superconducting electrodes, separated by a thin insulating film (with typical thickness of \unit[1-3]{nm}) in a sandwich structure, giving rise to an intrinsic capacitance $C_{\rm J}$. Two bound electrons with opposite spins, known as a Cooper pair, in the superconducting electrodes, can tunnel coherently through the insulating barrier one by one. These tunneling Cooper pairs form the so-called supercurrent, governed by the dc Josephson effect~\cite{Josephson1962}: $I = I_{\rm c} \sin \phi$, where $\phi$ is the gauge-invariant phase difference between the two superconducting electrodes and the critical current $I_{\rm c}$ is determined by material properties. The phase difference $\phi$ is connected to the voltage $V$ across the junction through the relation $ \partial \phi / \partial t = 2e V / \hbar$, where $e$ is the elementary charge. 

From these two equations, it can be shown that the Josephson junction has an energy $-E_{\rm J} \cos\phi$, where $E_{\rm J} = \hbar I_{\rm c} / (2e)$ is called the Josephson energy. Furthermore, the current-voltage relations reveal that the Josephson junction functions like a nonlinear inductance with $\abs{L} = \hbar / (2e I_{\rm c} \cos\phi)$. It is this nonlinearity of the Josephson junction that brings about the anharmonicity of SQCs. In a given SQC, we can thus select the two lowest-energy levels from the non-equally spaced energy spectrum. These two levels form a \textit{quantum bit} (qubit) for quantum-information processing. 

When an ac voltage $V$ is applied to the two electrodes of the Josephson junction, the supercurrent $I$ is periodically modulated as $I = I_{\rm c} \sin(\omega t + \phi)$ with the Josephson frequency $\nu = \omega / (2\pi) = 2eV / h$. This is called the ac Josephson effect. The energy $h\nu$ equals the energy change of a Cooper pair transferred across the junction. The voltage applied to a Josephson junction is typically on the order of a few microvolts. Thus, SQCs are usually operating at frequencies in the microwave regime.

\subsection{Three basic SQC types}
\label{sec:ThreeSQCTypes}

\begin{figure}
\centering
\includegraphics[width=\linewidth]{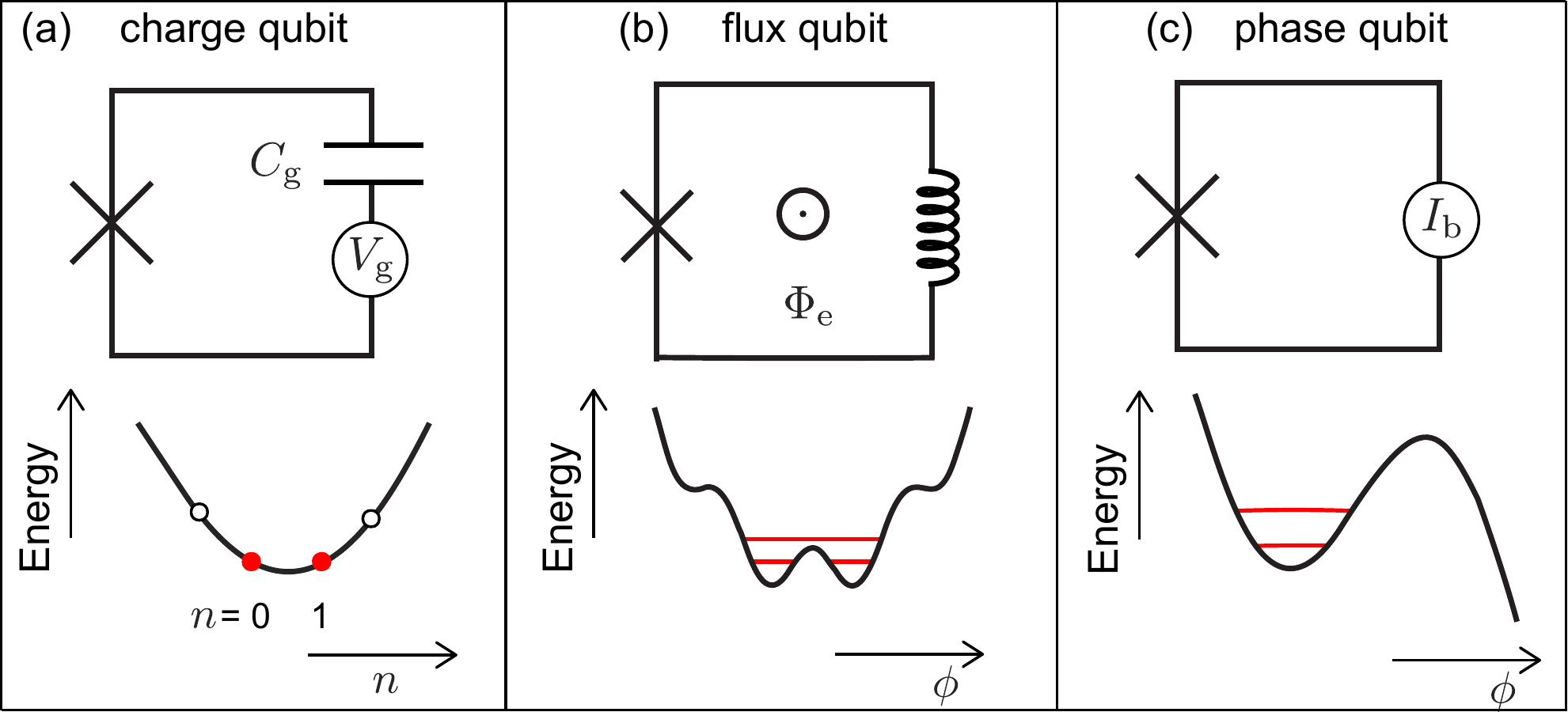}
\caption{Schematic diagrams of the three basic superconducting quantum circuits and their potential energies. (a) Charge qubit, (b) flux qubit, and (c) phase qubit.
\label{fig:BasicQubits}}
\end{figure}

There are three basic types of SQCs~\cite{Devoret2004, Wendin2005} with Josephson junctions, classified by the ratio $E_{\rm J} / E_{\rm C}$, where $E_{\rm C} = e^2 / (2C)$ is the electrostatic Coulomb energy (single-electron charging energy). The relevant capacitance $C$ is either the capacitance $C_{\rm J}$ of the Josephson junction or the capacitance of a superconducting island connected to the Josephson junction, depending on the circuit. Quantization of the SQCs, which reveals their discrete energy levels, can be realized by defining a pair of canonically conjugate variables: $\phi$, the gauge-invariant phase difference, and $n$, the number of Cooper pairs on the superconducting island. This pair of conjugate variables satisfies the commutation relation $\comm{\phi}{n} = i$ (or, rather, $\comm{e^{i\phi}}{n} = e^{i\phi}$~\cite{GerryKnight}) and obeys the Heisenberg uncertainty relation $\Delta\phi \,\Delta n \geq 1$. When $E_{\rm J} > E_{\rm C}$, the Josephson behavior of the junction dominates and the charge number $n$ has large quantum fluctuations. Conversely, when $E_{\rm C} > E_{\rm J}$ the charging behavior of the capacitance dominates; the charge number $n$ is well defined and $\phi$ has large quantum fluctuations.

Just like natural atoms, SQCs are \textit{multi-level systems}. When we limit our study to only the two lowest-energy levels, a two-state system, i.e., a qubit, can be defined. Using the ideas described here, SQCs based on Josephson junctions can basically be divided into three kinds of circuits, called \textit{charge}, \textit{flux}, and \textit{phase qubits}, which are shown in \figref{fig:BasicQubits} and discussed in more detail below.

\subsubsection{Superconducting charge-qubit circuits}
\label{sec:SuperconductingChargeQubits}

A \textit{charge-qubit circuit}, also called a single Cooper-pair box, can be defined when $E_{\rm C} / E_{\rm J} \sim 10$. As schematically shown in Fig.~\ref{fig:BasicQubits}(a), the circuit consists of a superconducting island (with $n$ excess Cooper pairs) connected to ground via one or two small Josephson junctions (with Josephson energy $E_{\rm J}$ and capacitance $C_{\rm J}$)~\cite{Bouchiat1998, Nakamura1999, Pashkin2009}. In the latter case, the two Josephson junctions form a superconducting quantum interference device (SQUID). A gate voltage $V_{\rm g}$, applied to the island through a gate capacitance $C_{\rm g}$, is used to control the energy-level spacing of the qubit circuit. For a single-junction circuit, the single-electron charging energy is $E_{\rm C} = e^2 / [2(C_{\rm J} + C_{\rm g})]$. 

If the superconducting energy gap $\Delta E$ of the island is much larger than the thermal energy $k_{\rm B} T$, where $k_{\rm B}$ is Boltzmann's constant and $T$ is the temperature of the qubit environment, then only Cooper pairs can tunnel coherently through the Josephson junction. In this case, the system Hamiltonian can be written as
\be
H = 4 E_{\rm C} (n - n_{\rm g})^2 - E_{\rm J} \cos \phi,
\label{eq:HamiltonianCPB}
\ee
where $n_{\rm g} = C_{\rm g} V_{\rm g} / (2e)$ is the gate-charge number. We recall that $n$ is the number of excess Cooper pairs on the island, and $\phi$ is the gauge-invariant phase difference across the Josephson junction. To quantize the circuit, we use $n$ and $\phi$ as a pair of quantum-mechanical conjugate operators. Given that the conditions
\be
\Delta E \gg E_{\rm C} \gg E_{\rm J} \gg k_{\rm B} T
\ee
are satisfied, the charge energy $E_{\rm C}$ dominates \eqref{eq:HamiltonianCPB} for most values of $n_{\rm g}$ and, then, a charge-qubit circuit can be defined. In the charge basis $\ket{n}$, the Hamiltonian in \eqref{eq:HamiltonianCPB} can be rewritten as
\be
H = \sum_n \left[ 4 E_{\rm C} (n - n_{\rm g})^2 \ketbra{n}{n} - \frac{1}{2} E_{\rm J} (\ketbra{n+1}{n} + \ketbra{n}{n+1}) \right],
\label{eq:HamiltonianCPBChargeBasis}
\ee
which shows that the charge-qubit circuit is a multilevel system. The energy-level spacing \textit{can be changed} by varying $n_{\rm g}$ through the external gate voltage $V_{\rm g}$, i.e., the gate voltage acts as a control parameter in such an artificial atom. 

We note that the value $n_{\rm g} = n + (1/2)$ plays a very special role in the circuit:
\begin{enumerate}[(i)]
\item The eigenstates of the system are symmetric around this point and have well-defined parities there.
\item The charging energies $4 E_{\rm C} (n - n_{\rm g})^2$ are degenerate at this point for any two adjacent charge states $\ket{n}$ and $\ket{n+1}$. This degeneracy can be broken by the Josephson energy.
\item The two lowest-energy levels of the circuit are well separated from the upper levels at this point, so the circuit \textit{can be reduced} to a two-state system by only considering the two lowest-energy levels. Thus, a charge qubit is formed, described by the Hamiltonian
\be
H_{\rm q} = - 2 E_{\rm C} (1 - 2 n_{\rm g}) \sz - E_{\rm J} \sx / 2,
\ee
where $\sz = \ketbra{1}{1} - \ketbra{0}{0}$ and $\sx = \ketbra{1}{0} + \ketbra{0}{1}$ are the Pauli operators in the charge basis. We note that the state $\ket{1}$ ($\ket{0}$) corresponds to the spin state $\ket{\uparrow}$ ($\ket{\downarrow}$).
\item Up to first order in perturbation theory, the qubit transition frequency is \textit{insensitive to offset-charge noise}, and thus the coherence time of the charge qubit is optimized. 
\end{enumerate}
Considering properties (ii) and (iv), $n_{\rm g} = n + (1/2)$ is usually called the degenerate or optimal point. For SQUID-based two-junction circuits, $E_{\rm J}$ and $C_{\rm J}$ in the Hamiltonian of a single-junction circuit are replaced by $E_{\rm J} \cos(\pi \Phi_{\rm e} / \Phi_0)$ and $2C_{\rm J}$, respectively, where $\Phi_{\rm e}$ is the external magnetic flux through the SQUID loop and $\Phi_0$ is the flux quantum. Thus, the gate voltage $V_{\rm g}$ and the external magnetic flux $\Phi_{\rm e}$ together enable \textit{full control} for these SQUID-based charge-qubit circuits.

\subsubsection{Superconducting flux-qubit circuits}

A \textit{flux-qubit circuit}, also called a persistent-current-qubit circuit, can be defined when $E_{\rm J} / E_{\rm C} \sim 50$~\cite{Mooij1999, VanderWal2000, Friedman2000, Chiorescu2003}. This qubit is analogous to a particle with an anisotropic mass moving in a two-dimensional periodic potential~\cite{Orlando1999}. As schematically shown in \figref{fig:BasicQubits}(b), a flux-qubit circuit consists of a superconducting loop interrupted by one, three, or more than three Josephson junctions. For a flux-qubit circuit with a single-junction loop~\cite{Leggett1980, Friedman2000}, a large self-inductance is needed to obtain enough quantum states in the local minimum of the circuit's potential energy. Such a large self-inductance requires a large-size loop, making the qubit sensitive to flux noise. However, flux-qubit circuits based on three junctions~\cite{Mooij1999, VanderWal2000, Chiorescu2004} enable a reduction of the loop size while retaining a large inductance; thus, these circuits have been extensively adopted for qubit devices. In such circuits, the size of one junction is $\alpha$ times smaller than those of the other two (identical) junctions. If $\alpha \leq 0.5$, the flux-qubit circuit only has a single potential well. To create a double-well potential, $\alpha > 0.5$ is used, and typically $0.6 < \alpha < 0.7$ to avoid effects of charge noise. A magnetic flux $\Phi_{\rm e}$ through the superconducting loop is used to adjust the potential energy of the circuit and, thus, to \textit{vary} its energy-level spacing. 

The Hamiltonian of the three-junction circuit can be written as~\cite{Orlando1999, Liu2005a}
\be
H = \frac{P_{\rm p}^2}{2M_{\rm p}} + \frac{P_{\rm m}^2}{2M_{\rm m}} + U(\varphi_{\rm p}, \varphi_{\rm m}),
\label{eq:HFluxQubit}
\ee
where $M_{\rm p} = 2 C_{\rm J} (\Phi_0 / 2 \pi)^2$ and $M_{\rm m} = M_{\rm p} (1 + 2 \alpha)$ can be considered effective masses, while $P_{\rm p} = - i \hbar \partial / \partial \varphi_{\rm p}$ and $P_{\rm m} = - i \hbar \partial / \partial \varphi_{\rm m}$ can be considered effective momenta. Moreover, $\varphi_{\rm p} = (\varphi_1 + \varphi_2) / 2$ and $\varphi_{\rm m} = (\varphi_1 - \varphi_2) / 2$ are defined by the phase drops $\varphi_1$ and $\varphi_2$ across the two larger junctions. The effective potential $U(\varphi_{\rm p}, \varphi_{\rm m})$ is given by
\be
U(\varphi_{\rm p}, \varphi_{\rm m}) = 2 E_{\rm J} (1 - \cos \varphi_{\rm p} \cos \varphi_{\rm m}) + \alpha E_{\rm J} [1 - \cos (2 \pi f + 2 \varphi_{\rm m})],
\ee
where $f = \Phi_{\rm e} / \Phi_0$ is the reduced magnetic flux. It is clear that the shape of the double-well potential energy $U(\varphi_{\rm p}, \varphi_{\rm m})$ can be changed from asymmetric to symmetric if $f$ is changed from $f \neq 0.5$ to $f = 0.5$ by adjusting the external magnetic flux $\Phi_{\rm e}$. Thus, $\Phi_{\rm e}$ is a \textit{control parameter} for various properties of these flux-qubit circuits.

In these SQCs, the two lowest energy levels are well isolated from the upper levels when $f$ is near the point $f = 0.5$. Thus, a two-level system (a qubit) can be realized by only considering these two lowest energy levels. By projecting the full Hamiltonian onto these two levels, we obtain the approximate Hamiltonian of the flux qubit,  
\be
H_{\rm q} = \frac{\varepsilon \sz + \delta \sx}{2},
\ee
with $\varepsilon = I_{\rm p} (2 \Phi_{\rm e} - \Phi_0)$, and Pauli operators $\sz = \ketbra{\circlearrowleft}{\circlearrowleft} - \ketbra{\circlearrowright}{\circlearrowright}$ and $\sx = \ketbra{\circlearrowleft}{\circlearrowright} + \ketbra{\circlearrowright}{\circlearrowleft}$ given in the circulating-current basis. Here, $\ket{\circlearrowleft}$ and $\ket{\circlearrowright}$ are, respectively, the anti-clockwise and clockwise supercurrents $I_{\rm p}$ circulating in the loop of the qubit circuit. The parameter $\delta$ is the \textit{tunneling-coupling strength} between the two current states, which correspond to different potential wells. We mention that the flux qubits are, up to first order in perturbation theory, insensitive to electromagnetic noise when $f = 0.5$. Thus, the point corresponding to $f = 0.5$ is called the \textit{optimal working point} for a flux qubit. Furthermore, the inversion symmetry of the potential energy $U(\varphi_{\rm p}, \varphi_{\rm m})$ for the variables $\varphi_{\rm p}$ and $\varphi_{\rm m}$ is well defined at this point. This symmetry implies that the parities of all the eigenstates of the system also are well defined. In this sense, superconducting flux-qubit circuits are the same as natural atoms.

\subsubsection{Superconducting phase-qubit circuits}

The study of superconducting \textit{phase-qubit circuits} can be dated back to the 1980s and the exploration of quantum-mechanical properties of macroscopic degrees of freedom~\cite{Martinis1985, Clarke1988}. As schematically shown in \figref{fig:BasicQubits}(c), these circuits consist of a large Josephson junction with $E_{\rm J} / E_{\rm C} \sim 10^6$~\cite{Ramos2001, Martinis2002, Yu2002, Martinis2009}. The ``tilted washboard'' shape of the junction's potential energy is controlled through an applied dc bias current $I_{\rm b}$. The magnetic flux threading the phase-qubit loop can also act as a control parameter for the washboard potential. 

The Hamiltonian of the phase-qubit circuit is given by
\be
H = \left( \frac{2e}{\hbar} \right)^2 \frac{p^2}{2C_{\rm J}} - \frac{\Phi_0}{2\pi} I_{\rm b} \varphi - E_{\rm J} \cos \varphi,
\ee
with the charge $Q = 2 e p / \hbar$ in the junction capacitance defining a ``momentum'' $p$. By similarly treating $\varphi$ as a ``position'', letting $p$ and $\varphi$ form a pair of canonically conjugate quantum operators with the commutation relation $\comm{\varphi}{p} = i$, the phase-qubit circuit can be quantized. In such a circuit, the bias current $I_{\rm b}$ is usually very close to the critical current $I_{\rm c} = 2 e E_{\rm J} / \hbar$ of the Josephson junction. Because of this, the energy levels have small anharmonicity. By projecting a given multi-level phase-qubit circuit onto its two-lowest-energy levels, a phase qubit and its Hamiltonian is obtained, just as for charge and flux qubits. However, there is \textit{no symmetry} for the potential energy of phase-qubit circuits and, thus, there are no well defined parities for the eigenstates. We mention that there is no optimal point to minimize the effect of electromagnetic noise on phase-qubit circuits, but the phase qubit is \textit{insensitive to offset-charge noise} since the ratio $E_{\rm J} / E_{\rm C}$ is so large.

\subsection{Various extensions of SQCs}
\label{sec:SQCExtensions}

\begin{figure}
\centering
\includegraphics[width=\linewidth]{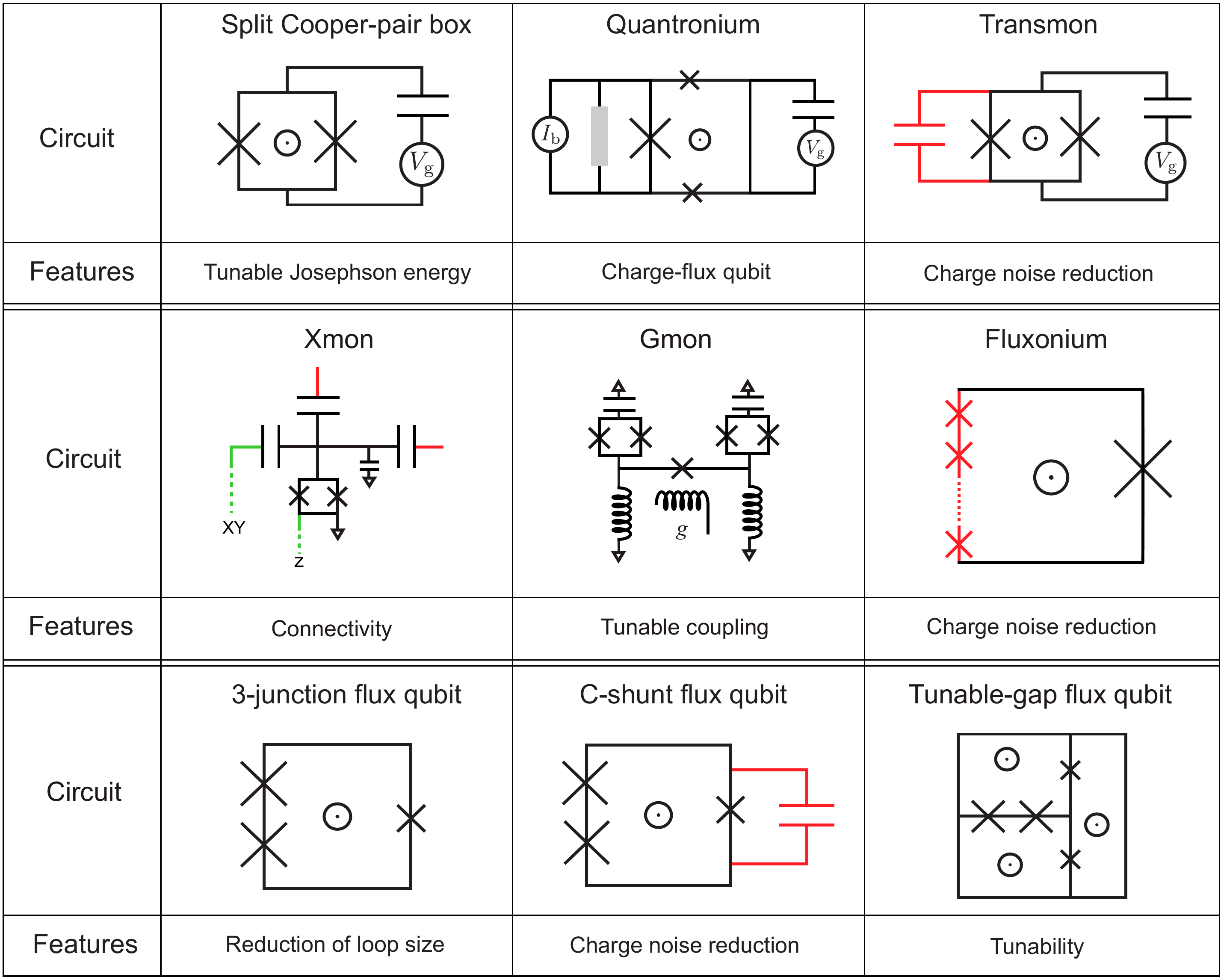}
\caption{Various extensions of the three basic superconducting qubits. The split Cooper-pair box and the 3-junction flux qubit shown in the left column were already discussed in the previous section. For the other circuits, explanations are given in the text. Parts marked in red are additions on earlier SQC designs. For the quantronium, when the qubit is in the excited state, the zero-voltage superconducting state is switched to a finite-voltage classical state, resulting in a resistance indicated by the grey box. For the xmon, the red lines show the new connections made possible by the cross-shaped capacitance. The green XY line excites the qubit, and the green Z line tunes the frequency of the qubit.
\label{fig:QubitExtensions}}
\end{figure}

To realize long coherence times, easy connectivity, as well as fast and full control of SQCs, the three basic SQCs discussed above have been further improved and modified in various kinds of superconducting-circuit designs. In this section, we briefly summarize the new generations of SQCs. Several representative examples of these new qubit circuits are summarized in \figref{fig:QubitExtensions}, where the main improvement for each circuit is emphasized.

In early charge and flux qubits, coherence times were limited to a few nanoseconds. The first significant improvement of coherence times was achieved in a so-called \emph{quantronium}~\cite{Vion2002, Cottet2002} circuit, in which the Josephson-junction energy is approximately equal to the charge energy, $E_{\rm J} / E_{\rm C} \sim 1$, such that the qubit operates in an intermediate regime of charge and flux. This circuit, which was initially called a charge-flux-qubit circuit, achieved coherence times on the order of \unit[500]{ns}. 

Reducing the loop size in a flux-qubit circuit will decrease its sensitivity to flux noise, but increases the sensitivity to charge noise instead. To circumvent this limitation, a shunt capacitor can be introduced~\cite{You2007} to the smaller junction in the three-junction flux qubit, leading to greatly improved coherence times (up to three orders of magnitude better!) in experiments~\cite{Steffen2010, Yan2016}. There is also a capacitor-shunted phase-qubit design~\cite{Steffen2006}. To eliminate current noise, arising due to the asymmetric design of the three-junction flux-qubit circuits, \textit{four-junction flux-qubit circuits} have been introduced~\cite{Bertet2005, Yoshihara2006, Gustavsson2012, Yan2013, Stern2014} and theoretically analyzed~\cite{Qiu2016}. 

Replacing the small Josephson junction in the three-junction flux-qubit circuit by a SQUID and adopting a gradiometric design, full control of the qubit can be realized. In this case, the minimum energy splitting $\delta$ of the flux qubit can be controlled by the magnetic flux through the SQUID loop. This is called a \textit{tunable-gap flux-qubit circuit}~\cite{Paauw2009, Zhu2010, Schwarz2013}.

To improve the coherence times of charge-qubit circuits, the \textit{transmon}-qubit circuit~\cite{Koch2007} was proposed. In the transmon design, a large capacitor is added to shunt the Josephson junctions such that the ratio $E_{\rm J} / E_{\rm C}$ is on the order of several tens up to several hundreds. This improvement overcomes the charge noise by making the energy levels flat as a function of $n_{\rm g}$, but comes at the expense of lower anharmonicity, which limits the speed of qubit operations.

\textit{Fluxonium}-qubit circuits~\cite{Manucharyan2009}, with a junction shunted by an array of large-capacitance tunnel junctions, are also insensitive to the fluctuations of offset charges, but maintain a large anharmonicity due to their large inductance. Experiments with fluxonium circuits show that the relaxation due to quasiparticle dissipation can be coherently suppressed by designing a $\pi$-phase difference across the Josephson junction, leading to coherence times exceeding \unit[1]{ms}~\cite{Pop2014}. 

The planar transmon circuits have been adapted to form \emph{xmon}-qubit circuits~\cite{Barends2013} with cross-shaped capacitors, which combine easy connectivity, fast control, and long coherence. By coupling two xmon-qubit circuits through a Josephson junction functioning as a tunable inductance, a \emph{gmon}-qubit circuit~\cite{Chen2014, Geller2014} is realized. This design incorporates the advantages of fast tunable coupling between the two qubits, long coherence, and minimal cross-talk. Recently, another SQC design, the \emph{gatemon}-qubit circuit, was demonstrated~\cite{Larsen2015, DeLange2015}. In this circuit, instead of the conventional sandwich structure of a Josephson junction, the two superconductors are now bridged by a semiconductor nanowire. By applying a gate voltage to this nanowire, the qubit resonance frequency can be adjusted. There are also other kinds of superconducting qubits beyond what we have mentioned here, e.g., phase-slip qubits, Andreev-level qubits, and d-wave qubits~\cite{Zagoskin2007}, as well as more proposals, e.g., toroidal qubits~\cite{Zagoskin2015}. In all extended quantum circuits, a qubit is defined by projecting a multi-level system onto its two lowest energy levels.

\subsection{Selection rules for microwave-induced transitions}
\label{sec:SelectionRules}

Superconducting artificial atoms, similar to natural atoms, can interact with microwave electric and magnetic fields, applied to either capacitors or inductors. These electromagnetic fields can induce transitions between different quantum states. In natural atoms, the transitions induced by electric-dipole interactions have \textit{selection rules} due to the well-defined parities of the dipole moment and the eigenstates of the system, with the symmetries of $SO(3)$ or $SO(4)$. A parity operator $\Pi$, with eigenvalues $\pm 1$, can be defined as space inversion~\cite{Cohen1991}. An eigenstate of a given system is said to be even (odd) if it corresponds to the eigenvalue $+1$($-1$) of $\Pi$. An operator $A$ that commutes with $\Pi$ (i.e., $\Pi A \Pi = A$) is called an even operator. Similarly, an odd operator anti-commutes with $\Pi$. The selection rules arise from the fact that the matrix elements of an even operator are zero for eigenstates of different parities, while the matrix elements of an odd operator are zero for eigenstates of equal parities~\cite{Cohen1991}. Since the dipole moment operator has odd parity, it can only link transitions between initial and final states of different parities.

However, in superconducting artificial atoms, the transitions induced by a microwave field \textit{can be engineered} by changing the parities of the system eigenstates and dipole moments through external parameters, which control the symmetry of the system's potential energy. For example, the dipole-like interaction between a three-junction flux-qubit circuit and a time-dependent magnetic flux $\Phi_{\rm a} (t)$ can be described by the Hamiltonian~\cite{Liu2005a}
\be
H_{\rm d} = - \frac{2\pi \alpha E_{\rm J}}{\Phi_0} \sin (2\pi f + 2 \varphi_{\rm m}) \Phi_{\rm a}(t).
\ee
The term $\sin(2\pi f + \varphi_{\rm m})$ plays the role of the dipole moment in an electric-dipole interaction. If $f = 0.5$, this term is an odd function of $\varphi_{\rm m}$. Moreover, for this value of $f$, the potential energy $U(\varphi_{\rm p}, \varphi_{\rm m})$ in \eqref{eq:HFluxQubit} is an even function of $\varphi_{\rm p}$ and $\varphi_{\rm m}$. Therefore, at this particular point, a transition, induced by the magnetic flux $\Phi_{\rm a}(t)$, between an initial state $\ket{i}$ and a final state $\ket{f}$, can only occur when these states have different parities. That is, the matrix element $\brakket{f}{H_{\rm d}}{i} \neq 0$ for $f = 0.5$ only if $\ket{i}$ and $\ket{f}$ have different parities. Otherwise $\brakket{f}{H_{\rm d}}{i} = 0$. This selection rule for such a microwave-field-induced transition is the same as that for an electric-dipole transition. However, when $f \neq 0.5$ there is no such selection rule, and transitions between any two states are possible. 

In superconducting artificial atoms, there are two types of microwave-induced transitions. One is the same as that seen in electric-dipole interactions, for which the initial and final states must have different parities, as, e.g., in flux-qubit circuits with $f = 0.5$ and charge-qubit circuits with $n_{\rm g} = 0.5$. In these cases, we say that there is a \textit{selection rule for microwave-induced transitions}. The other type of transitions occur, e.g., in phase qubits, and in flux- and charge-qubit circuits at the points $f \neq 0.5$ or $n_{\rm g} \neq 0.5$, respectively. Here, microwave-induced transitions between \textit{any} two states are possible. In these cases, the qubit circuit Hamiltonians all lack a well-defined symmetry and we say that there are no selection rules for these transitions.

Note that the transitions of the flux- and charge-qubit circuits can be changed, from having selection rules to lacking them, by continuously varying an external parameter. This provides a convenient way to control the interaction between SQCs and microwave fields. The selection rules and potential-energy symmetries of the three basic SQCs are summarized in~\tabref{tab:SelectionRules}.
\begin{table}
\centering
\renewcommand{\arraystretch}{1.2}
\renewcommand{\tabcolsep}{0.15cm}
\begin{tabular}{|c|c|c|c|}
\hline
\textbf{Qubits} & \textbf{Special values} & \textbf{Inversion symmetry} & \textbf{Selection rules} \\
\hline
Charge qubits & $n_{\rm g} = 0.5$ & Yes & Yes \\
\hline
Charge qubits & $n_{\rm g} \neq 0.5$ & No & No \\
\hline
Flux qubits & $f = 0.5$ & Yes & Yes \\
\hline
Flux qubits & $f \neq 0.5$ & No & No \\
\hline
Phase qubits & --- & No & No \\
\hline
\end{tabular}
\caption{Summary of selection rules and symmetries of potential energies for SQCs implementing charge qubits at the gate-charge numbers $n_{\rm g} = 0.5$ and $n_{\rm g} \neq 0.5$, flux qubits at the reduced magnetic fluxes $f = 0.5$ and $f \neq 0.5$, and phase qubits.
\label{tab:SelectionRules}}
\end{table}

\subsection{Three-level superconducting artificial atoms}
\label{sec:ThreeLevel}

\begin{figure}
\centering
\includegraphics[width=\linewidth]{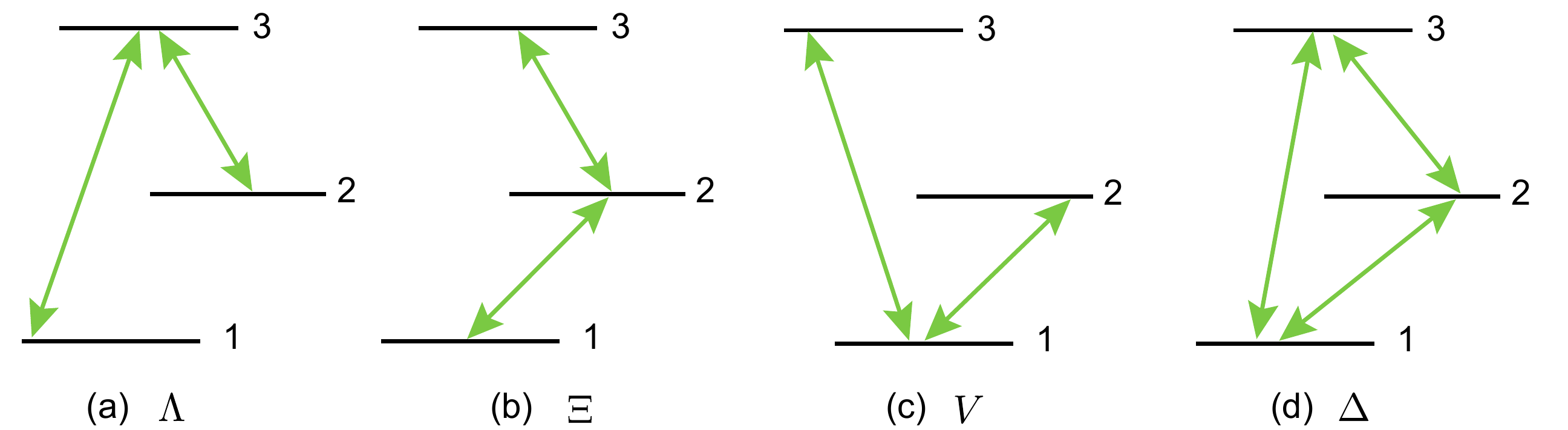}
\caption{Schematic diagrams for different transition configurations of three-level atoms. (a) $\Lambda$-type transitions. (b) Ladder-type transitions, also known as cascade-type, $\Xi$, or $\Sigma$ transitions. (c) $V$-type transitions. (d) $\Delta$-type transitions, which do not occur in natural atoms, but are possible in some superconducting artificial atoms.
\label{fig:EnergyLevel}}
\end{figure}

Having focused mostly on superconducting two-level systems (qubits) so far, we now briefly turn to multi-level systems (\textit{qudits}). The simplest multi-level system beyond qubits is a three-level system (\textit{qutrit}), which has been extensively studied in quantum optics and atomic physics. In natural atomic systems, three-level systems can be divided into three different configurations: $\Lambda$, $\Xi$, and $V$, shown in Figs.~\ref{fig:EnergyLevel}(a), (b), and (c), respectively. These configurations are the ones allowed by selection rules for transitions induced by electric-dipole interactions, where transitions between two states with the same parity are forbidden. However, in superconducting artificial atoms, the parities of the eigenstates can be engineered through external parameters such as magnetic fluxes or electric fields, as discussed in~\secref{sec:SelectionRules}. When there are no well-defined parities for the eigenstates of these systems, \textit{previously forbidden transitions become allowed}. For instance, in a three-level flux-qubit circuit~\cite{Liu2005a} with $f = 0.5$, the transitions induced by an external magnetic flux can only connect states with different parities. But when $f \neq 0.5$, transitions between \textit{any} two states are possible. In this case, as shown in \figref{fig:EnergyLevel}(d), three-level systems can have cyclic or $\Delta$-type transitions, which \textit{do not exist} in natural atomic systems. In such a $\Delta$-type system, a single-photon transition from the ground state $\ket{1}$ to the second excited state $\ket{3}$, and the two-photon transition from the ground state $\ket{1}$ to the first excited state $\ket{2}$, and then from $\ket{2}$ to the second excited state $\ket{3}$, can coexist~\cite{Liu2005a}. The coexistence of these single- and two-photon processes has been demonstrated in experiment~\cite{Deppe2008}. 

This type of transitions enable new applications including single-photon generation~\cite{You2007a, Astafiev2007, Ashhab2009}, amplifying signals without population inversion~\cite{Joo2010, Jia2010}, qubit cooling~\cite{You2008}, electromagnetically induced transparency~\cite{SunHuiChen2014}, downconversion~\cite{Marquardt2007b, Koshino2009, Sanchez-Burillo2016}, three-wave mixing~\cite{Liu2014a}, and quantum routing of single photons~\cite{Zhou2013}. Moreover, they can also be used for correlated microwave lasing~\cite{Peng2015a}, where the linewidth of an unconverted signal is smaller than the Schawlow-Townes limit.

All four types of three-level-system transitions shown in \figref{fig:EnergyLevel} \textit{can be engineered in SQCs} by tuning some external parameters. For example, a $\Delta$-type three-level system can be constructed by charge, transmon, xmon, and flux qubit circuits, operated away from their optimal points, or phase-qubit circuits. $\Xi$-type three-level systems can be obtained by applying a gate voltage $V_{\rm g}$ to the charge~\cite{Niemczyk2009}, transmon, and xmon qubit circuits, or by applying a magnetic bias $\Phi_{\rm e}$ to the flux-qubit loop, to satisfy the conditions $n_{\rm g} = V_{\rm g} C_{\rm g} / (2e) = 1/2$ or $\Phi_{\rm e} / \Phi_0 = 0.5$, respectively. The $\Xi$-type transition can also be obtained by minimizing the transition matrix element between the states $\ket{1}$ and $\ket{3}$ of the $\Delta$-type transitions. In the same way, three-level SQCs with $V$- or $\Lambda$-type transitions can be obtained by engineering SQCs to minimize the transition matrix elements between the states $\ket{2}$ and $\ket{3}$ or $\ket{1}$ and $\ket{2}$ in $\Delta$-type transitions. For example, $V$-type transitions have been realized using coupled charge qubits~\cite{Srinivasan2011}. $\Lambda$-type transitions can be observed by using flux qubits with single~\cite{Yang2004}, three, and four junctions, or the fluxonium qubit~\cite{Manucharyan2009, QuantumMachines2014}.

\subsection{Readout of qubit states}
\label{sec:Readout}

There are various methods to read out the state of superconducting qubits~\cite{Siddiqi2011}. These methods can be categorized with respect to, e.g., (1) specific measured observables, (2) whether the measurements enable the complete, or only an incomplete, reconstruction of a given state, (3) whether special types of measurements (e.g., weak measurements or quantum nondemolition measurements) are used, or (4) whether a measurement on a single copy of a given state is enough or not.

(1) Considering measured observables, we can distinguish three methods for measuring SQC qubit states. For charge qubits, the measured physical quantity is usually the \textit{electric charge on the superconducting island}. This charge can be detected by a single-electron transistor (SET), as in the experiment of Ref.~\cite{Nakamura1999}. The current in the SET serves as an indicator for the state of the charge qubit. For a flux qubit, the two basis states of the qubit are defined by \textit{persistent currents} circulating in the clockwise and counter-clockwise directions, which can be detected by a dc SQUID. By switching the bias current applied to the SQUID to a dissipative state, information about the qubit states is gained, as, e.g., in the experiment of Ref.~\cite{VanderWal2000}. In the case of a phase-qubit, its states are measured using the \textit{tunneling} out of the zero-voltage state of a current-biased Josephson junction~\cite{Martinis1985}. When the bias current is below the value of the critical current of the Josephson junction, the phase qubit energy levels are situated in a potential minimum corresponding to the zero-voltage state. Quantum tunneling of the phase, which is much more likely to occur in the qubit's excited state, will switch the zero-voltage state to a finite-voltage state. In the case of the quantronium-qubit circuit, its quantum state can be transferred to a phase qubit. One can then follow the procedure for measuring a standard phase qubit~\cite{Vion2002}.

(2) Qubit states may comprise many complementary features which cannot be observed simultaneously and precisely due to the Heisenberg uncertainty relations or the no-cloning theorem. Thus, a full characterization of a qubit state must include all its complementary aspects. This can be achieved by performing a series of measurements on a large number of identically prepared copies of a given quantum system. After this so-called \textit{quantum state tomography} (QST) was theoretically introduced to solid-state and superconducting qubit systems~\cite{Liu2004, Liu2005c}, it was first realized experimentally in superconducting phase-qubit systems and later also used in other types of SQCs~\cite{Steffen2006, Katz2008, Neeley2008, Filipp2009}. QST is now extensively applied to superconducting systems for characterizing quantum states of both qubits and microwaves (e.g., Refs.~\cite{Houck2007, Hofheinz2009, Mallet2011, Eichler2011}; see also~\secref{sec:PhotonDetectionCorrelation}).

(3) A few particular types of measurements, with related different quantum-mechanical interpretations, are sometimes applied in SQCs. Here, we mention three such measurements. The first is \textit{quantum nondemolition} (QND) measurements, which are discussed in more detail for photons in~\secref{sec:PhotonDetection}. Briefly put, a QND measurement preserves the state it projects the measured system into. For example, the above-mentioned measurements of charge, flux, phase, and quantronium qubits are not QND. In SQCs, a QND measurement is usually performed using either a dispersive coupling between a qubit and a cavity field~\cite{Wallraff2004, Johnson2010} or the coupling between a qubit and a nonlinear resonator~\cite{Lupascu2004, Siddiqi2006, Lupascu2006, Lupascu2007}. A more detailed description of dispersive coupling is given in~\secref{sec:JaynesCummingsDispersive}. In this coupling regime, the qubit frequency is far detuned from that of the cavity field. The corresponding effective Hamiltonian reveals that the cavity field experiences a frequency shift depending on the state of the qubit~\cite{Blais2004, Siddiqi2006}. Thus, the qubit state can be read out through this frequency shift of the cavity field. This method can be applied to any kind of SQCs that are dispersively coupled to a single-mode cavity field.

The states of a superconducting qubit can also be continuously monitored by weakly coupling the qubit to a low-frequency $LC$ circuit~\cite{Ilichev2003}. Such a \textit{weak measurement} only collects partial information about the qubit~\cite{Katz2008} and it is interesting to note that weak measurements can be used to prepare arbitrary qubit states~\cite{Ashhab2010a}. To resolve the measured weak signal, various amplifiers, such as the Josephson bifurcation amplifier~\cite{Siddiqi2004, Siddiqi2006, Lupascu2006, Vijay2009}, the Josephson ring amplifier~\cite{Bergeal2010}, or parametric amplifiers~\cite{Vijay2011} are used (a more extensive overview of amplifiers is given in~\secref{sec:PhotonDetectionHomodyne}). There are also experiments to perform \textit{weak-value measurements}~\cite{Tan2015}, which are quite different from weak measurements~\cite{Aharonov1988, Kofman2012}.

(4) Most of the above-mentioned measurements must be performed many times to collect statistics about the qubit state. However, there are also techniques to measure the state of a superconducting qubit in a single experimental run, which is called a \textit{single-shot measurement}~\cite{Cooper2004, Astafiev2004, Lupascu2005, Lupascu2006, Siddiqi2006, Mallet2010, Chow2010, Reed2010, Sete2013, Sun2014}.

\subsection{Resonators}
\label{sec:Resonators}

The two most fundamental components in cavity QED are natural atoms and optical cavities. In the previous sections, we showed how the natural atoms are replaced by superconducting artificial atoms in microwave photonics. In this section, we show how a variety of different \textit{resonators for microwave photons} can replace the optical cavities. These resonators, which usually are constructed from electrical circuits, are used to \textit{store} or \textit{guide} microwave photons. Such a resonator can also act as a \textit{quantum data bus}, transferring quantum information between superconducting artificial atoms.

The simplest resonator is the $LC$ circuit, consisting of a capacitor $C$ connected in series with an inductor $L$. Other superconducting devices can be described in terms of lumped-element $LC$ circuits, e.g., superconducting cavities~\cite{Pozar2011}, transmission-line resonators~\cite{Wallraff2004}, and waveguides~\cite{Astafiev2010}. An important difference is that the latter resonators support multiple modes, while the $LC$ circuit just has a single mode. In terms of the capacitor charge $Q$ and the inductor current $I$, the Hamiltonian of the $LC$ oscillator is written as
\be
H = \frac{Q^2}{2C} + \frac{\Phi^2}{2L},
\ee
where $\Phi = I L$ is the flux through the inductor. If we define a pair of conjugate variables $Q$ and $\Phi$ such that these satisfy the canonical commutation relation $\comm{Q}{\Phi} = i \hbar$, the above Hamiltonian gives a quantum version of the $LC$ oscillator. As such, the Hamiltonian can be rewritten as $H = \hbar \omega_0 (a^\dag a + 1/2)$ in terms of the creation and annihilation operators defined, respectively, by
\be
a^\dag = \frac{1}{\sqrt{2 \hbar \omega_0}} \left( \frac{Q}{\sqrt{C}} - i \frac{\Phi}{\sqrt{L}} \right),
\quad a = \frac{1}{\sqrt{2 \hbar \omega_0}} \left( \frac{Q}{\sqrt{C}} + i \frac{\Phi}{\sqrt{L}} \right),
\ee
where $\omega_0 = 1 / \sqrt{LC}$ is the oscillator frequency. It is clear that the $LC$ circuit behaves as a \textit{harmonic oscillator} with equally spaced energy levels. The frequency $\omega_0$ \textit{can be engineered} in a very large range of possible values using the parameters $L$ and $C$.

Transmission lines most commonly appear in superconducting quantum devices in the form of \textit{coplanar waveguides} (CPWs)~\cite{Wallraff2004, Frunzio2005, Goppl2008}. A CPW consists of a central conductor, inserted in a slot of the ground plane. By terminating a length of CPW with capacitors, functioning like Fabry-Perot cavity mirrors, a CPW resonator is formed. The capacitors control the quality factor of the CPW resonator. A low quality factor, corresponding to high loss, is suitable for fast readout. However, a high quality factor is needed for observing strong coupling between a qubit and a photon. The internal quality factors of CPW resonators can reach values above $10^6$~\cite{Megrant2012}.

Transmission lines can be described by a distributed circuit model with a series of parallel $LC$ oscillators~\cite{Collin2001, Pozar2011}. The voltages in the transmission line are governed by a standard wave equation and can vary between different points. In~\appref{app:Quantizing1DTransmissionLine}, we show how the $LC$-circuit model of an infinite transmission line is quantized. Finite-length transmission lines support standing waves with a discrete mode structure, while infinite transmission lines support traveling waves with a continuous spectrum of frequencies. Here, the term infinite means that the microwaves in a CPW are never reflected back from its end, i.e., the microwave energy is carried away. In other words, infinite transmission lines are dissipative. Grounding one end of an infinite transmission line is equivalent to inserting a \textit{mirror} in open space.

The frequency of a \textit{transmission-line resonator} can be engineered in a very large range of values. Intuitively, the resonance frequency is changed by tailoring the length of the resonator, and thus altering its standing wavelengths. For example, if both ends of a resonator are open, then the frequency of its $n$th eigenmode is $\omega_n = \pi v / (d n)$, where $d$ is the length of the resonator and $v$ is the wave velocity. If instead one end of the resonator is open, while the other is connected to ground, the eigenfrequencies are $\omega_n = \pi v / d (n + \frac{1}{2})$.

\textit{Frequency-tunable} transmission-line resonators have important applications in superconducting quantum information processing. Such a tunable frequency can be realized either by changing the resonator's inductance per unit length or by controlling its boundary conditions. Both these tasks can be achieved by incorporating a dc SQUID at a suitable place in the resonator. The Josephson energy of the SQUID can be tuned to be either small or large using the magnetic flux through the SQUID loop~\cite{Wallquist2006}. SQUID-based tuning of the frequency in a transmission-line resonator has been experimentally demonstrated both by altering the boundary conditions of the resonator~\cite{Palacios-Laloy2008, Yamamoto2008, Sandberg2008, Sandberg2008a} and by changing its inductance~\cite{Wang2013APL, Vissers2015}.

It is worth noting that if the resonance frequency of the resonator is tuned faster than the photon lifetime, the frequency of the photons stored in the resonator is changed when these photons leak out from the resonator~\cite{Sandberg2008, Sandberg2008a}. Frequency-tunable resonators are also studied for implementation of \textit{controllable coupling} between different quantum elements, which, e.g., can be used to create shaped photons (see~\secref{sec:ShapingPhotons}).

Another application of tunable resonators is the \textit{dynamical Casimir effect} (DCE), where, if a mirror is accelerated at a speed commensurate with the speed of light, photons can be generated from the vacuum~\cite{Moore1970}. It is very hard to realize the DCE by mechanical modulation of a mirror. However, the change of boundary conditions created by a moving mirror can be achieved with tunable resonators~\cite{Johansson2009, Johansson2010, Johansson2013}. By modulating a SQUID-terminated resonator,  an SQC experiment achieved the first ever observation of the DCE~\cite{Wilson2011} (for more details about this and other relativistic physics possible in SQCs, see the review in Ref.~\cite{Nation2012}). Tunable resonators can also be used to simulate the twin paradox~\cite{Lindkvist2014}, generate entangled states~\cite{Felicetti2014}, realize a beam-splitter operation~\cite{Chirolli2010}, demonstrate single-photon parametric frequency conversion~\cite{Zakka-Bajjani2011}, and achieve quantum interference with photons of two different frequencies~\cite{Nguyen2012}.

Expanding from separate resonators, a large \textit{array of resonators} can be connected together. Such coupled resonators are commonly realized using interconnecting capacitors~\cite{Schmidt2013}. A tunable coupling between resonators can be mediated by a qubit~\cite{Mariantoni2008, Reuther2010, Baust2015}, a SQUID~\cite{Peropadre2013, Wulschner2015}, and other quantum or classical elements. In these setups, photons can hop between coupled neighboring resonators, and transfer excitations between qubits. Arrays of coupled resonators enable photon-based simulation of many-body physics on a chip, as described, e.g., in Refs.~\cite{Koch2010, Houck2012, Schmidt2013, Anderson2016, Noh2017, Fitzpatrick2017} and in \secref{sec:QuantumSimulation}. These resonator arrays can also be used to create \textit{metamaterials}, which are discussed in~\secref{sec:Metamaterials}.

The aforementioned transmission lines are two-dimensional (2D) resonators, patterned on a chip with qubits by using well-established semiconductor fabrication technologies, such as lithography. These 2D resonators are small and scalable. However, the electromagnetic energy in these resonators is concentrated near the surface of the chip, leaving it susceptible to surface dielectric losses. To overcome this limitation, a three-dimensional (3D) transmon was introduced in the circuit-QED architecture~\cite{Paik2011}. The \textit{3D resonator} is a rectangular metal box, which has a larger mode volume and more energy  stored in the vacuum than a 2D resonator. The internal quality factor of a 3D resonator can exceed $10^8$~\cite{Reagor2013}, which is two orders of magnitude greater than what has been achieved in its planar counterpart~\cite{Megrant2012}. The cavity of the 3D resonator is usually a micromachined chamber in bulk materials, assembled from two halves. In the center of the bisected cavity, a chip with a transmon qubit is mounted. With the introduction of the 3D resonator, the coherence time of the transmon was improved by an order of magnitude over typical planar-resonator setups at that time~\cite{Paik2011}. Later, this coherence time was improved even further~\cite{Rigetti2012}. Instead of the complicated wirings in the planar resonator design, the qubit is now suspended in the middle of a cavity and has limited addressability. A two-cavity design has been developed~\cite{Kirchmair2013}, with the qubit sticking into both cavities. The second cavity is used for qubit readout. Apart from the 3D transmon design, a flux qubit in a 3D resonator has also been reported~\cite{Stern2014}.

While the 3D resonators offer longer coherence times, the 2D planar resonators are more compatible with the current microfabrication processes. Some research on multilayer structures has been devoted to combining the two approaches~\cite{Minev2016}. In the multilayer structure, the 3D resonators are micromachined~\cite{Brecht2015} in wafers of different planes, leaving vacuum gaps for energy storage. The planar qubit is patterned inside an aperture in the boundaries of a resonator.

Finally, it can be noted that there are a few experiments where superconducting qubits have been coupled to \textit{mechanical} resonators. In most of these experiments, charge, phase, or transmon qubits were coupled to small vibrating pieces of metal hosting localized phonons~\cite{LaHaye2009, OConnell2010, Pirkkalainen2013, Pirkkalainen2015, Rouxinol2016}. Very recently, transmons have also been coupled to resonators for surface acoustic waves~\cite{Manenti2017} and bulk acoustic waves~\cite{Chu2017} in piezoelectric substrates. In setups like these, the great ability to generate, manipulate, and read out non-classical \textit{photonic} states in circuit QED (see Secs.~\ref{sec:PhotonGeneration} and \ref{sec:PhotonDetection}) could now potentially be applied to \textit{phonons} as well. The short wavelength of phonons means that resonators can be made much smaller than for photons, which opens up new possibilities for miniaturized devices, multimode resonators and a new quantum-optics regime where the artificial atom is much larger than the wavelength~\cite{Kockum2014, Aref2016, Guo2016}. Furthermore, mechanical resonators can be used for \textit{transduction} between various quantum systems in hybrid setups~\cite{Schuetz2015} and to convert between microwave and optical light~\cite{Bochmann2013, Andrews2014, Shumeiko2016}.

\section{Other microwave components}
\label{sec:MicrowaveComponents}

In quantum optics, cavities and atoms are at the heart of every experimental setup. However, many additional components may be needed for proper preparation, control, and readout, depending on the details of the experiment. For microwave photonics with superconducting circuits, the situation is similar. In addition to the artificial atoms and resonators discussed in~\secref{sec:BasicConcepts}, the experimentalist's toolbox may also include mirrors, phase shifters, beam-splitters, circulators, switches, routers, mixers, amplifiers, and detectors. While these components all have implementations either in quantum optics or in classical microwave technology, they may be challenging to transfer to the domain of microwave photonics. For example, the low energy of microwave photons compared to optical photons makes single-photon detection harder for the former. Also, some components used in classical microwave technology are bulky, lossy and/or require the use of strong magnetic fields, all of which make them less than ideal for use with superconducting chips in a cryostat with limited space. Because of these limitations, new solutions specifically suited for microwave photonics have been developed or are under development. Just as for the superconducting artificial atoms, Josephson junctions play a vital role in almost all of these new solutions. 

In this section, we provide an overview of most of the components that make up the toolbox for microwave photonics. Detection schemes for microwave photons is too large a topic to fit here; it is instead treated in~\secref{sec:PhotonDetection}. Likewise, amplifiers are discussed in more detail in connection with homodyne and heterodyne detection in~\secref{sec:PhotonDetectionHomodyne} rather than here. When it comes to mirrors and phase shifters, they are more common in quantum optics than in microwave photonics. For mirrors, in addition to what was discussed in connection with resonators in the previous subsection, we only note that a termination of a transmission line constitutes one; the termination can be either through a capacitance~\cite{Hoi2013a}, a short-circuit~\cite{Hoi2015}, or through a SQUID~\cite{Wilson2011}. For photons in separate resonators, a relative phase shift could be realized by coupling each resonator dispersively to a qubit and then flipping (for a certain time) the state of one of the qubits~\cite{Peropadre2015}. A large cross-Kerr phase shift has been demonstrated for propagating photons scattering from a three-level atom~\cite{Hoi2013a}. Recently, a more general phase shifter, consisting of a hybrid coupler (see \secref{sec:MicrowaveComponentsBeamsplitters}) together with two flux-tunable resonators, was demonstrated~\cite{Naaman2016a}. In the rest of this section, we review beam-splitters (\secref{sec:MicrowaveComponentsBeamsplitters}), circulators (\secref{sec:MicrowaveComponentsCirculators}), switches and routers (\secref{sec:MicrowaveComponentsSwitchesAndRouters}), and mixers (\secref{sec:MicrowaveComponentsMixers}).

\subsection{Beam-splitters}
\label{sec:MicrowaveComponentsBeamsplitters}

A \textit{beam-splitter} takes two modes as inputs and mixes them into two output modes. It can be characterized by a $2\times2$ unitary matrix with determinant $+1$ or $-1$, depending on the convention used~\cite{Nielsen2000, Bouland2014}, and is usually implemented in quantum optics as a semitransparent mirror. Quantum theory for lossless beam-splitters was developed in the 1980's~\cite{Zeilinger1981, Fearn1987} and was later extended to include losses~\cite{Barnett1998}. In 1994, it was shown that any unitary $N\times N$ matrix can be implemented with an array of beam-splitters ($\leq N(N-1)/2$) and phase shifters~\cite{Reck1994}; recently this result was extended to only requiring a finite number of copies of one non-trivial beam-splitter~\cite{Bouland2014}. It is therefore no surprise that beam-splitters constitute basic building blocks for quantum optics~\cite{Braunstein2005}; for example, they are integral to schemes for linear optical quantum computing~\cite{Knill2001, Kok2007}, which could also be considered for microwave photonics~\cite{Adhikari2013}. 

\begin{figure*}
\centering
\includegraphics[width=\linewidth]{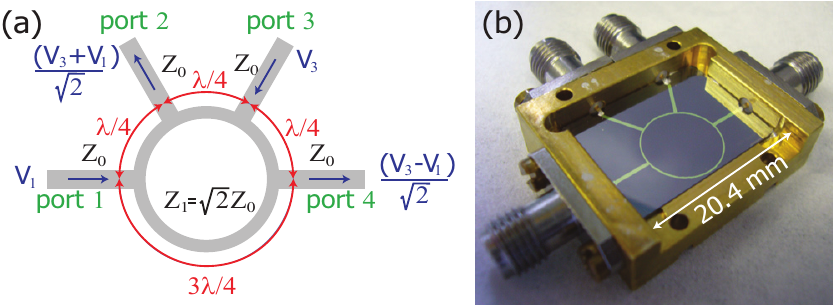}
\caption{A microwave beam-splitter implemented as a 180 degree hybrid coupler~\cite{Hoffmann2010}. (a) The hybrid coupler is a four-port device with distances between the ports chosen such that inputs on ports 1 and 3 will not give any output through those ports due to destructive interference, but will mix to give outputs through ports 2 and 4 due to constructive interference. (b) A photo of such a device, which demonstrated good coupling and isolation in a $\unit[2]{GHz}$ bandwidth around a center frequency of $\unit[6]{GHz}$. 
Reprinted figure from E.~Hoffmann et al., \href{http://dx.doi.org/10.1063/1.3522650}{Appl.~Phys.~Lett.~\textbf{97}, 222508 (2010)}, with the permission of AIP Publishing.
\label{fig:180DegreeHybridRing}}
\end{figure*}

An early version of a beam-splitter for microwave photonics simply put two transmission lines close to each other along a short distance~\cite{Gabelli2004}. In this case, the beam-splitter was used to enable correlation measurements on microwave photons, which has been a common application since then~\cite{Mariantoni2005, DaSilva2010, Mariantoni2010, Menzel2010, Bozyigit2011, Lang2011, Eichler2012a, Hoi2012, Menzel2012, Eichler2013PhD, Hoi2013NJP, Lang2013, Woolley2013, DiCandia2014} (see \secref{sec:PhotonDetectionCorrelation}). Another common application has been entanglement generation~\cite{Menzel2012, DiCandia2014, Ku2015, Fedorov2016, Narla2016}. However, many of the later experiments opted for using a so-called 180- or 90-degree \textit{hybrid coupler} (either on- or off-chip)~\cite{Mariantoni2005, Mariantoni2009, Hoffmann2010, Mariantoni2010, Menzel2010, Abdo2011, Bozyigit2011, Ku2011, Lang2011, Hoi2012, Menzel2012, Abdo2013PRL, Eichler2013PhD, Hoi2013NJP, Lang2013, DiCandia2014, Ku2015, Fedorov2016, Narla2016}, well-known from classical microwave technology~\cite{Cohn1968, Collin2001, Pozar2011}; an example is shown in \figref{fig:180DegreeHybridRing}.

Another microwave-beam-splitter design is the Wilkinson power divider~\cite{Wilkinson1960, Collin2001, Pozar2011}, which takes a single input and splits it into two; such a device has also been used in microwave photonics experiments~\cite{Mariantoni2009, Mariantoni2010}. However, a full quantum analysis reveals the presence of a second, internal port, which has the same effect as vacuum noise input on the second input port of an ordinary beam-splitter~\cite{Mariantoni2009}.

In contrast to the beam-splitters discussed above, which are designs from classical microwave technology taken to the quantum realm, a circuit-QED version of a three-wave mixer (see \secref{sec:MicrowaveComponentsMixers}), made from four resonators coupled via four Josephson junctions, has also been demonstrated to work as a beam-splitter for propagating microwave photons of different frequencies~\cite{Abdo2013PRL}. The parameters of this beam-splitter can be tuned \textit{in situ} all the way from full reflection to full transmission by changing the amplitude of a pump. There are also theoretical investigations of using an atom~\cite{Roulet2014}, or a cavity coupled to an atom~\cite{Oehri2015}, as a beam-splitter for single photons traveling in a transmission line.

Another approach to beam-splitting is to not look at propagating photons, but instead consider photons in two resonators (or two modes of one resonator), interacting through a coupling $\hbar g \left(a^\dag b + a b^\dag\right)$, where $g$ is the coupling strength and $a, a^\dag$ ($b, b^\dag$) are the annihilation and creation operators for the first (second) resonator/mode. This has been investigated for microwave photons both theoretically~\cite{Mariantoni2008, Chirolli2010, Adhikari2013, Haeberlein2013, Peropadre2013, Peropadre2015} and experimentally~\cite{Zakka-Bajjani2011, Haeberlein2013, Baust2015, Wulschner2015, Lecocq2016}. Tunable coupling has been demonstrated through a SQUID~\cite{Zakka-Bajjani2011, Wulschner2015}, a flux qubit~\cite{Baust2015}, and a mechanical resonator~\cite{Lecocq2016}. This could potentially be used to implement boson sampling~\cite{Aaronson2013} with microwaves~\cite{Peropadre2015}.

\subsection{Circulators}
\label{sec:MicrowaveComponentsCirculators}

A \textit{circulator} is a device with three or more ports, where input on one port is routed to become output from the next port in the sequence. For an ideal three-port circulator, the scattering matrix is~\cite{Pozar2011}
\bea
S = 
\begin{pmatrix}
	0 & 0 & 1 \\
	1 & 0 & 0 \\
	0 & 1 & 0
\end{pmatrix}.
\eea
Circulators are \textit{nonreciprocal} devices~\cite{Anderson1965, Anderson1966, Ranzani2014, Ranzani2015} (reciprocity means that the measured scattering does not change if the source and the detector are interchanged~\cite{Deak2012}), with many applications for routing signals in experiments in microwave photonics. For example, by terminating one port of the circulator in a matched load, a two-port \textit{isolator}~\cite{Jalas2013} is formed, only allowing signal propagation in one direction. This is often used in experiments to protect a circuit QED system from noise travelling back from an amplifier.

In classical microwave technology, circulators have been investigated and used for more than half a century~\cite{Allen1956, Auld1959, Fay1965, Tanaka1965, Collin2001, Pozar2011}. However, the typical implementation relies on Faraday rotation in ferrite materials~\cite{Allen1956, Auld1959, Fay1965, Collin2001, Pozar2011}, which requires strong magnetic fields and wavelength-size components, both of which are hard to integrate with superconducting circuits. There exist other circulator designs, e.g., based on transistors~\cite{Tanaka1965}, realized through three coupled modulated resonators~\cite{Estep2014}, designed for sound with a ring of circulating fluid in its center~\cite{Fleury2014}, or based on the quantum Hall effect~\cite{Viola2014, Mahoney2016, Placke2016}. There are also optical isolators or diodes based on modulation of the refractive index in a waveguide or photonic crystal~\cite{Lira2012, Wang2013PRL}, or using nonreciprocal transmission in waveguides coupled via two coupled microcavities (one lossy, one with gain)~\cite{Peng2014a} (the same setup with mechanical oscillators instead could realize a phonon diode~\cite{Zhang2015}). Furthermore, directional coupling of a quantum dot, or a natural atom in a microresonator, to photons with different polarization in a photonic-crystal waveguide, or an optical fiber, can be used for single-photon diodes and circulators~\cite{Sollner2015, Scheucher2016}. A directional coupler has also been demonstrated with superconducting transmission lines~\cite{Ku2011}, but this is not yet a circulator.

One way to circumvent, or at least reduce, the need for circulators in microwave photonics experiments is to make \textit{directional amplifiers}. After a theoretical study showed that SQUID amplifiers can display directionality~\cite{Kamal2012}, directional amplification has been realized with two \textit{Josephson parametric converters} (JPCs, four resonators coupled via a ring interrupted by four Josephson junctions) pumped with two microwave tones~\cite{Abdo2013PRX, Abdo2014}; the setup has been dubbed a \textit{Josephson directional amplifier} (JDA). Building on a graph-based theory, developed to analyze non-reciprocity in coupled systems~\cite{Ranzani2014, Ranzani2015}, directional amplification has also been experimentally realized with three pumps on pairs of ports of a single JPC~\cite{Sliwa2015} and with another similar three-mode circuit~\cite{Lecocq2017}. Furthermore, directional amplification has also been demonstrated with travelling-wave amplifiers, implemented with a large number of Josephson junctions forming a nonlinear lumped-element transmission line~\cite{Macklin2015, White2015}. Recently, a setup, with a waveguide running alongside a 2D square lattice of parametrically driven resonators in a synthetic magnetic field, was shown to work as a topologically protected directional travelling-wave amplifier, which also could function as a circulator~\cite{Peano2016}.

There are several proposals for how a circulator for microwave photonics could be built on-chip. For example, three resonators coupled via a ring containing three Josephson junctions can be made to work as a circulator by tuning the magnetic flux through the ring~\cite{Koch2010, Nunnenkamp2011}. A related proposal is to make a phononic circulator by introducing a complex coupling among three coupled mechanical oscillators~\cite{Habraken2012}. This is done by coupling two of the oscillators to two driven optical resonators. See also the recent optomechanics experiment of Ref.~\cite{Fang2016}, which demonstrated directional amplification and could be extended to a circulator, as well as the microwave optomechanics experiment of Ref.~\cite{Peterson2017}, where isolation of $\unit[20]{dB}$ was achieved. Another way to achieve directionality is to couple two resonators both through a coherent channel and through a common bath~\cite{Metelmann2015, Metelmann2016}. If the common bath is another lossy resonator, the scattering matrix of an ideal circulator can be recovered.

Another class of proposals involve rapid modulation of various couplings in a system. It has been shown that modulating the couplings between resonators in a square lattice, with proper phase off-set between the modulation signals, creates an effective magnetic field that gives unidirectional propagation~\cite{Fang2012}. Recently, this idea was implemented in a superconducting circuit where three transmon qubits were coupled in a ring and their couplings modulated in such a way~\cite{Roushan2016}. That experiment demonstrated unidirectional propagation of one and two photons in this setup, and it could potentially be extended to a larger lattice. With three harmonic oscillators connected via biharmonically modulated couplings, both circulation and directional amplification can be realized~\cite{Kamal2016}. A four-port circulator design, where four ports are connected via four tunable inductance bridges (which could be implemented using SQUIDs), has also been analyzed~\cite{Kerckhoff2015}.

The JDA mentioned above can also be configured to work as a circulator~\cite{Sliwa2015}. By making the three pump phases in the device add up to $\pm\pi/2$, clockwise or counterclockwise circulation can be achieved. In the experiment, $\unit[10.5]{dB}$ isolation and less than $\unit[1]{dB}$ insertion loss was demonstrated over a bandwidth of $\unit[11]{MHz}$. This was recently improved to $\unit[30]{dB}$ isolation in the related Josephson-amplifier circuit of Ref.~\cite{Lecocq2017}. These results open up for using circulators in microwave photonics experiments to achieve unidirectional propagation~\cite{Stannigel2012, Sathyamoorthy2014} (see \secref{sec:WaveguideQED}).

\subsection{Switches and routers}
\label{sec:MicrowaveComponentsSwitchesAndRouters}

\textit{Switches} and \textit{routers} are devices which control transport of a signal conditioned on some other control input. In the simplest case, the control determines whether the signal is transmitted or reflected; in a more complex device, the signal can be routed into one of multiple output ports. Going towards the quantum regime, many of the proposals and experiments reviewed in this subsection concern routing/switching with single photons as signal and/or control. In a few cases, these devices are also referred to as \textit{single-photon transistors}~\cite{Chang2007, Chen2013, Neumeier2013, Manzoni2014, Gonzalez-Ballestero2016, Kyriienko2016}. A truly quantum router or switch could have the control in a superposition state, giving rise to a superposition in the output~\cite{Lemr2013, Yuan2015}.

For microwave switching at cryogenic temperatures, there are promising implementations based on microelectromechanical systems (MEMS) ~\cite{Gong2009, Attar2014} or high-electron-mobility transistors (HEMTs)~\cite{Ward2013, Hornibrook2015}. However, since these components are hard to integrate with superconducting circuits, recently a few devices based on Josephson junctions have been developed~\cite{Chapman2016, Naaman2016, Pechal2016}. By tuning the magnetic flux through a SQUID that couples two resonators through a mutual inductance~\cite{Naaman2016}, or through an array of SQUIDs forming a tunable inductor bridge with four inductors connecting two ports~\cite{Chapman2016}, single-pole, single-throw switches have been demonstrated. Here, the control (the magnetic flux) determines whether an input is reflected or transmitted; the devices have a bandwidth of several GHz~\cite{Chapman2016, Naaman2016}. In Ref.~\cite{Pechal2016}, a single-pole, double-throw switch, where the input is routed into one of two possible outputs, was demonstrated in a setup consisting of two hybrid couplers (beam-splitters, see \secref{sec:MicrowaveComponentsBeamsplitters}) coupled via two transmission-line resonators. The resonator frequencies are tunable via magnetic flux through SQUIDs; an input on one of the first beam-splitter ports is either reflected through the other port of the same beam-splitter or transmitted out through a port of the second beam-splitter, depending on whether the input is on resonance with the resonators or not. The bandwidth of the device is limited to roughly $\unit[100]{MHz}$, but it can handle a photon flux of $10^5\:\mu\rm{s}$. Finally, we note that the theory proposal for a four-port circulator based on tunable inductance bridges~\cite{Kerckhoff2015} and the experimentally demonstrated JDA in its circulator mode~\cite{Sliwa2015}, both discussed in the previous subsection, can be regarded as routers for microwave photonics.

For single-photon routing and switching there are a plethora of theoretical proposals and experimental implementations based on one or a few atoms, sometimes in combination with a cavity. The simplest system is a single two-level atom, which will perfectly reflect a single photon in an open transmission line if it is on resonance (the photon is transmitted if it is off resonance, see \secref{sec:WaveguideQED}). If the atom frequency can be tuned, the atom functions as a \textit{single-photon switch}~\cite{Shen2005, Zhou2008, Zhou2009}. For a three-level $\Xi$ atom, driving the upper transition will change the transition frequency of the lower transition, giving the same effect; this has been demonstrated to work very well in experiments with superconducting circuits~\cite{Abdumalikov2010, Hoi2011, Li2012} (see also \secref{sec:EITandATS}). In quantum optics, three-level $\Lambda$ atoms, sometimes with transitions coupling to different polarizations, can be used for single-photon routing controlled by a single photon~\cite{Chang2007, Shomroni2014, Albrecht2016, Sipahigil2016}. Recently, it was proposed that coupling a superconducting flux qubit to such a $\Lambda$ system could be used to route optical photons~\cite{Xia2016}.

Increasing the complexity of the setups, routing and switching can also be realized by letting one or more atoms couple to multiple transmission lines; control of which transmission line the input exits through can then be realized by changing atomic frequencies or driving atomic transitions~\cite{Neumeier2013, Zhou2013, Lu2014, Yan2014, Gonzalez-Ballestero2016, Kyriienko2016, Sala2016}. Including a resonator as well, there are proposals~\cite{Bermel2006, Xia2013} and experiments~\cite{Dayan2008} routing or switching of single photons using a ring resonator coupled to an atom. Routing can also be done through a photon blockade effect~\cite{Rosenblum2011}, which can be realized through the Jaynes--Cummings nonlinearity (see \secref{sec:DressedStates}) with an atom in a resonator~\cite{Volz2012, Chen2013}. Other experiments in quantum optics have achieved routing or switching with electromagnetically induced transparency (EIT, see~\secref{sec:EITandATS}) controlling the transmission through an optical cavity containing $\Lambda$ atoms~\cite{Mucke2010}, with two fibers coupled to a resonator containing an atom~\cite{OShea2013}, or with a three-level atom in a cavity at the end of a transmission line~\cite{Tiecke2014}. The setup in Ref.~\cite{Mariantoni2008}, with tunable coupling between two resonators mediated by a qubit, can also be called a quantum switch.

\subsection{Mixers}
\label{sec:MicrowaveComponentsMixers}

A \textit{mixer} usually takes two signals at frequencies $\omega_1$ and $\omega_2$ as input and generates output at new frequencies, typically at $\omega_1 + \omega_2$ or $\abs{\omega_1 - \omega_2}$. Mixers are widely used in classical microwave technology~\cite{Pozar2011}, but for quantum applications in superconducting circuits a new design is needed~\cite{Flurin2014PhD}. Various mixing processes in quantum optics and microwave photonics are discussed in more detail in~\secref{sec:NonlinearProcess}; in this subsection we only briefly review some implementations based on Josephson junctions.

\begin{figure}
\centering
\includegraphics[width=\linewidth]{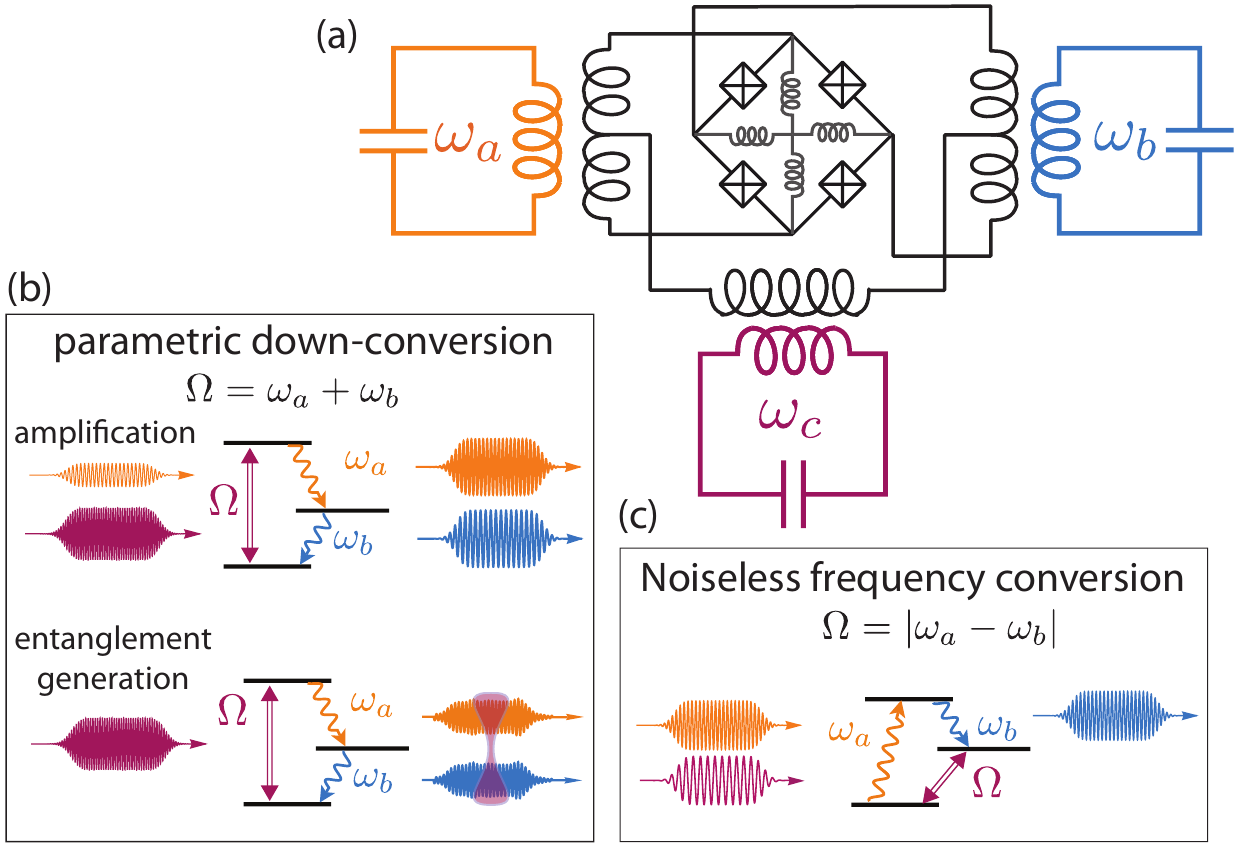}
\caption{The Josephson mixer~\cite{Flurin2014PhD}. (a) Circuit diagram of a Josephson mixer with an inductively shunted Josephson ring modulator coupled to three modes that are to be mixed. (b) One application of three-wave mixing is parametric down-conversion, which can be used for amplification (see Secs.~\ref{sec:NonlinearProcThreeWaveMixing} and \ref{sec:PhotonDetectionHomodyne}) and entanglement generation, as shown schematically here. (c) Another application is noiseless frequency conversion, which has been demonstrated in experiments with superconducting circuits. 
Reprinted figure from E.~Flurin, The Josephson mixer - A swiss army knife for microwave quantum optics, Ph.D.~thesis, Ecole Normale Superieure, Paris (2014), with permission from E.~Flurin.
\label{fig:JosephsonMixer}}
\end{figure}

The \textit{Josephson ring modulator} (JRM), first demonstrated and discussed in Refs.~\cite{Bergeal2010, Bergeal2010a}, is the device at the heart of on-chip mixers for microwave photonics in the quantum regime. The JRM is a ring interrupted by four Josephson junctions; when connected to resonators a JPC (discussed previously in \secref{sec:MicrowaveComponentsCirculators}) is formed. The nonlinearity of the Josephson junctions is the source of the three-wave mixing that can occur with three ports connected to the JRM~\cite{Bergeal2010a, Abdo2013PRX}. A circuit diagram of a \textit{Josephson mixer} with a shunted JRM~\cite{Roch2012, Flurin2012, Flurin2014PhD, Pillet2015} is shown in \figref{fig:JosephsonMixer} along with examples of applications of three-wave mixing. Josephson mixers have been used in experiments for amplification~\cite{Bergeal2010, Roch2012, Schackert2013, Flurin2014PhD, Pillet2015}, noiseless frequency conversion~\cite{Abdo2013PRL}, beam-splitting~\cite{Abdo2013PRL}, and entanglement generation/two-mode squeezing~\cite{Bergeal2012, Flurin2012, Flurin2014PhD, Flurin2015}. In the latest characterization of a Josephson mixer, amplification with a power gain of $\unit[20]{dB}$ over a bandwidth of $\unit[50]{MHz}$, quantum efficiency of $\unit[70]{\%}$, and saturation power of $\unit[-112]{dBm}$ was demonstrated~\cite{Pillet2015}. We also note that a variation of the Josephson-mixer design has been used to realize a vector and quadrature modulator for microwaves~\cite{Naaman2016a}.

\section{Circuit QED}
\label{sec:CircuitQED}

The interaction between an atomic system and electromagnetic fields at the quantum level has been intensively studied in cavity quantum electrodynamics (cavity QED)~\cite{Haroche2006}. Analogously to cavity QED, the superconducting artificial atoms reviewed in Secs.~\ref{sec:ThreeSQCTypes}-\ref{sec:SQCExtensions} can be coupled to quantized microwave fields in the transmission-line or 3D resonators reviewed in \secref{sec:Resonators}~\cite{Blais2003, Buisson2000, Marquardt2001, You2003, Yang2003, Yang2004, Zagoskin2004, Paik2011, Stern2014}. With flexible circuit designs, this new field, which is called \textit{circuit QED}~\cite{Blais2004, Wallraff2004}, has opened new directions for exploring microwave quantum optics on a superconducting chip. In this section, we first briefly review the coupling mechanisms used in circuit QED and the Hamiltonians they give rise to (\secref{sec:LightMatterCoupling}). We then discuss (Secs.~\ref{sec:CircuitQED_StrongCoupling}-\ref{sec:ultrastrong}) the physics that can be probed in circuit QED in various parameter regimes (e.g., strong, ultrastrong, and dispersive coupling), defined by the relations between coupling strength, decoherence rates, and transition frequencies in the system.

\subsection{Light-matter coupling in circuit QED}
\label{sec:LightMatterCoupling}

In a circuit QED setup, the coupling between a superconducting artificial atom and a resonator can be either \textit{inductive}~\cite{Liu2004, You2003} or \textit{capacitive}~\cite{You2003a, Blais2004}, analogous to \textit{magnetic} or \textit{electrical-dipole} couplings in quantum optics~\cite{Devoret2007}. In experimental realizations, charge (transmon) and phase qubits are usually coupled to a transmission-line resonator via a capacitor. Flux-qubit circuits can also be coupled to a resonator through a capacitor, but more often their coupling to the resonator is through a mutual inductance. For more details on the couplings between SQCs and resonators, we refer to the review of Ref.~\cite{Xiang2013} and references therein. We note that superconducting qubits also can be designed to couple both capacitively and inductively to a resonator field, as considered in, e.g., Ref.~\cite{Baksic2014}. These coupling mechanisms can lead to interesting phase transitions in quantum many-body physics~\cite{Meaney2010, Nataf2010, Nataf2010a, Baksic2014}.

It is known that the coupling strength between a superconducting qubit and a single-mode microwave field, in principle, can be engineered to be extremely large~\cite{Devoret2007}. For example, one can reach the \textit{ultrastrong-coupling} (USC) regime, where the strength of the coupling between the qubit and the resonator field is on the same order as the bare transition frequencies in the system (see~\secref{sec:ultrastrong}). Moreover, the coupling between a qubit and a resonator can be tunable \textit{in situ}~\cite{Gambetta2011, Srinivasan2011, Hoffman2011a, Whittaker2014, Shanks2013} (see \secref{sec:ShapingPhotons}). The general Hamiltonian describing the interaction between a superconducting qubit and a single-mode microwave field can be written as~\cite{QuantumMachines2014, Blais2004, Wallraff2004, Liu2014, Liu2014a, Zhao2015, Zhao2016},
\be
H = \omega_{\rm r} \left( a^\dag a + \frac{1}{2} \right) + \frac{\omega_{\rm q}}{2} \sz + g_z \sz \left(a + a^\dag \right) + g_x \sx \left( a + a^\dag \right),
\label{eq:general}
\end{equation}
which corresponds to breaking the inversion symmetry of the potential energy of the qubit. For simplicity, we set $\hbar=1$ here and throughout the following sections. Here $a$ ($a^\dag$) is the photon annihilation (creation) operator while $\sx$ and $\sz$ are Pauli matrices, which can be written as $\sx = \ketbra{e}{g} + \ketbra{g}{e}$ and $\sz = \ketbra{e}{e} - \ketbra{g}{g}$ in terms of the ground $\ket{g} \equiv \ket{\downarrow}$ and excited $\ket{e} \equiv \ket{\uparrow}$ states of the two-level qubit. The first (second) term in \eqref{eq:general} represents the free Hamiltonian of the resonator field (qubit) with frequency $\omega_{\rm r}$ ($\omega_{\rm q}$). The third (fourth) term describes the longitudinal (transverse) coupling between the qubit and the resonator field with coupling strength $g_z$ ($g_x$).

For charge and flux qubits operating at the degeneracy point, where the qubit has a well-defined inversion symmetry of the potential energy, the longitudinal coupling term vanishes ($g_z=0$). Then \eqref{eq:general} is reduced to the \textit{quantum Rabi Hamiltonian}~\cite{Rabi1936, Rabi1937}. However, for the phase qubit, the quantum Rabi Hamiltonian can only be obtained by neglecting the longitudinal coupling term, because no physical rule guarantees that $g_z = 0$. Under the \textit{rotating-wave approximation} (RWA)~\cite{Bloch1940}, corresponding to neglecting all terms which do not conserve the number of excitations, the quantum Rabi Hamiltonian is further simplified to the \textit{Jaynes--Cummings} (JC) Hamiltonian~\cite{JaynesCummings}. In the following, we will mainly focus on the JC and quantum Rabi Hamiltonians.

Note that the \textit{longitudinal coupling} also has attracted much research interest~\cite{Nakamura2001, Irish2005, Wilson2007, Liu2014}. One can design circuits with only the longitudinal-coupling term~\cite{Kerman2013, Billangeon2015, Didier2015, Richer2016}, for example, a flux qubit or a transmon inductively coupled to a resonator~\cite{Kerman2013, Didier2015, Billangeon2015}. By modulating such a longitudinal coupling at the resonator frequency, a fast and pure quantum nondemolition (QND) readout of the qubit can be realized~\cite{Didier2015}. Also, scalable quantum information processing architectures can also be extended from the transverse-coupling~\cite{Liu2006a, Blais2007} to the longitudinal-coupling case~\cite{Billangeon2015}. Moreover, longitudinal coupling can be used for fast preparation of microwave nonclassical states~\cite{Zhao2015, Zhao2016}, quantum switching of the coupling between the qubit and a transversely coupling field~\cite{Liu2014}, and fast entangling gates~\cite{Royer2016}.

To characterize the strength of the qubit-resonator coupling, it can be compared to either the bare transition frequencies of the qubit and the resonator, or to the decoherence rates of the system. Denoting the qubit-resonator coupling $g$, the resonator decay rate $\kappa$, and the qubit relaxation rate $\gamma$, 
\begin{enumerate}[(i)]
\item the \textit{weak-coupling} regime is defined by $g < \max \{ \gamma, \kappa \}$ and 
\item the \textit{strong-coupling} regime is reached when $g > \max \{ \gamma, \kappa \}$. 
\end{enumerate}
Recent studies have showed that SQCs can also reach 
\begin{enumerate}[(i)]
\setcounter{enumi}{2}
\item the USC regime, defined by $g \gtrsim 0.1 \omega_{\rm r/q}$, or even 
\item the \textit{deep-strong coupling} (DSC) regime, where $g \gtrsim \omega_{\rm r/q}$. 
\end{enumerate}
These coupling regimes have been summarized in~\tabref{tab:coupling}. The strong-coupling regime, which is easier to achieve with SQCs than in conventional quantum optics, and the USC regime, which has been reached with SQCs but not in conventional quantum optics, have both been extensively explored for superconducting quantum information processing and quantum physics. In the following, we therefore devote subsections to each of these regimes to summarize results from such work.

\begin{table}[ht]
\centering
\renewcommand{\arraystretch}{1.2}
\renewcommand{\tabcolsep}{0.15cm}
\begin{tabular}{ | c | c | c | c | c | }
\hline
 & \textbf{Weak coupling} & \textbf{Strong coupling} & \textbf{USC} & \textbf{DSC} \\ 
\hline
\textbf{Coupling strength} & $g < \max \{ \gamma, \kappa \}$ & $g > \max \{ \gamma, \kappa \}$ & $g \gtrsim 0.1 \omega_{\rm r/q}$ & $g \gtrsim \omega_{\rm r/q}$ \\
\hline
\textbf{Model} & JC & JC & Q.~Rabi & Q.~Rabi \\ 
\hline
\end{tabular}
\caption{Parameter regimes for light-matter coupling. All symbols are defined in the text. The last line of the table shows which models are usually needed to treat the physics in the different regimes. Note that the coupling $g$ is compared to different quantities in the different cases. Thus, it is technically possible for the coupling to be, e.g., weak and ultrastrong at the same time. However, most discussion of strong coupling, such as in \secref{sec:CircuitQED_StrongCoupling} below, usually assumes that $g < 0.1 \omega_{\rm r/q}$. Likewise, most work on the USC and DSC regimes assume $g > \max \{ \gamma, \kappa \}$.
\label{tab:coupling}}
\end{table}

\subsection{Strong-coupling regime}
\label{sec:CircuitQED_StrongCoupling}

When the coupling between a qubit and resonator is strong, their coherent interaction contributes more to the system dynamics than the decoherence in the system. If the coupling is strong but not ultrastrong, the system is well described by the JC Hamiltonian. In this subsection, we review properties of the JC Hamiltonian, including dressed states and the regime of dispersive coupling, and their ubiquitous applications in circuit QED.

\subsubsection{Jaynes--Cummings model}
\label{sec:JaynesCummings}

The Jaynes--Cummings model describes one of the most fundamental setups of quantum optics, in which a two-level system (TLS, a qubit) is coupled to a single-mode resonator (cavity) field. This model gives a mathematical framework for various quantum phenomena, including nonclassical-state generation, quantum state transfer, topological physics using photons, and many others. A comprehensive (although not recent) review~\cite{Shore1993} summarizes the first thirty years of research based on the JC model, which is described by the Hamiltonian
\be
H_{\text{JC}} = \omega_{\rm r} \left(a^\dag a + \frac{1}{2} \right) + \frac{\omega_{\rm q}}{2} \sz + \frac{\Omega_0}{2}(a \sp + a^\dag \sm),
\label{eq:JC}
\ee
where we introduce the \textit{vacuum Rabi frequency} $\Omega_0 = 2g$. Here, the ladder operators are defined as $\sp = \ketbra{e}{g} \equiv \ketbra{\uparrow}{\downarrow}$ and $\sm = \ketbra{g}{e} \equiv \ketbra{\downarrow}{\uparrow}$. The RWA used to arrive at the JC Hamiltonian from the quantum Rabi Hamiltonian is valid when $(\omega_{\rm q} + \omega_{\rm r}) \gg \{ g, \abs{\omega_{\rm q} - \omega_{\rm r}} \}$, i.e., when the coupling is below the threshold for the USC regime and the qubit and the resonator are not too far from resonance. Physics beyond the RWA is discussed in \secref{sec:ultrastrong}. 

\subsubsection{Dressed states}
\label{sec:DressedStates}

\begin{figure}
\centering
\includegraphics[width=\linewidth]{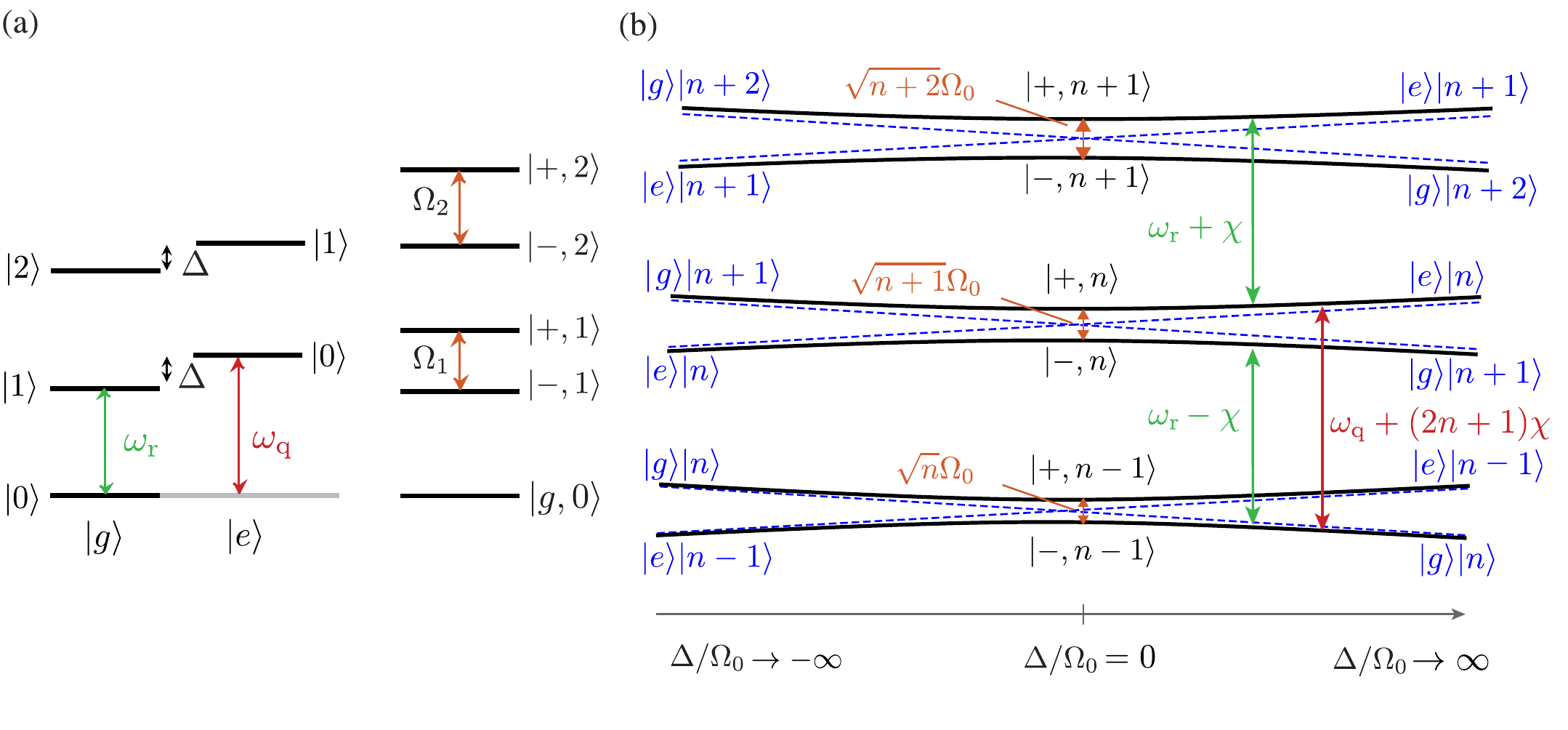}
\caption{Energy-level diagrams for the bare and dressed states in the Jaynes--Cummings model. (a) Bare and dressed states. To the left, the bare states are represented in two ladders for $\ket{g}$ and $\ket{e}$. The states $\ket{g}\!\ket{n+1}$ and $\ket{e}\!\ket{n}$ are close in energy, separated by $\Delta = \omega_{\rm q} - \omega_{\rm r}$, but without any coupling there are no possible transitions between these states. To the right, the coupling $g$ is switched on and the nearby states hybridize, forming a new ladder with level spacings $\Omega_n = \sqrt{\Delta^2 + (n+1)\Omega_0^2}$. (b) Energy levels as a function of detuning. The characteristic $\sqrt{n}$ nonlinearity of the Jaynes--Cummings ladder occurs at the resonance $\Delta = 0$ (orange). In the dispersive regime $\abs{\Delta} \gg \Omega_0$, the dressed states in the asymptotic limit correspond to the unperturbed states with shifted energies (blue). The effective resonator frequency (green) is $\omega_{\rm r} \pm \chi$, where $\chi = \frac{g^2}{\Delta}$, depending on the qubit state. The qubit (red) experiences a Stark shift $2n\chi$ due to the photons in the resonator and a Lamb shift $\chi$ due to the vacuum fluctuations in the resonator.
\label{fig:DressedStates}}
\end{figure}

The third term in the JC Hamiltonian, given in \eqref{eq:JC}, describes the electrical-dipole coupling between a qubit and a resonator field, where the qubit can be excited by absorbing a photon, or return to the ground state by emitting a photon. Thus, the coupling only connects the states $\ket{g}\!\ket{n+1}$ and $\ket{e}\!\ket{n}$, where $\ket{n}$ ($n = 0, 1, \ldots$) is the Fock state of the resonator field. The Hamiltonian in \eqref{eq:JC} can be diagonalized in the subspace $\{\ket{g}\!\ket{n+1}, \, \ket{e}\!\ket{n} \}$ with the eigenvalues
\be
E_n^{\pm} \equiv E(\ket{\pm,n}) = \omega_{\rm r} \left(n + \frac{1}{2} \right) \pm \frac{1}{2} \sqrt{\Delta^2 + \Omega_{n,\rm r}^2}
\label{eq:eigenf}
\ee
and the corresponding eigenstates
\bea
\ket{+,n} &=& \cos \left( \frac{\theta_n}{2} \right) \ket{e}\!\ket{n} + \sin \left( \frac{\theta_n}{2} \right) \ket{g}\!\ket{n+1}, \nn \\
\ket{-,n} &=& - \sin \left( \frac{\theta_n}{2} \right) \ket{e}\!\ket{n} + \cos \left( \frac{\theta_n}{2} \right) \ket{g}\!\ket{n+1},
\label{eq:dressed_states}
\end{eqnarray}
where the mixing angle $\theta_n$ is defined by $\tan \theta_n = \Omega_{n,\rm r} / \Delta$, with detuning $\Delta = \omega_{\rm q} - \omega_{\rm r}$, and $\Omega_{n,\rm r} = \Omega_0 \sqrt{n+1}$ is the $n$-photon Rabi frequency on resonance. The doublets of states given in Eq.~(\ref{eq:dressed_states}) are called \textit{dressed states} in quantum optics and atomic physics. Using these dressed states, the Hamiltonian in \eqref{eq:JC} can be rewritten in a diagonal form
\be
H = - \frac{\Delta}{2} \ket{g}\!\ket{0}\!\bra{0}\!\bra{g} + \sum_{n=0}^\infty \left(E_n^+ \ketbra{+,n}{n,+} + E_n^- \ketbra{-,n}{n,-} \right).
\ee
In \figref{fig:DressedStates}, we show detailed energy-level diagrams for the bare and dressed states of the JC Hamiltonian.

Dressed states have been essential for explaining experimental results in circuit QED ever since the beginning of the field~\cite{Wallraff2004, Wilson2007, Bishop2008, Fink2008, Fink2009, Wilson2010a} and it has been proposed that the dressed states themselves could be used for quantum information processing in SQCs~\cite{Liu2006a}. Adding a driving field to the qubit-resonator system can result in ``dressing of the dressed states''~\cite{Carmichael2008}. Such \textit{doubly dressed states} have also been studied in experiments with SQCs \cite{Bishop2008, Kockum2013}. By engineering the dressed states in a circuit-QED setup, an impedance-matched $\Lambda$-type three-level system can be realized~\cite{Koshino2013a}. Using such a system, microwave down-conversion~\cite{Inomata2014} and single-photon detection~\cite{Koshino2015, Koshino2016} can be implemented. Furthermore, three-state dressed states and multiphoton multi-level dressed states have been observed via higher-order Rabi sidebands in SQCs~\cite{Koshino2013b, Braumuller2015}. Also, collective dressed qubit states in the Tavis--Cummings model~\cite{TavisCummings}, the extension of the JC model to multiple qubits, have been demonstrated in SQCs~\cite{Fink2009a}.

\subsubsection{Resonant coupling}

In the resonant case, $\omega_{\rm r} = \omega_{\rm q}$, a quantum of energy bounces back and forth between the qubit and the resonator at a rate given by the Rabi frequency $\Omega_{n,\rm r}$. Note that the more photons there are in the cavity, the faster the single-photon exchange between a qubit and a resonator. The energy exchange is known as \textit{vacuum Rabi oscillations} when $n=0$; the splitting of $\Omega_0$ in the frequency domain due to the first doublet of JC eigenstates is called the \textit{vacuum Rabi splitting}. If the quantized resonator field is replaced by a classical field, then the normal Rabi oscillation (or splitting) can be observed. Observation of vacuum Rabi oscillations/splitting is a clear indicator of strong coupling and it has been used as such in many circuit-QED designs~\cite{Wallraff2004, Chiorescu2004, Johansson2006, Sillanpaa2007}. The Rabi oscillations can be stabilized for a very long time using a feedback-control method~\cite{Vijay2012, Cui2013}.

As shown in \eqref{eq:eigenf} and \figref{fig:DressedStates}, the anharmonicity of the JC ladder (i.e., the $n$th-doublet splitting) scales with $\sqrt{n}$. This can be used to reveal the quantization (graininess) of the resonator field, as has been demonstrated experimentally in SQCs~\cite{Bishop2008, Fink2008, Leek2010, Hofheinz2008}, semiconductors~\cite{Khitrova2006}, and Rydberg atoms~\cite{Brune1996, Schuster2008}. The resonant interaction between a superconducting qubit and a microwave resonator field can be used to deterministically generate Fock states~\cite{Hofheinz2008, Wang2008} and to systematically synthesize an arbitrary superposition of such states~\cite{Hofheinz2009} (see \secref{sec:OneCavity}); it can also be used to effectively blockade single microwave photons~\cite{Birnbaum2005, Lang2011} (see \secref{sec:PhotonBlockade}). Single-qubit states can also be mapped to another qubit via a resonator serving as a quantum bus~\cite{Sillanpaa2007}. Both the Fock-state generation above and this last example make great use of the \textit{tunability} that SQCs possess. By choosing to tune the qubit into resonance only for a certain time, some fraction of a Rabi oscillation period, the state that is transferred to the resonator can be exquisitely controlled.

\subsubsection{Dispersive coupling}
\label{sec:JaynesCummingsDispersive}

When $\omega_{\rm r} \neq \omega_{\rm q}$ and $g \ll \abs{\Delta} = \abs{\omega_{\rm q} - \omega_{\rm r}} $, the qubit and the resonator field are said to be in the \textit{dispersive regime} (large-detuning regime). In this case, there is no resonant photon absorption or emission. If the ratio $g/\Delta$ is small enough, the Hamiltonian in~\eqref{eq:JC} can be effectively written as
\be
H \approx \left[ \omega_{\rm r} + \frac{g^2}{\Delta} \sz \right] a^\dag a + \frac{1}{2} \left(\omega_{\rm q} + \frac{g^2}{\Delta} \right) \sz,
\label{eq:dispersive}
\ee
up to first order in $g/\Delta$, by applying the Schrieffer-Wolff transformation $U = \exp [g(a^\dag \sm - a \sp) / \Delta]$ to eliminate the linear-order term $a \sp + a^\dag \sm$~\cite{Wallraff2004, Blais2004, Liu2005b, Boissonneault2009} (see \appref{app:JCDispDerivation} for more details).

Equation~(\ref{eq:dispersive}) shows that the resonator field experiences a frequency shift of magnitude $\chi = g^2 / \Delta $, depending on the qubit state. This shift can be used to perform a QND measurement on the qubit by probing the transmission of the resonator~\cite{Blais2004, Gambetta2008}. As shown in~\figref{fig:DressedStates}, the resonator frequency is $\omega_{\rm r} + \chi$ ($\omega_{\rm r} - \chi$) if the qubit is in the excited (ground) state. Dispersive readout is not the only way to measure the qubit state; the JC nonlinearity can also be used for that purpose~\cite{Brune1996, Schuster2008, Bishop2008, Reed2010, Bishop2010, Boissonneault2010}.

Looking at \eqref{eq:dispersive} from the point of view of the qubit, also shown in~\figref{fig:DressedStates}, its transition frequency experiences a \textit{Stark shift} $2\chi n$ when there are $n$ photons inside the resonator, and a vacuum \textit{Lamb shift} $\chi$, which is present even when there are no photons inside the resonator~\cite{Schuster2005, Gambetta2006}. The Stark shift can be used to measure the number of photons in the resonator; see \secref{sec:PhotonDetectionQNDandCavity} for a more detailed discussion. The vacuum-induced Lamb shift $\chi$ has been observed experimentally in SQCs~\cite{Fragner2008}. 

This dispersive coupling between a qubit and a resonator can be used to simultaneously manipulate hundreds of photons, to create superpositions of coherent states~\cite{Vlastakis2013}, and to control photon states~\cite{Krastanov2015}. If two qubits are dispersively coupled to a resonator, an effective qubit-qubit interaction, mediated by \textit{virtual} photons in the resonator, is created when the qubits are on resonance with each other~\cite{Blais2007}. In this way, quantum states can be transferred between spatially separated qubits. Unlike the resonant coupling between a qubit and a resonator field, the dispersive coupling can avoid resonator losses. This type of qubit-qubit coupling has been demonstrated experimentally in SQCs~\cite{Majer2007}, also with multiple resonator modes~\cite{Leek2010, Filipp2011}. In the USC regime (see \secref{sec:ultrastrong}), three- and four-qubit interactions mediated by virtual photons can be engineered in a similar way~\cite{Stassi2017}.

\begin{figure}
\centering
\includegraphics[width=\linewidth]{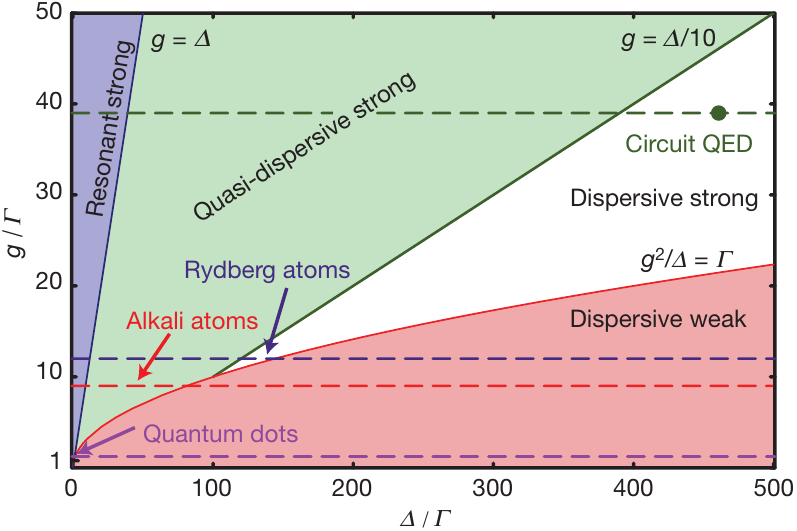}
\caption{A parameter-space diagram (or phase-like diagram) for cavity QED~\cite{Schuster2007}. The space is spanned by the atom-photon coupling strength $g$ and the detuning $\Delta = \omega_{\rm q} - \omega_{\rm r}$ between the atom ($\omega_{\rm q}$) and resonator ($\omega_{\rm r}$) frequencies, all given in units of $\Gamma = \max \{ \gamma, \kappa, 1/T \}$. Here $\gamma$ ($\kappa$) is the atom (resonator) decay rate and $1/T$ is the inverse of the atomic-transition time through the resonator in some cavity-QED setups. The dashed-horizontal lines indicate various cavity QED systems; from the top: circuit QED, Rydberg atoms, alkali atoms, and quantum dots. In the violet region, the atom and resonator are in resonance ($g > \Delta$) and strongly coupled ($g > \Gamma$). This is the regime where vacuum Rabi oscillations can be observed. Other regions correspond to different dispersive regimes: (i) weak dispersive (colored pink), (ii) strong dispersive (white), and (iii) strong quasi-dispersive (green). In the weak-dispersive region, the ac Stark shift $ 2\chi = 2 g^2 / \Delta < \Gamma$ is too small to resolve individual photons in qubit spectroscopy, but a QND measurement of the qubit state is still possible with many photons. In the strong-dispersive regime, QND measurements can be performed with demolition less than 1\%. This demolition is larger in the strong quasi-dispersive regime, where, however, a single-photon resolution remains feasible. 
Reprinted figure by permission from Macmillan Publishers Ltd: Nature, D.~I.~Schuster et al., \href{http://dx.doi.org/10.1038/nature05461}{Nature \textbf{445}, 515 (2007)}, copyright (2007). 
\label{fig:DispersiveRegime}}
\end{figure}

In \figref{fig:DispersiveRegime}, we display a parameter-space diagram for cavity QED, showing how coupling strength and detuning in terms of decoherence rates define different regimes. In the \textit{weak-dispersive} regime, the \textit{Purcell effect}~\cite{Purcell1946}, in which the spontaneous emission of a qubit can be enhanced or suppressed by modifying the density of states, has been observed in transmons~\cite{Houck2008}, and employed to generate single microwave photons~\cite{Houck2007} (see \secref{sec:OneCavity}). In the \textit{strong-dispersive} regime, the Stark shift per photon is larger than the decay rates of the resonator and the qubit, i.e., $2\chi > \max \{ \gamma, \kappa \}$. This regime is known as the number-splitting regime, since the qubit spectrum resolves individual photon-number states~\cite{Gambetta2006, Schuster2007}.

\subsection{Ultrastrong and deep-strong coupling regimes}
\label{sec:ultrastrong}

The Jaynes--Cummings model describes the interaction between light fields and matter under the condition of not-too-strong coupling strength, resonance or near resonance, and without strong driving~\cite{Berlin2004}. It has been the workhorse of quantum optics for decades. However, in the last few years a number of remarkable experiments have pushed past the boundaries of validity for the JC model, reaching the USC regime of light-matter interaction, where the coupling strength $g$ becomes comparable to the qubit and resonator transition frequencies ($\omega_{\rm q}$ and $\omega_{\rm r}$, respectively). These experiments have also inspired much theoretical work on the USC regime, which is a rapidly growing field of research. Below, we first review experimental setups reaching the USC regime and then discuss how the physics in this regime differ from that in the strong-coupling regime, which was treated in \secref{sec:CircuitQED_StrongCoupling}.

\subsubsection{Experimental implementations}
\label{sec:UltrastrongExp}

The experiments that have reached the USC regime are summarized in \tabref{tab:ultrastrongExperiment}. Although the USC regime has been realized in several types of systems, including semiconductor quantum wells~\cite{Anappara2009, Gunter2009, Todorov2010, Geiser2012}, THz meta-materials with a 2D electron gas (2DEG)~\cite{Scalari2012, Maissen2014, Zhang2016a}, organic molecules~\cite{Schwartz2011, Kena-Cohen2013, Gambino2014, Mazzeo2014, Gubbin2014, George2016a}, and a submillimeter-sized sphere in a magnetic-field-focusing resonator with photon-magnon and magnon-magnon couplings~\cite{Goryachev2014}, the experimental realizations using SQCs~\cite{Niemczyk2010, FornDiaz2010, Baust2016, Forn-Diaz2016, ChenZhen2016, Yoshihara2017, Forn-Diaz2017, Yoshihara2017a, Bosman2017} stand out in several respects. The SQC experiments are the only ones which have USC coupling to a \textit{single} (albeit artificial) atom, they offer unparalleled possibilities for probing the properties and dynamics of such an ultrastrongly coupled system, and a recent SQC experiment~\cite{Yoshihara2017} is the first ever to reach the DSC regime, where $g > \omega_{\rm q/r}$. Furthermore, driven SQCs in the strong-coupling regime have  recently been used to perform quantum simulations of the dynamics of an USC system~\cite{Langford2016, Braumuller2016}, building on earlier theoretical proposals~\cite{Ballester2012, Mezzacapo2014a}.

\begin{table}
\centering
\renewcommand{\arraystretch}{1.2}
\renewcommand{\tabcolsep}{0.15cm}
\begin{tabular}{ | c | c | c | c | c | }
\hline
\textbf{System} & \textbf{Experiments}  & \textbf{Top exp.} & $\omega_{\rm r} / 2\pi$ & $g / \omega_{\rm r}$ \\
\hline
SQCs & \cite{Niemczyk2010, FornDiaz2010, Baust2016, Forn-Diaz2016, ChenZhen2016, Yoshihara2017, Forn-Diaz2017, Yoshihara2017a, Bosman2017} & \cite{Yoshihara2017} & \unit[5.7]{GHz} & 1.34 \\ 
\hline
2DEG & \cite{Scalari2012, Maissen2014, Zhang2016a} & \cite{Maissen2014} & \unit[310]{GHz} & 0.87 \\
\hline
Organic molecules & \cite{Schwartz2011, Kena-Cohen2013, Gambino2014, Mazzeo2014, Gubbin2014, George2016a} & \cite{Gambino2014} & \unit[270]{THz} & 0.3 \\ 
\hline
Semiconductor polaritons & \cite{Anappara2009, Gunter2009, Todorov2010, Geiser2012} & \cite{Geiser2012} & \unit[670]{GHz} & 0.27 \\
\hline
Cavity-QED magnons & \cite{Goryachev2014} & \cite{Goryachev2014} & \unit[20]{GHz} & 0.1 \\
\hline
\end{tabular}
\caption{Experimental observations of the USC and DSC regimes for light-matter interaction. The values for qubit-resonator coupling strength $g$ and resonator frequency $\omega_{\rm r}$ refer to those experiments reporting the strongest (to our knowledge) coupling for a given class of systems. Since the experiments usually have a tunable matter frequency $\omega_{\rm q}$, but tune that frequency into or close to resonance, we only compare $g$ to $\omega_{\rm r}$.
\label{tab:ultrastrongExperiment}}
\end{table}

Circuit-QED setups can be engineered to realize a variety of coupling strengths and coupling mechanisms. One way to achieve a large inductive coupling strength is to make the central conductor of a transmission-line resonator thin and part of the loop of a flux qubit, possibly including a Josephson junction in the central-conductor part of the loop~\cite{Abdumalikov2008, Bourassa2009}. This type of setup was used in the first SQC experiment to reach the USC regime~\cite{Niemczyk2010} and also in later SQC experiments, which demonstrated USC to two resonators~\cite{Baust2016} and single-photon-driven high-order sideband transitions~\cite{ChenZhen2016}. Most other SQC experiments that have reached the SQC regime~\cite{FornDiaz2010, Forn-Diaz2016, Yoshihara2017, Yoshihara2017a}, including the only one to reach the DSC regime~\cite{Yoshihara2017}, have also used flux qubits, but coupled to a lumped-element $LC$ oscillator instead. Flux qubits have large anharmonicity, which means that they can be treated as two-level systems even when the coupling is ultrastrong. However, a recent experiment demonstrated USC between a weakly anharmonic transmon and a transmission-line resonator~\cite{Bosman2017}, using a careful design of coupling capacitances in the system~\cite{Bosman2017a}. 

Besides these experimentally implemented designs, many more ways to reach USC with SQCs have been proposed. The proposals include using a flux qubit with a rapidly controllable gap~\cite{Fedorov2010}, a switchable coupling~\cite{Peropadre2010}, a multimode coupling with a left-handed transmission line~\cite{Egger2013}, a transmon made part of a transmission line~\cite{Bourassa2012}, and a transmon coupled to an array of Josephson junctions~\cite{Andersen2017}. There is also a proposal for coupling a superconducting artificial atom simultaneously to both quadratures of the resonator using a combination of capacitive and inductive coupling~\cite{Baksic2014}. Such a setup could potentially realize the pure JC Hamiltonian with USC. Finally, we note that a recent experiment has demonstrated ultrastrong coupling of a flux qubit to an open one-dimensional (1D) transmission line~\cite{Forn-Diaz2017}. Such ultrastrong coupling to a \textit{continuum} of modes could potentially also be achieved with a transmon qubit coupled to propagating surface acoustic waves~\cite{Gustafsson2014, Aref2016}. For more discussion about USC to an open waveguide, see \secref{sec:WaveguideUSC}.

\subsubsection{Theory for USC and DSC}
\label{sec:UltrastrongTheory}

Since the JC Hamiltonian is not valid in the USC regime, analyzing the properties and dynamics of the qubit and the resonator in this regime requires using the full quantum Rabi Hamiltonian~\cite{Rabi1936, Rabi1937},
\be
H_{\text{Rabi}} = \omega_{\rm r} \left(a^\dag a + \frac{1}{2} \right) + \frac{\omega_{\rm q}}{2} \sz + g \sx \left(a + a^\dag \right),
\label{eq:Rabi}
\ee
or, for some SQC setups, the generalized version of this Hamiltonian given in \eqref{eq:general}. The last part of \eqref{eq:Rabi} contains the \textit{counter-rotating} terms $a \sm$ and $a^\dag \sp$, which oscillate with frequency $\omega_{\rm r} + \omega_{\rm q}$. These terms describe the simultaneous creation or annihilation of both atom and resonator excitations. It is the presence of these terms, which are neglected in the RWA to derive the JC Hamiltonian~\cite{JaynesCummings, Yoo1985, GerryKnight, Haroche2006} (see also \appref{app:RWA}), that result in various novel phenomena in the USC and DSC regimes. 

One immediate consequence of the presence of the counter-rotating terms is that the number of excitations in the system, $N = a^\dag a + \sp \sm$, is \textit{not} a conserved quantity, since $\comm{N}{H_{\text{Rabi}}} \neq 0$. In the JC Hamiltonian, $N$ is conserved ($\comm{N}{H_{\text{JC}}} = 0$); this is what makes that model easy to diagonalize. Since $N$ is not conserved in the quantum Rabi model, a wealth of new processes become possible in the USC regime. For example, the Rabi oscillations in the JC model can be extended to \textit{multiphoton} Rabi oscillations, where multiple photons excite the qubit~\cite{Ma2015, Garziano2015}, a single photon can excite \textit{multiple} qubits~\cite{Garziano2016}, and frequency conversion of photons can be realized if a qubit is coupled ultrastrongly to two resonator modes~\cite{Kockum2017}. In fact, analogues of almost \textit{all nonlinear-optics processes}, such as three- and four-wave mixing, are feasible~\cite{Anton2017}, as well as various many-body interactions between qubits~\cite{Stassi2017}. All these examples are higher-order processes where an effective coupling is created between two resonant system states, connected via transitions that do not conserve $N$.

While $N$ is not conserved in the quantum Rabi model, the \textit{parity} $\Pi = - \sz (-1)^{a^\dag a}$ is, since $\comm{\Pi}{H_\text{Rabi}} = 0$. This gives the quantum Rabi model a $\mathbb{Z}_2$ symmetry. The parity operator $\Pi$ has eigenvalues $p = \pm 1$, which divide the states of the model into two parity chains:
\bea
&& \ket{g}\!\ket{0} \leftrightarrow \ket{e}\!\ket{1} \leftrightarrow \ket{g}\!\ket{2} \leftrightarrow \ket{e}\!\ket{3} \leftrightarrow \cdots \quad (p=+1),\nn\\
&& \ket{e}\!\ket{0} \leftrightarrow \ket{g}\!\ket{1} \leftrightarrow \ket{e}\!\ket{2} \leftrightarrow \ket{g}\!\ket{3} \leftrightarrow \cdots \quad (p=-1).
\eea
Solving the quantum Rabi model is a difficult problem because the two parity subspaces are infinite-dimensional; by contrast, the JC model has two-dimensional subspaces. It was only a few years ago that an analytical solution was finally found, given in the form of transcendental functions~\cite{Braak2011, Solano2011}. This solution sheds more light not only on the USC regime, but also in the DSC regime~\cite{Casanova2010}, where the separate parity chains can exhibit collapses and revivals. Recently, analytical solutions have also been found for extensions of the quantum Rabi model to multiple qubits~\cite{Braak2013, Peng2013} and multiple resonator modes~\cite{Chilingaryan2015, Duan2015, Alderete2016}. Interestingly, in these extended models, there exist ``dark-like eigenstates'', which are superpositions of photonic and qubit excitations whose eigenenergies are independent of the coupling strength~\cite{Peng2016}.

Since the analytical solution of the quantum Rabi model is complicated, various approximate theoretical methods have been developed to analyze the dynamics including counter-rotating terms~\cite{Irish2007, Chen2008, Zueco2009, Ashhab2010, Hausinger2010, Hwang2010, Beaudoin2011, Hausinger2011, Rossatto2016}. For example, in the dispersive regime of the quantum Rabi model, when $g \ll \abs{\omega_{\rm q} - \omega_{\rm r}}$, one can apply the unitary transformation~\cite{Zueco2009, Klimov2009}
\be
U = \exp \left[ \frac{g}{\omega_{\rm r} - \omega_{\rm q}}(a^\dag \sm - a \sp) - \frac{g}{\omega_{\rm r} + \omega_{\rm q}}(a^\dag \sp - a \sm) \right] \ee
to \eqref{eq:Rabi} and expand to first order in $g / \abs{\omega_{\rm q} - \omega_{\rm r}}$. This gives the effective Hamiltonian
\be
H_{\text{Rabi}} \approx \omega_{\rm r} a^\dag a + \frac{\omega_{\rm q}}{2} \sz + \frac{g^2}{2} \left( \frac{1}{\omega_{\rm q} - \omega_{\rm r}} + \frac{1}{\omega_{\rm q} + \omega_{\rm r}} \right) \sz \left(a + a^\dag \right)^2,
\ee
where, in addition to the Stark shift $g^2 / (\omega_{\rm q} - \omega_{\rm r})$ known from the dispersive regime of the JC Hamiltonian reviewed in \secref{sec:JaynesCummingsDispersive}, the qubit frequency also gets a \textit{Bloch--Siegert shift}~\cite{Bloch1940} $g^2 / (\omega_{\rm q} + \omega_{\rm r})$ due to the counter-rotating terms~\cite{Klimov2009}. Several experiments with SQCs have observed the Bloch--Siegert shift and used it to show that they have reached physics beyond the JC model~\cite{FornDiaz2010, Baust2016, Forn-Diaz2016, Bosman2017}.

The ground state $\ket{G}$ of the quantum Rabi model in the USC regime is not the product of the photonic vacuum and the qubit ground state, $\ket{g}\!\ket{0}$, which is the case in the JC model. Instead, at high coupling strength the ground state is an \textit{entangled Schr\"odinger cat-like photon-qubit} state, which exhibits various nonclassical properties, including squeezing~\cite{Hines2004, Levine2004, Ashhab2010, Meaney2010, Nataf2010a, Garziano2014}. The ground state $\ket{G}$ contains a non-zero number of photons, $\brakket{G}{a^\dag a}{G} \neq 0$, which require careful physical interpretation. Standard input-output relations say that the output photon flux from the resonator is proportional to $\expec{a^\dag a}$, but here that would imply energy leaving a system which already is in its ground state~\cite{Ciuti2006, Ridolfo2012}. A correct treatment of decoherence, input-output, and correlation functions for an ultrastrongly coupled system requires considering how operators induce transitions between eigenstates of the system~\cite{Ashhab2010, Beaudoin2011, Ridolfo2012, Stassi2016}. This leads to interesting and surprising results, e.g., modification of photon blockade~\cite{Ridolfo2012}, both bunched and anti-bunched statistics for photons emitted from a thermal resonator~\cite{Ridolfo2013}, and bunched emission from a two-level atom ultrastrongly coupled to a resonator~\cite{Garziano2017}. 

Although the photons in the ground state of an ultrastrongly coupled qubit-resonator system cannot leak out spontaneously from the resonator, there are ways to release or probe them. For example, the spectrum of an ancillary probe qubit connected to a USC system will contain information about the virtual photons in $\ket{G}$~\cite{Lolli2015}. If the cavity is part of an opto-mechanical system, it may also be possible to detect the radiation pressure caused by the virtual ground-state photons~\cite{Cirio2017}.

To release photons from from $\ket{G}$, one approach is to non-adiabatically modulate either the qubit frequency or the coupling strength~\cite{Ciuti2005, Takashima2008, Werlang2008, Dodonov2008, Dodonov2009, DeLiberato2009, Beaudoin2011, Carusotto2012, Garziano2013, Shapiro2015a}; this is closely connected to the dynamical Casimir effect discussed in \secref{sec:Resonators}, where photons are generated from the vacuum by rapidly moving boundaries. Interestingly, the production of these photons can occur even in the weak-coupling regime ($g < \max\{\gamma, \kappa\}$)~\cite{DeLiberato2016}. If an auxiliary level is added to the qubit, such that a $\Xi$-type three-level atom with only the upper transition ultrastrongly coupled to the resonator is formed, a drive on the lower transition of the atom can periodically switch the ultrastrong interaction between the upper transition and the resonator on and off~\cite{Gunter2009, Ridolfo2011} and thus create photons~\cite{Carusotto2012}. Even without any outside drive, \textit{spontaneous conversion} from virtual to real photons can occur in such a three-level setup~\cite{Stassi2013}. Applying the drive to other transitions in the system, \textit{stimulated emission of virtual photons} can be realized~\cite{Huang2014}, which can be used to study how virtual photons virtual particles dress physical excitations in a quantum system~\cite{DiStefano2017}. We also note that ground-state photons can be released through electroluminescence if the photonic mode is coupled to an electronic two-level system~\cite{Cirio2016}.

An important question for USC physics and its implementation in SQCs is whether an $A^2$ term should be added to \eqref{eq:Rabi}. In atomic physics, this term originates from a coupling Hamiltonian $\propto (p - eA)^2$, where $p$ is the momentum of a charged particle, $e$ is the elementary charge, and $A \propto (a+a^\dag)$ is the electromagnetic vector potential. The $A^2$ term can usually be ignored for weaker couplings, but in the USC regime its inclusion determines whether the system can undergo a \textit{superradiant phase transition}~\cite{Hepp1973, Wang1973, Rzazewski1975, Bialynicki-Birula1979, Emary2003}. 

Whether the $A^2$ term is present in the Hamiltonians describing circuit-QED setups, what magnitude it then has, and how that affects the possibility of observing a superradiant phase transition, is the subject of ongoing debate and investigations~\cite{Nataf2010, Viehmann2011, Garcia-Ripoll2014, Jaako2016, Bamba2016, Bamba2017}. Another, somewhat surprising, consequence of the $A^2$ term is that it leads to an \textit{effective decoupling} of the qubit from the resonator if the system goes deep enough into the DSC regime~\cite{DeLiberato2014}. A striking consequence of this result is that the spontaneous-emission rate of the qubit, thought to increase monotonically with $g$ due to the Purcell effect, decreases dramatically when $g$ becomes large enough~\cite{DeLiberato2014}.

Ultrastrong coupling has several potential applications in quantum computing. For example, quantum gates in USC systems can operate extremely fast, at sub-nanosecond time scales~\cite{Romero2012, Wang2017}. Furthermore, various unique characteristics of the USC regime, e.g., the entangled ground state and the $Z_2$ symmetry, can be utilized to realize protected quantum computation with multiple resonators~\cite{Nataf2011}, quantum memory~\cite{Kyaw2014, Stassi2017a}, and holonomic quantum computation~\cite{Wang2016c}. Circuit-QED systems in the USC regime can also be used for simulations of the Jahn-Teller model~\cite{Larson2008, Bourassa2009, Dereli2012} and exploration of symmetry-breaking quantum vacuum, which is analogous to the Higgs mechanism~\cite{Garziano2014}. Finally, it can also be noted that many well-known phenomena from the strong-coupling regime change dramatically in the USC regime and thus need to be reexamined. Examples include the Purcell effect~\cite{DeLiberato2014}, the spontaneous emission spectrum~\cite{Cao2011a}, the Zeno effect~\cite{Lizuain2010, Cao2012}, and photon transfer in coupled cavities~\cite{Felicetti2014a}.

\section{Waveguide QED}
\label{sec:WaveguideQED}

\begin{figure}
\centering
\includegraphics[width=\linewidth]{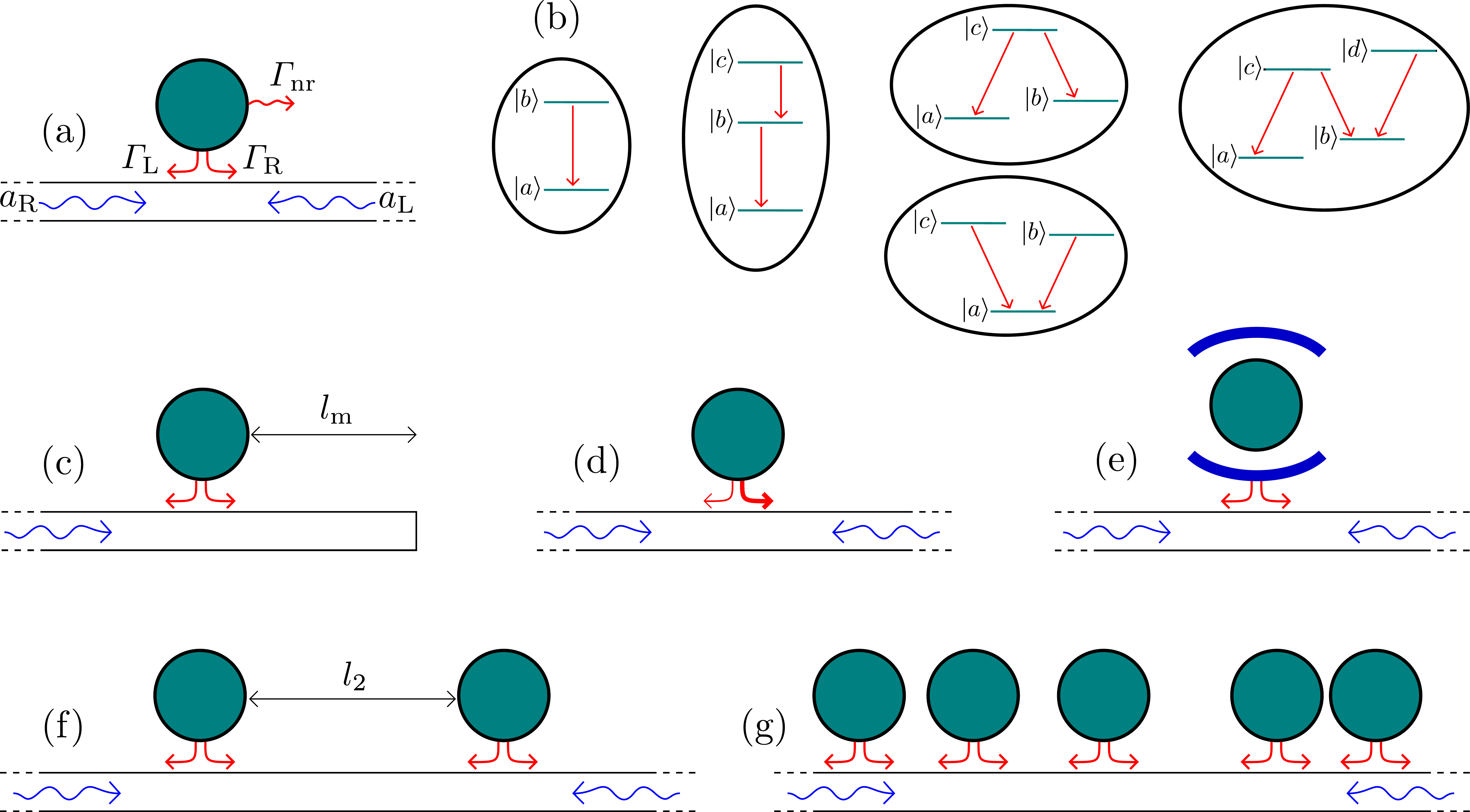}
\caption{An overview (not exhaustive) of waveguide QED setups. (a) The most basic structure is a 1D waveguide with a continuum of right- and left-propagating modes $a_{\rm R}$ and $a_{\rm L}$, respectively, coupled to a single atom. The atom can relax, with rates $\Gamma_\text{R}$ and $\Gamma_\text{L}$, to the modes in the transmission lines. Usually, $\Gamma_\text{L} = \Gamma_\text{R}$. The atom can also relax via other channels, sometimes called nonradiative decay, with a rate $\Gamma_\text{nr}$. In many cases with superconducting artificial atoms, this decay is negligible, i.e., $\Gamma_\text{nr} \ll \Gamma_{1\text{D}} = \Gamma_\text{L} + \Gamma_\text{R}$. (b) The atom(s) coupled to the waveguide can have various level structures, e.g., two levels, three levels in the $\Xi$, $\Lambda$, and $V$ configurations, and four levels in an $N$ configuration. (c) The waveguide can be semi-infinite, terminated by a mirror at some distance $l_\text{m}$ from the atom(s). (d) The waveguide can be chiral, i.e., $\Gamma_\text{L} \neq \Gamma_\text{R}$, possibly with one these rates being zero. (e) The atom(s) can be coupled to a resonator, which then in turn is coupled to the waveguide. (f) There can be two atoms, separated by a distance $l_2$, coupled to the waveguide. (g) There can be many atoms coupled to the waveguide.
\label{fig:WaveguideQEDSetups}}
\end{figure}

Quantum optics has for a long time been concerned with either atoms in free (3D) space, or atoms interacting with single electromagnetic modes in cavities. However, in the last decade experimental investigations have begun on atoms, both natural and artificial, coupled to various types of 1D open waveguides. In these setups, dubbed \textit{waveguide QED}, the absence of a cavity means that the atoms will interact with a \textit{continuum} of propagating photonic modes, but the confinement to 1D instead of 3D is enough to still achieve a strong coupling. As shown in~\figref{fig:WaveguideQEDSetups}, a great many configurations can be envisioned for waveguide QED. Much of the work in the field concerns using the atoms to mediate effective interactions between photons propagating in the waveguide. Another important topic is collective effects, such as super- and subradiance~\cite{Dicke1954}, for multiple atoms interacting via the waveguide.

In this section, we review experimental and theoretical work in waveguide QED with a focus on superconducting artificial atoms. However, we also mention much work done with other systems in mind, since it often can be transferred to the realm of microwave photonics. We begin in~\secref{sec:WaveguideQEDExpSCAtoms} with an overview of experiments done with superconducting systems and then compare to experiments in other types of systems in~\secref{sec:WaveguideQEDExpOther}. In~\secref{sec:WaveguideQEDTheory}, we then review theoretical work dealing with the multitude of possible waveguide QED setups shown in \figref{fig:WaveguideQEDSetups}. For another recent review of the field, see Ref.~\cite{Roy2017}.

\subsection{Experiments}

Atoms in free space have been studied experimentally for a long time, but truly 1D waveguide systems have only been realized in the last few years. Arguably, the first such experiment was performed in 2007 when CdSe quantum dots were coupled to surface plasmons propagating in a silver nanowire~\cite{Akimov2007}. The first experiment with a superconducting artificial atom coupled to propagating microwave photons was realized in 2010~\cite{Astafiev2010}. Here, we review and compare the experiments done with superconducting artificial atoms and other systems.

\subsubsection{Superconducting artificial atoms}
\label{sec:WaveguideQEDExpSCAtoms}

\begin{figure}
\centering
\includegraphics[width=\linewidth]{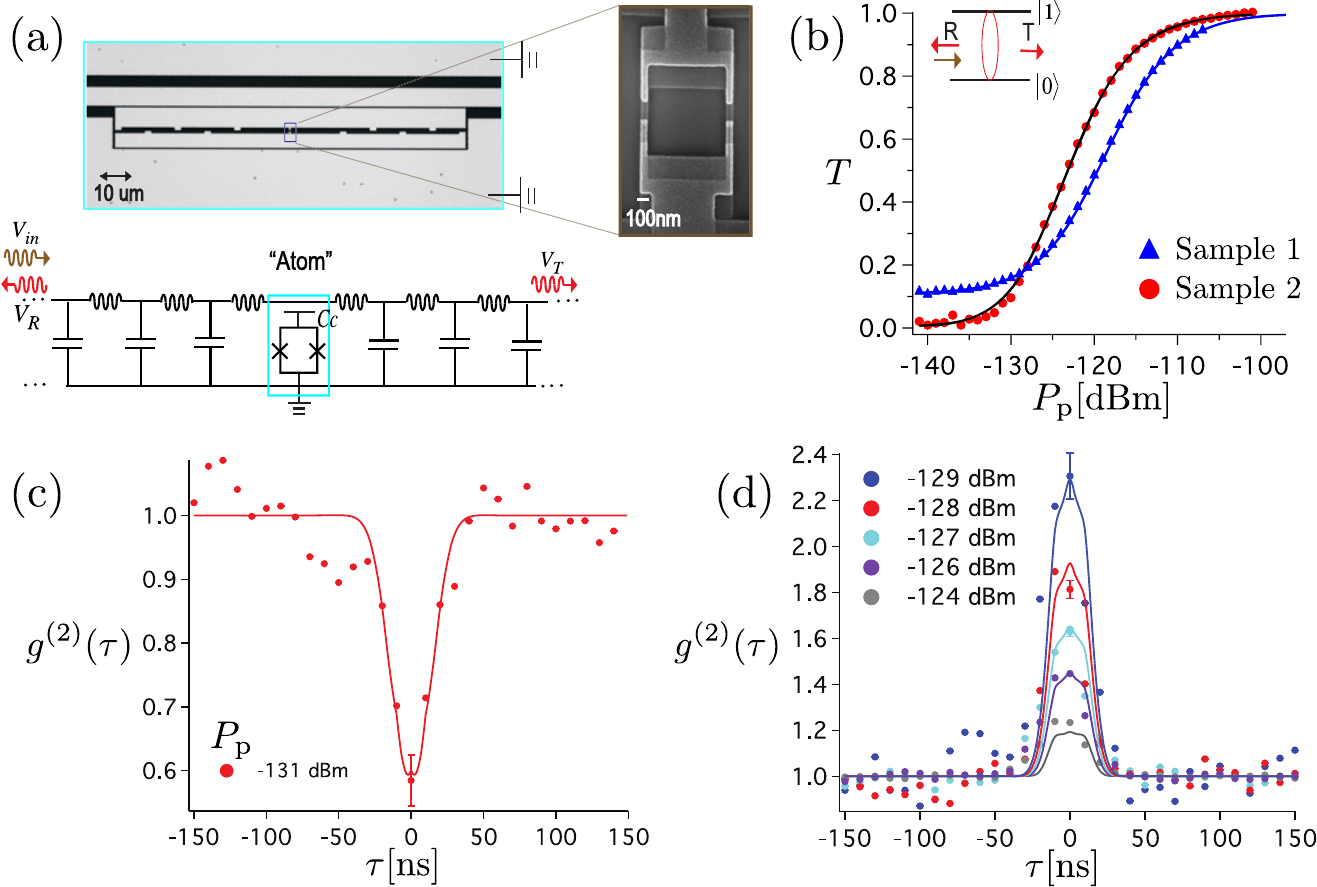}
\caption{Experiments with a superconducting artificial atom in an open 1D transmission line~\cite{Hoi2012, Hoi2013NJP}. (a) The transmon circuit placed close to the transmission line. Zoom: The SQUID connecting the two superconducting islands of the transmon. (b) Measured power transmission $T$ as a function of the input power $P_{\rm p}$ (on resonance with the artificial atom) for two samples. The solid lines are theory. Sample 2 has a smaller pure dephasing rate than sample 1 and therefore achieves a larger transmission extinction. (c) Second-order correlation function $g^{(2)}(\tau)$ for the reflected microwave signal, showing a clear anti-bunching dip. The solid lines are theory calculations including temperature, finite detector bandwidth, and trigger jitter. (d) The same as (c), but for the transmitted signal at a number of different input powers. Reprinted figure from I.-C.~Hoi et al., \href{http://iopscience.iop.org/1367-2630/15/2/025011}{New J.~Phys.~\textbf{15}, 25011 (2013)}. 	
\label{fig:ReflectionAndG2ExpSCAtom}}
\end{figure}

In the seminal work of Ref.~\cite{Astafiev2010}, a superconducting flux qubit was coupled to an open microwave transmission line, realizing the setup illustrated in \figref{fig:WaveguideQEDSetups}(a). Besides demonstrating the \textit{Mollow triplet}~\cite{Mollow1969} (see also \secref{sec:EITandATS}) in resonance fluorescence measurements of the atom (further explored in subsequent work~\cite{Abdumalikov2011}; Mollow sidebands had been measured earlier with a transmon qubit inside a resonator~\cite{Baur2009}), the experiment also clearly measured a fundamental effect unique to the 1D geometry. This is the \textit{power-dependent reflection} of a coherent probe signal on resonance with the atom. When the atom is excited, it emits equally in both directions of the transmission line. This emission interferes destructively with the probe signal and the result is perfect reflection. However, the atom can only handle one photon per relaxation time before it becomes \textit{saturated}, so when the probe power increases, the transmission approaches unity. In Ref.~\cite{Astafiev2010}, a $94\%$ extinction of the transmission was measured at low probe power. This can be compared to experiments with natural atoms in 3D free space, where the many possible emission directions for the atom weakens the interference effect. Despite using lenses for extensive focussing, the best such experiments have shown transmission extinction of $7$--$12\%$~\cite{Vamivakas2007, Gerhardt2007, Tey2008, Wrigge2008, Hwang2009}, although higher numbers are possible according to theory~\cite{Zumofen2008}. Another experiment with a superconducting transmon qubit in a 1D transmission line improved the record to $99.6\%$ in 2011 [see Figs.~\ref{fig:ReflectionAndG2ExpSCAtom}(a) and \ref{fig:ReflectionAndG2ExpSCAtom}(b)] by optimizing the setup to reduce pure dephasing and minimizing relaxation to other channels beside the transmission line~\cite{Hoi2011, Hoi2013NJP, Hoi2013PhD}.

The perfect reflection of a single photon from a single two-level atom in an open 1D waveguide entails that the reflected part of a coherent drive on the atom will be \textit{anti-bunched}, while the transmitted part will be \textit{bunched}. This was confirmed in experiments~\cite{Hoi2012, Hoi2013NJP} with a transmon qubit where the second-order correlation function $g^{(2)}(\tau)$ was measured, demonstrating $g^{(2)}(0) < 1$ for the reflected signal and $g^{(2)}(0) > 2$ for the transmitted signal, as shown in Figs.~\ref{fig:ReflectionAndG2ExpSCAtom}(c) and \ref{fig:ReflectionAndG2ExpSCAtom}(d).

In other experiments, the superconducting artificial atom in the waveguide has had a three-level $\Xi$ structure with ground state $\ket{0}$ and excited states $\ket{1}$ and $\ket{2}$. For such a system with low anharmonicity (the transition frequencies $\omega_{01}$ and $\omega_{12}$ are similar), three-state dressed states have been observed when driving at $(\omega_{01} + \omega_{12})/2$~\cite{Koshino2013b}. In another experiment, the $\ket{0} \leftrightarrow \ket{2}$ transition was driven to create \textit{population inversion}, resulting in a small amplification (by stimulated emission) of a signal resonant with the $\ket{0} \leftrightarrow \ket{1}$ transition~\cite{Astafiev2010a}. 

By instead driving the $\ket{1} \leftrightarrow \ket{2}$ transition with an amplitude $\Omega$, an \textit{Autler--Townes splitting}~\cite{Autler1955} (see also \secref{sec:EITandATS}) of the $\ket{0} \leftrightarrow \ket{1}$ transition (resulting in resonances at $\omega_{01}\pm\Omega/2$ instead of $\omega_{01}$) has been achieved. Such a splitting had been measured earlier with superconducting qubits in resonators~\cite{Baur2009, Sillanpaa2009}, but here it was used to control the transmission of a weak probe on $\omega_{01}$ (compare the transmission extinction above), yielding a high on/off ratio~\cite{Abdumalikov2010, Hoi2011} that can be used for \textit{single-photon routing}~\cite{Hoi2011}. When the signals close to resonance with $\ket{0} \leftrightarrow \ket{1}$ and $\ket{1} \leftrightarrow \ket{2}$ transitions both are at the single-photon level, an effective photon-photon interaction in the form of a large \textit{cross-Kerr} phase shift (10--20 degrees per photon) has been observed for coherent signals~\cite{Hoi2013a}. This is much larger than what has been achieved in systems with single natural atoms (a phase shift on the order of 1 radian per photon was very recently demonstrated using an atomic ensemble~\cite{Beck2016}) and could be utilized for single-photon detection (see~\secref{sec:PhotonDetectionItinerant} for details).

Just as ultrastrong coupling has been reached in circuit QED, as described in~\secref{sec:ultrastrong}, it has now also been reached in waveguide QED using a superconducting flux qubit coupled to an open transmission line. In the experiment of Ref.~\cite{Forn-Diaz2017}, the decay rate $\Gamma_{1\text{D}}$ exceeded the qubit transition frequency. This opens up for experimental studies of the spin-boson model~\cite{Leggett1987} in hitherto unexplored parameter regimes and a number of interesting theoretical predictions reviewed in~\secref{sec:WaveguideUSC} below.

By alternating coplanar-waveguide sections with different impedances, a recent experiment created a microwave photonic-crystal waveguide with a \textit{photonic bandgap} and coupled a transmon strongly to the upper band~\cite{Liu2016a}. Tuning the transmon transition frequency into the bandgap, a bound polariton state was observed. The bound state gives rise to a transmission peak inside the bandgap even though the photonic part of the polariton is exponentially localized around the transmon.

Increasing the number of atoms, two transmons have been coupled to an open 1D transmission line at a distance $l_2$ from each other [the setup in \figref{fig:WaveguideQEDSetups}(f)]~\cite{VanLoo2013}. The relevant wavelength for interaction between the atoms via the transmission line is set by the transition frequencies of the atoms, which can be tuned to change $l_2$ in units of wavelengths. The experiment demonstrated distance-dependent exchange interaction between the atoms as well as collective decay (\textit{super-} and \textit{subradiance}). The latter was also studied recently with two transmons in a bad cavity~\cite{Mlynek2014}, which is a setup somewhere in between waveguide QED and normal cavity QED.

A few experiments with superconducting circuits have explored the setup with a semi-infinite waveguide, i.e., a waveguide terminated by a mirror, shown in \figref{fig:WaveguideQEDSetups}(c). An example without any artificial atom is the experimental demonstration of the dynamical Casimir effect, where a SQUID at the end of a transmission line was used to implement a rapidly tunable boundary condition (an effective moving mirror)~\cite{Wilson2011}. Another example is the cross-Kerr photon-photon interaction discussed above, which was enhanced when the atom was placed at the end of a waveguide, only allowing the atom to emit in one direction instead of two~\cite{Hoi2013a}. In Refs.~\cite{Koshino2013a, Inomata2014, Inomata2016}, a flux qubit placed in a resonator realized an effective three-level $\Lambda$ system [see \figref{fig:WaveguideQEDSetups}(b)], which was placed at the end of a transmission line. Engineering the two relaxation rates from the highest excited state $\ket{c}$ to be equal, an interference effect similar to that for a two-level atom in an open waveguide causes the system to fully absorb an incoming weak signal at $\omega_{ac}$ and down-convert it by only decaying to $\ket{b}$. As discussed further in~\secref{sec:PhotonDetectionItinerant}, this can be used for single-photon detection by detecting when the system has switched to $\ket{b}$.

\begin{figure*}
\centering
\includegraphics[width=\linewidth]{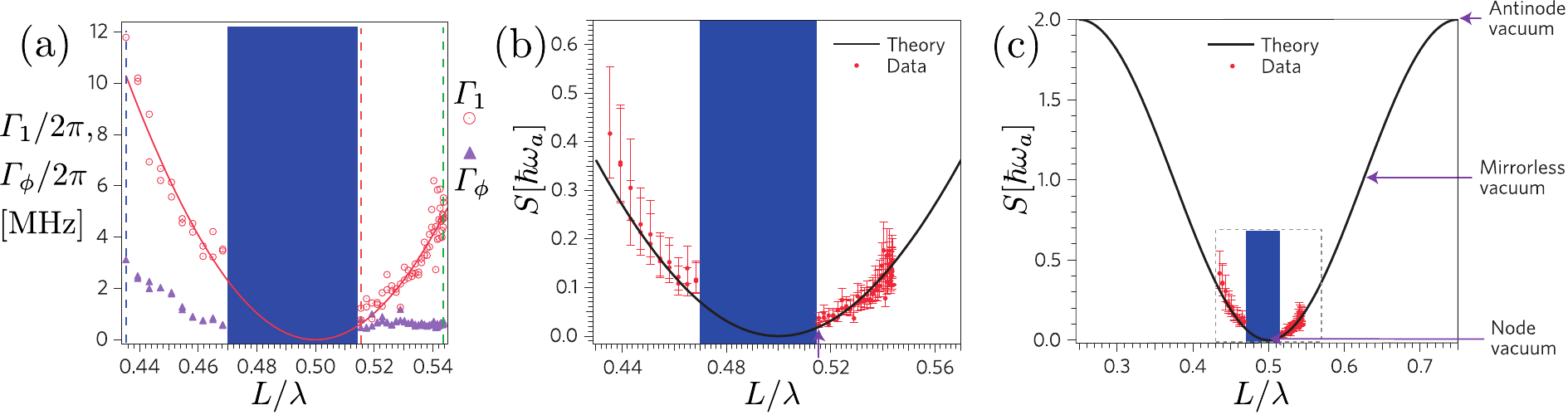}
\caption{Measuring vacuum fluctuations with a superconducting artificial atom at a distance $L$ from an effective mirror in a semi-infinite transmission line~\cite{Hoi2015}. (a) Measured relaxation rate $\Gamma_1$ and pure dephasing rate $\Gamma_\phi$ for the atom as a function of $L$. Varying the atom transition frequency $\omega_{\rm a}$ changes the distance to the mirror in terms of wavelengths $\lambda = 2\pi v/\omega_{\rm a}$, where $v$ is the wave velocity in the transmission line. Close to the node for the electromagnetic field, around $L = \lambda/2$, the rates are too weak to measure (shaded area). Note that the measured relaxation rate changes by around an order of magnitude. The solid line is theory. (b) Power spectral density $S$ for the vacuum fluctuations of the electromagnetic field at $\omega$ in units of $\hbar\omega_{\rm a}$, extracted from the measured relaxation rate. (c) Zoom out from (b) showing the theoretical predictions for a larger span of $L$, including antinodes for the field. For an open transition line (mirrorless vacuum), $S(\omega_{\rm a}) = \hbar\omega_{\rm a}$, with the right- and left-propagating modes contributing half a quantum each to the fluctuations (see \appref{app:Quantizing1DTransmissionLine}). 
Reprinted figure by permission from Macmillan Publishers Ltd: Nature Physics, I.-C.~Hoi et al., \href{http://dx.doi.org/10.1038/nphys3484}{Nat.~Phys.~\textbf{11}, 1045 (2015)}, copyright (2015). 
\label{fig:SCAtomMirrorVacuum}}
\end{figure*}

In the experiment of Ref.~\cite{Hoi2015}, a transmon qubit was placed at a distance from the end of a transmission line. Just as in Ref.~\cite{VanLoo2013} discussed above, the relevant measure of distance is in terms of the wavelength set by the atom transition frequency. It is vacuum fluctuations at this frequency that causes the atom to relax from its excited state. Due to the presence of the mirror, the atom can be placed at a node of these fluctuations, lowering the relaxation rate. By tuning the atom frequency, and thus the effective distance to the mirror, a decrease in the relaxation rate by one order of magnitude was observed, as shown in \figref{fig:SCAtomMirrorVacuum}(a). Measuring the relaxation rate was then in turn used to measure the spectral density of the vacuum fluctuations, showing that at the node they are well below half a quantum [see Figs.~\ref{fig:SCAtomMirrorVacuum}(b) and \ref{fig:SCAtomMirrorVacuum}(c)].

Finally, a recent experiment coupled a superconducting artificial atom (a transmon) not to propagating microwave photons, but to propagating \textit{phonons} in the form of surface acoustic waves (SAWs) on a piezoelectric GaAs substrate~\cite{Gustafsson2014}. In this version of 1D waveguide QED (or rather QAD, quantum \textit{acoustodynamics}), the phonons propagate five orders of magnitude slower than the photons discussed so far, and since their frequency is still in the microwave range the wavelength is correspondingly shorter, on the order of \unit[1]{$\mu$m}. This is significantly smaller than the size of the transmon, so the experiment realizes a ``giant artificial atom''~\cite{Kockum2014, Guo2016}, which couples to the waveguide at several points, wavelengths apart. The setup offers many interesting possibilities for waveguide QED experiments in new regimes~\cite{Aref2016}, including ultra-strong coupling, and can potentially be used to interface with other artificial and natural atoms~\cite{Schuetz2015, Schutz2015} as well as optical light~\cite{Shumeiko2016}.

\subsubsection{Other systems}
\label{sec:WaveguideQEDExpOther}

The experimental realizations of waveguide QED not using superconducting circuits mainly follow two approaches, either coupling quantum dots to photonic crystal waveguides or coupling natural atoms to optical fibers, although there are more setups. In the seminal experiment of Ref.~\cite{Akimov2007}, quantum dots were coupled to surface plasmons in a nanowire, with around 60\% of the emission from the dots going into this waveguide. The loss of the remaining emission to free space, i.e., a relatively large $\Gamma_\text{nr}$, is a problem which superconducting circuits do not suffer from, but it has proven a challenge to reduce it in other systems.

Theoretical studies have shown that quantum emitters can couple efficiently to waveguides for plasmons made from nanowires~\cite{Chang2007a}, graphene~\cite{Koppens2011, Christensen2012}, and carbon nanotubes~\cite{Martin-Moreno2015}, but later experiments with quantum dots and natural atoms have seen them coupled to photonic crystal waveguides instead~\cite{Lund-Hansen2008, Laucht2012, Arcari2014, Goban2015, Reithmaier2015, Sollner2015, Hood2016}, where theory predicts that $\Gamma_\text{nr}$ can be made small~\cite{MangaRao2007, Lecamp2007}. Indeed, in Ref.~\cite{Arcari2014} 98.4\% of the quantum dot emission went into the waveguide. For an in-depth review of quantum dots coupled to photons in various configurations, see Ref.~\cite{Lodahl2015}; a review on various emitters coupled to propagating surface plasmons is given in Ref.~\cite{Huck2016}.

Several experiments have coupled natural atoms, trapped and cooled, to optical fibers~\cite{Bajcsy2009, Vetsch2010, Goban2012, Mitsch2014, Goban2014, Beguin2014, Corzo2016, Sorensen2016} or to resonators that are in turn coupled to fibers~\cite{Volz2014, Shomroni2014}. These experiments also struggle to reduce $\Gamma_\text{nr}$. For a review on photon-photon interaction mediated by natural atoms, see Ref.~\cite{Chang2014}.

Other experimental realizations of waveguide QED include dye molecules coupled to a waveguide for optical light made from an organic crystal~\cite{Faez2014}, a gold nanoparticle coupled to a photonic waveguide made from a silica nanofiber~\cite{Petersen2014}, and an NV center coupled to a plasmonic groove waveguide~\cite{Bermudez-Urena2015}.

Comparing with superconducting circuits, waveguide QED with other systems suffer from larger $\Gamma_\text{nr}$, although quantum dots have almost caught up in this respect. While superconducting circuits operate at microwave frequencies, most other systems work at optical frequencies, where, e.g., photon detection is easier to perform (see~\secref{sec:PhotonDetection}). We also note that recently several non-superconducting systems have realized \textit{chiral} waveguides~\cite{Mitsch2014, Petersen2014, Sollner2015, Young2015} [see \figref{fig:WaveguideQEDSetups}(d)] using spin-orbit interaction of the confined light~\cite{Bliokh2015, Bliokh2015a, Bliokh2015b}. Chiral waveguides should also be possible to implement with superconducting circuits, using circulators (see~\secref{sec:MicrowaveComponentsCirculators}) in a setup like that of \figref{fig:SathyamoorthyPhotonDetection} in~\secref{sec:PhotonDetectionItinerant}.

Going beyond pure 1D waveguide QED, we note that experiments have been performed earlier with trapped ions in free space interacting with focussed laser light. In this kind of setup, super- and subradiance has been observed from two trapped ions separated by a tunable distance~\cite{DeVoe1996} and a few experiments have investigated a trapped ion in front of a mirror~\cite{Eschner2001, Wilson2003, Bushev2004, Dubin2007}, showing that the distance to the mirror affects both the relaxation rate and the Lamb shift~\cite{Lamb1947} of the ion due to interference with the ion's mirror image. While the Lamb shift of a superconducting qubit in a resonator has been measured~\cite{Fragner2008}, it remains to be done in a waveguide.

\subsection{Theory}
\label{sec:WaveguideQEDTheory}

The experimental progress in waveguide QED, both with superconducting circuits and in other systems, has inspired a wealth of theoretical work in the field. Below, we review this rapidly growing body of literature, dividing the work mainly into open and semi-infinite waveguides, which couple to one, two, or many atoms. We also consider chiral waveguides, coupling to cavities containing atoms, and ultrastrong coupling. However, a detailed account of analytical and numerical tools such as (continuous) matrix product states~\cite{Fannes1992, White1992, Schollwock2005, Haegeman2010, Verstraete2010, Prior2013}, formalisms for cascaded quantum systems~\cite{Gardiner1993, Carmichael1993, Gardiner2004, Gough2009, Gough2009a, Tezak2012, Zhang2013, Zhang2014a}, Fock state input to open quantum systems~\cite{Gheri1998, Baragiola2012, Gough2012, Baragiola2014PhD}, etc., is beyond the scope of this review. For some overview, see, e.g., Refs.~\cite{Zhang2014a, Combes2016} and the introductions of Refs.~\cite{Nysteen2015, Xu2015, Pichler2015}.

The continuum of modes in the 1D waveguide, which sometimes is modeled as a chain of coupled cavities for analytical and numerical convenience~\cite{Zhou2008, Zhou2008a, Zhou2009, Shi2009, Longo2010, Liao2010a, Longo2011, Roy2011a, Longo2012, Moeferdt2013, Martens2013, Peropadre2013a, Lu2014, Lombardo2014, Wang2014, Sanchez-Burillo2014, Schneider2015, Sanchez-Burillo2015, Calajo2015, Kocabas2015, Sanchez-Burillo2016b}, can be viewed as a bath for the atom(s) coupling to the waveguide, thus realizing the well-studied \textit{spin-boson model}~\cite{Leggett1987} and making it possible to derive a master equation for the atom(s)~\cite{Gardiner2004}. However, much of the work in the field of waveguide QED is more concerned with effects on the photons propagating in the waveguide.

\subsubsection{Open waveguide}
\label{sec:WaveguideQEDTheoryOpen}

Theoretical work on 1D waveguides coupled to atoms really took off in the middle of the 2000s', but there are some earlier works that either eliminated cavities from cavity QED or took the step from 3D to 1D. Transmission in the ``bad-cavity'' limit for a cavity QED setup with a single atom was studied~\cite{Rice1988, Turchette1995, Turchette1995a} and Kimble coined the term ``one-dimensional atom'' for such systems~\cite{Turchette1995a}. There were also papers on 1D waveguides with periodic scattering potentials~\cite{Bendickson1996} or coupling to harmonic oscillators~\cite{Konik1998, Xu2000}.  

\paragraph{One atom}

The simplest waveguide QED system consists of a single atom coupled to the waveguide, where some of the earliest efforts are Refs.~\cite{LeClair1997, Domokos2002, Kojima2003, Shen2005}. Most papers on the subject so far are concerned with two-level atoms~\cite{LeClair1997, Domokos2002, Kojima2003, Shen2005, Shen2007a, Koshino2007, Chang2007, Shen2007, Zhou2008, Longo2009, Zhou2009, Shi2009, Zheng2010, Dzsotjan2010, Rephaeli2010, Fan2010, Longo2010, Longo2011, Wang2011, Derouault2012, Longo2012, Kocabas2012, Witthaut2012, Valente2012b, Moeferdt2013, Rephaeli2013, Zheng2013b, Huang2013PRA, Roulet2014, Lombardo2014, Li2014, Lindkvist2014NJP, Fang2014, Zang2015, Kocabas2015, Nysteen2015, Nysteen2015a, Pletyukhov2015, Schneider2015, Bera2016, Guo2016, Nurdin2016, Roulet2016, Sanchez-Burillo2016b}, but atoms with three levels~\cite{Chang2007, Koshino2009, Witthaut2010, Roy2011, Kolchin2011, Valente2012a, Zheng2012, Witthaut2012, Martens2013, Fan2013, Peropadre2013NJP, Lu2014, Roy2014, Wang2014, Fang2015a, Xu2016a} and more~\cite{Zheng2011, Zheng2012, Kockum2014, Sanchez-Burillo2016, Sanchez-Burillo2016a} have also been studied. The question usually asked is how the atom affects the transport of photons, either in Fock states with one~\cite{Domokos2002, Kojima2003, Shen2005, Shen2007a, Zhou2008, Koshino2009, Longo2009, Zhou2009, Shi2009, Zheng2010, Fan2010, Witthaut2010, Roy2011a, Zheng2011, Zheng2012, Derouault2012, Martens2013, Rephaeli2013, Huang2013PRA, Li2014, Kocabas2015, Schneider2015, Guo2016, Nurdin2016, Sanchez-Burillo2016, Sanchez-Burillo2016a, Sanchez-Burillo2016b}, two~\cite{Kojima2003, Shen2007a, Koshino2007, Shen2007, Shi2009, Zheng2010, Fan2010, Longo2011, Roy2011a, Roy2011, Zheng2011, Longo2012, Zheng2012, Moeferdt2013, Rephaeli2013, Xu2013a, Kocabas2015, Nysteen2015, Nysteen2015a, Schneider2015, Guo2016, Sanchez-Burillo2016a, Xu2016a}, or many~\cite{Longo2009, Shi2009, Zheng2010, Zheng2012, Xu2015, Roulet2016} excitations, or in coherent states~\cite{Domokos2002, Zheng2010, Kocabas2012, Zheng2012, Zheng2013b, Peropadre2013NJP, Lindkvist2014, Zang2015, Pletyukhov2015, Bera2016}. 

For the case of a single photon, one can also ask how well it can excite the atom~\cite{Rephaeli2010, Longo2011, Wang2011}. It turns out that the atom can be \textit{perfectly excited}, provided that the photon wavepacket shape is a rising exponential, which is the time-reverse of decay from the atom~\cite{Rephaeli2010, Wang2011}. If the atom-waveguide coupling can be modulated, a photon of arbitrary shape can be perfectly absorbed~\cite{Nurdin2016}. A related problem is stimulated emission from an excited atom induced by a single photon~\cite{Rephaeli2012, Valente2012b}. The probabilities of the two outgoing photons traveling in the same or opposite directions depends on the wavepacket shape of the incoming photon~\cite{Rephaeli2012}. Other aspects of single-photon transport that have been studied include the effect of the waveguide dispersion relation~\cite{Zhou2008, Roy2011a, Sanchez-Burillo2016b} and of rapidly modulating the atom frequency~\cite{Zhou2009}.

For a single photon it does not matter if the system in the waveguide is a two-level atom or a harmonic oscillator, since only one transition is involved, but for two or more photons a single atom can be saturated and thus mediate an \textit{effective photon-photon interaction}. A clear signature of the atom-induced nonlinearity in the multi-photon transport is provided by the second-order correlation function $g^{(2)}(\tau)$~\cite{Kojima2003, Chang2007, Zheng2010, Kocabas2012, Zheng2012, Moeferdt2013, Peropadre2013NJP, Roy2014, Fang2014, Fang2015a, Pichler2015}, which, as we saw above in~\secref{sec:WaveguideQEDExpSCAtoms}, has also been studied in experiments with superconducting circuits. A single photon on resonance with the atom will be perfectly reflected, but it will also saturate the atom such that any additional photons arriving at the same time are transmitted; this causes the reflected light to become anti-bunched while the transmitted light becomes bunched. Another effect of this saturation is that a two-level atom in an open waveguide cannot be a perfect analog of a beam-splitter~\cite{Longo2012, Roulet2014, Nysteen2015a}; the Hong-Ou-Mandel effect~\cite{Hong1987}, where two photons arriving from opposite directions (opposite input ports) always end up traveling in the same direction, is not perfectly reproduced.

Driving a three-level atom can give rise to electromagnetically induced transparency (EIT), affecting photon transport~\cite{Witthaut2010, Roy2011, Zheng2011, Martens2013, Roy2014, Sanchez-Burillo2016a}, as discussed further in~\secref{sec:EITandATS}, or to single-photon frequency conversion~\cite{Wang2014, Sanchez-Burillo2016}. A single photon can also be used to switch the state of a three-level atom, thereby determining whether a subsequently arriving photon will be reflected or transmitted, i.e., realizing a \textit{single-photon transistor}~\cite{Chang2007, Witthaut2010}. Actually, the transmission of a single photon via a multi-level atom can both enhance (photon-induced tunneling) and decrease (photon blockade) the transmission of a following photon~\cite{Zheng2011, Zheng2012}. Driving one transition in a three-level atom can regulate single-photon transport, implementing a \textit{single-photon router}~\cite{Hoi2011, Lu2014}.

If two photons scatter off separate transitions in a $\Xi$-type three-level atom, one of them will get a $\pi$ phase shift if they are both reflected~\cite{Kolchin2011}. Similarly, the arrival of a single photon resonant with the first transition can give rise to a large phase shift for a coherent probe resonant with the second transition~\cite{Hoi2013a, Fan2013}, but this is not enough to enable detection of the single photon (measuring the shift of the probe)~\cite{Fan2013}.

A single atom could couple to the waveguide at several points, forming a ``giant atom'' ~\cite{Kockum2014}. Interference between these coupling points give rise to a frequency-dependent relaxation rate, which can markedly alter system properties~\cite{Kleppner1981, Lewenstein1988, Wilhelm2004}. If the travel time between the coupling points becomes comparable to or large than the timescale set by the coupling strength, new effects occur, such as polynomial decay, additional single-photon reflection peaks beside the one at the atomic transition frequency, and switching of $g^{(2)}(\tau)$ between anti-bunching and bunching~\cite{Guo2016}.

Other work with a single atom in an open waveguide include using the atom as a single-photon source~\cite{Zheng2011, Lindkvist2014}, considering the effect of transversal modes (quasi-1D)~\cite{Huang2013PRA, Li2014}, looking at full counting statistics for scattered coherent light~\cite{Pletyukhov2015}, and considering entanglement generated between photons scattered off the atom~\cite{Nysteen2015, Nysteen2015a, Pletyukhov2015}. Recently, methods have been developed for investigating \textit{non-Markovian} regimes with time delays and feedback~\cite{Zhang2013, Fang2014, Pichler2015, Grimsmo2015}. There are also some general results for an arbitrary quantum system coupled to the waveguide~\cite{Koshino2007, Xu2013a, Pichler2015, Grimsmo2015}.

\paragraph{Two atoms}

Introducing a second atom into the open waveguide allows for atom-atom interaction and more complicated photon scattering processes. Just as for a single atom, most studies consider two-level atoms~\cite{Stephen1964, Lehmberg1970a, Milonni1974, Ficek1990, Ordonez2004, LeKien2005, Rist2008, Zhou2008a, Dzsotjan2010, Kim2010, Chen2010, Chen2011, Gonzalez-Tudela2011, Rephaeli2011, Martin-Cano2011, Cheng2012, Huidobro2012, Zheng2013a, Huang2013a, Gonzalez-Ballestero2013, Lalumiere2013, Zang2013, Derouault2014, Gonzalez-Ballestero2014, Fang2014, Fratini2014, Redchenko2014, Laakso2014, Pichler2015, Facchi2016}, but there are some results for atoms with three~\cite{Fang2015a, Douglas2015} and more levels~\cite{Lalumiere2013, Shahmoon2014}. Single-photon~\cite{Zhou2008a, Kim2010, Chen2010, Chen2011, Rephaeli2011, Cheng2012, Huang2013a, Zang2013, Derouault2014, Gonzalez-Ballestero2014, Fang2014} and two-photon~\cite{Rephaeli2011, Huang2013a, Gonzalez-Ballestero2014, Laakso2014} transport, as well as scattering of coherent light~\cite{Ficek1990, Rist2008, Zheng2013a, Lalumiere2013, Fang2014}, has been studied, often by calculating $g^{(2)}(\tau)$~\cite{Rist2008, Zheng2013a, Huang2013a, Fang2014, Laakso2014, Fang2015a}, but there is also much interest in how the interaction of the two atoms affects their decay~\cite{Stephen1964, Lehmberg1970a, Milonni1974, Ordonez2004, LeKien2005, Dzsotjan2010, Huidobro2012, Lalumiere2013, Redchenko2014, Derouault2014, Facchi2016} and Lamb shifts~\cite{Stephen1964, Ordonez2004, Dzsotjan2010}.

\begin{figure}
\centering
\includegraphics[width=1\columnwidth]{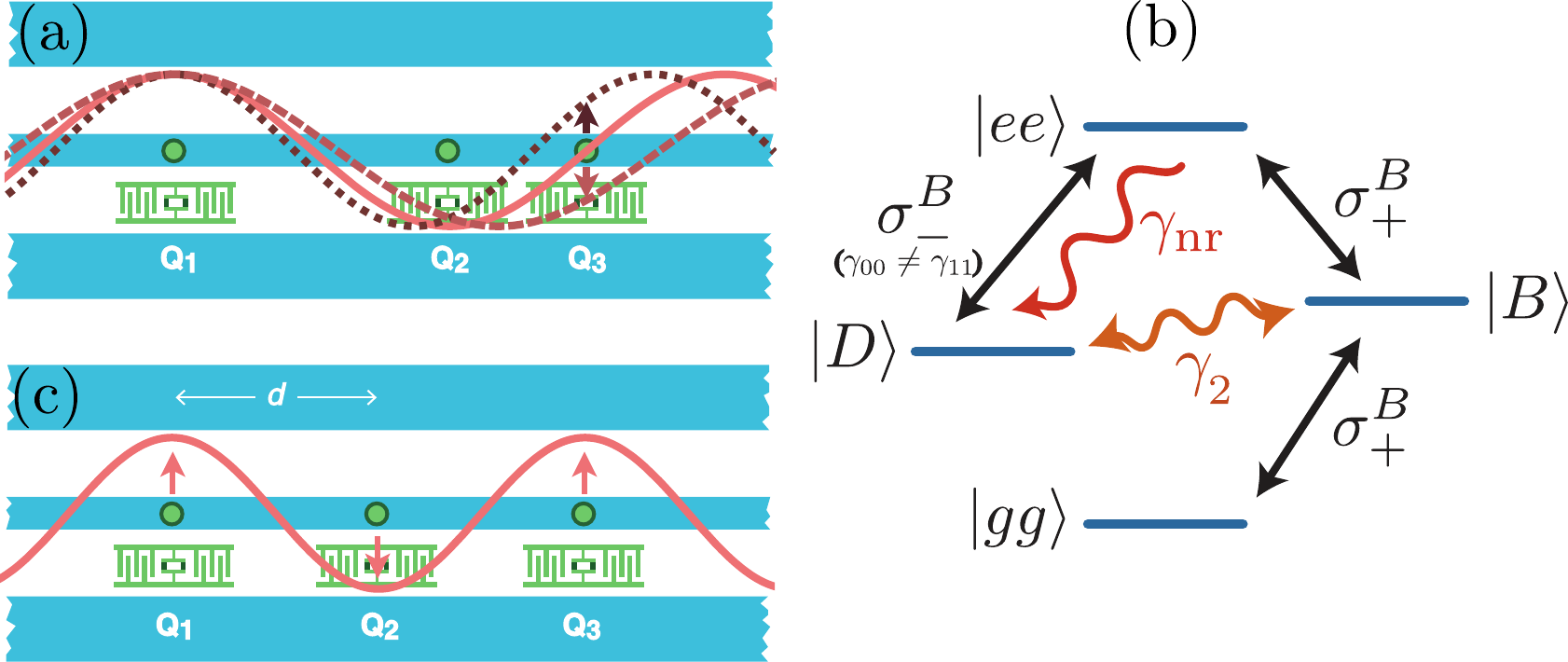}
\caption{Interference effects for two atoms in an open waveguide~\cite{Lalumiere2013}. (a) The identical artificial atoms $Q_1$ and $Q_2$ are separated by a distance $d=\lambda_0/2$, where $\lambda_0$ is the wavelength of a photonic drive (solid line) on resonance with the atoms. The drive is only able to excite antisymmetric superpositions of the atoms. Conversely, only the antisymmetric superposition $\ket{B} = \frac{\ket{ge} - \ket{eg}}{\sqrt{2}}$ (the bright state) can decay into the waveguide, while the symmetric superposition $\ket{D} = \frac{\ket{ge} + \ket{eg}}{\sqrt{2}}$ (the dark state) is protected from decay. The decay rate of $\ket{B}$ is $2\Gamma$, which is termed superradiance since it exceeds the decay rate $\Gamma$ for an excitation in a single atom. The decay rate of $\ket{D}$ is $0$, which is referred to as subradiance. If the two atoms are instead $Q_1$ and $Q_3$, separated by a distance $d=\lambda_0$, the bright and dark states become $\ket{B} = \frac{\ket{ge} + \ket{eg}}{\sqrt{2}}$ and $\ket{D} = \frac{\ket{ge} - \ket{eg}}{\sqrt{2}}$, respectively. (b) Level structure for two atoms with bright and dark states. A dressed raising operator $\sigma_+^B$ can be defined by diagonalizing the dissipation term in \eqref{eq:MasterEqNAtomsOpenWaveguide}. The dark state $\ket{D}$ can be populated if the atoms are not identical, if there is decay to other reservoirs except the waveguide, or if there is dephasing $\gamma_2$. (c) The exchange interaction is found by integrating over all the continuous modes in the waveguide except at the atomic transition frequency. The expression being integrated is proportional to $1/\delta$, where $\delta$ is the detuning between the mode frequency and the atoms. In the figure, $Q_1$ and $Q_3$ are separated by $3\lambda_0/4$. The two modes shown as dashed lines will then have different signs at $Q_3$, but their detunings will also have different signs. These double sign differences cancel, making all modes contribute constructively to the exchange interaction, which reaches its maximum here. On the other hand, $Q_1$ and $Q_2$ are separated by $\lambda_0/2$, such that the two modes have the same sign here. However, their detunings still differ in sign and their contributions cancel, making the exchange interaction zero for this interatomic separation. 
Reprinted figure with permission from K.~Lalumi{\`{e}}re et al.,~\href{http://dx.doi.org/10.1103/PhysRevA.88.043806}{Phys.~Rev.~A~\textbf{88}, 043806 (2013)}. \textcircled{c} 2013 American Physical Society.
\label{fig:TwoAtomsInterference}}
\end{figure}

Early work on two atoms in 3D free space include Refs.~\cite{Stephen1964, Lehmberg1970a, Milonni1974, Ficek1990} (a review on two-atom systems can be found in Ref.~\cite{Ficek2002}), where relaxation, Lamb shifts and exchange interactions for the atoms were investigated. For $N$ identical atoms, placed at positions $x_n$ in an open waveguide, one can derive the Markovian master equation (assuming negligible travel time between the atoms)~\cite{Lehmberg1970, Pichler2015a}
\bea
\dot{\rho} = -i \comm{H_{\rm atoms} + \frac{\Gamma}{2} \sum_{n,m}^N \sin \left( k \abs{x_n - x_m} \right) \sigma_+^{(n)} \sigma_-^{(m)}}{\rho} \nn\\
+ \Gamma \sum_{n,m}^N \cos \left( k \abs{x_n - x_m} \right) \left( \sigma_-^{(n)} \rho \sigma_+^{(m)} - \frac{1}{2} \left\{\sigma_+^{(m)} \sigma_-^{(n)}, \rho \right\} \right), 
\label{eq:MasterEqNAtomsOpenWaveguide}
\eea
where $\Gamma = \Gamma_\text{L} + \Gamma_\text{R}$ is the total relaxation rate for a single atom coupled to the waveguide (we neglect relaxation to other environments, $\Gamma_\text{nr}$), $H_{\rm atoms}$ is the Hamiltonian for the individual atoms (with Lamb shift included), $\sigma_+^{(n)}$ ($\sigma_-^{(n)}$) is the raising (lowering) operator for atom $n$, and $k$ is the wavenumber for photons resonant with the atoms. The second term in the commutator is an exchange interaction between the atoms, mediated by virtual photons in the waveguide. The last term is the decay of the atoms, which contains collective parts. These effects, including their dependence on the interatomic distance and their connection to dark and bright states, are explained further in \figref{fig:TwoAtomsInterference}. Non-Markovian effects, which arise when the travel time between the atoms is no longer negligible and require modifications of \eqref{eq:MasterEqNAtomsOpenWaveguide}, have become an increasingly popular topic in the last few years~\cite{Milonni1974, Rist2008, Zheng2013a, Gonzalez-Ballestero2013, Fang2014, Redchenko2014, Laakso2014, Pichler2015}. Some studies have also considered lossy waveguides~\cite{Dzsotjan2010, Chen2011, Gonzalez-Tudela2011, Paulisch2015}.

The distance-dependent interaction between the two atoms makes it relevant to study how they can be entangled by external drives~\cite{Chen2011, Gonzalez-Tudela2011, Martin-Cano2011, Zheng2013a, Gonzalez-Ballestero2013, Gonzalez-Ballestero2014} or by relaxation~\cite{Facchi2016}; it can also be used for a phase gate between the atoms~\cite{Dzsotjan2010} or photon blockade~\cite{Huang2013a}. Other photonic transport properties studied include EIT~\cite{Cheng2012, Douglas2015}, special (anti)bunching features~\cite{Fang2014}, and the possibility of creating molecular-like potentials for photons~\cite{Douglas2015}. Also, two atoms coupled to a transmission line may experience a large Casimir force~\cite{Shahmoon2014}. Finally, since a single atom can perfectly reflect a single photon on resonance, two atoms can form a sort of \textit{effective cavity} in the waveguide~\cite{Rist2008, Zhou2008a, Zang2013, Fratini2014}, which could be used for photon rectification, i.e., as a \textit{diode}~\cite{Fratini2014}.

\paragraph{Many atoms}

Increasing the number of atoms in the waveguide above two, the result is mostly various generalizations of phenomena already seen in some form with two atoms, though some important novelties arise. As before, most studies have focussed on two-level atoms~\cite{Lehmberg1970, Friedberg1973, Manassah1983, Yudson2008, Tsoi2008, Rakhmanov2008, Stannigel2012, Chang2012, Shvetsov2013, Roy2013, Gonzalez-Tudela2013, Fang2014, Ramos2014, Zoubi2014, Asai2015, Caneva2015, Fang2015, Pichler2015a, Sanchez-Burillo2015, Shi2015, Paulisch2015, Calajo2015, Li2015, Haakh2015, Asenjo-Garcia2016, Brod2016, Brod2016a, EkinKocabas2016, Guimond2016a, Liao2016, Mirza2016, Mirza2016a, Munro2016, Ruostekoski2016}, but there are results for systems with three~\cite{Leung2012, Fan2013, Sathyamoorthy2014, Shi2015, Li2015, Albrecht2016, Gonzalez-Tudela2016} and more levels~\cite{Chang2008, Hafezi2012, Gonzalez-Tudela2013, Lalumiere2013}. Transport of one~\cite{Tsoi2008, Roy2013, Fang2014, Asenjo-Garcia2016, Liao2016, Brod2016a}, two~\cite{Hafezi2012, Roy2013, Fang2015, Albrecht2016, Brod2016a}, and more photons~\cite{Yudson2008, Hafezi2012, Sanchez-Burillo2015, Shi2015}, as well as coherent light~\cite{Fang2014, Munro2016, Ruostekoski2016}, has been studied, usually including calculations of $g^{(2)}(\tau)$~\cite{Hafezi2012, Roy2013, Fang2014, Fang2015, Munro2016}, but makes up a smaller part of the total body of work than for one and two atoms. While most studies of two or more atoms consider the case where the atoms are identical, Ref.~\cite{Liao2016} treats the case of nonidentical atoms.

Just as for two atoms, early work with many atoms in 3D free space~\cite{Lehmberg1970, Friedberg1973, Manassah1983}, as well as some later 1D work~\cite{Gonzalez-Tudela2013, Haakh2015}, was mainly concerned with relaxation, Lamb shifts and exchange interaction for the atoms along the lines of \eqref{eq:MasterEqNAtomsOpenWaveguide}. Later, the spontaneous emission of a single excited atom among many others~\cite{Tsoi2008}, and non-Markovian effects~\cite{Fang2015, Shi2015, Guimond2016a}, have also been studied. Topics related to the interaction between the atoms are decoherence-free subspaces~\cite{Paulisch2015}, multiphoton generation~\cite{Gonzalez-Tudela2016}, as well as the use of photonic drives to generate entanglement~\cite{Stannigel2012, Gonzalez-Tudela2013, Mirza2016, Mirza2016a} and dark states~\cite{Stannigel2012, Ramos2014, Pichler2015a}.

With many atoms, the effective cavity formed by two atoms mentioned in the previous section can be extended to contain a single atom, either by using just two surrounding atoms as mirrors~\cite{Li2015}, or by using two surrounding ensembles of atoms to form Bragg mirrors~\cite{Chang2012, Guimond2016a}. The Rabi-oscillation period in such a Bragg-mirror cavity is set by the cavity length, the number of atoms in the Bragg mirrors, and the strength with which these couple to the waveguide~\cite{Guimond2016a}. The setup with single-atom mirrors could potentially be used for (destructive) single-photon detection, by increasing the probability of the intra-cavity $\Lambda$ atom to absorb an incoming photon~\cite{Li2015}. Nondestructive photon detection can also be achieved by accumulating a cross-Kerr phase shift for a coherent probe beam passing several $\Xi$ atoms together with the single photon~\cite{Sathyamoorthy2014}, but circulators are needed to make the propagation unidirectional; in an open transmission line reflections will prevent a large enough phase shift from accumulating~\cite{Fan2013} (see~\secref{sec:PhotonDetectionItinerant} for details).

Other applications with many atoms in the open waveguide include quantum memory~\cite{Leung2012}, a single-photon switch for single-photon transport~\cite{Albrecht2016}, nonreciprocal photon transport~\cite{Roy2013}, a CPHASE gate between photons propagating in two transmission lines interacting via pairs of coupled atoms~\cite{Brod2016}, and the formation of a quantum metamaterial~\cite{Rakhmanov2008, Shvetsov2013, Asai2015} (see \secref{sec:Metamaterials}).

\paragraph{Cavities with atoms}

A cavity with atoms can either be side-coupled~\cite{Bermel2006, Shen2009, Shen2009a, Rephaeli2010, Liao2010, Shi2011, Hughes2012, Yan2012, Shi2013, Xu2016a} to the waveguide, as in \figref{fig:WaveguideQEDSetups}(e), or directly coupled~\cite{Auffeves-Garnier2007, Zhou2008, Shen2009, Shi2011}, i.e., interrupting the waveguide. Most studies of such systems consider the simplest setup with a single atom in the cavity~\cite{Bermel2006, Auffeves-Garnier2007, Zhou2008, Shen2009, Shen2009a, Rephaeli2010, Shi2011, Hughes2012, Yan2012, Shi2013, Xu2016a}, but two atoms in two separate cavities have also been investigated~\cite{Zeeb2015}, as well as a nonlinear cavity without any atoms~\cite{Liao2010} and optomechanical cavities~\cite{Chang2011}. The atoms usually have two levels~\cite{Auffeves-Garnier2007, Zhou2008, Shen2009, Shen2009a, Rephaeli2010, Shi2011, Hughes2012, Shi2013, Zeeb2015}, but three~\cite{Yan2012} and more~\cite{Bermel2006, Xu2016a} levels have also been considered.

\begin{figure}
\centering
\includegraphics[width=0.5\linewidth]{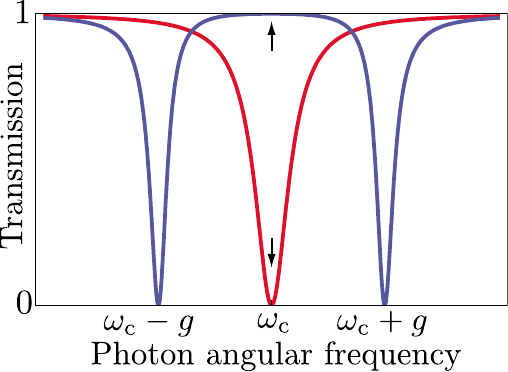}
\caption{Single-photon transmission in a waveguide with a side-coupled cavity containing a single two-level atom~\cite{Shen2009}. The atom can either be far detuned from the cavity frequency $\omega_{\rm c}$ (red curve) or on resonance with the cavity (blue curve). In the first case, many photons can be reflected at $\omega_{\rm c}$. In the second case, single photons will be reflected at $\omega_{\rm c} \pm g$ ($g$ is the coupling strength in the Jaynes--Cummings Hamiltonian), which correspond to the transition frequencies for the dressed states formed by the cavity and the atom. As indicated by the arrows in the figure, the cavity can be switched from full transmission to full reflection by tuning the atom transition frequency. 
Reprinted figure with permission from 
J.-T.~Shen et al., \href{http://dx.doi.org/10.1103/PhysRevA.79.023837}{Phys.~Rev.~A \textbf{79}, 23837 (2009)}. \textcircled{c} 2009 American Physical Society. 	
\label{fig:TransmissionCavityAtom}}
\end{figure}

Most of these works study single-photon~\cite{Bermel2006, Zhou2008, Shen2009, Shen2009a, Rephaeli2010, Yan2012, Shi2013, Xu2016a}, two-photon~\cite{Liao2010, Shi2011, Yan2012, Shi2013, Xu2016a}, or coherent-light transport~\cite{Auffeves-Garnier2007, Hughes2012}, sometimes calculating $g^{(2)}(\tau)$~\cite{Shi2011, Shi2013}. Note that, unlike the atoms in the open waveguide, the cavity can absorb several photons. However, the cavity+atom system can form dressed states (see~\secref{sec:JaynesCummings}) that give rise to several transmission/reflection peaks and nonlinear behavior, as shown in \figref{fig:TransmissionCavityAtom}. Other topics studied include the optimal wavepacket shape of an incoming photon for excitation of the atom in the cavity~\cite{Rephaeli2010}.

\subsubsection{Semi-infinite waveguides}

Terminating the waveguide by a mirror as in \figref{fig:WaveguideQEDSetups}(c), sometimes referred to as a ``half-cavity'', introduces a number of new features compared to the open-waveguide case. Photons can now bounce off the mirror and return to atoms they interacted with previously, giving rise to \textit{interference effects}. Placing the atom(s) at the mirror is equivalent to having an open, unidirectional waveguide, and can therefore be used for some setups discussed in~\secref{sec:ChiralWaveguides} below.

Even without any atoms, the semi-infinite waveguide hosts interesting physics in the form of the dynamical Casimir effect if the mirror can be moved rapidly, which can be realized for superconducting circuits~\cite{Johansson2009, Johansson2010, Johansson2013} (see~\secref{sec:Resonators}; other relativistic effects could potentially also be modeled with SQUIDs in waveguides~\cite{Nation2012}). Recently, a setup with two semi-infinite waveguides connected through a parametric coupling has also been studied~\cite{Wang2016a}. However, most of the work in the field concerns a single atom placed at some distance from the mirror. Usually, the atom is a two-level system~\cite{Hinds1991, Dorner2002, Kojima2004, Koshino2006, Dong2009, Glaetzle2010, Chen2011a, Koshino2012, Witthaut2012, Valente2012, Wang2012, Peropadre2013NJP, Bradford2013, Tufarelli2013, Kockum2014, Lindkvist2014, Tufarelli2014, Pichler2015, Shi2015, Song2016}, but atoms with three~\cite{Beige2002, Peropadre2011, Valente2012a, Witthaut2012, Fang2015a} and more levels~\cite{Meschede1990, Hinds1991, Zheng2013, Koshino2016a} have also been studied.

\begin{figure}
\centering
\includegraphics[width=0.85\linewidth]{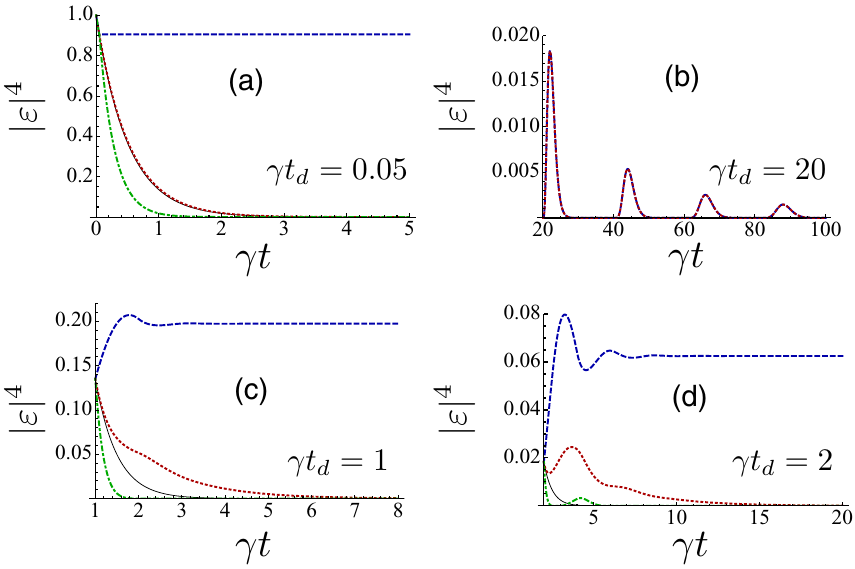}
\caption{Excitation amplitude $\varepsilon(t)$ as a function of time for an atom in front of a mirror in (a) the Markovian regime, (b) the strongly non-Markovian regime and (c), (d) the crossover between Markovian and non-Markovian regimes ($\gamma$ is the atomic linewidth and $t_d$ the round-trip travel time)~\cite{Tufarelli2014}. All plots show the cases $\phi = 0$ (blue dashed line), $\phi = \pi/2$ (red dotted line), $\phi = \pi$ (green dot-dashed line), and, for comparison, the result for an open waveguide (black solid line). In (b)-(d), the plots start from $t_d$; at previous times the behavior is identical to that in (a). Note the $\phi$ dependence on the relaxation rate in (a) and the revivals of the atomic population in (b). 
Reprinted figure with permission from 
T.~Tufarelli et al., \href{http://dx.doi.org/10.1103/PhysRevA.90.012113}{Phys.~Rev.~A \textbf{90}, 12113 (2014)}. \textcircled{c} 2014 American Physical Society. 	
\label{fig:AtomMirrorNonMarkov}}
\end{figure}

The first studies of an atom in front of a mirror were done for the 3D~\cite{Meschede1990, Hinds1991} and quasi-1D~\cite{Dorner2002, Beige2002} cases (recently the quasi-1D setup with a rectangular waveguide has also been investigated~\cite{Song2016}). Here, and also in several later 1D studies, the focus was on the modification of the atom relaxation rate ~\cite{Meschede1990, Dorner2002, Beige2002, Glaetzle2010, Koshino2012, Wang2012, Bradford2013, Tufarelli2013, Kockum2014, Tufarelli2014, Pichler2015} and Lamb shift~\cite{Meschede1990, Hinds1991, Dorner2002, Koshino2012, Kockum2014, Pichler2015} due to the presence of the mirror. In fact, the situation is rather similar to the case of two atoms in an open waveguide discussed above, since the atom can be thought of as interacting with its mirror image. The result is a relaxation rate $\Gamma(1+\cos \phi)$ and a shift of the transition frequency by $\frac{\Gamma}{2}\sin\phi$, where $\phi$ is the phase acquired by a photon traveling from the atom to the mirror and back. If the round-trip travel time is not negligible, the system becomes non-Markovian~\cite{Dorner2002, Glaetzle2010, Bradford2013, Tufarelli2013, Tufarelli2014, Pichler2015} and the atom can exhibit revivals. A comparison between the open and semi-infinite waveguide cases, including both Markovian and non-Markovian regimes, is shown in \figref{fig:AtomMirrorNonMarkov}. The picture can be complicated further by adding mirror oscillations, which give rise to sideband photons~\cite{Glaetzle2010}.

The relaxation rate of the atom can be \textit{tuned}, either by changing the atom transition frequency or by moving the mirror. This could be used to create a quantum memory, using a high, possibly time-dependent, coupling to absorb an incoming photon and then trap it by turning off the relaxation~\cite{Wang2012}. Optimization of photon absorption has been studied further~\cite{Chen2011a, Peropadre2011} (it could be used for photon detection if the atom has a $\Lambda$ structure~\cite{Peropadre2011, Witthaut2012}), and so has optimization of stimulated emission~\cite{Valente2012}, which could be used for photon cloning~\cite{Valente2012a}. 

Similar to two atoms in an open waveguide, a single atom in front of a mirror can form an effective cavity for a single photon~\cite{Dong2009}. The atom can also be used for single-photon generation~\cite{Lindkvist2014}, to entangle photons~\cite{Shi2015} or to squeeze them~\cite{Koshino2006}. Four-level atoms can be used to implement a photonic phase gate~\cite{Zheng2013} or various two-qubit gates between an atom and a photon~\cite{Koshino2016a}.

With two atoms in front of the mirror, a photon can perform a controlled-Z gate between them~\cite{Ciccarello2012}. If the two atoms are placed in front of separate mirrors, they can be used to sort photon numbers~\cite{Witthaut2012}. Recently, methods have also been develop to study many atoms in front of a mirror, which allow for, e.g., $g^{(2)}(\tau)$ calculations and exploration of the non-Markovian regime~\cite{Fang2015, Shi2015}. There is also an example where a cavity with an atom inside is placed in front of the mirror~\cite{Bradford2013}.

\subsubsection{Chiral waveguides}
\label{sec:ChiralWaveguides}

Chiral waveguides, where, as shown in \figref{fig:WaveguideQEDSetups}(d), the relaxation rates to left- and right-moving photons differ, have only started to attract attention in the last decade. There are now a few studies on single~\cite{Koshino2009, Roy2010, Fan2010, Yan2011, Gonzalez-Ballestero2016, Guimond2016}, two~\cite{Zang2013}, and multiple~\cite{Pletyukhov2012, Stannigel2012, Ramos2014, Sathyamoorthy2014, Ringel2014, Pichler2015a, Mirza2016, Mirza2016a, Ramos2016} atoms in a chiral waveguide, considering atoms with two~\cite{Roy2010, Fan2010, Yan2011, Stannigel2012, Zang2013, Ramos2014, Ringel2014, Pichler2015a, Mirza2016, Mirza2016a, Ramos2016}, three~\cite{Koshino2009, Sathyamoorthy2014, Gonzalez-Ballestero2016, Guimond2016}, and more~\cite{Pletyukhov2012} levels. For recent reviews of chiral quantum optics and spin-orbit interactions of light, see Refs.~\cite{Lodahl2016, Bliokh2015a, Bliokh2015b}.

An important feature of the chiral waveguide is that the \textit{asymmetry} in the relaxation rates destroys the interference that gave rise to perfect single-photon reflection with a single atom in an open waveguide. With this in mind, transport of one~\cite{Koshino2009, Roy2010, Fan2010, Yan2011, Zang2013, Ringel2014, Gonzalez-Ballestero2016}, two~\cite{Roy2010, Fan2010, Ringel2014, Gonzalez-Ballestero2016}, and more~\cite{Pletyukhov2012} photons has been studied anew, suggesting applications like a photon diode~\cite{Roy2010, Gonzalez-Ballestero2016}, a single-photon router~\cite{Gonzalez-Ballestero2016}, a single-photon transistor~\cite{Gonzalez-Ballestero2016} and a single-photon-transmission switch~\cite{Zang2013}. In the limit of unidirectional relaxation, which for a single atom is equivalent to placing the atom at the end of a semi-infinite waveguide, the interference that led to single-photon reflection in the open waveguide can be harnessed in a different way if the atom is of the $\Lambda$ type with equal relaxation rates from the upper level. Then, a photon exciting one of the transitions will always relax via the other transition, which can be used for frequency conversion~\cite{Koshino2009}.

Other applications with multiple atoms in a chiral waveguide are the generation of dark and entangled states~\cite{Stannigel2012, Ramos2014, Pichler2015a, Mirza2016, Mirza2016a, Ramos2016} (a dark state can also form with a single $V$-type atom in front of a mirror~\cite{Guimond2016}) and photon detection~\cite{Sathyamoorthy2014}. Recently, non-Markovian effects for multiple atoms have been studied in detail~\cite{Ramos2016}. Some possible implementations of chiral waveguides are superconducting waveguides with circulators~\cite{Stannigel2012, Sathyamoorthy2014} and systems with spin-orbit coupling of evanescent light~\cite{Bliokh2015, Bliokh2015a, Bliokh2015b} (see also~\secref{sec:WaveguideQEDExpOther}).

\subsubsection{Ultrastrong coupling}
\label{sec:WaveguideUSC}

Superconducting artificial atoms can not only couple ultrastrongly to cavities~\cite{Devoret2007, Bourassa2009, Peropadre2010, Bourassa2012}, as discussed in~\secref{sec:ultrastrong}, but also to an open waveguide~\cite{Bourassa2009, LeHur2012, Peropadre2013a}, as recently demonstrated in an experiment~\cite{Forn-Diaz2017}. A few studies has explored this setting for one~\cite{LeHur2012, Peropadre2013a, Goldstein2013, Sanchez-Burillo2014, Diaz-Camacho2015, Snyman2015, Bera2016, Gheeraert2017}, two~\cite{Diaz-Camacho2015}, and many~\cite{Sanchez-Burillo2015} two-level atoms, looking at both single-~\cite{LeHur2012, Peropadre2013a, Goldstein2013, Sanchez-Burillo2014, Diaz-Camacho2015}, multi-photon~\cite{Sanchez-Burillo2015}, and coherent-state~\cite{LeHur2012, Bera2016} transport. One interesting result is that a single incoming photon interacting with a single two-level atom can now be converted to another frequency~\cite{Goldstein2013, Sanchez-Burillo2014}.

In the USC regime with a cavity, the rotating-wave approximation can no longer be used and the ground state contains virtual photons. In the open waveguide case, there are also virtual photons in the ground state, but these are now present in a continuum of modes~\cite{Peropadre2013a}. Furthermore, at strong enough coupling, the Markov approximation breaks down~\cite{Peropadre2013a}. Other new phenomena in this regime include a modified Lamb shift~\cite{Sanchez-Burillo2014, Diaz-Camacho2015}, a decrease in the spontaneous emission rate as the coupling increases~\cite{Diaz-Camacho2015}, spontaneous emission in the form of Schr\"{o}dinger-cat states~\cite{Gheeraert2017}, and a modified distance dependence for waveguide-mediated atom-atom interaction~\cite{Sanchez-Burillo2015}. There is also a close connection to Kondo physics~\cite{LeHur2012, Goldstein2013, Snyman2015}.

\section{Quantum optics and atomic physics on a superconducting chip}
\label{sec:QuantumOptics}

Many phenomena in quantum optics and atomic physics have been demonstrated in the microwave domain using circuit-QED systems. In this section, we review some of these experimental achievements. We first discuss Autler--Townes splitting, the Mollow triplet, and electromagnetically induced transparency in \secref{sec:EITandATS} and then cover topics such as sideband transitions (\secref{sec:Sideband}), various multiphoton processes (\secref{sec:multiphoton-transitions}), lasing (\secref{sec:Lasing}), squeezing (\secref{sec:SqueezedStates}), photon blockade (\secref{sec:PhotonBlockade}), and quantum jumps (\secref{sec:QuantumJumps}). We focus on explaining the basic mechanisms of these phenomena and their experimental realizations in SQCs. The treatment in this section is not exhaustive. As shown in Table~\ref{tab:Experiments} and the table of contents, quantum-optics and atomic-physics experiments with microwaves are discussed in many sections of this review.

\begin{table}
\centering
\renewcommand{\arraystretch}{1.2}
\renewcommand{\tabcolsep}{0.15cm}
\begin{tabular}{|l|l|}
\hline
\textbf{Phenomena} & \textbf{References}\\
\hline
Autler--Townes splitting & Secs.~\ref{sec:WaveguideQEDExpSCAtoms} and \ref{sec:EITandATS}\\
\hline
Bell, GHZ and W states & \secref{sec:DiVincenzoCriteria} \\
\hline
Correlated-emission lasing & \secref{sec:ThreeLevel}\\
\hline
Dressed states & \secref{sec:DressedStates} \\
\hline
Electromagnetically induced transparency & \secref{sec:EITandATS} \\
\hline
Even and odd coherent states & \secref{sec:OneCavity} \\
\hline
Landau--Zener transitions & \secref{sec:multiphoton-transitions} \\
\hline
Lasing with and without population inversion & \secref{sec:Lasing} \\
\hline
Microwave entangled states & \secref{sec:TwoOrMoreCavities} \\
\hline
Microwave quantum state engineering & Secs.~\ref{sec:NoCavity}, \ref{sec:OneCavity}, \ref{sec:PhotonDetectionQNDandCavity}, and \ref{sec:PhotonDetectionCorrelation} \\
\hline
Nonlinear optics & \secref{sec:NonlinearProcess} \\
\hline
Photon blockade & \secref{sec:PhotonBlockade} \\
\hline
Photon generation & \secref{sec:PhotonGenerationSCCircuits} \\
\hline
Quantum contextuality & Refs.~\cite{Wei2010, Jerger2016} \\
\hline
Quantum jumps & \secref{sec:QuantumJumps} \\
\hline
Quantum nondemolition (QND) measurements & Secs.~\ref{sec:Readout} and \secref{sec:PhotonDetectionQNDandCavity} \\
\hline
Quantum Zeno effect & Refs.~\cite{Gambetta2008, Wang2008a, Zhou2009, Helmer2009, Lizuain2010, Cao2010, Sabin2011, Feng2011, Cao2012, Chantasri2013, Li2014a, Mirrahimi2016, Cohen2016} \\
\hline
Rapid adiabatic passage and STIRAP & Secs.~\ref{sec:PhotonGenerationQuantumOpticsApproaches} and \ref{sec:OneCavity} \\
\hline
Reconstruction of microwave quantum states & Secs.~\ref{sec:SqueezedStates}, \ref{sec:PhotonDetectionHomodyne}, and \ref{sec:PhotonDetectionCorrelation}, Ref.~\cite{Shalibo2013} \\
\hline 
Resonance fluorescence & \secref{sec:multiphoton-transitions} \\
\hline
Sideband transitions & \secref{sec:Sideband} \\
\hline
Squeezing and squeezed states & \secref{sec:SqueezedStates} \\
\hline
Ultrastrong light-matter coupling & \secref{sec:ultrastrong} \\
\hline
\end{tabular}
\caption{A summary of phenomena from atomic physics and quantum optics that have been experimentally realized in circuit-QED systems. Most of these experiments are reviewed in the present section, but, as shown in the table, others are discussed in other parts of the review.
\label{tab:Experiments}}
\end{table}

\subsection{Mollow triplet, Autler--Townes splitting, and electromagnetically induced transparency}
\label{sec:EITandATS}

\begin{figure}
\centering
\includegraphics[width=\linewidth]{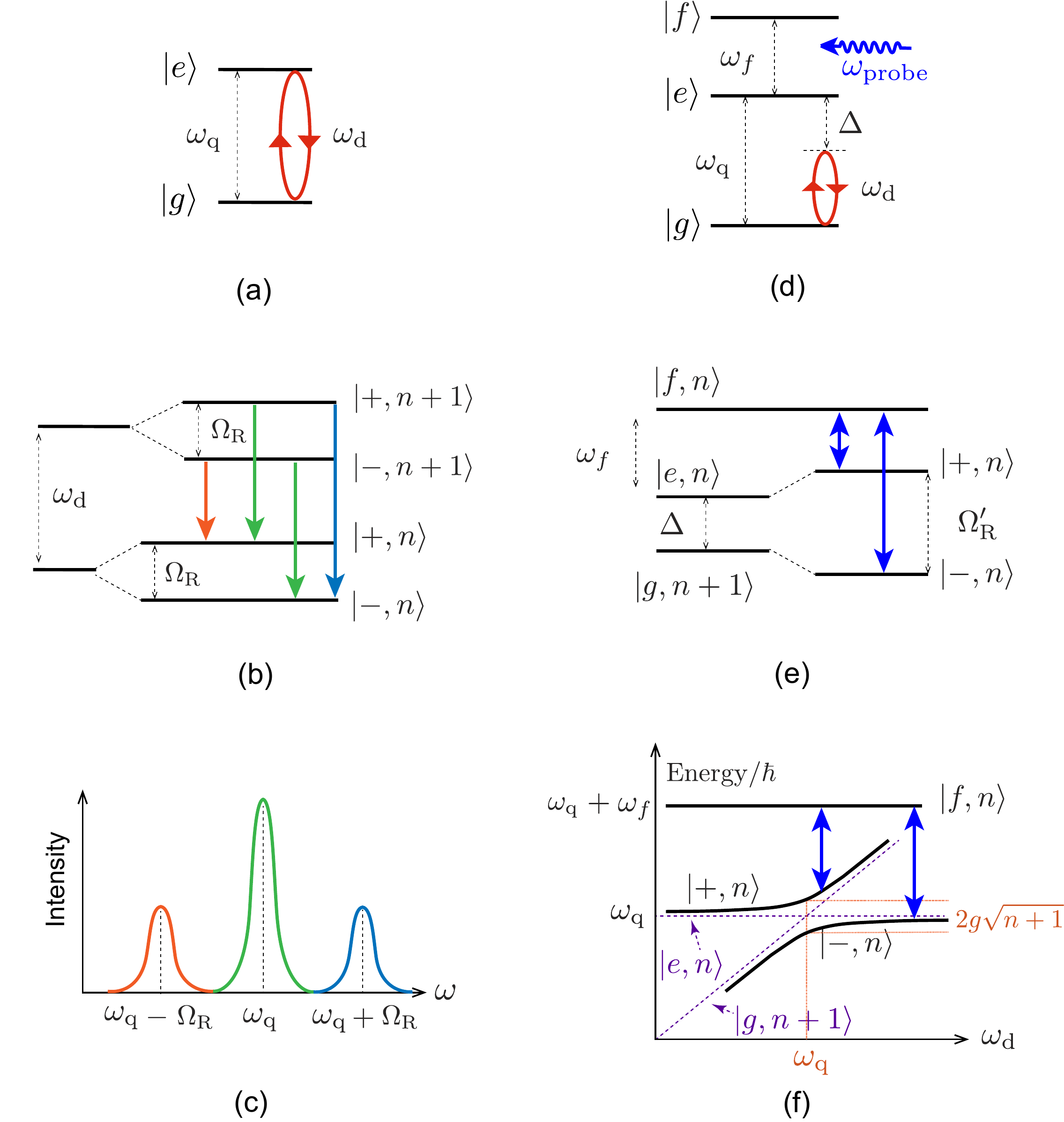}
\caption{Mollow triplet (a, b, c) and Autler--Townes doublet (d, e, f). (a) To create the Mollow triplet, a two-level atom is driven resonantly (red arrows, drive frequency $\omega_{\rm d}$ equal to the transition frequency $\omega_{\rm q}$). (b) The strong drive dresses the states of the atom, creating four possible transitions (at three different transition frequencies) between two adjacent doublets $\ket{\pm,n+1}$ and $\ket{\pm,n}$ in the dressed-states basis. (c) These transitions give rise to the Mollow triplet in the spectrum of the resonance fluorescence from the atom. (d) In one incarnation of the ATS, the transition $\ket{e} \leftrightarrow \ket{g}$ in a three-level atom is driven by a strong field at frequency $\omega_{\rm d}$, with detuning $\Delta = \omega_{\rm q} - \omega_{\rm d}$. A weak field probes the $\ket{e} \leftrightarrow \ket{f}$ transition. (e) In the dressed-states basis, the $\ket{e} \leftrightarrow \ket{f}$ transition splits into two, indicated by blue arrows. The new transitions are separated by $\Omega_{\rm R}' = \sqrt{\Omega_{\rm R}^2 + \Delta^2}$. (f) The energies of the dressed states change with the frequency of the strong driving field $\omega_{\rm d}$. The uncoupled states (dashed lines) cross at $\omega_{\rm d} = \omega_{\rm q}$. An anticrossing is formed by the dressed states $\ket{+,n}$ and $\ket{-,n}$; the minimum splitting is a measure of the coupling strength.
\label{fig:ATS}}
\end{figure}

One of the simplest quantum-optics experiments is to coherently drive a single atom. In the case where the atom only has two levels and the drive is resonant (the drive frequency $\omega_{\rm d}$ equals the atomic transition frequency $\omega_{\rm q}$) we know from~\secref{sec:DressedStates} that a Jaynes--Cummings ladder will form. The pairs of states in the ladder are separated by splittings $\Omega_{\rm R} = 2g \sqrt{n}$, where $g$ is the coupling between the atom and the drive field and $n$ is the number of drive photons. If the drive is strong enough, $\sqrt{n} \approx \sqrt{n+1}$ and the splittings become uniform. As shown in \figref{fig:ATS}(b), this leads to a situation where transitions can occur at three different frequencies: $\omega_{\rm q}$ and $\omega_{\rm q} \pm \Omega_{\rm R}$. If the splitting $\Omega_{\rm R}$ is larger than the atom linewidth, three lines at these frequencies can be seen in the resonance-fluorescence spectrum of the atom, as shown in \figref{fig:ATS}(c). This is the well-known \textit{Mollow triplet}~\cite{Mollow1969}.

In superconducting circuits, the Mollow triplet was first observed using a transmon qubit in a resonator~\cite{Baur2009}. Shortly thereafter, the Mollow triplet was also demonstrated in the resonance fluorescence from a flux qubit in an open 1D transmission line~\cite{Astafiev2010}. In a subsequent work~\cite{Abdumalikov2011}, the same setup was used to investigate the dynamics of both coherent and incoherent emission in resonance fluorescence. Other SQC experiments studying resonance fluorescence have demonstrated nonclassical effects like photon antibunching~\cite{Lang2011} and quantum features characteristic of weak values in the interference between past and future quantum states~\cite{Campagne-Ibarcq2014}. Recently, resonance fluorescence in a squeezed vacuum was observed using a transmon qubit~\cite{Toyli2016}.

Closely related to the Mollow triplet is \textit{Autler--Townes splitting} (ATS)~\cite{Autler1955, Cohen1996autler}, also known as the ac Stark effect. In both cases, an atomic transition is strongly driven. However, the ATS refers to the splitting of a different atomic transition in a multilevel atom due to this driving, as shown in Figs.~\ref{fig:ATS}(d)-(f). In that illustration, the drive is on the $\ket{g} \leftrightarrow \ket{e}$ transition in a three-level atom, which leads to a probe field of frequency $\omega_{\rm p}$ seeing a split of the $\ket{e} \leftrightarrow \ket{f}$ transition, whose resonance frequency changes from $\omega_f$ to $\omega_f \pm \Omega_{\rm R}'/2$. Since a weak probe resonant with an atomic transition is strongly absorbed by the atom, the ATS can turn off such absorption by changing the resonance frequency of the transition.

The reduced absorption of a weak probe due to a strong drive in the ATS resembles \textit{electromagnetically induced transparency} (EIT)~\cite{Harris1990, Boller1991, Harris1997, Fleischhauer2005}. Also in EIT, strong absorption of a resonant weak probe field by a three-level atom is switched off by driving the atom. EIT has many potential applications in nonlinear optics, e.g., cross-phase modulation, coherent population transfer, lasing without inversion, slowing light, and light storage~\cite{Harris1997, Marangos1998, Fleischhauer2005}. In conventional EIT, there is only one transparency window for the probe, but recent studies in optomechanical systems have shown that multiple transparency windows can exist~\cite{WangHui2014}.

While both ATS and EIT can give rise to a dip in the absorption spectrum of a weak probe field, the origin of this dip is different in the two cases~\cite{Abi-Salloum2010a, Anisimov2011}, as we will explain in more detail in \secref{sec:ATSvsEIT} below. In the ATS case, the drive is strong enough to shift the transition frequency. In the EIT case, the drive is not very strong and the dip is instead mainly due to \textit{Fano interference}~\cite{Fano1961, Imamoglu1989, Fleischhauer2005}, where destructive interference between different excitation pathways leads to partial or complete reduction of the absorption. The drive strength has a threshold value, determined by the decay rates of the three-level atom, that distinguishes EIT from ATS. To determine whether an experimentally observed absorption dip is due to EIT or ATS, the so-called Akaike information criterion can be applied~\cite{Anisimov2011}. In this way, it has been shown that EIT thought to have been observed using a flux qubit~\cite{Abdumalikov2010} actually was ATS~\cite{Anisimov2011}. Experiments in other systems, such as cold atoms~\cite{Giner2013} and whispering-gallery-mode resonators~\cite{Peng2014}, have demonstrated the transition from EIT to ATS by varying the drive strength.

\begin{figure}
\centering
\includegraphics[width=\linewidth]{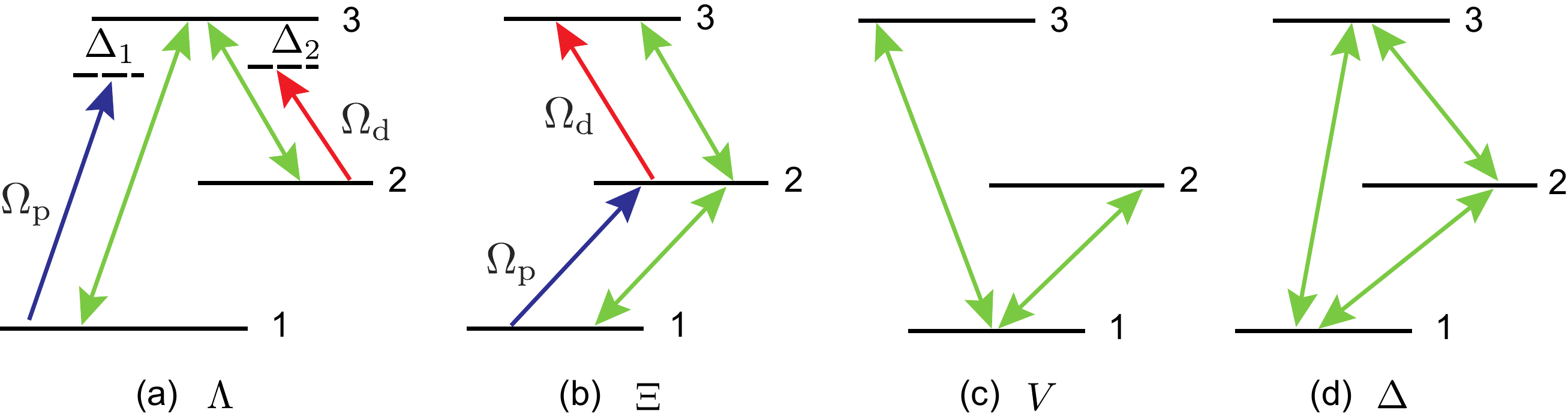}
\caption{Configurations of three-level systems. Green arrows indicate the allowed transitions in each configuration. As shown in \secref{sec:ThreeLevel}, natural three-level atoms can only have the (a) $\Lambda$, (b) $\Xi$, and (c) $V$ configurations, but artificial atoms can be designed also in the (d) $\Delta$ configuration. (a) The setup analyzed in more detail in \secref{sec:ATSvsEIT}. The $\Lambda$ atom interacts with a probe field (red arrow) of strength (Rabi frequency) $\Omega_{\rm p}$ and detuning $\Delta_1$ from the $\ket{1} \leftrightarrow \ket{3}$ transition and a drive field (blue arrow) of strength $\Omega_{\rm d}$ and detuning $\Delta_2$ from the $\ket{2} \leftrightarrow \ket{3}$ transition. (b) Driving the upper transition of a $\Xi$ atom is the only configuration in natural atoms, in addition to the $\Lambda$ one, that allows perfect cancellation of the probe absorption through Fano interference.
\label{fig:EIT_ThreeLevel}}
\end{figure}

The configuration of the three-level atom interacting with the probe and drive fields also affects whether ATS and EIT can occur~\cite{Abi-Salloum2010a, Lee2000}. Although the four different types of three-level atoms ($\Lambda$, $\Xi$, $V$, and $\Delta$, shown in~\figref{fig:EIT_ThreeLevel}) offer many ways to apply the drive and the probe to different transitions, it has been shown that, for natural atoms, cancellation of absorption solely due to Fano interference is only possible for $\Lambda$ [\figref{fig:EIT_ThreeLevel}(a)] and upper-driven $\Xi$ [\figref{fig:EIT_ThreeLevel}(b)] atoms~\cite{Lee2000}. However, as discussed in \secref{sec:ThreeLevel}, three-level SQCs can be in a $\Delta$ configuration [\figref{fig:EIT_ThreeLevel}(d)]. In that configuration, the response can be engineered by varying the strength of the driving field~\cite{SunHuiChen2014} and the transparency window can be sandwiched between the absorption and amplification bands~\cite{Joo2010}.

\subsubsection{ATS versus EIT: the details}
\label{sec:ATSvsEIT}

To clarify the physics governing ATS and EIT, we analyze a driven $\Lambda$ system in the setup shown in ~\figref{fig:EIT_ThreeLevel}(a). An analysis of other setups can be found in Ref.~\cite{Abi-Salloum2010a}. In the interaction picture, we can write the Hamiltonian of the driven $\Lambda$ system as~\cite{Fleischhauer2005}
\be
H = - \frac{1}{2} \left( \Omega_{\rm p} \ketbra{3}{1} e^{i \Delta_1 t} + \Omega_{\rm d} \ketbra{3}{2} e^{i \Delta_2 t} + \text{H.c.} \right),
\label{eq:HDrivenLambda}
\ee
where the Rabi frequency $\Omega_{\rm p}$ ($\Omega_{\rm d}$) describes the strength of the probe (drive), and $\Delta_1 = \omega_{31} - \omega_{\rm p}$ ($\Delta_2 = \omega_{32} - \omega_{\rm d}$) is the detuning of the probe (drive) from the $\ket{1} \leftrightarrow \ket{3}$ ($\ket{2} \leftrightarrow \ket{3}$) transition. By applying the transformation $U = \ketbra{2}{2} \exp[-i (\Delta_1 - \Delta_2) t] + \ketbra{3}{3} \exp(-i \Delta_1 t)$, the Hamiltonian in \eqref{eq:HDrivenLambda} can be rewritten as
\be
\tilde{H} = -\frac{1}{2}
\begin{bmatrix}
0 & 0 & \Omega_{\rm p} \\
0 & - 2 \delta & \Omega_{\rm d} \\
\Omega_{\rm p} & \Omega_{\rm d} & - 2 \Delta
\end{bmatrix},
\ee
in a rotating reference frame, where we introduced the two-photon detuning $\delta = \Delta_1 - \Delta_2$ and set $\Delta_1 = \Delta$. When $\delta=0$, the eigenvalues of the Hamiltonian $\tilde{H}$ are simplified to
\be
\epsilon_0 = 0, \qquad \epsilon_\pm = \frac{1}{2} \left(\Delta \pm \sqrt{\Delta^2 + \Omega_{\rm p}^2 + \Omega_{\rm d}^2} \right),
\ee
with the corresponding eigenstates
\bea
\ket{\Psi_0} &=& \cos\Theta \ket{1} - \sin\Theta \ket{2}, \\
\ket{\Psi_+} &=& \sin\Theta \sin\Phi \ket{1} + \cos\Phi \ket{3} + \cos\Theta \sin\Phi \ket{2}, \\
\ket{\Psi_-} &=& \sin\Theta \cos\Phi \ket{1} - \sin\Phi \ket{3} + \cos\Theta \cos\Phi \ket{2},
\label{eq:EigenstatesEIT}
\eea
where $\tan\Theta = \Omega_{\rm p} / \Omega_{\rm d}$ and $\tan (2\Phi) = \sqrt{\Omega_{\rm p}^2 + \Omega_{\rm d}^2} / \Delta$. The eigenstate $\ket{\Psi_0}$, which corresponds to the zero eigenvalue, is known as a \textit{dark state}. Once the system is in this dark state, there is no population in the upper state $\ket{3}$. The dark state can be applied for coherent population trapping (CPT) and adiabatic population transfer~\cite{Marangos1998}.

\begin{figure}
\centering
\includegraphics[width=\linewidth]{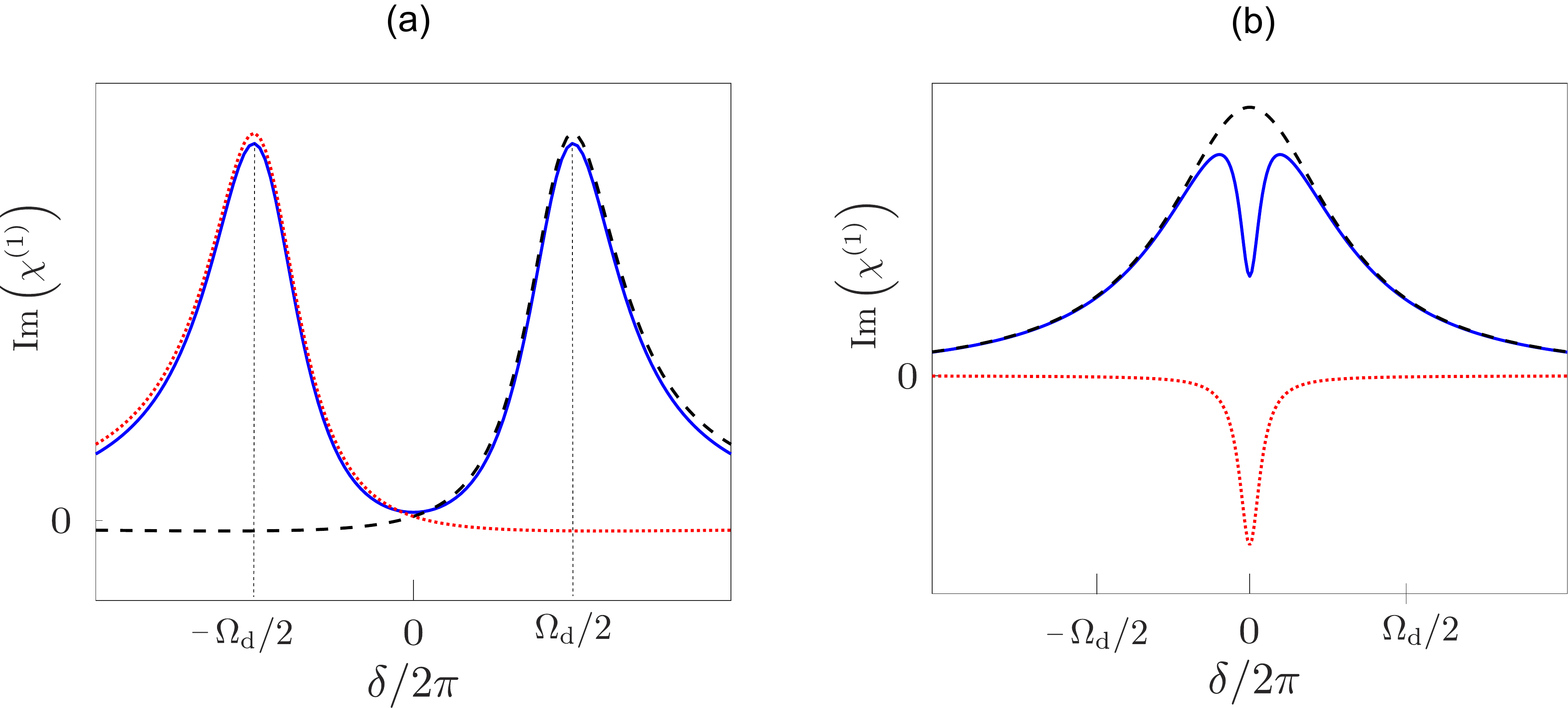}
\caption{Spectra of absorption (blue solid curves), corresponding to the imaginary part of $\chi^{(1)}$, for (a) ATS and (b) EIT. ATS is induced by a strong driving characterized by $\Omega_{\rm d}/[(\gamma_{31} - \gamma_{21})/2] > 1$, while EIT is due to a weak drive, $\Omega_{\rm d}/[(\gamma_{31} - \gamma_{21})/2] < 1$. The ATS spectrum is composed of two positive Lorentzians (dashed black curve and dotted red curve) of equal width, separated by $\Omega_{\rm d}$. The EIT spectrum is composed of a broad positive Lorentzian (dashed black curve) and a narrow negative Lorentzian (dotted red curve), both centered at $\delta = 0$. The dip in the EIT absorption spectrum is due to the interference between these two contributions, while the dip in the ATS absorption spectrum is due to the separation between the two Lorentzians.
\label{fig:Decomposition}}
\end{figure}

The difference between ATS and EIT can be understood by looking at the linear response of the $\Lambda$ system to the weak probe field. This linear response is characterized by the first-order susceptibility $\chi^{(1)}(-\omega_{\rm p},\omega_{\rm p})$. The imaginary part of $\chi^{(1)}$ represents absorption, while the real part determines refraction. The susceptibility can be found by solving the standard master equation (including dissipation) for the density matrix $\rho$ of the $\Lambda$ system. Assuming $\rho_{11} = 0$, one finds that $\chi^{(1)}(-\omega_{\rm p},\omega_{\rm p}) \propto \rho_{31}$, leading to~\cite{Fleischhauer2005}
\be
\chi^{(1)} (-\omega_{\rm p},\omega_{\rm p}) = \frac{ \frac{i\Omega_{\rm p}}{2} \left(\delta - \frac{i\gamma_{21}}{2} \right) }{ \left(\delta + \Delta -\frac{i\gamma_{31}}{2} \right) \left(\delta - \frac{i\gamma _{21}}{2} \right) - \frac{\Omega_{\rm d}^2}{4} },
\ee
where $\gamma_{31} = \Gamma_3 + \gamma_{3\,\text{deph}}$ and $\gamma_{21} = \gamma_{2\,\text{deph}}$, with $\Gamma_3$ the total spontaneous emission rate from the state $\ket{3}$ and $\gamma_{2\,\text{deph}}$ ($\gamma_{3\,\text{deph}}$) the dephasing rate of the state $\ket{2}$ ($\ket{3}$). This equation can be decomposed into two parts~\cite{Agarwal1997, Anisimov2008}:
\be
\chi^{(1)} = \frac{\chi_+}{\delta - \delta_+} + \frac{\chi_-}{\delta - \delta_-}.
\ee
When $\omega_{\rm p} = \omega_{31}$, i.e., $\Delta = 0$, the poles $\delta_{\pm}$, corresponding to two resonances, are given by
\be
\delta _{\pm} = i \frac{\gamma_{31} + \gamma_{21}}{4} \pm \sqrt{\Omega_{\rm d}^2 - \frac{1}{4} (\gamma _{31} - \gamma _{21})^2}.
\ee
It is the sign of the expression under the square root that determines whether EIT or ATS occurs. When $\Omega_{\rm d} > (\gamma_{31} - \gamma_{21})/2$, the imaginary part of $\chi^{(1)}$ is composed of two positive Lorentzians width equal widths and peaks separated by $\Omega_{\rm d}$~\cite{Anisimov2011, Abi-Salloum2010a, Peng2014}, as shown in \figref{fig:Decomposition}(a). This is the ATS. When $\Omega_{\rm d} < (\gamma_{31} - \gamma_{21})/2$, the imaginary part of $\chi^{(1)}$ is composed of a broad positive Lorentzian and a narrow negative Lorentzian, both centered at $\delta = 0$. The interference between these two contributions gives rise to a dip in the absorption spectrum~\cite{Anisimov2011, Abi-Salloum2010a, Peng2014}, as shown in \figref{fig:Decomposition}(b). This is the EIT, occurring for a lower drive strength than the ATS.

\subsubsection{ATS and EIT in superconducting quantum circuits}
\label{sec:ATSandEITinSQCs}

In SQCs, EIT was first theoretically introduced for flux qubits in the $\Lambda$ configuration as a sensitive probe of decoherence~\cite{Murali2004, Dutton2006}. Early on, it was also proposed to use the dark state of the EIT scheme [see discussion below \eqref{eq:EigenstatesEIT}] for quantum-information transfer and entanglement generation in two SQUID-based qubits inside a cavity~\cite{Yang2004}. Later, it has been shown that a three-level system formed by dressing a superconducting qubit with a single-mode resonator field not only can be used for EIT, but also for enhancing the absorption of a probe, i.e., \textit{electromagnetically induced absorption} (ETA)~\cite{Ian2010}. Other studies have shown that a combination of EIT and lasing without inversion (LWI, see \secref{sec:LWI}) can be realized by controlling the relative phase in a $\Delta$ system~\cite{Joo2010} and that EIT can be used for on-demand storage and retrieval of a microwave pulse in a linear array of fluxonium qubits coupled to a transmission line~\cite{Leung2012}. The Autler--Townes effect and dark states in two-tone driving of a three-level superconducting system have also been theoretically studied~\cite{Li2011a}. There is also an EIT proposal based on longitudinal coupling between a qubit and a resonator~\cite{Wang2016d}.

Many experiments have demonstrated ATS using various three-level superconducting artificial atoms~\cite{Sillanpaa2009, Baur2009, Abdumalikov2010, Kelly2010, Hoi2011, Li2012, Suri2013, Novikov2013}.  Moreover, coherent population trapping has been observed in three-level phase-qubit circuits~\cite{Kelly2010} and adiabatic population transfer has been experimentally demonstrated in transmon and phase-qubit circuits~\cite{Xu2016}. As discussed above, the strength $\Omega_{\rm d}$ of the drive, compared to relevant dissipation rates in the system, separates ATS and EIT. For EIT, stringent requirements on these dissipation rates must be satisfied, which is hard to achieve with SQCs~\cite{SunHuiChen2014}. It was therefore only recently that EIT was finally realized in two experiments with a hybrid $\Lambda$-type three-level system formed by a transmon qubit and a single-mode microwave field~\cite{Novikov2015, Liu2016b}. Based on a theoretical proposal~\cite{Gu2016}, EIT has also been experimentally demonstrated by engineering the relevant decay rates of doubly dressed superconducting qubits~\cite{Long2017}.

Other interesting studies include replacing the strong classical drive field with a quantized cavity mode. An early study showed that an empty cavity can induce transparency for the probe field~\cite{Field1993}. This vacuum-induced transparency~\cite{Field1993} was demonstrated in \textit{an ensemble} of atoms, where the control field was enhanced by an optical cavity~\cite{Tanji-Suzuki2011}, and vacuum-induced ATS was recently observed in a \textit{single} superconducing artificial atom~\cite{Peng2017}. Also, it has been shown that a system with EIT gives rise to a group-velocity delay that depends on the number of photons in the driving field interacting with an ensemble of atoms~\cite{Nikoghosyan2010}. The effect of the number of photons in the driving field for both EIT and ATS was recently further studied in the context of superconducting circuits~\cite{Ding2017}.

\subsection{Sideband transitions}
\label{sec:Sideband}

\begin{figure}
\centering
\includegraphics[width=0.85\linewidth]{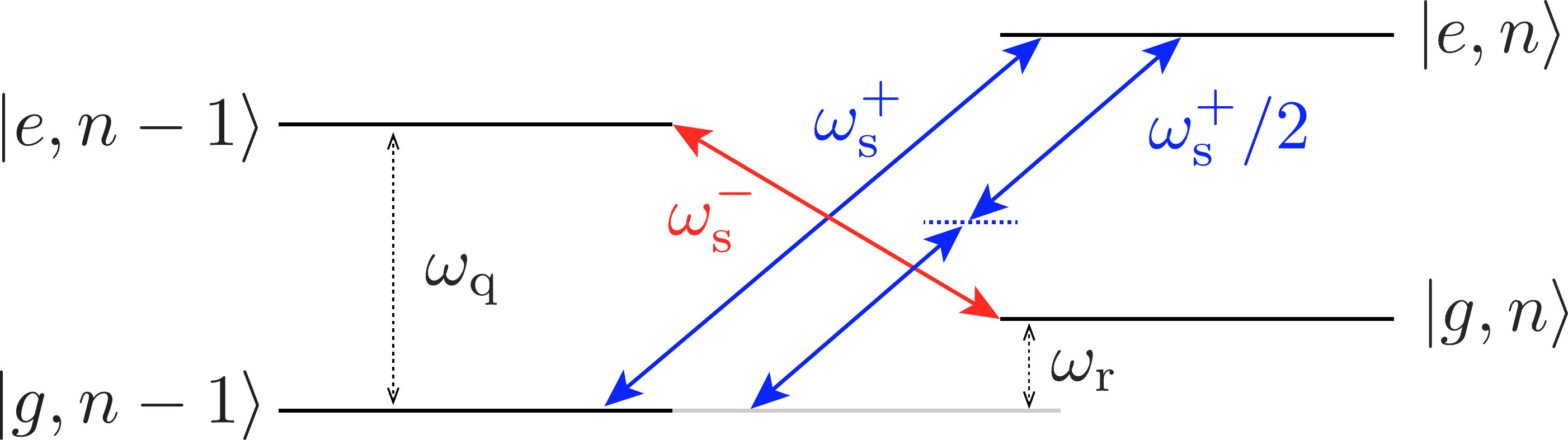}
\caption{Sideband transitions. A red sideband excitation (red arrow) excites the qubit (resonator) by combining an external drive photon at $\omega_{\rm s}^-$ with an excitation from the resonator (qubit). A blue sideband excitation (blue arrows) corresponds to exciting both the qubit and the resonator simultaneously. Due to selection rules that apply to charge and flux qubits at their degeneracy points (see \secref{sec:SelectionRules}), single-photon blue sideband transitions can be forbidden. However, the transition can then still be accessed by using two (or more) drive photons at a lower frequency.
\label{fig:sideband}}
\end{figure}

Sideband transitions are transitions that involve excitation or de-excitation of both a qubit and a resonator, as sketched in \figref{fig:sideband}. The \textit{blue-sideband} transition shown corresponds to exciting both the qubit and the resonator simultaneously at the frequency $\omega_{\rm s}^+ = \omega_{\rm r} + \omega_{\rm q}$. In the \textit{red-sideband} transition shown, a signal at $\omega_{\rm s}^- = \omega_{\rm q} - \omega_{\rm r}$ excites the qubit by taking one photon from the resonator.

Both red- and blue-sideband transitions have been extensively studied for quantum information processing in ion traps~\cite{LeibfriedRMP2003, Blatt2008, BlattNatPhys2012}, where the frequency of the vibrational motion is much smaller than the transition frequency between the two electronic energy levels chosen to form a qubit. Inspired by such experiments with trapped ions, a scalable superconducting circuit, in which superconducting qubits can be controlled and addressed as trapped ions, was proposed in Ref.~\cite{Liu2007}. In this proposal, superconducting qubits are coupled to the ``vibrational" mode provided by a superconducting $LC$ circuit or a cavity field. This allows for single-qubit rotations and selective tuning of the coupling between a qubit and the $LC$ circuit by adjusting the frequencies of time-dependent external fields to match the condition of sideband excitations. A trapped-ion model in SQCs can also be built by engineering circuit-QED systems with dressed states~\cite{Liu2006a, Blais2007}. 

Sideband transitions have been observed in many SQC experiments. In addition to the overview given here, more examples can be found in Secs.~\ref{sec:multiphoton-transitions} and \ref{sec:Lasing}, including sidebands in resonance fluorescence and Rabi sideband lasing. Already in the early days of circuit QED, both red- and blue-sideband transitions were observed, and exploited to generate entangled states, using a flux qubit coupled to an $LC$ circuit~\cite{Chiorescu2004}. Another experiment, with a charge-phase qubit coupled to a low-frequency $LC$ resonator, showed the analogy between sideband transitions in their system and in a vibrating diatomic molecule~\cite{Gunnarsson2008}. A two-photon sideband transition~\cite{Wallraff2007} has been used to create entanglement and implement two-qubit operations in SQCs~\cite{Leek2009}. In order to realize fast two-qubit gates, it has been proposed~\cite{Beaudoin2012}, and demonstrated in a circuit-QED experiment~\cite{Strand2013}, that modulating the qubit frequency can make sideband transitions a first-order process. Sideband transitions have also been used to generate multi-photon Fock states in SQCs~\cite{Leek2010} and it has been shown theoretically that simultaneously driving multiple sideband transitions makes it possible to create an arbitrary Fock-state superposition~\cite{Strauch2012}. With longitudinal coupling (see \secref{sec:LightMatterCoupling}), \textit{multiphoton sideband transitions}, which can be used for fast preparation of Fock states and their superpositions, become possible~\cite{Zhao2015}.

\subsection{Strongly driven multiphoton processes}
\label{sec:multiphoton-transitions}

A two-level atom subjected to a monochromatic driving field will absorb or emit photons at the drive frequency if the drive is weak. Viewed in the time domain, the atom will exhibit Rabi oscillations between its ground and excited states. However, the behavior of the system becomes more complicated if the drive strength increases to a point where its associated Rabi frequency exceeds the atomic linewidth. At such drive strengths, \textit{higher-order interactions} between the driving field and the two-level system, or their sidebands, can occur. For example, an early experiment demonstrated excitation of, and Rabi oscillations with, a charge qubit through one-, two-, and three-photon processes~\cite{Nakamura2001}. Note that such multiphoton Rabi oscillations are different from the multiphoton \textit{vacuum} Rabi oscillations that are possible with ultrastrong coupling~\cite{Garziano2015} (see \secref{sec:ultrastrong}). In the spectral domain, multiphoton absorption driving transitions between neighboring levels has been demonstrated in flux~\cite{Saito2004, Saito2006a, Deppe2008} and phase~\cite{Wallraff2003, Shnyrkov2006} qubits. Recently, multiphoton sideband transitions were also probed for a flux qubit ultrastrongly coupled to multiple modes of a transmission-line resonator~\cite{ChenZhen2016}.

For a qubit described by a Hamiltonian $H = \omega_{\rm q} \sz/2$, varying an external bias in $\sx$ takes the qubit through an avoided level crossing. For example, at the optimal working point of a flux qubit, when the amplitude of the driving field is large~\cite{Ashhab2007}, the qubit can be driven periodically through the avoided crossing and \textit{Landau--Zener--St\"{u}ckelberg--Majorana} (LZSM) transitions can occur~\cite{Oliver2009, Shevchenko2010, Shevchenko2012, Shevchenko2012a, Sun2012}. Such LZSM transitions can be multiphoton processes (e.g., up to 17 photons contributed in the experiment of Ref.~\cite{Stehlik2012}). There have been several experiments studying LZSM physics in strongly driven SQCs and similar systems~\cite{Oliver2005, Sillanpaa2006, Wilson2007, Izmalkov2008, LaHaye2009, Wilson2010a, Leppakangas2013a, Neilinger2016}. Extensions to a biharmonic drive~\cite{Satanin2014} and multiple qubits~\cite{Satanin2012} have also been studied.

If the two-level system has a broken inversion symmetry of its potential energy, the Hamiltonian with a classical driving field of strength $\Omega_{\rm d}$ and drive frequency $\omega_{\rm d}$ becomes
\be
H = \frac{\epsilon}{2} \sz + \frac{\Delta}{2} \sx + \frac{\Omega_{\rm d}}{2} \sz \cos \omega_{\rm d} t,
\ee
where $\Delta$ is the qubit gap and $\epsilon$ the tunneling energy. This Hamiltonian can be rewritten as
\be
H = \frac{\omega_{\rm q}}{2} \tilde{\sigma}_z + \frac{\Omega_{\rm d}}{2} \left( \cos\theta \tilde{\sigma}_z - \sin\theta \tilde{\sigma}_x \right) \cos\omega_{\rm d} t
\label{eq:HLongitudinalTransverseDrive}
\ee
in the diagonalized qubit basis with $\omega_{\rm q} = \sqrt{\epsilon^2 + \Delta^2}$ and $\sin\theta = \Delta/\omega_{\rm q}$. Here, $\tilde{\sigma}_z$ and $\tilde{\sigma}_x$ are defined by the eigenstates of the Hamiltonian $(\epsilon \sz + \Delta \sx) / 2$. Equation~(\ref{eq:HLongitudinalTransverseDrive}) describes a qubit coupled to an external driving field both transversely and longitudinally. It has been shown that this longitudinal coupling can induce coexistence of multiphoton transitions~\cite{Liu2014}, which do not exist in natural atomic systems. For more discussion on longitudinal coupling, see \secref{sec:LightMatterCoupling}.

Going beyond the two-level approximation, multiphoton transitions have also been studied in many-level quantum circuits~\cite{Braumuller2015}. While that study focused on a single artificial atom, multiphoton transitions between dressed resonator-qubit states have also been studied to reveal the characteristic $\sqrt{n}$ nonlinearity in the Jaynes--Cummings model (see \secref{sec:DressedStates}). In the resonant case, this nonlinearity has been demonstrated experimentally with SQCs~\cite{Fink2008, Bishop2008, Kockum2013}.

\subsection{Lasing}
\label{sec:Lasing}

Lasing is perhaps the most well-known, and most widely applied, quantum-optics process. In the case of SQCs with microwave photons, micromasers based on Josephson junctions were proposed in the 1990s~\cite{Free1995, Hatakenaka1996}. The recent progress in SQCs has revived interest in implementing lasing on a superconducting chip. There have been various theoretical studies~\cite{You2007a, Zhirov2008, Marthaler2008, Ashhab2009, Marthaler2011a, Jia2011, Navarrete-Benlloch2014} paving the way for achieving a tunable micromaser that can generate nonclassical light. In this section, we review lasing, with and without population inversion, in SQCs. For generation of other nonclassical states of light in SQCs, see Secs.~\ref{sec:SqueezedStates} and \ref{sec:PhotonGeneration}.

\subsubsection{Lasing with population inversion}

Lasing occurs when stimulated emission exceeds absorption. The ratio between the rates for stimulated emission and absorption is proportional to $\rho_{\text{ee}}B_{\text{stem}}/(\rho_{\text{gg}}B_{\text{abs}})$, where $\rho_{\text{gg}}$ ($\rho_{\text{ee}}$) is the population of the ground (excited) state and $B_\text{abs}$ ($B_\text{stem}$) is the Einstein coefficient for absorption (stimulated emission). Since $B_\text{abs} = B_\text{stem}$, \textit{population inversion} ($\rho_{\text{ee}} > \rho_{\text{gg}}$) is usually required for lasing. Population inversion has been demonstrated in SQCs~\cite{Berns2008}.

Conventional lasers use many atoms and operate in the weak-coupling regime. However, in the regime of strong coupling between atoms and photons, \textit{single-atom lasing}, where a single atom is excited and used to generate many photons, is possible~\cite{Mu1992}. Single-atom lasing was first demonstrated in a cavity-QED experiment~\cite{McKeever2003}, which also showed that the single-atom laser differs from conventional many-atom lasers in several respects. For example, the single-atom laser does not have a sharp pumping threshold for the onset of lasing and the statistical properties of the output light are different~\cite{Ashhab2009}.

In SQCs, the strong-coupling regime is relatively easy to reach (see \secref{sec:LightMatterCoupling}). An early proposal showed how persistent single-photon generation could be generated using lasing with population inversion in a flux-qubit circuit~\cite{You2007}. Single-artificial-atom lasing was first demonstrated in an experiment with a charge qubit coupled to a transmission-line resonator~\cite{Astafiev2007}. There, the population inversion was created by single-electron-tunneling events~\cite{Rodrigues2007, Ashhab2009}. For such lasing, operating in the regime of strong coupling and low temperature, the quantum fluctuations of the photon number dominate over thermal noise. This leads to phase diffusion and the linewidth of the lasing field is also strongly influenced~\cite{Andre2009, Andre2009a, Andre2010}. Recently, microwave lasing was also demonstrated using a Cooper-pair transistor embedded in a high-$Q$ superconducting microwave cavity~\cite{Chen2014b}; this can be used to generate amplitude-squeezed light.

As discussed in \secref{sec:WaveguideQED}, strong (and even ultrastrong) coupling of a superconducting qubit to photons has not only been achieved with a resonator, but also with an open transmission line. In such a setup, a flux qubit in 1D open space has been demonstrated to work as a broadband-tunable quantum amplifier~\cite{Astafiev2010a}, which is a step towards single-emitter lasing. Recent work demonstrating coupling between a superconducting qubit and a continuum of surface acoustic waves~\cite{Gustafsson2014} could also potentially be extended to realize lasing (or rather, phasing, since microwave \textit{phonons} would be emitted)~\cite{Kockum2014, Aref2016}.

\begin{figure}
\centering
\includegraphics[width=\linewidth]{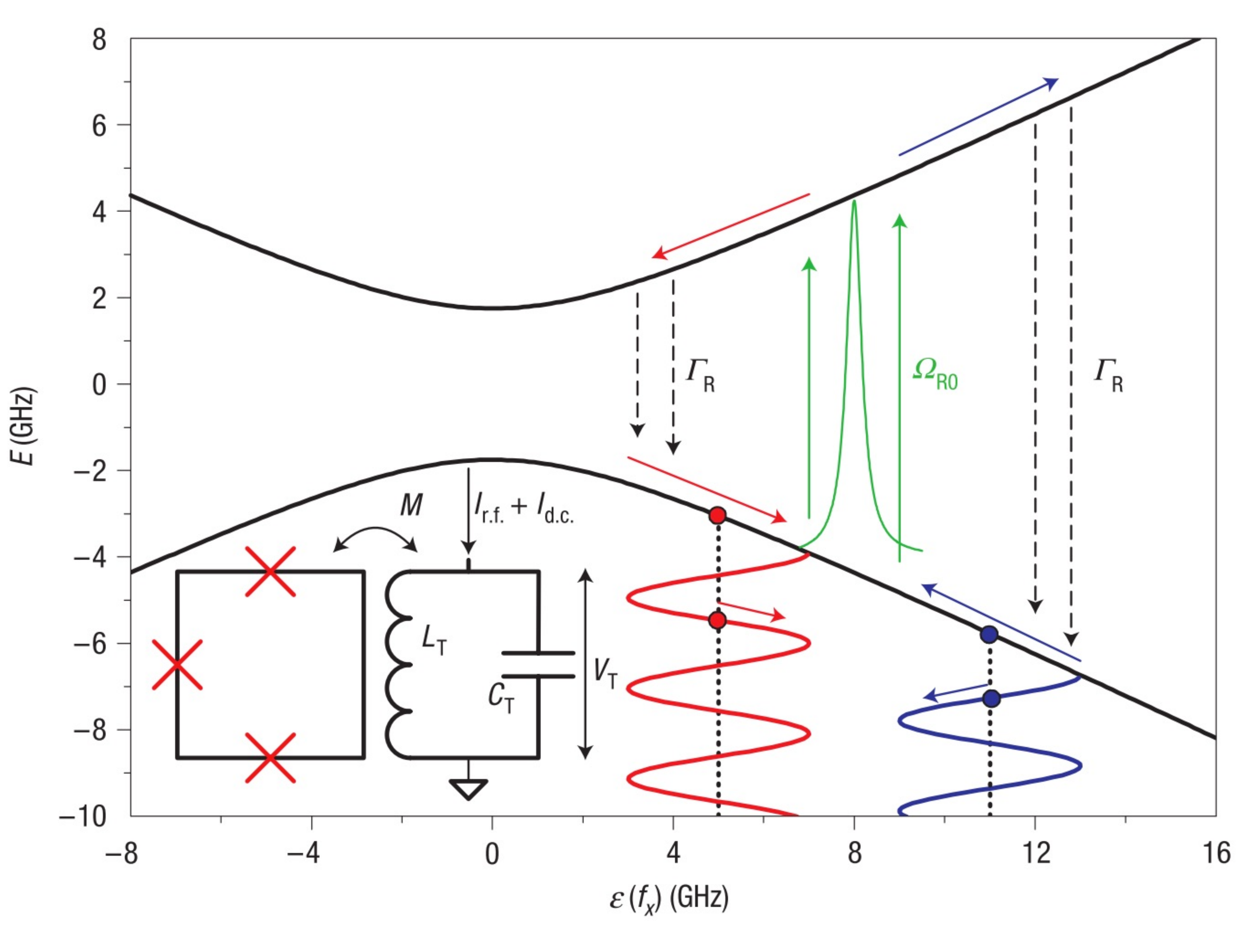}
\caption{Sisyphus cooling and amplification of an $LC$ oscillator~\cite{Grajcar2008, Nori2008}. The oscillator, with inductance $L_{\rm T}$ and capacitance $C_{\rm T}$, is coupled to a flux qubit (the loop with three red crosses representing Josephson junctions) through a mutual inductance $M$. The frequency of the oscillator, $\omega_{\rm r}$, is much lower than the transition frequency of the qubit, $\omega_{\rm q}$. The energy levels of the qubit (solid black curves) change as a function of the energy bias $\varepsilon (f_x)$, which depends on the flux $f_x$. This flux is set by the slowly oscillating current $I_{\rm r.f.}$ in the $LC$ oscillator and a dc bias $I_{\rm d.c.}$ of the qubit. The qubit is also driven by a high-frequency signal at $\omega_{\rm d} \approx \unit[8]{GHz}$ (green curve) with amplitude $\Omega_{\rm R0}$. In the Sisyphus analogy, the qubit plays the role of the boulder and $I_{\rm r.f.}$ of the $LC$ oscillator corresponds to Sisyphus. In Sisyphus cooling, shown by blue arrows and curves, $I_{\rm d.c.}$ is set such that $\omega_{\rm q} > \omega_{\rm d}$. If the qubit starts in the ground state somewhere to the right of $\omega_{\rm d}$, the slow oscillation of $I_{\rm r.f.}$ will push it towards a point where $\omega_{\rm q} \approx \omega_{\rm d}$ (blue arrow pointing uphill to the left). At that point, the drive becomes resonant and excites the qubit (green vertical arrow). The $LC$ oscillator then pushes the qubit away from the resonance (blue arrow pointing uphill to the right). At some point, the qubit relaxes to the ground state (dashed black vertical arrows) and the cycle begins anew. If the relaxation rate $\Gamma_{\rm R}$ is of the same size as $\omega_{\rm r}$, the qubit is pushed uphill almost all the time. This pushing takes energy from the oscillator and dissipates it to the environment through the relaxation process, resulting in cooling of the oscillator. Reversing the process with $\omega_{\rm q} < \omega_{\rm d}$ (red arrows and curves), the qubit rolls downhill most of the time and the oscillator gains energy, which results in amplification. Reprinted figure by permission from Macmillan Publishers Ltd: Nature Physics, M.~Grajcar et al.,~\href{http://dx.doi.org/10.1038/nphys1019}{Nat.~Phys.~\textbf{4}, 612 (2008)}, copyright (2008).
\label{fig:Sisyphus}}
\end{figure}

\textit{Sisyphus cooling}~\cite{Wineland1992} and \textit{amplification} is another mechanism that can be used to approach lasing in SQCs. It has been experimentally demonstrated in a circuit-QED setup with a flux qubit coupled to an $LC$ oscillator~\cite{Grajcar2008, Nori2008}, building on a theoretical proposal for cooling a mechanical resonator coupled to a controllable charge qubit~\cite{Zhang2005prl}. The experiment and the principle for Sisyphus cooling and amplification are sketched in \figref{fig:Sisyphus} (see also Ref.~\cite{Shevchenko2012} for a comprehensive review). Both cooling and amplification (a trend toward lasing) of the $LC$ oscillator were observed. Later experimental work increasing the quality factor of the oscillator brought the system very close to the theoretical lasing limit~\cite{Skinner2010}. In these experiments, the resonator frequency $\omega_{\rm r}$ is in the MHz range, slow compared to the qubit frequency $\omega_{\rm q}$, which is in the GHz range. This situation is also seen with a nanomechanical resonator capacitively coupled to a charge qubit, a setup where cooling of the mechanical degree of freedom has been demonstrated and which can be used to create macroscopic nonclassical states~\cite{Naik2006}. In all these setups, and also in another experiment with a flux qubit coupled to an $LC$ resonator~\cite{Ilichev2003}, the qubit is subjected to an external drive at $\omega_{\rm d}$ with amplitude (Rabi frequency) $\Omega_{\rm R}$. As explained in \figref{fig:Sisyphus}, lasing (or at least amplification) of the resonator occurs when $\omega_{\rm d} > \omega_{\rm q}$ (blue detuning), while cooling of the resonator can be realized when $\omega_{\rm d} < \omega_{\rm q}$ (red detuning).

An experiment has also demonstrated cooling of a flux qubit itself through a microwave drive~\cite{Valenzuela2006, Chiorescu2006}. The basic idea of this experiment is to cool the qubit through a process that can be considered the inverse of population inversion. In this case, a drive is used to transfer population from the first excited state of the qubit to its second excited state. The flux-qubit circuit has been engineered such that relaxation from the second excited state to the ground state is much faster than other transitions in the system, which results in the external drive effectively transferring population from the first excited state to the ground state. This result has been theoretically extended to cooling of nearby two-level systems~\cite{You2008}.

Further theoretical studies~\cite{Hauss2008a, Hauss2008} have shown that \textit{two-photon lasing} can be achieved with a single superconducting artificial atom. When the drive is blue-detuned, the number of photons released in the lasing can be controlled by tuning $\Omega_{\rm R}$ such that either $\Omega_{\rm R} \approx \omega_{\rm r}$, which leads to single-photon lasing, or $\Omega_{\rm R} \approx 2\omega_{\rm r}$, which leads to two-photon lasing. At the optimal point of a flux qubit, the two-photon process dominates~\cite{Hauss2008a, Hauss2008} due to a selection rule~\cite{Liu2005a}. Later experiments have demonstrated single-photon amplification~\cite{Oelsner2013} and strong evidence for both single-photon and two-photon lasing with a single qubit~\cite{Neilinger2015}.

The above lasing scenario is closely related to \textit{dressed-state lasing} or \textit{Rabi sideband lasing}, also known as hidden inversion in dressed states; these types of lasing are often discussed in the context of lasing without population inversion~\cite{Mompart2000} (see \secref{sec:LWI}). As discussed in \secref{sec:EITandATS} above, driving a qubit strongly can give rise to a Mollow triplet~\cite{Mollow1969} with peaks at $\omega_{\rm d}$ and $\omega_{\rm d} \pm \Omega_{\rm R}$ (see \figref{fig:ATS}). The latter two peaks are the Rabi sidebands. If $\omega_{\rm d} > \omega_{\rm q}$, a probe field with frequency $\omega_{\rm p}$ experiences absorption at $\omega_{\rm d} + \Omega_{\rm R}$ and amplification at $\omega_{\rm d} - \Omega_{\rm R}$ due to population inversion in the dressed-state basis~\cite{Mompart2000}. In optical systems, experiments have demonstrated such Rabi sideband amplification~\cite{Wu1977} and lasing~\cite{Khitrova1988}. In such optical experiments, the high-frequency transitions between adjacent doublets are often used. Corresponding circuit-QED experiments, on the other hand, have coupled an oscillator to the low-frequency transition within a doublet of a driven flux qubit, i.e., $\omega_{\rm p} = \Omega_{\rm R}$~\cite{Oelsner2013, Shevchenko2014}.

In experiments using superconducting qubits with a weak anharmonicity, e.g., a transmon qutrit, a strong drive can excite both the second and third levels of the system. The dressed states that then form involve six Rabi sidebands. Both amplification and attenuation of a probe signal for such a system has been observed in an experiment~\cite{Koshino2013b}.

\subsubsection{Lasing without population inversion}
\label{sec:LWI}

When lasing is realized through population inversion, the pump power needed is proportional to (at least) the fourth power of the lasing frequency. Therefore, \textit{lasing without inversion} (LWI) is desirable for implementations of short-wavelength lasers. LWI can be realized through a variety of different mechanisms, as summarized in the comprehensive review of Ref.~\cite{Mompart2000}. These mechanisms include recoil-induced lasing, hidden inversion in a coherent-population-trapping basis, hidden inversion in a dressed-state basis, and lasing without hidden inversion. The last process can occur in $V$- and $\Xi$-type three-level systems due to quantum interference~\cite{Mompart2000}.

In natural atoms, the recoil effect gives different frequency shifts to the spectra of stimulated emission and absorption. This asymmetry between the emission and absorption processes leads to the existence of a frequency range where stimulated emission exceeds absorption even without population inversion~\cite{Mompart2000}. Applying this idea of asymmetry between stimulated emission and absorption to circuit QED, it has been proposed that LWI can be realized with a driven qubit coupled to a transmission-line resonator in the presence of a dissipative electromagnetic environment~\cite{Marthaler2011}. In this setup, LWI arises because the coupling to the dissipative environment can, in a similar way as the atomic recoil effect, enhance photon emission as compared to absorption. LWI has also been studied theoretically for a superconducting $\Delta$-type three-level system where a relative phase between driving fields is controlled~\cite{Jia2010, Joo2010}.

In a recent experiment with a superconducting qubit placed at the end of a transmission line, the amplitude of a weak coherent probe was amplified by $\unit[7]{\%}$ using an amplification mechanism without population inversion~\cite{Wen2017}. In contrast to the Rabi sideband lasing discussed above, here the drive was resonant with the qubit, i.e., $\omega_{\rm d} = \omega_{\rm q}$, which means that there is no population inversion even among the dressed states. As predicted by Mollow in 1972, a weak probe will nevertheless be amplified when $\omega_{\rm p}$ lies in-between the frequencies of the Mollow triplet~\cite{Mollow1972}. This amplification can be explained in terms of an irreversible four-photon process~\cite{Friedmann1987}.

\subsection{Squeezed states}
\label{sec:SqueezedStates}

Quantization of the electromagnetic field, where two non-commuting observables obey the Heisenberg uncertainty relation, directly leads to vacuum fluctuations. These fluctuations can be revealed in, e.g., the Casimir effect and the Lamb shift, and set the quantum limit for measurement sensitivity~\cite{Ma2011}. Looking at a single mode of the quantized electromagnetic field, it can be decomposed into two quadrature components. In a coherent state, the quantum fluctuations in the two quadratures are equally large and minimize the uncertainty product in the Heisenberg uncertainty relation. It is also well known that the quantum fluctuations in a coherent state are randomly distributed in phase and that they are equal to the zero-point fluctuations of the vacuum.

In a \textit{squeezed state}, the quantum fluctuations are not equally distributed between the two quadratures. In general, a squeezed state may have less noise than a coherent state in one quadrature, but this comes at the expense of increased fluctuations in the other quadrature. Squeezed states have many applications (see Refs.~\cite{Dodonov2002, Ma2011} for reviews), including gravitational force detection, precision measurements~\cite{Ma2011}, and continuous-variable quantum information processing~\cite{Braunstein2005}. Squeezed states may also modify the coherence time of an atom~\cite{Gardiner1986}.

Due to the tunable nonlinearity and low losses that can be realized for microwaves, SQCs are promising devices for producing squeezed states~\cite{Yurke1988, Yurke1989, Hu1996, Everitt2004}. For example, theoretical studies~\cite{Zagoskin2008, Zagoskin2012a} have shown that a superconducting resonant tank circuit~\cite{Ilichev2003} can be used to produce squeezed states, and that the generated squeezed states can be further applied to minimize quantum fluctuations. There are many theoretical proposals for generating squeezing using circuit-QED systems~\cite{Moon2005, Zhang2009, Navarrete-Benlloch2014, Elliott2015} and other superconducting microwave devices~\cite{Ojanen2007, Kronwald2013, Didier2014}. Multimode squeezed vacuum could potentially be generated by dissipation in a circuit-QED system~\cite{Porras2012} or with a Josephson traveling-wave
amplifier~\cite{Grimsmo2016}. It has also been pointed out that a superconducting qubit can act as a probe to detect microwave squeezing~\cite{Boissonneault2014, Ong2013}.

The most popular way to produce microwave squeezing is to use a \textit{Josephson parametric amplifier}. For example, amplification and squeezing of quantum noise has been demonstrated both with a tunable Josephson metamaterial~\cite{Castellanos-Beltran2008} and with a Josephson parametric amplifier~\cite{Mallet2011, Zhong2013}. Experiments have also shown that quadrature-squeezed electromagnetic vacuum, generated by a Josephson parametric amplifier, can be used to reduce the radiative decay of superconducting qubits~\cite{Murch2013a} and to modify resonance fluorescence~\cite{Toyli2016}. As discussed further in Secs.~\ref{sec:MicrowaveComponentsMixers} and \ref{sec:PhotonDetectionHomodyne}, parametric amplifiers have recently been explored to improve the fidelity of the dispersive readout of superconducting qubit states~\cite{Barzanjeh2014}, to detect microwave states~\cite{Eichler2011}, and to generate entanglement~\cite{Flurin2015} or other two-mode nonclassical correlations and squeezed states~\cite{Bergeal2012, Flurin2012}.

\subsection{Photon blockade}
\label{sec:PhotonBlockade}

Reminiscent of the Coulomb blockade for electrons, \textit{photon blockade} means that subsequent photons are prevented from resonantly entering a cavity. Photon blockade originates from anharmonic energy levels of the cavity light field. The anharmonicity is usually due to photon-photon interactions induced by a nonlinear medium or atoms in the cavity~\cite{Tian92, Leonski94, Miran96, Imamoglu1997}. One of the basic conditions for observing photon blockade is that the decay rate of the cavity field should be less than the photon-photon interaction strength. If that is the case, a nonlinear medium or an atom in a cavity can work as a \textit{turnstile device}, where photons pass one by one. Thus, photon blockade could be used as a microwave single-photon source~\cite{Wang2016b} or as a single-photon transistor.

Photon blockade can be revealed by measuring nonclassical photon-counting statistics, e.g., photon antibunching or sub-Poissonian photon number statistics, in the second-order correlation function $g^{(2)}(\tau)$ (see \secref{sec:PhotonDetectionCorrelation}). Specifically (see, e.g., Ref.~\cite{Miran10}), antibunching is characterized by $g^{(2)}(\tau)> g^{(2)}(0)$, while sub-Poissonian photon statistics (confusingly also often called photon antibunching) correspond to $g^{(2)}(0)<1$. Note that these two effects are distinctly different~\cite{Zou90} and that classical light (e.g., thermal light) cannot exhibit neither sub-Poissonian statistics nor photon antibunching.

Following experimental observations in cavity QED~\cite{Birnbaum2005} and in a photonic-crystal cavity containing a quantum dot~\cite{Faraon2008}, photon blockade has been observed in circuit-QED setups with a superconducting qubit coupled either resonantly~\cite{Lang2011} or dispersively~\cite{Hoffman2011} to a microwave resonator. In the dispersive case, where the qubit and the resonator field were far detuned, the power spectrum of the transmitted light field as a function of photon bandwidth displayed a staircase character~\cite{Hoffman2011}. 

Further theoretical study has shown that photon blockade can be changed to \textit{transparency} in a circuit-QED system if the intensity of the probe field is increased~\cite{Liu2014c}. Indeed, if a Jaynes--Cummings system is driven strongly enough, a \textit{first-order dissipative phase transition} can occur due to the breakdown of photon blockade~\cite{Carmichael2015}. This phase transition was recently demonstrated in a circuit-QED experiment~\cite{Fink2017}.

The standard single-photon blockade has also been generalized to \textit{multiphoton blockade} (for multiphoton transitions)~\cite{Leonski1996, Miran96, Leonski1997, Shamailov2010} (see also early reviews in Refs.~\cite{Miranowicz2001,Leonski2001}), which has been the subject of several recent theoretical studies~\cite{Miranowicz2013, Miranowicz2014, Hovsepyan2014, Wang2015a, Miran16} and a cavity-QED experiment~\cite{Hamsen2017}. Considering that up to five-photon transitions in the Jaynes--Cummings ladder have been observed in a circuit-QED system~\cite{Bishop2008}, multiphoton blockade should be possible to see in such experiments. Other extensions of photon blockade include its behavior in the ultrastrong-coupling regime~\cite{Ridolfo2012}, photon blockade due to coherent feedback~\cite{Liu2013}, photon blockade with a single multi-level atom in an open waveguide, i.e., without a cavity~\cite{Zheng2011, Zheng2012} (see \secref{sec:WaveguideQEDTheoryOpen}), and ``unconventional photon blockade" in a setup with two coupled nonlinear resonators~\cite{Leonski2004, Miranowicz2006, Liew2010, Ferretti2010, Bamba2011, Bamba2011a, Flayac2013, Ferretti2013, Xu2013b, Lemonde2014, Flayac2015} (see \secref{sec:TwoOrMoreCavities}), a setup which has been realized in circuit QED~\cite{Eichler2014a}.

Photon blockade can also be used~\cite{Didier11} to detect the closely related phenomenon of \textit{phonon} blockade, which can occur in nanomechanical resonators~\cite{Liu10}. This phonon blockade is a useful indicator of the quantumness (or nonclassicality) of a mechanical resonator coupled to a superconducting qubit~\cite{Liu10, Miran16, Wang16a}.

\subsection{Quantum jumps}
\label{sec:QuantumJumps}

Quantum jumps are associated with the exchange of a quantum of energy between two system. The notion of quantum jumps goes back to the early days of quantum mechanics, when Einstein worried about wave-function collapse due to detection of a photoelectron. Bohr suggested that the interaction of light and matter occurs in such a way that the internal state of an atom undergoes an instantaneous transition upon the emission or absorption of a light quantum. These sudden transitions have become known as the Bohr-Einstein quantum jumps. In quantum optics, this quantum-jump approach has been developed and applied to solve dissipation problems. It is related to many concepts, including Monte-Carlo simulations, quantum trajectories, conditional density matrices, and collapse or reduction of a state vector~\cite{Carmichael1993a, Plenio1998, Wiseman2010}. Quantum jumps have also found important applications in quantum measurements, time and frequency standards, and precision spectroscopy.

Quantum jumps have been observed in various microscopic systems, first with trapped ions~\cite{BergquistPRL1986}. Photonic quantum jumps were first achieved in a one-electron cyclotron oscillator \cite{Peil1999} and later in cavity QED~\cite{Gleyzes2007, Guerlin2007}. The first experimental demonstration of quantum jumps between macroscopic quantum states used an SQC where a phase qubit coupled to two-level systems in a Josephson junction~\cite{Yu2008}.

Circuit QED provides an excellent platform for studying quantum jumps and other types of \textit{measurement backaction}~\cite{Gambetta2008} (see also Secs.~\ref{sec:PhotonDetectionItinerant} and \ref{sec:PhotonDetectionHomodyne}). Quantum jumps between states of a superconducting qubit have been observed in real time by coupling the qubit to a microwave readout cavity~\cite{Vijay2011}. In similar setups, such jumps have also been used to investigate quasiparticle excitations limiting the coherence of a fluxonium qubit~\cite{Vool2014} and to test measurement backaction~\cite{Hatridge2013}. Quantum jumps between photonic states have also been measured in a circuit-QED setup implementing quantum nondemolition (QND) parity measurements on a resonator using a transmon qubit~\cite{Sun2014}.
 
In qubit systems, it is usually quantum jumps between the two energy eigenstates, also called the longitudinal pseudospin components, that is observed. However, a recent circuit-QED experiment has shown that quantum jumps also can occur between \textit{transverse} superpositions of these eigenstates~\cite{Vool2016}. This and the other experiments mentioned above further settle the dispute about whether the probabilistic predictions of quantum mechanics can be used to describe the dynamics of a single quantum system. Furthermore, realtime monitoring of a superconducting qubit and the observation of quantum jumps are very important steps toward implementing quantum feedback control~\cite{Zhang2014a} and quantum error-correction codes (see \secref{sec:QEC}) in quantum information processing.

\section{Nonlinear processes}
\label{sec:NonlinearProcess}

In nonlinear optics, the response of a medium to an applied optical field depends \textit{nonlinearly} on the strength of the applied field~\cite{Kielich1981, Shen1984, Boyd2008}. In such a medium, the usual linear relation $P(t) = \epsilon_0 \chi^{(1)}E(t)$ between the induced polarization $P(t)$ and the applied electric field $E(t)$ is replaced by 
\be 
P(t) = \epsilon_0 \left( \chi^{(1)} E(t) + \chi^{(2)} E^2(t) + \chi^{(3)} E^3(t) + \ldots \right),
\label{eq:nonlinear}
\ee
where $\epsilon_0$ is the vacuum permittivity and $\chi^{(i)}$ is the $i$th-order nonlinear susceptibility~\cite{Boyd2008}. When the field $E(t)$ is sufficiently strong, the effects of the higher-order terms become important. These effects include the Kerr effect and nonlinear wave-mixing processes that can be applied for harmonic generation, frequency conversion, and amplification.

In this section, we briefly review quantum nonlinear processes in SQCs. We treat second-order nonlinearities in \secref{sec:NonlinearProcThreeWaveMixing} and third-order ones in \secref{sec:NonlinearProcKerr}. We mostly focus on nonlinear wave mixing and related phenomena, such as strong photon-photon interactions arising from third-order nonlinear processes. Photon-photon interactions at the single-photon level has many important applications in microwave photonics, e.g., single-photon switches, single-photon transistors, all-microwave quantum logic gates, and quantum control solely via microwave fields.

\subsection{Second-order nonlinearity and three-wave mixing}
\label{sec:NonlinearProcThreeWaveMixing}

The second-order nonlinear susceptibility $\chi^{(2)}$ enables \textit{three-wave mixing}, which is widely used for frequency up- and down-conversion. To see this, consider an input electric field consisting of two parts, a \textit{pump} at frequency $\omega_{\rm p}$ with amplitude $E_{\rm p}$ and a \textit{signal} at frequency $\omega_{\rm s}$ with amplitude $E_{\rm s}$, such that the total electric field can be written as $E(t) = E_{\rm p} \exp({-i \omega_{\rm p} t}) + E_{\rm s} \exp({-i \omega_{\rm s} t}) + \text{c.c.}$, where c.c.~denotes complex conjugate. Inserting this expression in the second-order term in \eqref{eq:nonlinear} results in components oscillating at the sum frequency $\omega_{\rm p} + \omega_{\rm s}$ and the difference frequency $\abs{\omega_{\rm p} - \omega_{\rm s}}$ (for more detailed calculations of this and other wave-mixing processes, see, e.g., Refs.~\cite{Boyd2008, Anton2017}).

\begin{figure}
\centering
\includegraphics[width=0.8\linewidth]{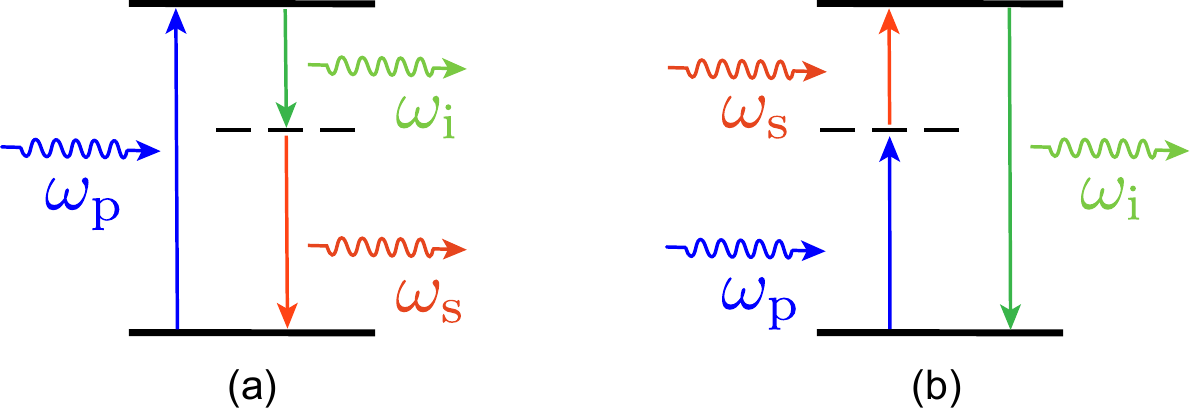}
\caption{Three-wave-mixing processes. (a) Difference-frequency generation. The second-order nonlinearity makes it possible for a high-frequency pump at $\omega_{\rm p}$ to be down-converted into a signal mode at $\omega_{\rm s}$ and an idler mode at $\omega_{\rm i} = \omega_{\rm p} - \omega_{\rm s}$, aided by stimulated emission generated by a signal at $\omega_{\rm s}$. (b) Sum-frequency generation. The pump at $\omega_{\rm p}$ and the signal at $\omega_{\rm s}$ together excite the medium, leading to emission of up-converted photons in the idler mode at $\omega_{\rm i} = \omega_{\rm p} + \omega_{\rm s}$.
\label{fig:ThreeWaveMixing}}
\end{figure}

To achieve \textit{difference-frequency generation}, a nonlinear medium is pumped into a higher-energy level by a strong field with frequency $\omega_{\rm p}$, as sketched in \figref{fig:ThreeWaveMixing}(a). Due to the nonlinearity, the presence of a signal at $\omega_{\rm s} < \omega_{\rm p}$ then stimulates the generation of outputs at $\omega_{\rm s}$ and $\omega_{\rm i} = \omega_{\rm p} - \omega_{\rm s}$. This results in amplification of both the signal at $\omega_{\rm s}$ and the \textit{idler} mode at $\omega_{\rm i}$. The case $\omega_{\rm i} = \omega_{\rm s}$ is known as \textit{degenerate amplification}; it can be used to generate \textit{squeezed states} (see \secref{sec:SqueezedStates}). \textit{Non-degenerate amplification}, i.e., $\omega_{\rm i} \neq \omega_{\rm s}$, does not result in squeezing of the individual signal and idler modes, but it nevertheless creates correlations between the two modes, which is called \textit{two-mode squeezing}~\cite{Walls2008}. In \textit{sum-frequency generation}, shown in \figref{fig:ThreeWaveMixing}(b), the pump and the signal combine to excite the medium to a higher energy level, resulting in emission into the idler mode at $\omega_{\rm i} = \omega_{\rm p} + \omega_{\rm s}$. In this case, neither amplification~\cite{Tien1958} nor squeezing is generated.

A prominent example of an SQC that can be used for three-wave mixing is the \textit{Josephson parametric converter} (JPC)~\cite{Bergeal2010}, which also is discussed in \secref{sec:MicrowaveComponentsMixers} and illustrated in \figref{fig:JosephsonMixer} of that section. A JPC consists of four identical large-area Josephson junctions in a superconducting ring, which couples to three modes exploiting a Wheatstone-bridge symmetry. When these three modes (pump, signal, and idler) satisfy the relation $\omega_{\rm p} - \omega_{\rm s} = \omega_{\rm i}$, the JPC operates as a nondegenerate parametric amplifier near the quantum limit~\cite{Bergeal2010a}. For JPCs in this mode of operation, two-mode squeezing of the signal and idler modes has been observed~\cite{Bergeal2012}, even when these two modes were spatially separated~\cite{Flurin2012}. The time-reversed process, where the signal and idler modes are attenuated to generate pump photons, has also been demonstrated~\cite{Schackert2013}. We note that two-mode squeezing in circuit QED can also be generated by a Kerr-type interaction (see \secref{sec:NonlinearProcKerr}), as observed in the experiment of Ref.~\cite{Eichler2011a}.

The JPC can be used for more tasks than just simple amplification. When the three modes of the JPC satisfy the relation $\omega_{\rm p} = \abs{\omega_{\rm s} - \omega_{\rm i}}$, the JPC can also operate as a noiseless tunable three-wave mixer~\cite{Roch2012}, a beam-splitter enabling fully coherent frequency conversion, or a photon combiner at the single-photon level~\cite{Abdo2013PRL}. Furthermore, by coupling two JPCs, one can construct a \textit{directional amplifier}, which eliminates the need for circulators~\cite{Abdo2013PRX}. A JPC can also used as a quantum node to store microwave photons in a memory and entangle them with a propagating output mode~\cite{Flurin2015}.

In nonlinear crystals, wave-mixing processes are restricted by the crystal symmetry~\cite{Boyd2008}. For example, second-order nonlinear processes can only occur in noncentrosymmetric media, i.e., in media without inversion symmetry. In centrosymmetric media, interference extinguishes the second-order processes such that wave-mixing only occurs through third- and higher odd-order processes. In SQCs, the intrinsic nonlinearity of an artificial atom enables \textit{symmetry breaking} (as discussed in \secref{sec:SelectionRules}) and thus three-wave mixing. There are various theoretical proposals for generating such second-order processes in SQCs. For instance, degenerate parametric down-conversion and generation of squeezing can be implemented by using two levels of a charge qubit operated away from its charge degeneracy point~\cite{Moon2005}. In such a system, the gate charge controls the energy-level spacing and, thus, a qutrit (i.e., three-level system) can be tailored for parametric down-conversion, assuming appropriate transition frequencies~\cite{Marquardt2007b, Koshino2009, Sanchez-Burillo2016}. Such a qutrit needs to have cyclic, $\Delta$-type transitions, which is possible in a variety of superconducting qubits, as discussed in \secref{sec:ThreeLevel}. Using a $\Delta$-type configuration, controllable three-wave mixing can be achieved~\cite{Liu2014a, Wang2014FeqConversion}.

Closely connected to the discussion in the previous paragraph is the issue of when \textit{multiphoton} excitation processes can be used for frequency conversion. In natural atomic systems, the parity of a single-photon excitation is odd, but a two-photon excitation is forbidden by selection rules (see \secref{sec:SelectionRules}) due to the inversion symmetry of the corresponding Hamiltonian. However, for some superconducting qubits this symmetry can be broken, allowing excitation by both one- and two-photon processes~\cite{Liu2014}. In experiments with a flux qubit coupled to a resonator, frequency up-conversion of microwave photons through such multiphoton processes has been demonstrated~\cite{Deppe2008, Niemczyk2009, Niemczyk2011}, using that the symmetry of the potential well for the flux qubit can be broken by applying an external magnetic flux. The experiment in Ref.~\cite{Deppe2008} also demonstrated coexistence of one- and two-photon processes, which was first predicted theoretically in Ref.~\cite{Liu2005a}. We also note that another experiment has shown that transitions in a pair of coupled flux qubits can be selectively excited or suppressed by engineering the corresponding selection rules~\cite{DeGroot2010}.

Several SQC experiments have demonstrated frequency conversion based on other mechanisms as well. For example, coherent frequency conversion between the two oscillation modes of a dc SQUID phase-qubit circuit with two internal degrees of freedom has been observed~\cite{Lecocq2012}. Furthermore, using an impedance-matched $\Lambda$-type system (formed by a qubit coupled to a resonator) with equal decay rates from the highest energy level~\cite{Koshino2013, Koshino2013a}, efficient microwave down-conversion through Raman transitions has been demonstrated~\cite{Inomata2014}. Frequency conversion has also been achieved in setups with coupled tunable resonators~\cite{Chirolli2010, Zakka-Bajjani2011}.

Finally, we note that the generalized Rabi Hamiltonian, describing the coupling between a superconducting flux qubit and a resonator, as discussed in Secs.~\ref{sec:LightMatterCoupling} and \ref{sec:UltrastrongTheory}, allows for processes which do not conserve the number of excitations in the system. These higher-order processes include \textit{multiphoton} vacuum Rabi oscillations, where two or more photons in the resonator are converted into a single qubit excitation and back~\cite{Ma2015, Garziano2015}, and a \textit{single} photon exciting \textit{multiple} qubits in a similar fashion~\cite{Garziano2016}. It was recently shown that almost \textit{any} analogue of nonlinear-optics processes, including all three- and four-wave mixing processes, the Kerr effect, etc., can be realized in circuit-QED setups with one or more resonators ultrastrongly coupled to one or more qubits in this way~\cite{Anton2017}. These analogues work \textit{without} any external drive, deterministically converting \textit{single} photons confined in resonators. Examples of such analogues include qubit-controlled frequency conversion processes between photons in two resonator modes coupled to a single tunable qubit~\cite{Kockum2017}.

\subsection{Third-order nonlinearity and Kerr interaction}
\label{sec:NonlinearProcKerr}

Photons usually do not interact with each other. This makes them ideal information carriers, but also makes it difficult to coherently manipulate their states. One way to make photons interact is to utilize the \textit{Kerr} or \textit{cross-Kerr} effects, which arise in media which have a third-order nonlinear susceptibility $\chi^{(3)}$. In the Kerr effect, the phase of a light field passing through such a medium is changed by an amount proportional to the intensity of the field \textit{itself}. In the cross-Kerr effect, the phase of the propagating field is changed by an amount proportional to the intensity of \textit{another} field. Since the $\chi^{(3)}$ nonlinearity in conventional materials is negligibly small at single-photon light intensities, photon-photon interaction is only possible in such setups when high-intensity light is used. However, there are many applications in quantum information science for quantum nonlinear optics that only uses single quanta of light~\cite{Chang2014}.

To realize a Kerr interaction in SQCs, one method is to use the inherent nonlinearity of a Josephson junction. As discussed in \secref{sec:SQCsAndJJs}, this nonlinearity stems from the Josephson-energy term proportional to $\cos\phi$, where $\phi$ is the gauge-invariant phase difference between the two superconducting electrodes of the junction. By expanding this term and introducing bosonic operators $a$ and $a^\dag$, one of the terms that one obtains is the Kerr interaction $K (a^\dag a)^2$, where $K$ is called the Kerr coefficient.

Kerr interaction based on a Josephson junction has been widely used for the amplification, bifurcation, and squeezing of quantum signals~\cite{Vijay2009} (see also \secref{sec:PhotonDetectionHomodyne}). In such Josephson parametric and bifurcation amplifiers, the Kerr nonlinearity is usually weaker than the photon decay rate $\kappa$. However, theoretical studies have shown that embedding the Josephson junction in a transmission line makes it possible to reach a wide range of values for $K$~\cite{Bourassa2012, Nigg2012, Leib2012}, including $K>\kappa$. Such setups have also been used to enhance qubit-resonator coupling, which can reach the ultrastrong~\cite{Bourassa2009, Niemczyk2010} and even deep-strong~\cite{Yoshihara2017} coupling regimes, as discussed in \secref{sec:ultrastrong}.

\begin{figure}
\centering
\includegraphics[width=\linewidth]{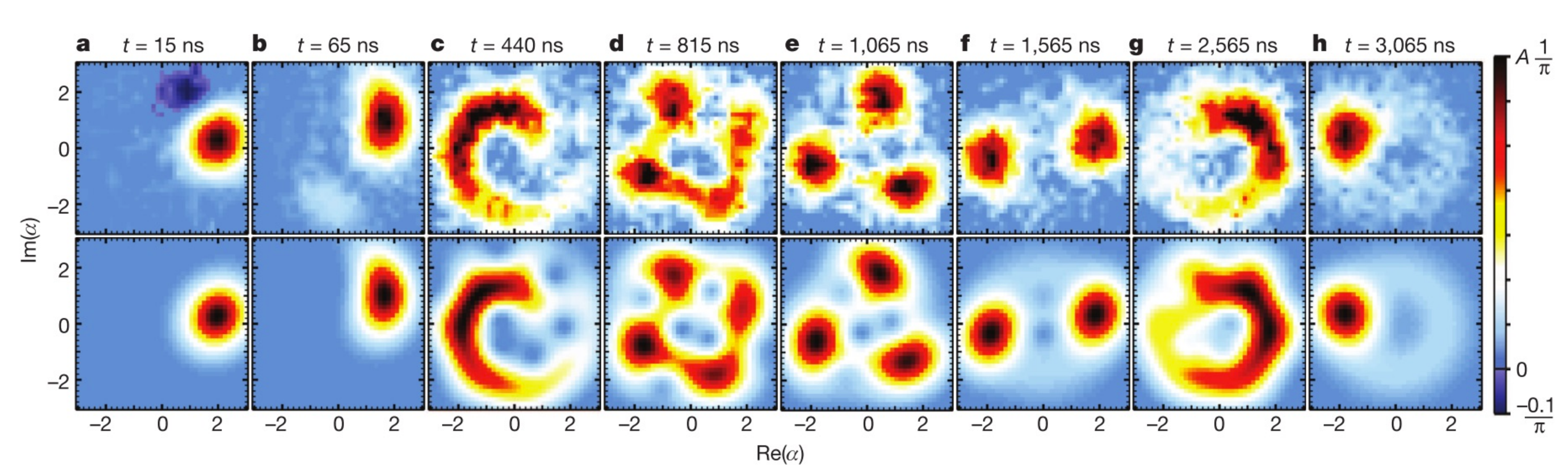}
\caption{Collapse and revival of a coherent state due to the single-photon Kerr effect~\cite{Kirchmair2013}. The system, a cavity with a Kerr nonlinearity induced by a qubit, starts in a coherent state with an average photon number of four. From left to right, the time evolution of the Husimi Q function for the cavity is shown. The upper row is experimental data; the lower row is a theoretical simulation. During the time evolution, it is clearly seen that superpositions of two, three, and even four coherent states appear before the system returns to a single coherent state again, as predicted in Ref.~\cite{Miran1990}. Such two- and multi-component
macroscopically-distinct superpositions are often referred to as Schr\"odinger cat and kitten states, respectively. 
Reprinted figure by permission from Macmillan Publishers Ltd: Nature, G.~Kirchmair et al.,~\href{http://dx.doi.org/10.1038/nature11902
}{Nature \textbf{495,} 205 (2013)}, copyright (2013). 
\label{fig:Kerr}}
\end{figure}

Another way to achieve Kerr interaction in SQCs is to couple the microwave field in a resonator dispersively to a qubit~\cite{Boissonneault2009, Boissonneault2010, Zhu2013a}. A more complicated setup, with two coupled charge qubit forming an $N$-type four-level system coupled to a resonator, has also been predicted to give rise to a large Kerr interaction~\cite{Rebic2009}. In the first experimental demonstration of the single-photon Kerr regime ($K>\kappa$), a 3D microwave resonator was dispersively coupled to a transmon qubit~\cite{Kirchmair2013}. This experiment confirmed that a coherent state in a Kerr medium experiences \textit{collapses} and \textit{revivals}, as shown in~\figref{fig:Kerr}. During this evolution, macroscopically distinct superpositions of two~\cite{Yurke1986} and more~\cite{Miran1990, Tanas1991} coherent states can be observed. These superpositions are usually interpreted as Schr\"odinger cat and cat-like (kitten) states, respectively.

The cross-Kerr effect has been experimentally demonstrated in SQCs both with resonator modes coupled dispersively to a qubit~\cite{Holland2015} and for two propagating microwave fields in an open transmission line interacting with separate transitions in a three-level transmon~\cite{Hoi2013a}. In the setup with resonators, the single-photon-resolved regime of the cross-Kerr interaction was reached. In the setup with propagating fields, a remarkably large phase shift of about 20 degrees per photon was observed. This latter setup has been studied theoretically to evaluate its potential as a microwave single-photon detector, as discussed further in \secref{sec:PhotonDetectionItinerant}. Somewhat surprisingly, it turns out that the cross-Kerr effect mediated by a single artificial three-level atom cannot be used to detect a single photon~\cite{Hoi2013a}. However, a chain of such three-level atoms, interspersed with circulators, can be used for this purpose~\cite{Sathyamoorthy2014}.

\section{Photon generation}
\label{sec:PhotonGeneration}

The generation of nonclassical states of light, such as single photons, is essential to many protocols for quantum computation and quantum communication. In quantum optics, single-photon \textit{generation} has posed a greater challenge than single-photon \textit{detection}. As we will see in this section and in~\secref{sec:PhotonDetection}, for microwave photonics with superconducting artificial atoms the situation is rather the opposite.

In this section, we review experimental and theoretical work on photon generation with superconducting circuits. We begin in~\secref{sec:PhotonGenerationQuantumOpticsApproaches} with a brief overview of the methods that have been used for photon generation in quantum optics. We then look at photon generation schemes for superconducting circuits in~\secref{sec:PhotonGenerationSCCircuits}, dividing them into setups without cavities, i.e., open transmission lines, setups with one cavity, and setups with two cavities (e.g., generation of NOON states). For lasing in SQCs, see \secref{sec:Lasing}.

\subsection{Approaches to photon generation in quantum optics}
\label{sec:PhotonGenerationQuantumOpticsApproaches}

Ideally, a system for single-photon generation should be \textit{deterministic} (able to produce photons ``on-demand''), not suffer from losses or accidental emission of multiple photons, have arbitrarily high repetition rate, and be able to \textit{shape} the photon wavepacket. As reviewed in Refs.~\cite{Lounis2005, Buller2010, Santori2010, Eisaman2011, Migdall2013, Chunnilall2014, Takeuchi2014}, deterministic sources investigated in quantum optics have usually been based on single quantum emitters, which are first excited by a classical pulse and then emit a photon when relaxing to a lower state. Examples of such quantum emitters (giving just a few references) are natural atoms~\cite{Brattke2001, McKeever2004, Hijlkema2007, Dayan2008, Nisbet-Jones2011, Mucke2013, Reiserer2015}, ions~\cite{Keller2004, Barros2009, Higginbottom2016}, molecules, quantum dots~\cite{Yuan2002, Laucht2012, He2013, Versteegh2014, Lodahl2015, Ding2016, Daveau2016, Loredo2016}, and color centers in diamond. Another approach is to drive a nonlinear medium to induce spontaneous parametric down-conversion (SPDC)~\cite{Soujaeff2007, Kolchin2008} or four-wave mixing~\cite{Gulati2014}, producing photon pairs. These methods are not deterministic, but \textit{probabilistic}; however, the photon production can be heralded by detecting one of the photons in the pair.

\begin{figure}
\centering
\includegraphics[width=0.7\linewidth]{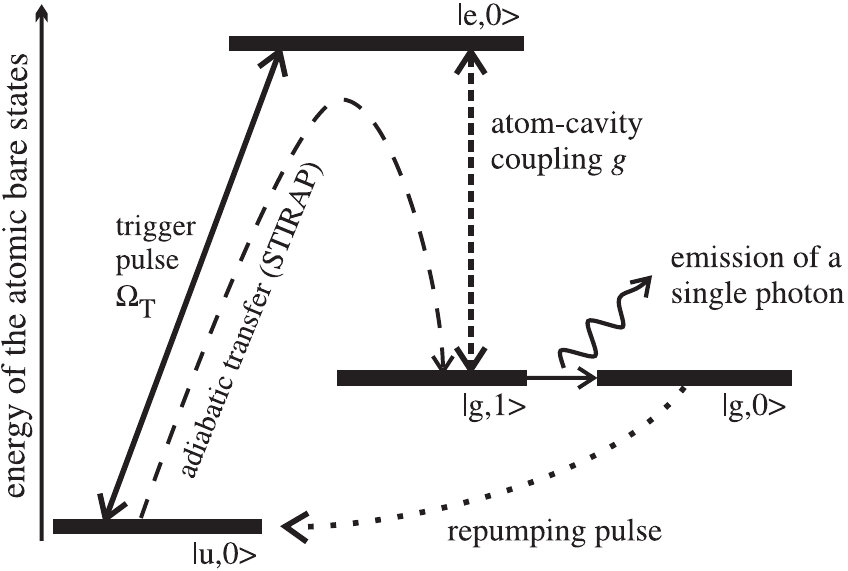}
\caption{A sketch of the level structure used for single-photon generation through a Raman process~\cite{Kuhn1999}. A classical drive with amplitude $\Omega_{\rm T}(t)$ is applied to the transition between the ground state $\ket{u}$ and the excited state $\ket{e}$ of a $\Lambda$-type three-level atom. A cavity is coupled to the $\ket{g} \leftrightarrow \ket{e}$ transition (but far detuned from $\ket{u} \leftrightarrow \ket{e}$) such that the excited atom can relax by emitting a cavity photon. In the vSTIRAP scheme, outlined in the main text, the drive is changed adiabatically in a way that generates a photon without populating $\ket{e}$ in the process. 
Reprinted figure from 
A.~Kuhn et al., \href{http://dx.doi.org/10.1007/s003400050822}{Appl.~Phys.~B \textbf{69}, 373 (1999)}, with permission from Springer Science and Business Media.	
\label{fig:STIRAP}}
\end{figure}

For deterministic photon generation from a single quantum emitter, the setup illustrated in \figref{fig:STIRAP} has been the subject of much theoretical~\cite{Parkins1993, Law1997, Kuhn1999, Brown2003, Vasilev2010, Mucke2013} and experimental~\cite{Keller2004, Hijlkema2007, Barros2009, Nisbet-Jones2011, Mucke2013} work. Here, a three-level atom in a $\Lambda$ configuration has one transition driven by a classical pulse and the other transition coupled to a cavity, such that a Raman process can generate a single photon in the cavity. However, if the atom starts in its ground state and the classical drive is increased adiabatically, the system can be kept in a dressed state involving only the two lower levels, resulting in the generation of a single photon without ever populating the excited state of the atom. This scheme is called \textit{vacuum-stimulated Raman adiabatic passage} (vSTIRAP). By more advanced control of the drive, the shape of the photon wavepacket can be designed~\cite{Keller2004, McKeever2004, Vasilev2010, Nisbet-Jones2011}. For a recent review of STIRAP, see Ref.~\cite{Vitanov2016}. Photon shaping has also been implemented with SPDC~\cite{Kolchin2008} and an atomic-ensemble quantum memory~\cite{Farrera2016}. Heralded photon shaping can be achieved by modulating one photon from a pair and then detecting it; the second photon can then acquire the same shape~\cite{Sych2016, Averchenko2016}.

Another interesting approach to photon generation is through \textit{projective measurements}. By measuring the phase shifts of Rydberg atoms passing through a microwave cavity, the photonic Fock states in the cavity can be distinguished~\cite{Gleyzes2007} (see \secref{sec:PhotonDetectionQNDandCavity}). Initializing the cavity in a coherent state using a classical drive, the measurement can then be used to project the cavity into a Fock state~\cite{Guerlin2007, Deleglise2008}. In this case, the Fock state generated is random, but by adding feedback to the system, either with excitations in the atoms or with coherent pulses, a specific Fock state can be created~\cite{Geremia2006, Dotsenko2009, Sayrin2011, Zhou2012, Peaudecerf2013}.

\subsection{Photon generation with superconducting circuits}
\label{sec:PhotonGenerationSCCircuits}

The strong coupling, tunability, and controllability of superconducting artificial atoms that make them promising for quantum computation also makes them well suited for photon generation. As we will see below, this has allowed for impressive engineering of Fock states in cavities, going well beyond what has been achieved in quantum optics previously.

\subsubsection{No cavity}\label{sec:NoCavity}

Neither a cavity nor an atom is actually necessary for photon generation. This is exemplified by the \textit{dynamical Casimir effect} (see \secref{sec:Resonators}), where a rapidly oscillating mirror produces pairs of photons whose frequencies sum to the oscillation frequency of the mirror (note that the photon generation is not deterministic). With superconducting circuits, the moving mirror can be implemented as a flux-tunable SQUID at the end of a transmission line~\cite{Johansson2009, Johansson2010, Johansson2013} and photon pair production has been demonstrated in such an experiment~\cite{Wilson2011}. Photons have also been produced in this manner in a resonator with a SQUID at one end~\cite{Wilson2010b, Johansson2010} and could also be created in two resonators connected through a SQUID~\cite{Felicetti2014}. As reviewed in Ref.~\cite{Nation2012}, a number of other photon-generating relativistic effects, e.g., Hawking radiation, could potentially be realized with superconducting circuits.

\begin{figure}
\centering
\includegraphics[width=\linewidth]{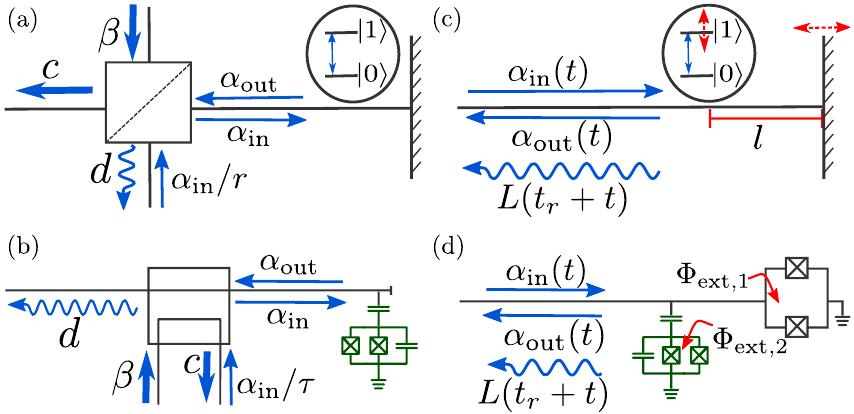}
\caption{Two setups for photon generation with a single two-level superconducting artificial atom at the end of a semi-infinite transmission line~\cite{Sathyamoorthy2015b}. (a) A coherent $\pi$-pulse $\alpha_{\rm in}$ excites the atom. The coherent part $\alpha_{\rm out}$ of the resulting output is removed by destructive interference with another coherent drive $\beta$ sent in through a beam-splitter with nearly perfect reflection, leaving only a single photon as output $d$. (b) A superconducting circuit for the setup in (a), with a directional coupler as the beam-splitter and a transmon qubit as the atom. (c) Alternatively, the atom can be placed at a tunable distance from the mirror. After the incoming pulse excites the qubit (weakly coupled to the transmission line), the distance is changed such that the coupling becomes zero. At a later time, the distance is changed again and the coupling is increased to release a photon. (d) The effective distance to the mirror can be changed either by tuning the qubit frequency or by terminating the transmission line with a SQUID. 
Reprinted figure with permission from 
S.~R.~Sathyamoorthy et al., \href{http://dx.doi.org/10.1103/PhysRevA.93.063823}{Phys.~Rev.~A \textbf{93}, 063823 (2016)}. \textcircled{c} 2016 American Physical Society.	
\label{fig:SathyamoorthyGenerationScheme}}
\end{figure}

Other cavity-less approaches involve either a driven Josephson junction~\cite{Leppakangas2014, Leppakangas2015, Leppakangas2016} or a superconducting artificial atom in an open or semi-infinite transmission line. While it has been shown that an $N$-type four-level atom in an open transmission line can give rise to photon blockade and increased probability of transmitting single-photon states of a coherent drive~\cite{Zheng2011}, most works focus on a two-level atom. As discussed in~\secref{sec:WaveguideQEDExpSCAtoms}, a two-level atom in an open transmission line can only reflect one photon at a time from a coherent drive (the antibunching of the reflected signal has been measured in an experiment~\cite{Hoi2012}). However, there is still a large probability of the reflected part of a coherent pulse being a zero-photon state, which means this will not be a deterministic photon source~\cite{Lindkvist2014}. More promising is to excite a two-level atom, strongly coupled to a semi-infinite transmission line, via a weakly coupled excitation line such that the atom relaxes by releasing a photon in the single direction of the strongly coupled line~\cite{Lindkvist2014, Peng2015b}. An experimental demonstration of this concept, yielding an efficiency above 50\% over a wide frequency range with a frequency-tunable flux qubit, was recently given~\cite{Peng2015b}. There are also proposals~\cite{Sathyamoorthy2015b} which do not require a separate excitation line, as shown in~\figref{fig:SathyamoorthyGenerationScheme}. A scheme for multiphoton generation using a large number of driven multi-level atoms in a waveguide has also been proposed~\cite{Gonzalez-Tudela2016}. Finally, traveling Schr\"odinger cat states could potentially be implemented in circuit QED by combining a Josephson traveling-wave amplifier, a microwave beam-splitter, and a single-photon detector to subtract a single photon from a squeezed state~\cite{Joo2016}.

\subsubsection{One cavity}
\label{sec:OneCavity}

With a single resonator, usually coupled to an artificial atom, there are many possible schemes for generating photons. We therefore sub-divide this section into parts dealing with the generation and control of Fock states in a cavity, shaping of the photon wavepacket (using various tunable couplings), photon generation using ultrastrong coupling, and other schemes.

\paragraph{Fock-state engineering and control}

\begin{figure*}
\centering
\includegraphics[width=1\linewidth]{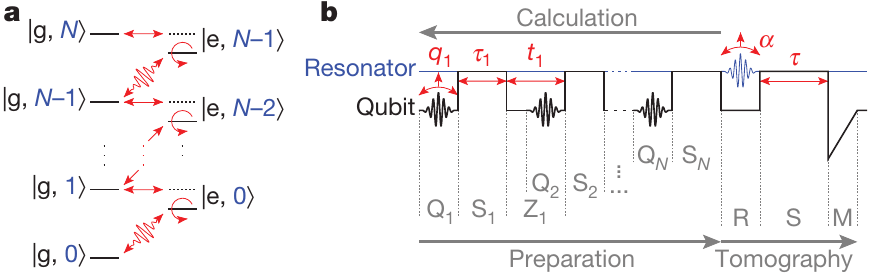}\vspace{0.5cm}
\includegraphics[width=1\linewidth]{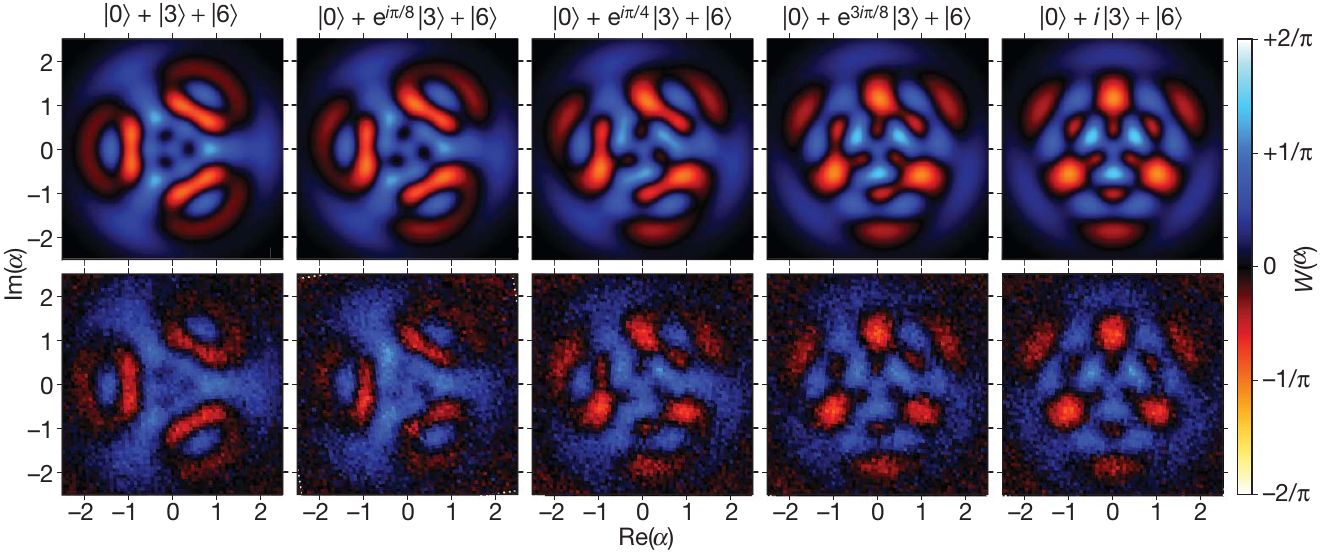}
\caption{Generation of arbitrary Fock-state superpositions in a cavity~\cite{Hofheinz2009}. (a) Energy levels for the qubit-resonator system with arrows representing the three types of operations used: undulating lines for driving the qubit (Q), straight horizontal lines for swapping states between the qubit and the resonator by bringing them on resonance for a certain time (S), and circles for phase rotations of the qubit (Z). (b) Pulse sequence for the state preparation, showing the qubit frequency in black, the resonator frequency in blue, and the pulses in red. The pulse sequence is calculated by starting from the desired final state and calculating backwards to the ground state. After the state has been prepared, it is measured with quantum tomography. (Bottom) Wigner tomography of several superposition states, showing theoretical simulations in the upper row and experimental results in the lower row. 
Reprinted figure by permission from Macmillan Publishers Ltd: Nature, M.~Hofheinz et al.,~\href{http://dx.doi.org/10.1038/nature08005}{Nature \textbf{459}, 546 (2009)}, copyright (2009).	
\label{fig:FockStatePreparationSchemeAndResults}}
\end{figure*}

A photon generation scheme well suited for superconducting circuits is to excite an artificial atom with a $\pi$-pulse, tune it into resonance with a resonator for half a Rabi oscillation period, and then tune it out of resonance again. In this way, a single photon is \textit{deterministically} generated in the resonator. By exciting the qubit again, a second photon can be added to the resonator, and so on. Generalizing the scheme to include all qubit rotations and varying interaction times, \textit{arbitrary} superpositions of Fock states can be generated, which was shown theoretically in the 90's~\cite{Vogel1993, Law1996} and then suggested for implementation with superconducting circuits~\cite{Liu2004}; see also \figref{fig:FockStatePreparationSchemeAndResults}. Later, it has been shown that arbitrary Fock states can be generated more rapidly, either by simultaneously driving multiple sideband transitions in the Jaynes--Cummings Hamiltonian with the qubit resonant with the resonator during the whole procedure~\cite{Strauch2012}, or by including longitudinal coupling to the qubit that makes multi-photon processes possible~\cite{Zhao2015}. It is also possible to generate Schr\"odinger cat states in a resonator coupled to a single qubit~\cite{Liu2005}.

The first photon-generation experiment with superconducting circuits used a $\pi$-pulse to excite a qubit which then emitted a photon as it relaxed (randomly) through Purcell decay~\cite{Houck2007}. In the following years, experiments using the scheme outlined above deterministically transferred a single photon to a resonator~\cite{Sillanpaa2007}, created single Fock states containing several photons~\cite{Hofheinz2008, Wang2008}, created arbitrary superpositions of Fock states in the resonator~\cite{Hofheinz2009, Wang2009} (see \figref{fig:FockStatePreparationSchemeAndResults} for details), and generated single photons leaving their resonators for further measurements~\cite{Bozyigit2011, Eichler2011, Lang2013}. Sideband transitions have also been used to generate multi-photon Fock states~\cite{Leek2010}. Also, the generation of large \textit{cat states} has been demonstrated~\cite{Wang2009, Vlastakis2013}.

The next step beyond generation of arbitrary superpositions of Fock states is \textit{universal control}, i.e., the ability to perform arbitrary unitary operations on Fock states in the resonator~\cite{Santos2005, Krastanov2015}. This can be achieved in superconducting circuits through a selective number-dependent arbitrary phase (SNAP) gate, which imparts phases $\theta_n$ on the different number states $\ket{n}$~\cite{Krastanov2015, Heeres2015}. An experimental demonstration, including the generation of the Fock state $\ket{1}$, has been given with a qubit dispersively coupled to a resonator strongly enough that the number-dependence of the dispersive shift to the qubit frequency can be resolved~\cite{Heeres2015}.

\paragraph{Shaping photons}
\label{sec:ShapingPhotons}

The wavepacket of a generated photon can be \textit{shaped} by modulating the coupling strength between the generating and receiving systems~\cite{Gough2012}. This has applications for the transfer of photons (and thus quantum information) between separate resonators~\cite{Cirac1997, Jahne2007, Korotkov2011, Sete2015} since an exponentially decaying wavepacket will be partly reflected by the receiving resonator. With superconducting circuits, an experiment has demonstrated absorption with efficiency above 99\% of a shaped coherent pulse arriving at a resonator~\cite{Wenner2014}.

The coupling that is modulated can be either between a qubit and the resonator or between the resonator and an outgoing transmission line. Tunable qubit-resonator coupling has been implemented experimentally both using a tunable mutual inductance, coupling a phase qubit and a lumped-element resonator~\cite{Allman2010}, using a three-island transmon qubit with a tunable dipole moment~\cite{Srinivasan2011, Gambetta2011, Hoffman2011a}, and using a three-level transmon where a tunable drive couples the states $\ket{g,1}$ and $\ket{f,0}$~\cite{Pechal2014, Zeytinoglu2015}. In the second setup, the coupling has been tuned by a factor of 1500~\cite{Hoffman2011a}. The last scheme has been used to control both amplitude, phase, and shape of single microwave photons when generating them~\cite{Pechal2014}.

Tunable coupling between a resonator and a transmission line has been demonstrated in two experiments. In the first, an inductive coupler including a flux-tunable SQUID was used to rapidly tune the resonator decay rate by more than two orders of magnitude and shape outgoing photons~\cite{Yin2013}. In the second experiment, a second, frequency-tunable resonator containing a SQUID was placed between the first resonator and the transmission line. By tuning the second resonator in and out of resonance with the first, the decay rate of the first resonator was modified by three orders of magnitude~\cite{Pierre2014}.

\paragraph{Ultrastrong coupling}

In the regime of ultrastrong coupling (see \secref{sec:ultrastrong}) between a qubit and a resonator, the ground state is not $\ket{g,0}$, but instead contains contributions from \textit{virtual} photons and qubit excitations. By modulating the coupling strength, photon pairs can be generated in a way that is reminiscent of the dynamical Casimir effect~\cite{Ciuti2005, DeLiberato2007, DeLiberato2009}. In fact, even an unmodulated thermalized ultrastrongly coupled qubit-resonator system displays interesting photon emission statistics. Usually, thermal radiation is bunched with $g^{(2)}(0) = 2$. However, at low temperatures and ultrastrong coupling, $g^{(2)}(0)$ approaches zero, which corresponds to single-photon emission~\cite{Ridolfo2013}.

If an auxiliary third level $\ket{s}$, having lower energy than $\ket{g}$ and with its transitions not ultrastrongly coupled to the resonator, is added to the qubit, additional ways to generate photons from the ultrastrongly coupled system's ground state become possible. Simple spontaneous relaxation can then cause the system to sometimes end up in $\ket{s,2}$, not only $\ket{s,0}$~\cite{Stassi2013}. The transition to $\ket{s,2}$ can be stimulated with an external drive to make the process deterministic~\cite{Huang2014}; photons can also be reabsorbed back into the vacuum using such a drive~\cite{DiStefano2017}.

\paragraph{Other schemes}

Beyond the photon-generation methods mentioned above, there are several more that have either been experimentally implemented with or suggested for superconducting circuits. Theoretical proposals include using a Raman scheme to generate single photons from a $\Lambda$ system formed by a flux qubit~\cite{Mariantoni2005}, realizing parametric down-conversion to produce photon pairs with a three-level Cooper-pair box in a resonator~\cite{Marquardt2007b}, implementing lasing by configuring a flux qubit as a three-level delta system~\cite{You2007a}, using a superconducting single-electron transistor in a weakly anharmonic resonator to squeeze the photon number distribution in the resonator~\cite{Marthaler2008}, and pulsed driving of a nonlinear oscillator~\cite{Gevorgyan2012}. STIRAP (see \secref{sec:PhotonGenerationQuantumOpticsApproaches}) and other adiabatic passage schemes have been considered in a number of theoretical works~\cite{Wei2008, Falci2013, Chen2015, DiStefano2016, Vepsalainen2016, Falci2017}.

On the experimental side, photons have been generated through adiabatic passage from $\ket{e,0}$ to $\ket{g,1}$ by tuning the frequency of a transmon qubit~\cite{Johnson2010} and through driving the sideband transition $\ket{g,0} \to \ket{e,1}$ in a setup where the cavity decays much faster than the qubit~\cite{Kindel2016}. Some more recent adiabatic-passage experiments can be found in Refs.~\cite{Xu2016, Kumar2016, Premaratne2017}; in Ref.~\cite{Premaratne2017}, Fock states with up to three photons were generated. Furthermore, photon blockade~\cite{Imamoglu1997} resulting from the nonlinearity of the Jaynes--Cummings Hamiltonian (see \secref{sec:JaynesCummings}) has been used to realize antibunched single-photon transmission through a resonator coupled to a transmon qubit~\cite{Lang2011}. Fock states have also been manipulated with a resonator resonant with the second transition in a transmon qubit~\cite{Juliusson2016}. Finally, several theoretical~\cite{Armour2013, Gramich2013, Dambach2015, Souquet2016} and experimental~\cite{Hofheinz2011, Chen2014b} studies have explored a setup with a voltage-biased Josephson junction coupled to a resonator. When a Cooper pair tunnels across the junction, one (or more) photons can be created in the cavity, provided that the Josephson frequency matches the resonator frequency.

\subsubsection{Two or more cavities}
\label{sec:TwoOrMoreCavities}

\begin{figure}
\centering
\includegraphics[width=\linewidth]{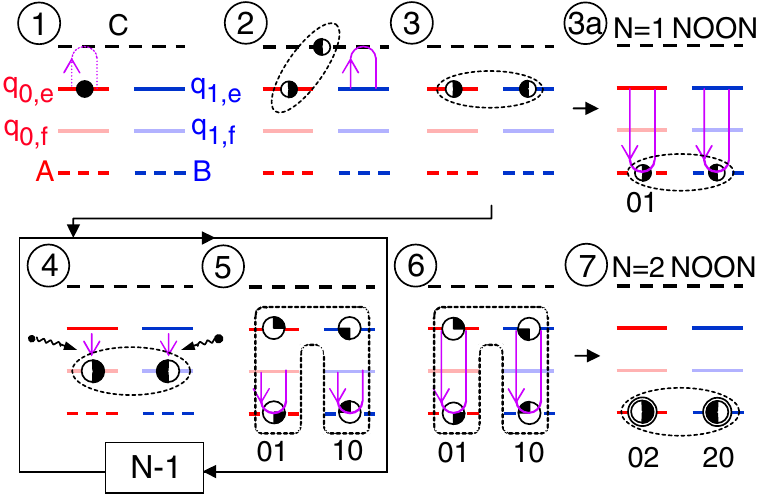}
\caption{An experimental protocol for the generation of NOON states~\cite{Wang2011PRL}. Three resonators, $A$, $B$, and $C$, are coupled to two artificial atoms, $q_0$ and $q_1$, with levels $\ket{g}$, $\ket{e}$, and $\ket{f}$. In step 1, $q_0$ is excited to $\ket{e}$ by a $\pi$-pulse. In step 2, $q_0$ is tuned into resonance with $C$ for a quarter of a Rabi oscillation period, forming the state $(\ket{0,e} + \ket{1,g})/\sqrt{2}$. Thereafter $C$ is tuned into resonance with $q_1$ for half a Rabi period, swapping its excitation to create the Bell state $(\ket{e,g} + \ket{g,e})/\sqrt{2}$ in step 3. To form the NOON state with $N=1$, $q_0$ and $q_1$ are then tuned into resonance with $A$ and $B$ to swap their respective states (step 3a). To generate NOON states with higher $N$, the atoms are instead excited from $\ket{e}$ to $\ket{f}$ (step 4) and then swap their higher excitations into $A$ and $B$ (step 5). The process is repeated until there are $N-1$ photons in the cavities. Then the excitations from the $\ket{e}$ states are swapped into $A$ and $B$ (steps 6-7). 
Reprinted figure with permission from H.~Wang et al.,~\href{http://dx.doi.org/10.1103/PhysRevLett.106.060401}{Phys.~Rev.~Lett.~\textbf{106}, 060401 (2011)}. \textcircled{c} 2011 American Physical Society.	
\label{fig:NOON}}
\end{figure}

Highly entangled states of photons in \textit{two} modes, e.g., the NOON state $(\ket{N,0} + \ket{0,N})/\sqrt{2}$, can improve measurement sensitivity~\cite{Boto2000} and have been realized in quantum optics experiments~\cite{Mitchell2004, Nagata2007, Vergyris2016}. In the last few years, a number of protocols have been proposed for the generation of NOON states in two superconducting resonators~\cite{Merkel2010, Strauch2010, Strauch2012, Su2014, Xiong2015, Zhao2016}, aided by one or two multi-level artificial atoms. In an experiment with three superconducting resonators connected to two phase qubits, the creation of NOON (and MOON, $(\ket{M,0} + \ket{0,N})/\sqrt{2}$, $M \neq N$) states with $N \leq 3$ has been demonstrated~\cite{Wang2011PRL}. The experimental protocol is explained in \figref{fig:NOON}. In a similar experimental setup, one and two photons created according to the protocol of \figref{fig:FockStatePreparationSchemeAndResults} have also been shuffled between the resonators via the qubits in a ``photon shell game'' and a ``tower of Hanoi game''~\cite{Mariantoni2011}. Entangled photons can also be generated with a SQUID implementing parametric frequency conversion between two cavity modes~\cite{Nguyen2012} and in a setup with an array of resonators coupled to a voltage-biased Josephson junction~\cite{Dambach2016, Dambach2016a}. The latter method is an extension of the scheme discussed above for a single resonator; here, the resonator frequencies sum to the Josephson frequency.

Schemes featuring two resonators can also be used for single-photon generation. A coherent drive on the first of two coupled nonlinear resonators (a setup sometimes referred to as a \textit{Bose--Hubbard dimer}) has been shown to generate antibunched photons from the first resonator~\cite{Liew2010}. Actually, only the second resonator needs to be nonlinear (the nonlinearity can be due to an atom coupling to the resonator) and the effect persists even if the nonlinearity is no larger than the resonator linewidth~\cite{Bamba2011}. This ``unconventional photon blockade", which was first investigated
in Refs.~\cite{Leonski2004,Miranowicz2006}, has been studied further in a number of theoretical works~\cite{Ferretti2010, Bamba2011a, Flayac2013, Ferretti2013, Xu2013b, Flayac2015}, with an explanation in terms of Gaussian squeezed states being given in Ref.~\cite{Lemonde2014}. The setup has been realized in an experiment with superconducting circuits, although in this instance it was used for quantum amplification instead of photon generation~\cite{Eichler2014a}. 

In a recent experiment, two resonators coupled to a superconducting transmon qubit realized a single-photon-resolved cross-Kerr interaction~\cite{Holland2015}. This allowed for the implementation of a protocol, based on driving both resonators, to stabilize a single-photon Fock state in one of the resonators. The same experimental setup, with the addition of a readout resonator for the qubit, has also been used to realize a two-mode cat state distributed over the two resonators~\cite{Wang2016}. Two resonators coupled to a superconducting qubit with both longitudinal and transversal couplings can also be used to realize photon blockade and single-photon generation~\cite{Wang2016b}.

\section{Photon detection}
\label{sec:PhotonDetection}

For most experiments in microwave photonics with superconducting artificial atoms, it is necessary to measure and characterize the photons in some manner. For a long time, this was an area where microwave photonics seemed to be at a disadvantage when compared to traditional quantum optics. At optical frequencies, there are good single-photon detectors~\cite{Hadfield2009, Buller2010, Eisaman2011, Migdall2013}. Furthermore, these detectors can be combined in different setups to realize homodyne and heterodyne detection, and also various correlation measurements. In microwave photonics, on the other hand, realizing a good single-photon detector is hard since the energy of a single microwave photon is roughly \textit{five orders of magnitude lower} than that of an optical photon. Consequently, detection in microwave photonics have so far mostly focussed on using amplifiers and linear detectors. In this way, not only homodyne and heterodyne detection, but also correlation measurements, have been realized. Using Josephson-junction based amplifiers, the available experimental technology is rapidly approaching ultimate quantum limits. In the last few years, there has also been considerable theoretical progress in the area  of single-photon detection for microwaves, which is now beginning to be implemented in experiments.

In this section, we review the current theoretical and experimental status for a number of detection schemes in microwave photonics. We first consider photon-number detection of both photons in cavities (\secref{sec:PhotonDetectionQNDandCavity}) and itinerant photons (\secref{sec:PhotonDetectionItinerant}). We then turn to homodyne and heterodyne detection implemented with amplifiers and linear detectors in~\secref{sec:PhotonDetectionHomodyne}. Finally, in~\secref{sec:PhotonDetectionCorrelation}, we look at how correlation measurements are realized. For the different kinds of measurements, we also discuss the form of measurement back-action they give rise to, as well as the concept of quantum nondemolition (QND) measurements~\cite{Braginsky1996}. Note that we here only consider measurements on \textit{photons}, not on atoms (though atoms are often used to realize measurements of photons).

\subsection{Quantum nondemolition measurements and detection of cavity photons}
\label{sec:PhotonDetectionQNDandCavity}

In general, detecting photons confined to a resonator is easier than detecting itinerant (traveling) photons. We therefore review the former case first. However, before we begin it is pertinent to briefly review the concept of QND measurements.

Measuring an observable $O$ in a QND way implies that subsequent measurements of the same variable will give the same result as the first measurement~\cite{Braginsky1980, Braginsky1996}, i.e., once the measurement has projected the system into an eigenstate of $O$, the subsequent measurement results are \textit{completely predictable}. It is far from always that an observable can be measured in a QND way. A simple example is the case of a free particle. Measuring the momentum $p$ of the particle will unavoidably introduce some uncertainty in the position $x$, but the momentum is conserved and subsequent measurements will give the same result. Thus, the momentum can be measured in a QND way. However, measuring $x$ cannot be done in a QND way, because any measurement of $x$ will introduce an uncertainty in $p$, which makes it impossible to predict the result of a subsequent measurement of $x$.

For the case of photon detection, a measurement is QND if it \textit{preserves the photon number}. For example, a QND measurement of photons in a resonator can project a coherent state into a Fock state, but subsequent measurements should give the same Fock state as a result (barring resonator decay).

QND measurements were first considered in the 1970s for detection of gravitational waves~\cite{Braginsky1975, Thorne1978, Unruh1979, Braginsky1980}, but were soon expanded to more general back-action evading measurements~\cite{Caves1980}. There have been many proposals and implementations in traditional quantum optics~\cite{Milburn1983, Yurke1985, Roch1992, Grangier1998} and lately also in quantum optomechanics and electromechanics~\cite{Clerk2008, Suh2014}.

The basic idea for measuring microwave photons in a resonator is to let them interact with an atom and then make a measurement on the atom. When the atom is on resonance with the resonator, we know from the Jaynes--Cummings model (see~\secref{sec:JaynesCummings}) that there will be Rabi oscillations with a frequency $\Omega_n \propto \sqrt{n+1}$ that depends on the photon number $n$. Letting the atom interact for a fixed time with the resonator and then measuring the state of the atom will thus give information about $n$. This scheme is \textit{not} QND since excitations are exchanged between the atom and the resonator. This type of measurement was first done with Rydberg atoms passing through a microwave cavity~\cite{Brune1996}. It has also been realized in circuit QED with phase qubits and used to study the generation and decay of Fock states in the resonator~\cite{Hofheinz2008, Wang2008, Hofheinz2009, Wang2009}.

A less invasive measurement can be realized by instead operating in the \textit{dispersive regime} of the Jaynes--Cummings model (see~\secref{sec:JaynesCummingsDispersive}). There, the photons shift the frequency of the atom by $2n\chi$ without any exchange of excitations taking place. Thus, a Ramsey experiment will reveal different phase shifts of the qubit depending on the photon number $n$. The drawback of the method is that it still requires multiple measurements to measure these phase shifts. Again, this method was first proposed~\cite{Brune1990} and realized experimentally~\cite{Nogues1999} for Rydberg atoms passing through a microwave resonator. The same setup has later been used for progressively more advanced measurements of the evolution of photon states in a resonator~\cite{Gleyzes2007, Guerlin2007, Bernu2008, Brune2008} and also for feedback on these states~\cite{Dotsenko2009, Sayrin2011, Zhou2012, Peaudecerf2013}. An overview of the experiments with Rydberg atoms can be found in Ref.~\cite{Haroche2013}.

\begin{figure}
\centering
\includegraphics[width=0.55\linewidth]{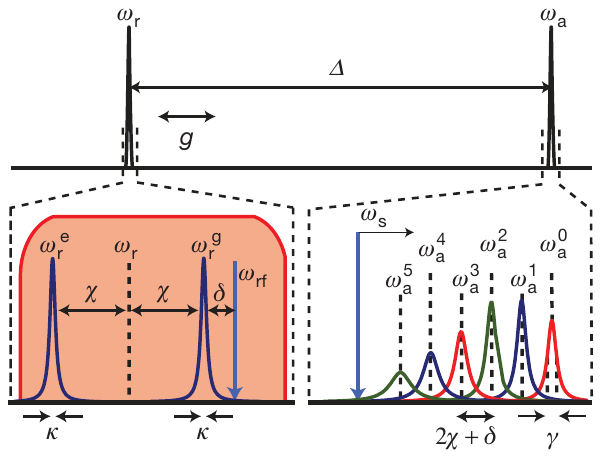}
\includegraphics[width=0.55\linewidth]{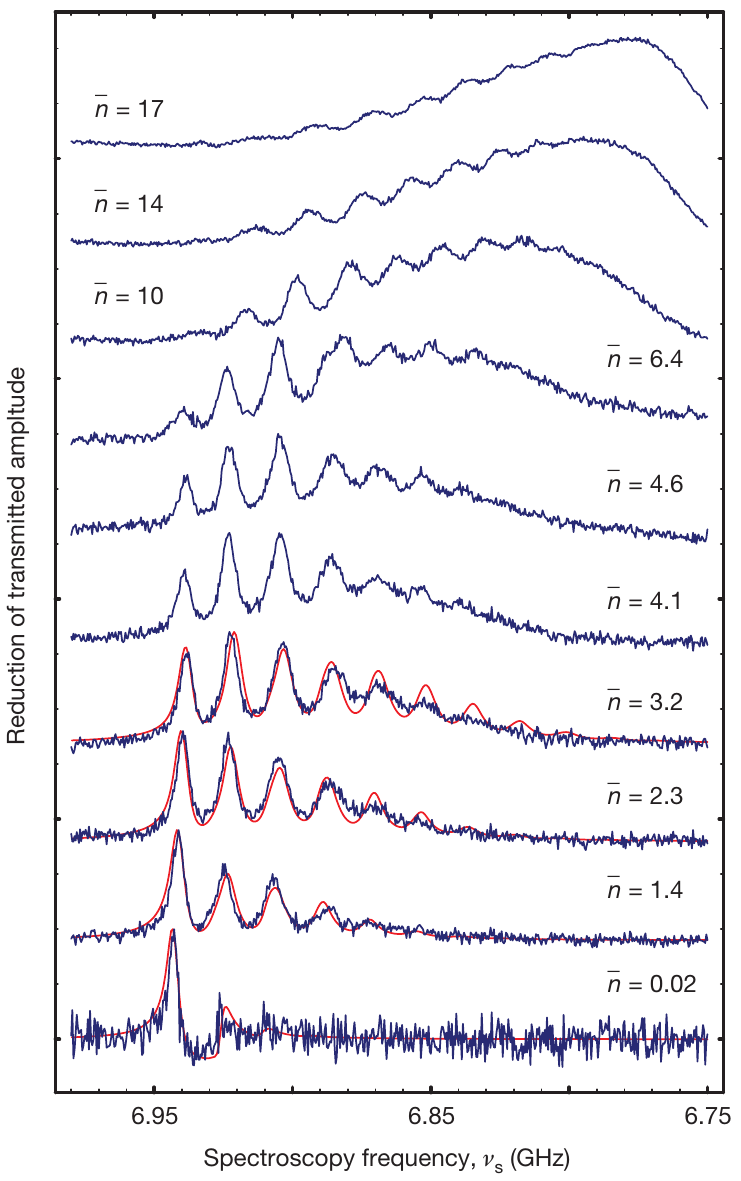}
\caption{A circuit QED measurement of photon number in a resonator~\cite{Schuster2007}. Top: Frequency diagram showing the resonator frequency $\omega_{\rm r}$ and the qubit frequency $\omega_{\rm a}$ together with their shifted resonances in the dispersive regime. Bottom: Results from two-tone spectroscopy (driving the qubit at $\nu_{\rm s}$ and measuring the transmission of the resonator at $\omega_{\rm rf}$) for different average photon numbers $\bar{n}$ in the resonator. Since $\chi \gg \gamma, \kappa$, several peaks, corresponding to different photon numbers, are clearly visible. 
Reprinted figure by permission from Macmillan Publishers Ltd: Nature, D.~I.~Schuster et al.,~\href{ http://dx.doi.org/10.1038/nature05461}{Nature \textbf{445,} 515 (2007)}, copyright (2007).	
\label{fig:SchusterPhotonDetection}}
\end{figure}

In circuit QED, the dispersive regime of the Jaynes--Cummings model has also been used to measure the photon number in a resonator with a charge qubit~\cite{Schuster2007}. There, both the dispersive shift of the qubit ($2n\chi $) and of the cavity ($\chi\sz$) were utilized, as shown in \figref{fig:SchusterPhotonDetection}. The transmission of the resonator at a frequency $\omega_{\rm rf}$ close to the resonance corresponding to the qubit ground state was measured while a second tone at frequency $\nu_{\rm s}$ was applied directly to the qubit. When $\nu_{\rm s}$ was on resonance with the shifted qubit frequency, the qubit was excited and a change in transmission at $\omega_{\rm rf}$ was observed since the resonator frequency had then been shifted by the qubit. In this way, by operating in the regime where $\chi$ is much larger than both the cavity decay $\kappa$ and the qubit decay $\gamma$, several Fock states could be distinguished. However, while the interaction between the qubit and the resonator is QND, the measurement is strictly speaking \textit{not} QND since the first probe tone affects the cavity.

A \textit{true} QND measurement of a single photon in a resonator was realized in circuit QED in 2010~\cite{Johnson2010, Johnson2011}. This time, a transmon qubit was coupled to a photon storage resonator in the quasi-dispersive regime, and also to a second resonator, used for qubit readout. By applying a $\pi$-pulse to the qubit at the frequency corresponding to a shift of $n$ photons, and then reading out the qubit state via the second cavity, a measurement of the type ``are there exactly $n$ photons in the storage resonator?'' was realized for $n = 0$ and $n=1$. The quasi-dispersive regime was used to realize a trade-off between QND interaction and large dispersive shifts, needed for the $\pi$-pulse to be frequency-selective. It should be possible to extend the scheme to higher $n$.

Recently, a refined version of the same setup (one transmon qubit and two resonators) was used to measure the \textit{photon-number parity} rather than the photon number in a QND way~\cite{Sun2014}. The procedure of that experiment is to start with the qubit in the ground state, apply a $\pi/2$-pulse, wait for a time $t = \pi/2\chi$, apply a second $\pi/2$-pulse, and finally read out the qubit via the second resonator. During the waiting time, the qubit acquires a phase $\exp (i n \pi)$. This means that the final measured state of the qubit ($\ket{e}$ or $\ket{g}$) will correspond to an even or odd number of photons in the storage resonator. There is also a proposal for measuring the phase of a superposition of Fock states in a resonator coupled to two qubits~\cite{Ng2010}.

\subsection{Detecting itinerant photons}
\label{sec:PhotonDetectionItinerant}

In traditional quantum optics, there are efficient detectors for \textit{itinerant} photons, based on, e.g., photomultipliers, avalanche photodiodes, superconducting transition-edge sensors, and superconducting nanowires~\cite{Lita2008, Hadfield2009, Buller2010, Eisaman2011, Marsili2013, Renema2014, Dauler2014, Chunnilall2014}. What all these detectors have in common is that the \textit{absorption} of the incoming photon gives rise to a signal that can be amplified and read out. Thus, these detectors are \textit{not} QND. 

In the case of microwave photonics, the detection schemes mentioned above do not work due to the low energy of the microwave photons. In the last few years, several theoretical proposals have been put forward to overcome this obstacle and a few experiments have been carried out. This research field was recently reviewed in Ref.~\cite{Sathyamoorthy2016} and there is also an earlier review in Ref.~\cite{Nation2012}, as well as a brief overview of photon generation and detection in Ref.~\cite{Yamamoto2013}. While the experimentally available microwave-photon detectors are not yet as good as their optical counterparts, the use of superconducting artificial atoms could allow the microwave-photon detectors to go further and realize QND detection.

Since microwave photons trapped in a cavity can be detected, as we saw in the previous section, one idea for the detection of itinerant photons is to catch them in a cavity and perform the detection there. Recently, experiments employing resonators with tunable coupling have demonstrated the capture of microwave photons with high efficiency~\cite{Yin2013, Wenner2014, Flurin2015}. However, this requires detailed knowledge of the arrival time and pulse shape of the photon to be captured, which makes it an impractical scheme for photon detection. There is a theoretical proposal for detecting an itinerant photon entering a cavity which is dispersively coupled to a second cavity~\cite{Helmer2009}. A strong coherent probe on the second cavity, read out by homodyne detection, can then realize a QND detection, but the efficiency is limited by a conflict between interaction strength and reflection of the photon to be detected. Finally, we also note that a recent experiment realized QND detection of an itinerant optical photon by letting it reflect off a cavity containing a single three-level atom prepared in a superposition state~\cite{Reiserer2013}. The reflection of a photon causes a phase flip of the atom state, which can be read out by a rotation followed by a projective measurement. Note that in this case the arrival time of the photon will not be known.

In \textit{cavity-free} photon-detection schemes, the basic ingredient is usually two- or three-level systems coupled directly to the open transmission line where the photon is propagating (see~\secref{sec:WaveguideQED} for more general information about such waveguide QED systems). Early on, it was proposed that microwave photons emitted in a quantum point contact could be measured using a nearby double quantum dot (DQD)~\cite{Aguado2000} and this experiment was later realized~\cite{Gustavsson2007}. The detection scheme is \textit{not} QND since the photon is absorbed in the DQD. 

\begin{figure}
\centering
\includegraphics[width=0.7\linewidth]{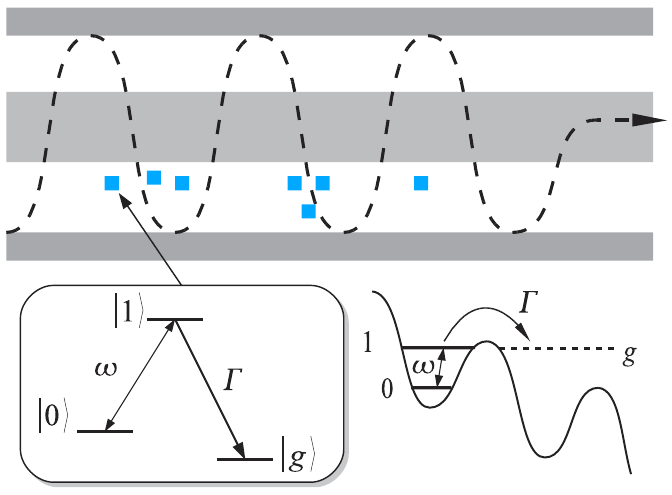}
\caption{A proposed setup for destructive detection of itinerant microwave photons~\cite{Romero2009a}. Top: A number of CBJJs (blue squares) are placed along the transmission line where the photon propagates. Bottom: Each CBJJ is tuned to have a tilted washboard potential with two states in a potential well (right), realizing an effective three-level system (left). When a photon with frequency $\gamma$ is absorbed by the CBJJ, it will rapidly tunnel from the $\ket{1}$ state to the continuum $\ket{g}$ (at a rate $\Gamma$), leading to a voltage drop over the junction that can be read out. 
Reprinted figure with permission from G.~Romero et al.,~\href{http://dx.doi.org/10.1103/PhysRevLett.102.173602}{Phys.~Rev.~Lett.~\textbf{102}, 173602 (2009)}. \textcircled{c} 2009 American Physical Society.
\label{fig:RomeroPicPhotonDetector}}
\end{figure}

In superconducting circuits, the first promising proposal suggested the use of current-biased Josephson junctions (CBJJs) along a transmission line as shown in \figref{fig:RomeroPicPhotonDetector}~\cite{Romero2009a, Romero2009}. A CBJJ can be made to realize a washboard potential with two states $\ket{0}$ and $\ket{1}$ in one of the potential wells as shown in the figure. The tunneling rate $\Gamma$ from $\ket{1}$ to the continuum is much greater than that from $\ket{0}$ and thus the CBJJ can be thought of as a three-level system as the figure shows. The idea is to let the CBJJ start in $\ket{0}$ and let the itinerant photon be resonant with the $\ket{0} \rightarrow \ket{1}$ transition. When the photon is absorbed, a tunneling event will occur with high probability and this leads to a voltage drop over the JJ which is easy to detect. If only a single CBJJ is used, the probability that it will absorb the photon is 50\%, but if $N$ CBJJs are placed along the transmission line instead, the probability that one of them will absorb and detect the photon quickly approaches 100\% as $N$ increases. Using more CBJJs also increases the bandwidth of the detector, but it was actually shown later that 100\% absorption can be achieved with a \textit{single} CBJJ if it is placed in front of a mirror~\cite{Peropadre2011}, which could possibly be formed by another artificial atom~\cite{Li2015}. Possible limitations to the efficiency of this type of photon detector includes tunneling from the $\ket{0}$ state and the relaxation process $\ket{1} \rightarrow \ket{0}$. The detector absorbs the photon and is thus \textit{not} QND.

An experiment with CBJJs placed at the end of a transmission line achieved a detection efficiency $\eta = 0.7$~\cite{Chen2011PRL}. In that experiment, two detectors were employed at the two outputs of a microwave beam-splitter to realize correlation measurements (a Hanbury-Brown--Twiss experiment~\cite{HanburyBrown1956}). In this way, $g^{(2)}(t)$ was measured for both coherent and thermal microwave photons. After this experiment, CBJJs as photon detectors in cavities have been analyzed further~\cite{Poudel2012, Govia2012, Schondorf2016}, sometimes being referred to as \textit{Josephson photomultipliers} (JPMs) to distinguish them from phase qubits which are operated in a different parameter regime.

Itinerant photons can also be absorbed efficiently by a $\Lambda$ system, which has been proposed and realized recently within circuit QED using dressed states of a resonator and a driven superconducting qubit~\cite{Koshino2013a, Koshino2013, Inomata2014}. Placing this system at the end of a transmission line and reading out the qubit via a second, open transmission line, it can work as a (destructive) detector for itinerant microwave photons in the first transmission line~\cite{Koshino2015}. A recent experimental implementation of this setup reached a single-photon detection efficiency close to $0.7$~\cite{Inomata2016}.

Another approach to detecting itinerant photons is to induce an interaction between them and a coherent probe, resulting in an effect on the probe that can then be read out with \textit{homodyne detection} (which is available for microwave photons, see~\secref{sec:PhotonDetectionHomodyne} below). Usually, this interaction is of the cross-Kerr type,
\be
H = \chi a^\dag a b^\dag b,
\label{eq:GenericCrossKerr}
\ee
with $\chi$ the interaction strength, $a$ the annihilation operator for the signal mode, and $b$ the annihilation operator for the probe mode. The presence of a photon in the signal mode will then give rise to a phase shift of photons in the probe mode, and vice versa. Photon detection based on this principle has been theoretically investigated in quantum optics~\cite{Imoto1985, Grangier1998, Munro2005, Greentree2009}, but experiments had not been able to reach larger phase shifts than a milliradian per photon~\cite{Matsuda2009, Venkataraman2012, Perrella2013} until very recently a phase shift of about 1 radian per photon was demonstrated with an atomic ensemble~\cite{Beck2016}.

However, in circuit QED an experiment with a \textit{single} three-level artificial atom (a transmon) in an open transmission line recently reported phase shifts of 20 degrees per photon~\cite{Hoi2013a}. In that experiment, the probe is close to resonance with the $\ket{0} \rightarrow \ket{1}$ transition of the transmon, while the signal is close to resonance with the $\ket{1} \rightarrow \ket{2}$. In the limit where both the signal and the probe are far detuned from the respective transitions, the photon-photon interaction mediated by the superconducting artificial atom reduces to the form of \eqref{eq:GenericCrossKerr}, but to account for the large phase shifts in the experiment the dynamics of the atom has to be included in the description.

\begin{figure}
\centering
\includegraphics[width=0.85\linewidth]{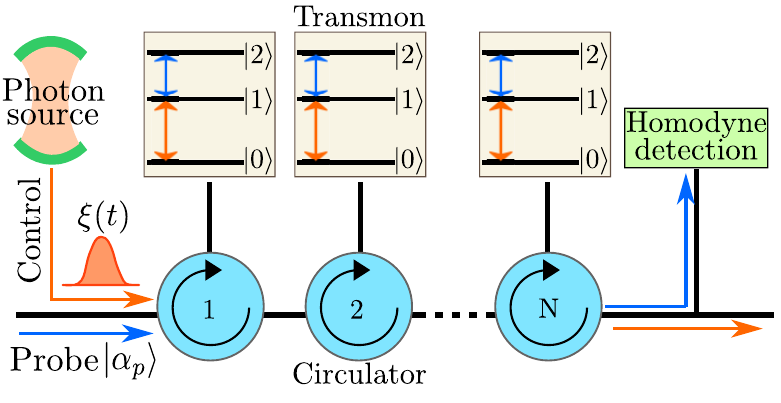}
\caption{A proposed setup for QND detection of itinerant photons~\cite{Sathyamoorthy2014}. A control (signal) photon with a wavepacket shape $\xi(t)$ is guided by circulators to interact sequentially with the $\ket{0} \rightarrow \ket{1}$ transition of $N$ three-level artificial atoms, while a coherent probe field interacts with the $\ket{1} \rightarrow \ket{2}$ transition. The three-level atoms induce an effective photon-photon interaction, resulting in a large phase shift of the probe which can be read out with homodyne detection. Since each atom sits at the end of a transmission line (i.e., in front of a mirror) the circulators ensure unidirectional propagation of all the photons. 
Reprinted figure with permission from S.~R.~Sathyamoorthy et al.,~\href{http://dx.doi.org/10.1103/PhysRevLett.112.093601}{Phys.~Rev.~Lett.~\textbf{112}, 093601 (2014)}. \textcircled{c} 2014 American Physical Society.
\label{fig:SathyamoorthyPhotonDetection}}
\end{figure}

The possibility of using the system from the experiment in Ref.~\cite{Hoi2013a} as a detector was investigated theoretically in Ref.~\cite{Fan2013}. There, it was found that the displacement of the probe induced by a single photon is \textit{not} large enough to overcome the inherent quantum noise in the probe. This negative result was in line with earlier theoretical investigations of the cross-Kerr effect~\cite{Shapiro2006, Shapiro2007, Gea-Banacloche2010}. However, a following study revealed that photon detection is nevertheless possible if several three-level atoms are cascaded after each other, using circulators to ensure unidirectional propagation as shown in \figref{fig:SathyamoorthyPhotonDetection}~\cite{Sathyamoorthy2014}. In this case, the phase shift of the probe \textit{accumulates} and rises above the quantum noise. By coupling a resonator to each atom and probing each resonator separately, the detector efficiency can be increased even further~\cite{Fan2014}. Since the signal photon propagates on after interacting with the atoms in the above setups, the detection is QND, unlike all existing detectors at optical frequencies (although there are proposals to achieve QND detection there also~\cite{Witthaut2012, Xia2015}). 

In addition to the proposals above, hot-electron nanobolometers are also being investigated as a possible microwave-photon detector~\cite{Govenius2014, Gasparinetti2015, Govenius2015a} and there may be more methods with Rydberg atoms that can be transferred to circuit QED setups~\cite{Penasa2016}. We also note that a single-electron transistor~\cite{Schoelkopf1998} has been used to detect propagating microwave \textit{phonons} (in the form of surface acoustic waves) with single-phonon sensitivity after averaging~\cite{Gustafsson2012}. Thus, there are now several promising schemes for detection of itinerant microwave photons, although some of them are still waiting for experimental implementation. Consequently, the use of photon detectors in circuit QED has seen an increased theoretical interest lately, with, e.g., qubit readout by a JPM~\cite{Govia2014}, multi-qubit parity measurements based on photodetection~\cite{Govia2015, Govenius2015}, and even dark-matter axion detection~\cite{Zheng2016} being considered. Since earlier, it is well-known from quantum optics that photon detection is crucial for applications such as optical quantum computing~\cite{Knill2001, Kok2007}, quantum key distribution~\cite{Jennewein2000, Lo2014}, Bell-inequality tests~\cite{Weihs1998, Giustina2013}, and quantum teleportation~\cite{Bouwmeester1997}.

To conclude this section, we briefly review how photon detection can be treated theoretically with \textit{stochastic master equations} (SMEs). Consider a cavity with Hamiltonian $H_\text{sys}$ containing photons (with annihilation operator $a$), leaking out to the environment at a rate $\kappa$. The master equation describing the time evolution for the density matrix $\rho$ of the cavity is then~\cite{Carmichael1999, Breuer2002, Gardiner2004}
\be
\dot{\rho} = - i \comm{H_\text{sys}}{\rho} + \kappa\lind{a}\rho,
\label{eq:MENoMeasurement}
\ee
where $\lind{a}\rho = a \rho a^\dag - \frac{1}{2} a^\dag a \rho -  \frac{1}{2} \rho a^\dag a$ is the Lindblad term~\cite{Lindblad1976} describing the cavity decay. This equation describes the behavior of the system averaged over many experiments. If we now place a photon detector outside the cavity, detecting a fraction $\eta$ of the photons that escape the cavity ($\eta$ is the measurement efficiency), the time evolution of the system, taking into account the information from the detector, follows the SME~\cite{Wiseman2010}
\be
\rd\rho = \left( -i\comm{H_\text{sys}}{\rho} + (1-\eta)\kappa\lind{a}\rho \right) \rd t + \measg{a}\rho\rd N(t) - \frac{1}{2}\eta\kappa\meas{a^\dag a}\rho \rd t,
\label{eq:SMEPhoton}
\ee
where we have introduced the notation $\measg{a}\rho = \frac{a\rho a^\dag}{\expec{a^\dag a}} - \rho$ and $\meas{a}\rho = a\rho + \rho a^\dag - \expec{a + a^\dag}\rho$. Here, $\rd N(t)$ is the stochastic increment in a stochastic process $N(t)$, which counts the number of photons that have been detected up until time $t$. The stochastic increment has the property $\rd N(t)^2 = \rd N(t)$, since in a small time interval one can only detect 0 or 1 photons, and its expectation value is $\expecc{\rd N(t)} = \eta\kappa\expec{a^\dag a} \rd t$.

Equation~(\ref{eq:SMEPhoton}) can be interpreted as the detector allowing us to follow the \textit{quantum trajectory}~\cite{Carmichael1993, WisemanPhD1994, Carmichael2008, Wiseman2010} of a single experiment. For every click in the detector, i.e., every time $\rd N(t) = 1$, the term $\measg{a}\rho$ removes a photon from the cavity to reflect our updated knowledge of the system. If $\eta < 1$, some photons escape the cavity without being detected, and this evolution can still only be captured on average by the Lindblad term in \eqref{eq:SMEPhoton}. Note that if we average over many experiments, the expectation value for $\rd N(t)$ ensures that we recover \eqref{eq:MENoMeasurement}.

\subsection{Homodyne and heterodyne detection}
\label{sec:PhotonDetectionHomodyne}

Homodyne and heterodyne detection is used to measure one or both quadratures of incoming light, giving information about \textit{amplitude} and \textit{phase} instead of just the number of photons. Due to the challenges of microwave photon detection described above, homodyne and heterodyne detection has been the prevailing type of measurement used in experiments in microwave photonics with superconducting artificial atoms, albeit with a slightly different implementation than in traditional quantum optics~\cite{Mariantoni2005, Eichler2012a} (a review of the latter is given in Ref.~\cite{Lvovsky2009}).

\begin{figure}
\centering
\includegraphics[width=0.7\linewidth]{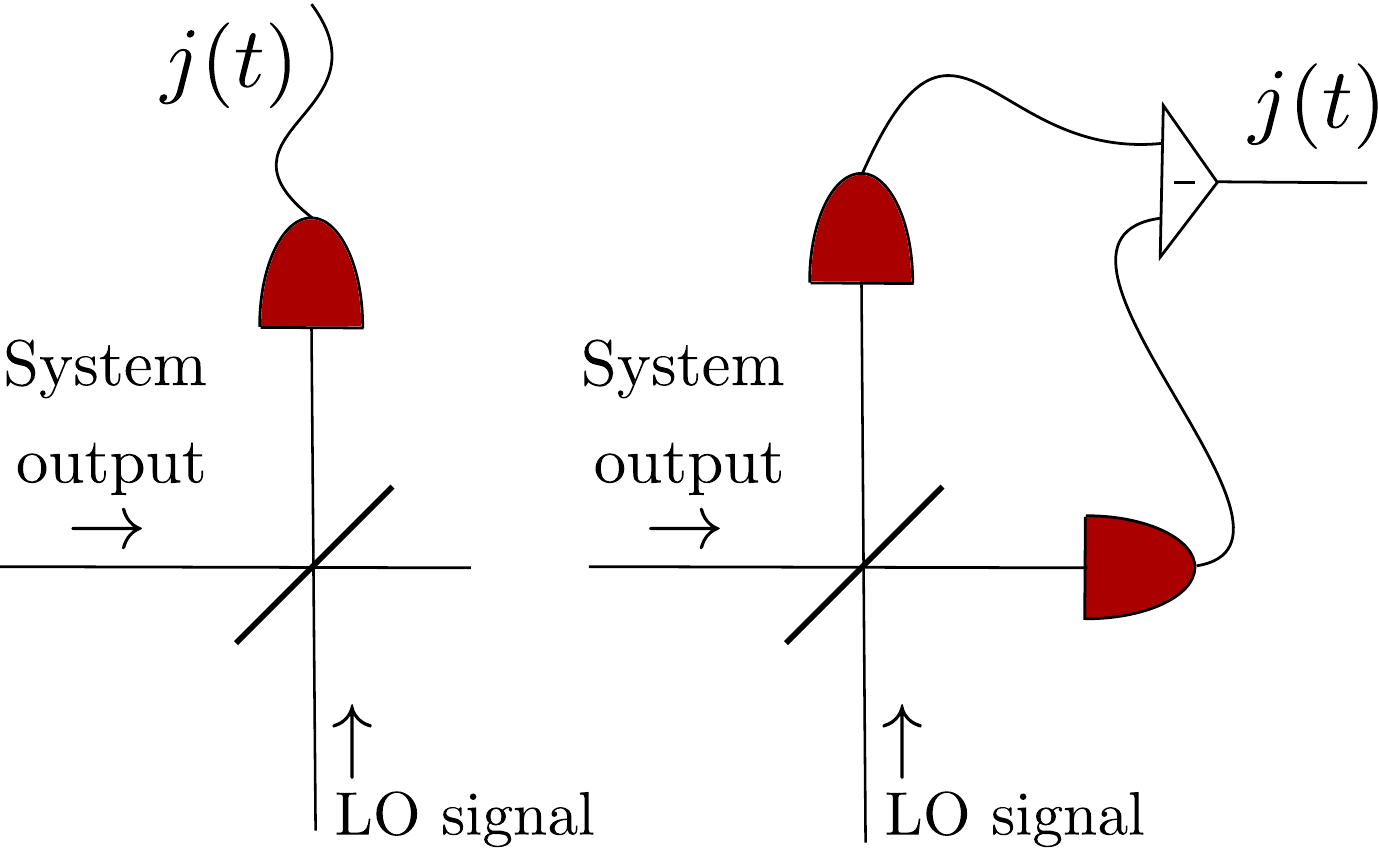}
\caption{Two setups for homodyne detection (equivalent in the limit of strong local oscillator). Left: Simple homodyne detection. The system output and the LO are mixed at a high-reflectivity beam-splitter, which reflects almost all the system output. This reflected signal is measured by a photon detector to give the homodyne current $j(t)$. Right: Balanced homodyne detection, which has been implemented in quantum optics~\cite{Smithey1993}. The system output is mixed with a strong LO in a 50:50 beam-splitter and photon detectors are placed at both of the beam-splitter outputs. The homodyne current is given by the difference of the photon detector outputs. 
\label{fig:TwoHomodyneSetups}}
\end{figure}

Two setups for \textit{homodyne} detection are sketched in \figref{fig:TwoHomodyneSetups}. The basic principle is to mix the system output to be measured with a strong coherent signal from a local oscillator (LO) and record the resulting photocurrent. The phase of the LO determines which quadrature is measured. Just as for photon detection, we can write down an SME for homodyne detection~\cite{Carmichael1993, Wiseman1993, Breuer2002, Wiseman2010},
\be
\rd \rho = -i\comm{H_\text{sys}}{\rho}\rd t+ \kappa\lind{a}\rho \rd t + \sqrt{\kappa\eta} \meas{ae^{-i\phi}} \rho \rd W(t), \label{EqSMEHomodyne}
\ee
where $\rd W(t)$ is a Wiener increment and $\phi$ is the phase set by the LO. The Wiener increment is a random variable with $\expecc{\rd W(t)} = 0$ and variance $\rd t$. Using the first property, we see that averaging over many quantum trajectories given by \eqref{EqSMEHomodyne} once again lets us recover the ordinary Lindblad master equation, \eqref{eq:MENoMeasurement}.

The measurement signal associated with the SME of \eqref{EqSMEHomodyne} is the \textit{homodyne current}
\be
j(t)\rd t = \sqrt{\kappa\eta} \expec{ae^{-i\phi} + a^\dag e^{i\phi}} \rd t + \rd W(t),
\label{EqHomodyneCurrent}
\ee
which has two notable features. Firstly, changing $\phi$ indeed determines which quadrature is measured; if $\phi=0$, we obtain information about $\expec{a+a^\dag}$, while if $\phi=\pi/2$, we obtain information about $i\expec{a^\dag - a}$. Secondly, the signal will be noisy, even for a vacuum bath, due to the stochastic increment $\rd W(t)$.

A \textit{heterodyne} measurement, which measures both quadratures simultaneously, can be realized by first splitting the system output at a 50:50 beam-splitter and then doing separate homodyne measurements, with a phase difference of $\pi/2$ between the LOs, on the two outputs from the beam-splitter. An SME similar to \eqref{EqHomodyneCurrent} can be written down for this case, but the measurement efficiency $\eta$ for each quadrature cannot exceed $1/2$ due to the splitting of the original output~\cite{Wiseman2010}.

Since the setups above are based on \textit{photon detectors}, they cannot be used straightforwardly for microwave photonics. Instead, in experiments in this field, the system output is first \textit{amplified} at cryogenic temperatures and only then measured at room temperature, usually with a vector network analyzer (VNA). The measurement efficiency is limited by the noise of the amplifier, and the best commercially available cryogenic amplifiers such as high-electron-mobility transistors (HEMTs) have noise temperatures of a few Kelvin~\cite{Bradley1999, Wadefalk2003, Pospieszalski2005}, which corresponds to upwards of at least 10 noise photons per signal photon. Since this degrades measurement efficiency considerably, great efforts have been made in the last decades to build \textit{quantum-limited amplifiers} using superconducting circuits. A thorough review on quantum amplification is given in Ref.~\cite{Clerk2010}, and some overview on the experimental developments can be found in Refs.~\cite{Nation2012, Schackert2013PhD, Flurin2014PhD, Roy2016a}. Below, we provide an overview of this field, which has seen much progress in the last few years.

\begin{figure}\centering
\includegraphics[width=\linewidth]{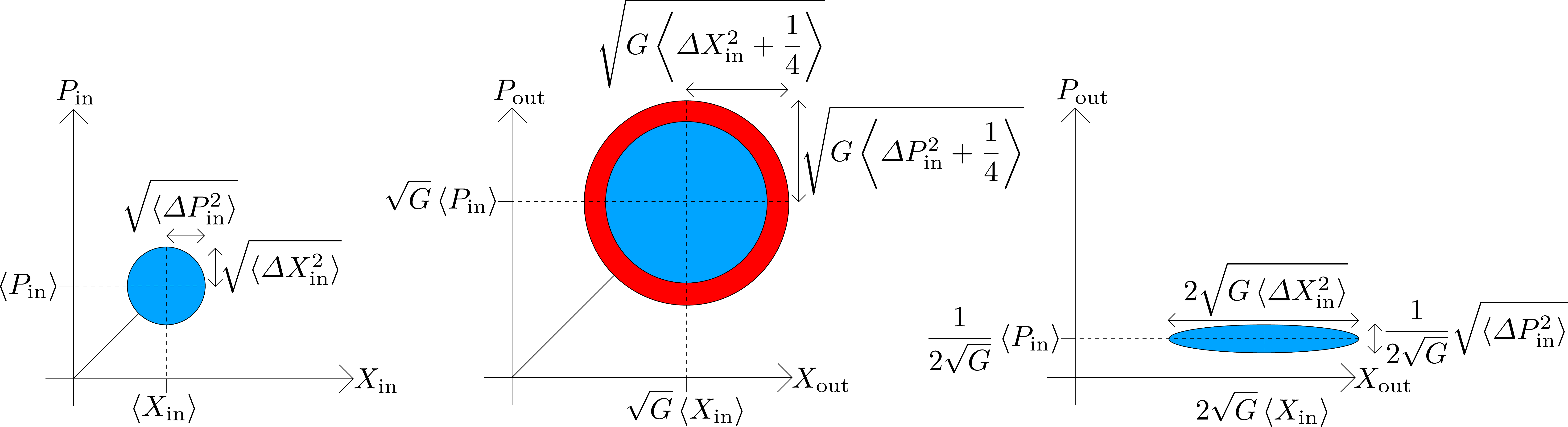}
\caption{Left: The input state with quadratures defined by $a_{\rm in} = X_{\rm in} + i P_{\rm in}$. Center: The output for phase-preserving amplification. The unavoidable extra added noise is shown in red. Right: The output for phase-sensitive amplification with $\phi = 0$ in the limit of large gain, $G \gg 1$. The $X$ quadrature has been amplified, without any extra noise added, while the $P$ quadrature has been deamplified. 
Figure adapted from E.~Flurin, The Josephson mixer - A swiss army knife for microwave quantum optics, Ph.D. thesis, Ecole Normale Superieure, Paris (2014).
\label{fig:PhaseSensitivePreservingAmplification}}
\end{figure}

To preserve the commutation relations $\comm{a_{\rm in}}{a_{\rm in}^\dag} = 1 = \comm{a_{\rm out}}{a_{\rm out}^\dag}$ for the input/output modes $a_{\rm in/out}$, an amplifier cannot simply give a pure amplitude gain $\sqrt{G}$ via $a_{\rm out} = \sqrt{G} a_{\rm in}$, but has to obey~\cite{Haus1962, Caves1982}
\be
a_{\rm out} = \sqrt{G} a_{\rm in} + \sqrt{G-1}b_{\rm in}^\dag,
\ee
where $b_{\rm in}$ represents another input mode. Amplifying both quadratures equally in this way, i.e., \textit{phase-preserving} (\textit{phase-insensitive}) amplification which is needed for heterodyne detection, will thus add at least half a photon of noise (the vacuum noise of $b_{\rm in}$). However, by choosing $b_{\rm in} = e^{i\phi}a_{\rm in}$, it is possible to instead amplify one of the quadratures \textit{noiselessly} at the expense of deamplifying the other quadrature. Such \textit{phase-sensitive} amplification is used to implement homodyne detection. The two amplifying schemes are illustrated in \figref{fig:PhaseSensitivePreservingAmplification}. We note that another scheme, with non-deterministic heralded noiseless phase-preserving amplification, has been realized in optics~\cite{Xiang2010}. 

Almost all cryogenic amplifiers trying to reach the quantum limit are based on Josephson junctions that provide a nonlinearity for parametric amplification, where a strong pump at frequency $\omega_{\rm p}$ amplifies a signal at frequency $\omega_{\rm s}$ and also produces additional photons at the ``idler'' frequency $\omega_{\rm i}$ (see \secref{sec:NonlinearProcThreeWaveMixing}). The amplifier is called \textit{degenerate} when signal and idler are in the same mode, and \textit{non-degenerate} when they are not. Various Josephson-junction-based amplifiers were investigated already several decades ago~\cite{Zimmer1967, Feldman1975, Wahlsten1977, Silver1981, Sweeny1985, Yurke1988, Yurke1989, Movshovich1990, Yurke1996}, but the field has seen renewed interest with the advent of circuit QED.

One of the first amplifiers developed in this new wave was the \textit{Josephson bifurcation amplifier} (JBA), based on a Josephson junction in a resonator~\cite{Siddiqi2004}, reviewed in Ref.~\cite{Vijay2009}. Here, the focus has mostly been on using the bifurcation for qubit readout~\cite{Siddiqi2006, Lupascu2006, Boulant2007, Lupascu2007, Manucharyan2007, Metcalfe2007, Naaman2008, Mallet2010}, but the JBA can also be used as a linear amplifier~\cite{Siddiqi2005, Manucharyan2007}.

The JBA is part of a larger family of amplifiers, \textit{Josephson parametric amplifiers} (JPAs), which are usually based on SQUID(s) in some form of resonator~\cite{Castellanos-Beltran2007, Castellanos-Beltran2008, Yamamoto2008, Castellanos-Beltran2009, Kamal2009, Castellanos-Beltran2010PhD, Eichler2011a, Mallet2011, Murch2013, Mutus2013, Wustmann2013, Zhong2013, DeLange2014, Eichler2014, Mutus2014, Simoen2014, Zhou2014, Roy2015}. There are additional SQUID-based amplifiers which could also be included under the JPA umbrella~\cite{Muck1998, Muck2003, Spietz2008, Spietz2009, Spietz2010, Hatridge2011, Vijay2012}. The JPA is normally operated as a degenerate, phase-sensitive amplifier, but it has also been used for phase-preserving amplification~\cite{Eichler2011a}. The JPA can be pumped not only through the resonator, but also by modulating the flux through the SQUID~\cite{Yamamoto2008, Wustmann2013, Mutus2013, Simoen2014, Zhou2014}.

Another type of amplifier is the \textit{Josephson ring modulator} (JRM), also called \textit{Josephson parametric converter} (JPC), which is a non-degenerate, phase-preserving amplifier, first proposed in Ref.~\cite{Bergeal2010}. This amplifier is based on a design with four Josephson junctions in a loop and has seen much investigation in the last few years~\cite{Bergeal2010, Abdo2011, Roch2012, Abdo2013PRB, Hatridge2013, Schackert2013, Schackert2013PhD, Campagne-Ibarcq2013, Flurin2014PhD, Pillet2015}. Its uses as a mixer are discussed in \secref{sec:MicrowaveComponentsMixers}. For directional amplification with JPCs and other Josephson amplifiers~\cite{Abdo2013PRX, Abdo2014, Sliwa2015, Macklin2015, White2015, Lecocq2017}, see \secref{sec:MicrowaveComponentsCirculators}.

There are also \textit{traveling-wave amplifiers}~\cite{HoEom2012, Yaakobi2013, OBrien2014, White2015, Macklin2015, Adamyan2016}, which have a large bandwidth and can be used to create multi-mode squeezing~\cite{Grimsmo2016}, as well as amplifiers based on SLUGs (superconducting low-inductance undulatory galvanometers)~\cite{Hover2012}, amplifiers with nonlinear resonators without Josephson junctions~\cite{Tholen2007}, an amplifier using cavity electromechanics~\cite{Ockeloen-Korppi2016, Ockeloen-Korppi2016a}, and the \textit{Josephson parametric dimer} (JPD), consisting of two coupled SQUID-based resonators~\cite{Eichler2014a}.

An amplifier that is to be used for homodyne or heterodyne detection should ideally have a low noise, i.e., high $\eta$, together with high gain (to overcome the noise of a following HEMT amplifier), large bandwidth (at least exceeding the bandwidth of the photons to be detected), and frequency tunability (to detect photons at different frequencies). For the phase-preserving JRM, the best reported $\eta$ is 0.82~\cite{Campagne-Ibarcq2013, Flurin2014PhD}, which comes with a gain above $\unit[20]{dB}$ and a bandwidth of a few MHz. For the phase-sensitive JPA, the highest reported $\eta$ is 0.68~\cite{Murch2013}, with a gain of $\unit[10]{dB}$ and a bandwidth of $\unit[20]{MHz}$. Traveling-wave amplifiers have bandwidths of several GHz, and very recently an $\eta$ of $0.75$ was reported~\cite{Macklin2015}.

\subsection{Correlation measurements}
\label{sec:PhotonDetectionCorrelation}

In traditional quantum optics, single-photon detectors are not only used for the various measurements discussed in the previous sections, but also to measure \textit{correlation functions} of photons. The first-order correlation function $G^{(1)}(t,\tau) = \expec{a^\dag(t) a(t+\tau)}$ is important for interference experiments such as the double-slit experiment, but to decisively distinguish quantum effects from classical ones measurements of the second-order correlation function $G^{(2)}(t,\tau) = \expec{a^\dag(t) a^\dag(t+\tau) a(t+\tau) a(t)}$, or its normalized version
\be
g^{(2)}(t, \tau) = \frac{G^{(2)}(t,\tau)}{(G^{(1)}(t,0))^2} = \frac{ \expec{a^\dag(t) a^\dag(t+\tau) a(t+\tau) a(t)} }{ \expec{a^\dag(t) a(t)}^2 },
\ee
are needed~\cite{Walls2008}. Following the Hanbury-Brown--Twiss (HBT) experiment~\cite{HanburyBrown1956}, where a setup with a beam-splitter followed by two intensity detectors was used to measure such correlations for the first time, a full quantum theory for photon correlations was developed in the early 1960s~\cite{Glauber1963a, Glauber1963, Mandel1965}. 

For coherent light $g^{(2)}(t, \tau) = 1$ for all $\tau$, while thermal (chaotic) light has $g^{(2)}(t, 0) = 2$, but approaches $g^{(2)}(t, \tau) = 1$ for long $\tau$ (all types of light do). In the latter case, the photons are \textit{bunched}, i.e., there is a higher probability for the photons to arrive in pairs. Classical fields can never have $g^{(2)}(t, \tau) < 1$. Truly quantum systems such as single-photon emitters in the form of atoms or molecules, on the other hand, can give rise to \textit{anti-bunching}, $g^{(2)}(t, \tau) > g^{(2)}(t, 0)$, with $g^{(2)}(t, 0)$ approaching 0 in the absence of photon pairs~\cite{Paul1982}. This has been shown in several quantum optics experiments employing photon detectors in an HBT setup~\cite{Kimble1977, Diedrich1987, Basche1992, Kuhn2002}.

\begin{figure}
\centering
\includegraphics[width=\linewidth]{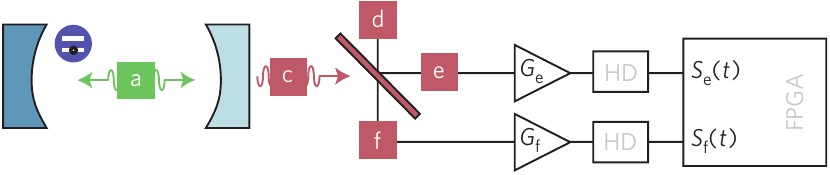}
\caption{An example of a Hanbury-Brown--Twiss setup modified for microwave photonics~\cite{Bozyigit2011}. The system to be measured, here a cavity with mode $a$ coupled to an artificial atom, releases photons to mode $c$, which is fed into a beam-splitter. The beam-splitter output modes $e$ and $f$ are fed through linear amplifiers before undergoing heterodyne detection. Reprinted figure by permission from Macmillan Publishers Ltd: Nature Physics, D.~Bozyigit et al.,~\href{http://dx.doi.org/10.1038/nphys1845}{Nat.~Phys.~\textbf{7}, 154 (2011)}, copyright (2011).	
\label{fig:CorrelationMsmt}}
\end{figure}

With single-photon detection at microwave frequencies being challenging, experiments in microwave photonics rely on another way to measure correlation functions, replacing the photon detectors after the beam-splitter with homodyne or heterodyne detection, implemented using linear amplifiers as discussed in the previous section. The first HBT experiments with microwaves used square-law detectors after the amplifiers~\cite{Gabelli2004, Zakka-Bajjani2010}, but after an experiment at optical frequencies replaced the single-photon detectors with homodyne detection~\cite{Grosse2007}, the theory was refined for this type of setup at microwave frequencies~\cite{DaSilva2010}. It is now understood how this kind of setup, shown in \figref{fig:CorrelationMsmt}, can be used to extract \textit{arbitrary} moments $\expec{(a^\dag)^n a^m}$~\cite{Menzel2010, Eichler2012a, Eichler2013PhD} and cumulants~\cite{Virally2016}, and experiments have been performed measuring these quantities for microwaves~\cite{Bozyigit2011, Menzel2010, Mariantoni2010, Eichler2011, Virally2016, Goetz2016}. Measuring moments of several orders allows for \textit{tomography} of microwave photons, reconstructing the Wigner function~\cite{Buzek1996, Eichler2011, Eichler2012a}.

Correlation measurements have also been used to show the anti-bunching (bunching) of the reflected (transmitted) signal from a superconducting artificial atom in a transmission line~\cite{Hoi2012}, to observe photon blockade~\cite{Lang2011}, and to demonstrate the Hong--Ou--Mandel (HOM) effect~\cite{Hong1987} (two photons that are mixed at a beam-splitter are emitted as a pair from one of the outputs) for microwaves~\cite{Lang2013, Woolley2013} (the HOM effect for microwaves was also observed in the experiment of Ref.~\cite{Nguyen2012}). However, it should be noted that an HBT experiment has been performed with microwave single-photon detectors as well~\cite{Chen2011} and that this setup in general is faster if the photon detectors are good enough. This is because measuring moments $\expec{(a^\dag)^n a^m}$ requires averaging over many measurements to filter out noise. The number of measurements needed is proportional to $(1+N_0)^{n+m}$, where $N_0$ is the number of noise photons added in the amplification chain~\cite{DaSilva2010, Eichler2013PhD}. This exponential increase makes measuring higher order moments impractical.

\section{Applications of large-scale superconducting quantum circuits}
\label{sec:Applications}

In the previous sections, we mainly reviewed how superconducting quantum circuits (SQCs) can be used to explore quantum optics and atomic physics in the microwave domain. However, these microwave-photonic devices can have many applications beside fundamental physics if they are scaled up to large numbers. In this section, we review the progress on three important potential applications of microwave SQCs based on Josephson junctions: quantum simulation (\secref{sec:QuantumSimulation}), quantum information processing (\secref{sec:QIP}), and metamaterials (\secref{sec:Metamaterials}).

\subsection{Quantum simulation}
\label{sec:QuantumSimulation}

More than three decades ago, Feynman proposed, in his pioneering work in Ref.~\cite{Feynman1982}, to simulate one quantum system by using another quantum system. Such \textit{quantum simulation} is sometimes considered an intermediate goal on the road toward universal quantum computation. Due to this, and also in view of the latest advances in coherent manipulation and state control for quantum systems, quantum simulation has recently been attracting growing interest in many areas of physics. In this section, we give an overview of some experimental achievements with SQCs for quantum simulation of condensed-matter physics. More comprehensive reviews of quantum simulators in various physical systems can be found in Refs.~\cite{Buluta2009, Nation2012, Houck2012, Georgescu2014}. We note that SQCs are particularly promising for quantum simulation, since they can be designed in a plethora of configurations and can have many parameters tuned during an experiment. They also have long coherence times and possess a strongly nonlinear element in Josephson junctions, which is useful for simulating nonlinear crystals.

In the context of quantum simulations, it is possible to get quite a lot of mileage out of just a single superconducting artificial atom. Since a two-level system can be considered a pseudospin-$1/2$ system, a superconducting qubit can be used to simulate the dynamics of a spin-$1/2$ system. For example, by geometrically manipulating a superconducting qubit with a microwave field such that the qubit state completes a closed path on the Bloch sphere, a geometric Berry's phase has been demonstrated~\cite{Leek2007, Berger2012, Berger2013}.

If the artificial atom supports $d > 2$ levels, it forms a \textit{qudit}, an extension of qubits that can speed up certain computing tasks and enable simpler implementation of some quantum algorithms~\cite{Muthukrishnan2000, Bullock2005, Lanyon2009}.
A qudit can also be used to simulate the dynamics of a large-spin system. In the experiment of Ref.~\cite{Neeley2009}, a superconducting qudit with up to five energy levels (which can be called a quintit) was used to emulate the dynamics of single spins with the principal quantum numbers $S = 1/2$, $1$, and $3/2$~\cite{Nori2009}. The experiment measured the Berry's phase for closed-path rotations of such spins and demonstrated the even (odd) parity of integer (half-integer) spins under $2\pi$-rotation. In another experiment, a superconducting qutrit (i.e., a quantum three-level system) was used to demonstrate non-Abelian and non-adiabatic geometric effects, which could be important for future simulations of synthetic-gauge fields and non-Abelian anyon statistics~\cite{Abdumalikov2013}.

A fully controllable superconducting qubit can also be used to simulate condensed-matter physics, e.g., \textit{topological quantum physics}, a topic which has attracted great interest recently due to the discovery of topological insulators in two and three dimensions~\cite{Hasan2010, Qi2011}. It has been shown that certain robust topological invariants, such as a Chern number, can be used to classify physical phenomena in such topological quantum systems~\cite{Chiu2016}. A change in the value of such a number, e.g., an increment in the filling factor of a integer quantum Hall state, can correspond to a nontrivial topological transition in the quantum system. The first Chern number is equivalent to the Berry phase, which can be obtained by integrating the Berry curvature over a closed surface. In an experiment with a transmon qubit, topological transitions from Chern number $1$ to $0$ were observed by modifying a drive that controlled parameters in the qubit Hamiltonian~\cite{Schroer2014}.

Topological phenomena were further explored in an experiment with a superconducting phase qubit, where the momentum space of the topological Haldane model was mapped to the parameter space of the single-qubit Hamiltonian~\cite{Roushan2014}. In this experiment, the microscopic spin texture of the associated states and their evolution across a topological phase transition were measured. The setup was then extended to two interacting qubits, where the emergence of an interaction-induced topological phase was observed~\cite{Roushan2014}. Adding another qubit to form a triangular loop where photons can hop between the qubits, synthetic magnetic fields, strong photon-photon interactions, and chiral ground-state currents for the interacting photons have been realized~\cite{Roushan2016}. This setup can be used as a building block for creating fractional quantum Hall states~\cite{Roushan2016}. We also note that the fractional statistical behavior of anyons, the use of which has been proposed for implementing topological quantum computation, recently was emulated in a circuit-QED systems with four qubits and one resonator~\cite{Zhong2016}, building on the theory in Ref.~\cite{You2010}. There is also a proposal for using superconducting qubits to realize Majorana qubits~\cite{You2014}.

Superconducting qubits can also be used to simulate other phenomena in condensed matter physics. Examples that have been demonstrated include weak localization~\cite{Chen2014c} and fermionic models~\cite{Barends2015}. Recently, a scalable quantum simulator with superconducting qubits was used to perform electronic-structure calculations for a hydrogen molecule~\cite{OMalley2016}. An array of superconducting qubits can also be used for many other simulation purposes, e.g., for exploring quantum phases in a Dicke-Ising model~\cite{Zhang2014}.

Another venue for quantum simulations with SQCs is resonators connected in \textit{large lattices}, in which each resonator interacts with a superconducting qubit following the Jaynes--Cummings Hamiltonian. Quantum simulation in photonic lattices is a large field with possible implementations in several types of systems~\cite{Carusotto2013, Noh2017}; circuit-QED lattices constitute one promising platform for such experiments~\cite{Tsomokos2010, Houck2012, Schmidt2013, Zhu2013, Anderson2016}. These lattices can be used to study, e.g., strongly correlated systems, both in and out of equilibrium, with broken time-reversal symmetry~\cite{Koch2010}, the quantum phase transition from a superfluid to a Mott insulator in the Bose--Hubbard model~\cite{Koch2009}, and artificial gauge fields.

Large circuit-QED lattices have already been implemented in several experiments and are being scaled up rapidly. In a minimal such lattice, a Jaynes--Cummings dimer with two coupled resonators connected to one qubit each, a dissipation-driven localization transition was observed~\cite{Raftery2014}. Other experiments have characterized the disorder in a 12-site lattice (forming a Kagome star)~\cite{Underwood2012} and demonstrated, in a 49-site Kagome lattice, that scanning defect microscopy can be used to image the photon lattice states~\cite{Underwood2016}. Recently, a dissipative phase transition was observed in a driven 72-site 1D circuit-QED lattice~\cite{Fitzpatrick2017}. Some other related experiments, used for simulating quantum annealing problems, are summarized below in \secref{sec:AdiabaticQCompQAnnealing} in the context of adiabatic quantum computing.

\subsection{Quantum-information processing}
\label{sec:QIP}

Early progress in SQCs was stimulated by their potential applications for \textit{quantum-information processing} (QIP). SQCs have numerous features that are advantageous for QIP, including, e.g., scalability, low loss, high nonlinearity, tunability, easy connectivity, and switching. However, to judge the potential of any platform for quantum computation, the gold standard is the  \textit{DiVincenzo criteria}~\cite{DiVincenzoFP2000}: scalable qubits, initialization, long coherence times, measurements, and universal gates, plus the two additional requirements of error correction and quantum memory. For SQCs, the road towards fulfilling all these criteria has been summarized into seven stages of development~\cite{Devoret2013}. In \secref{sec:DiVincenzoCriteria}, we review several (but not all) of the achievements of SQCs in satisfying the DiVincenzo criteria. We also discuss adiabatic quantum computing and quantum annealing with SQCs in \secref{sec:AdiabaticQCompQAnnealing}, and devote the separate \secref{sec:QEC} to quantum error correction with SQCs.
For a recent, more comprehensive review of QIP (including quantum simulation) with SQCs, see Ref.~\cite{Wendin2016}.

\subsubsection{Five criteria for scalable quantum computing with SQCs}
\label{sec:DiVincenzoCriteria}

Out of the five main DiVincenzo criteria, \textit{qubit readout} has already been discussed in \secref{sec:Readout}, where it became apparent that there are many well developed measurement methods. Likewise, methods for \textit{qubit initialization}, beyond simply letting the qubit relax to its ground state, have been developed and brought to high efficiency. These methods include initializations by measurements~\cite{Riste2012, Johnson2012, Govia2015a}, through a drive~\cite{Geerlings2013}, and by modifying the coupling to an environment~\cite{Reed2010a, Tuorila2016}. Methods for resonator reset have also been developed~\cite{Bultink2016, McClure2016, Boutin2016}. In the following, we go into some more detail about coherence, switchable coupling, and implementations of universal gates.

\paragraph{Coherence}

For QIP, long qubit \textit{coherence} times $T_1$ and $T_2$ are crucial. The \textit{energy-relaxation} time $T_1$ characterizes how fast a qubit decays from its excited state to its ground state. The \textit{dephasing} time $T_2$ is the time during which a qubit effectively retains its phase coherence. In general, $1/T_2=1/(2T_1)+1/T_{\varphi}$, where $T_{\varphi}$ is the time scale for pure dephasing, describing the relative-phase loss for a superposition state.

\begin{figure}
\centering
\includegraphics[width=\linewidth]{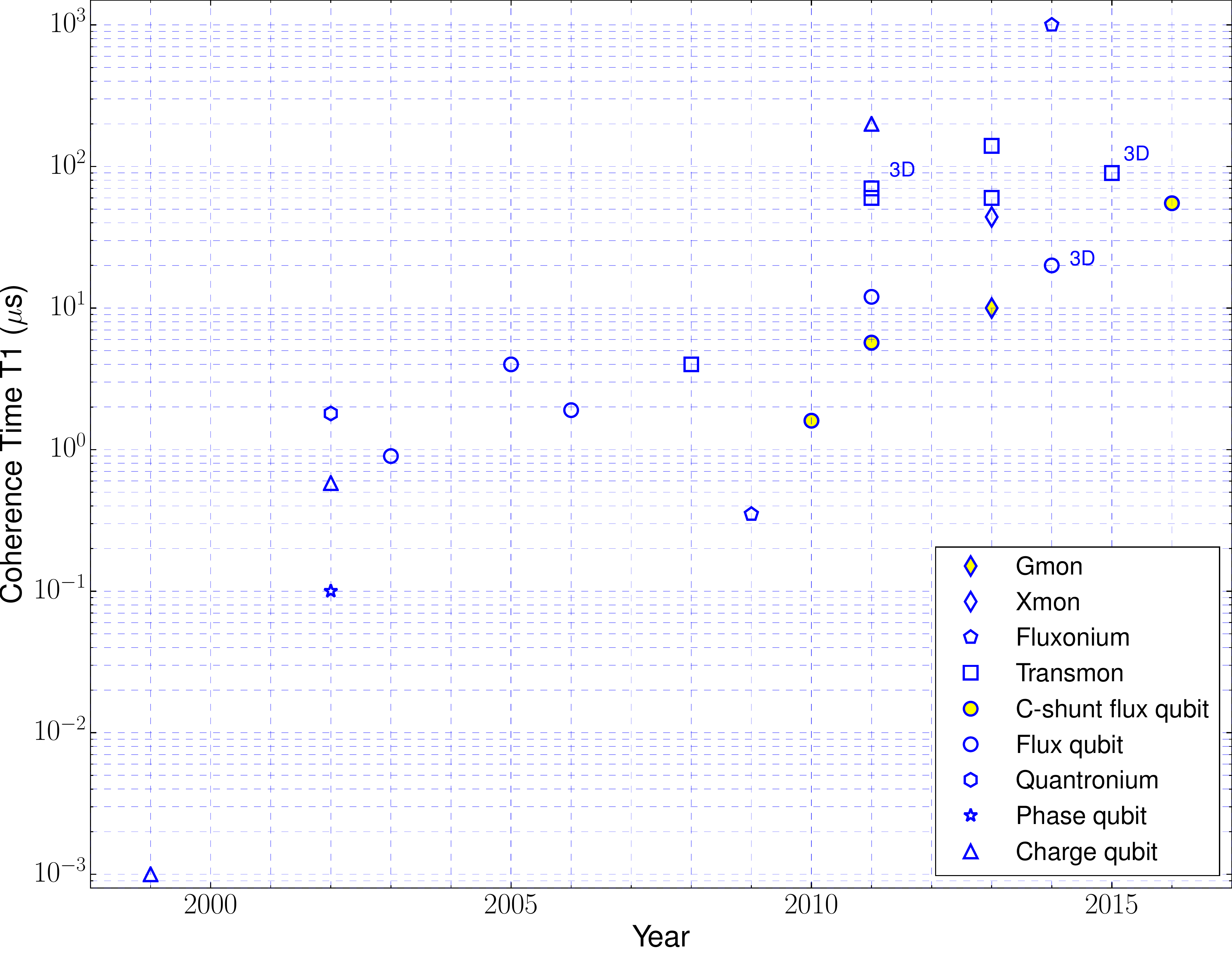}
\caption{Similar to the famous Moore's law for semiconductor industry, the coherence times of superconducting qubits also appear to exhibit exponential scaling. The graph shows, on a logarithmic scale, the energy-relaxation time $T_1$ achieved through the years in some state-of-the-art superconducting qubits. $T_1$ has increased by six orders of magnitude since 1999. This remarkable progress has been driven by novel circuit design, material and fabrication improvements~\cite{Oliver2013}, and understanding of dissipation mechanisms~\cite{Riste2013a, Pop2014}. 
\label{fig:Moore}}
\end{figure}

Spurred by the potential of QIP, the field of superconducting qubits has seen a remarkable development of coherence times in recent years. As shown in~\figref{fig:Moore}, the distressingly short decoherence times of a few nanoseconds observed in the earliest experiments have now been extended to around a hundred microseconds. This is about a thousand times longer than demonstrated time scales for initialization, read-out, and universal logic in circuit QED. These advances have been achieved through improvements of materials and qubit designs~\cite{Oliver2013}. However, the improvement of coherence times is still one of the main tasks for superconducting QIP. New materials and new methods for device engineering on a superconducting chip will likely be required to eliminate the remaining noise sources.

\paragraph{Coupling}

When it comes to scaling up superconducting circuits to many qubits, an important question is how to \textit{connect} the qubits. Connection can either be achieved through a quantum data bus, e.g., a single-mode microwave cavity field or an $LC$ oscillator, or by using a direct coupling between qubits. In order to implement universal gates and scale up the circuit for QIP, it is crucial that the coupling is \textit{switchable}. To couple and decouple two superconducting quantum elements connected by an always-on coupling, a common solution is to tune their frequencies in or out of resonance~\cite{Berkley2003, McDermott2005, Blais2003, Cleland2004, Sillanpaa2007, Majer2007, Eichler2012b}. This method has been widely adopted, even in the most recent implementations of universal gates~\cite{Fedorov2011, Reed2012, Corcoles2015, Barends2016}. However, there are limitations to this approach. For example, it becomes difficult to avoid problems with frequency-crowding in large-scale circuits. Also, tuning the qubit frequencies will move the qubits away from their optimal working points, and when the tuning is fast, the result is non-adiabatic information leakage.

Another way to switch off qubit-qubit coupling is to directly tune the coupling strength using an external magnetic flux~\cite{Makhlin2001, You2002}. This approach has been explored both theoretically~\cite{Rigetti2005, Liu2006, Grajcar2006} and experimentally~\cite{Hime2006, Niskanen2007, VanDerPloeg2007, Harris2007, Yamamoto2008a, Bialczak2011, Allman2014}. A highly coherent and rapidly tunable coupling circuit was recently demonstrated~\cite{Chen2014}. However, these quantum or classical couplers increase the complexity of SQCs and introduce new decoherence sources. Thus, the development of efficient quantum switches for the coherent coupling between different elements remains a big challenge for scalable SQCs. Recent studies~\cite{Liu2014, Wu2016} have shown that the coupling between two superconducting quantum elements can be perfectly switched on and off by a longitudinal control field. In this proposal, no auxiliary circuits or tunable frequencies for any quantum elements are required, which means that it could be a good candidate for realizing scalable superconducting QIP. Additional proposals for switchable coupling can be found in Refs.~\cite{Wei2005a, Ashhab2006a, Ashhab2007a, Ashhab2008}.

\paragraph{Gates}

Arbitrary \textit{universal gates} can be constructed by combining any nontrivial two-qubit gate [e.g., a controlled-NOT (CNOT) gate, a controlled-phase (CPHASE) gate, or an iSWAP gate] and single-qubit rotations. A comparison of speeds for various quantum gates can be found in Ref.~\cite{Ashhab2012}. In SQCs, the first implementation of a conditional gate used two coupled charge qubits~\cite{Yamamoto2003}. Later, the CNOT gate was demonstrated in a flux-qubit circuit~\cite{Plantenberg2007}. However, in these experiments the coupling between the two qubits was always on. Using a controllable coupling instead, an iSWAP-like gate for two flux qubits was realized around the same time~\cite{Niskanen2007}. In setups with two transmon qubits coupled through a microwave resonator, both the $\sqrt{\text{iSWAP}}$ gate~\cite{Majer2007}, and later the CPHASE gate~\cite{DiCarlo2009a}, have been demonstrated.

The three-qubit \textit{Toffoli gate}, which inverts the state of a target qubit conditioned on the state of two control qubits, can be used to construct a universal entangling gate. This gate can also be used for universal reversible classical computation and is a key element for the realization of some quantum error-correction codes. The standard implementation of the Toffoli gate, using only single- and two-qubit gates, requires six CNOT gates and ten single-qubit operations~\cite{Barenco1995}. Since the CNOT gate itself usually requires a sequence of operations in SQCs, it would seem that realizing the Toffoli gate for superconducting qubits would require very long qubit coherence times. However, by cleverly making use of auxiliary energy levels in three transmon qubits coupled via a microwave resonator, the Toffoli gate has been implemented much more efficiently in experiments~\cite{Fedorov2011, Reed2012}. This achievement opens up for realizing more complex quantum operations, useful for, e.g., error correction, in SQCs~\cite{Reed2012}.

Two-qubit gates have been used in SQC experiments for a variety of purposes, e.g., for coherent storage and transfer of quantum states~\cite{Sillanpaa2007}, and to demonstrate a violation of Bell's inequality~\cite{Ansmann2009}, following early proposals in Refs~\cite{Wei2005, Wei2006}. Quantum gates have also been applied to prepare the entangled Greenberger--Horne--Zeilinger (GHZ) and W states in three phase qubits~\cite{Neeley2010}, in three transmon qubits coupled via a transmission-line resonator~\cite{DiCarlo2010}, and with two phase qubits and a resonator mode~\cite{Altomare2010}; there have been many proposals for creating such states in SQCs~\cite{Wei2006PRL, Matsuo2007, Kim2008, Galiautdinov2008, Bishop2009, Helmer2009a, Wang2010, Feng2011, Yang2013, Garziano2015, Yang2016b, Liu2016c, Stassi2017}. In the context of QIP, two-qubit gates have been used to demonstrate simple quantum protocols and algorithms in two-qubit circuits. Recently, quantum algorithms have also been implemented in more complex circuits. For example, a circuit with seven superconducting quantum elements used $\sqrt{\text{iSWAP}}$ and CPHASE gates together with single-qubit operations to demonstrate a quantum von Neumann architecture by implementing the quantum Fourier transform and a three-qubit Toffoli-like gate~\cite{Mariantoni2011}. Another experiment with a quantum processor, consisting of nine elements, implemented Shor's algorithm to find the prime factors of 15~\cite{Lucero2012}. Recently, $\sqrt{\text{iSWAP}}$ gates have also been applied to generate $10$-qubit entanglement~\cite{Song2017} and to solve linear equations~\cite{Zheng2017}.

Quantum gates for SQCs are continuously being refined. Recent experiments have demonstrated impressive gate fidelities: above 99.9\% for single-qubit gates~\cite{Barends2014, Sheldon2016} and above 99\% for two-qubit gates~\cite{Barends2014, Sheldon2016a}. Furthermore, different gate implementations, e.g., the so-called cross-resonance~\cite{Rigetti2010, Chow2011, Sheldon2016a} and resonator-induced CPHASE~\cite{Cross2015, Puri2016, Paik2016} gates, as well as other schemes~\cite{Leek2009, Chow2013, McKay2016}, have been developed that allow for gates between fixed-frequency qubits, which can have longer coherence times than tunable qubits.

In conclusion, it can be said that the impressive progress for SQCs includes demonstrations showing that the five DiVincenzo for scalable QIP have been fulfilled at a reasonable level of performance in small-scale circuits. Thus, there does not seem to be any \textit{fundamental} roadblock hindering the construction of a quantum computer in a large-scale quantum circuits. In the following sections, we therefore review some more aspects of large-scale circuits (in addition to what we already treated when discussing quantum simulation in \secref{sec:QuantumSimulation}): adiabatic quantum computing and quantum error correction.

\subsubsection{Adiabatic quantum computing and quantum annealing}
\label{sec:AdiabaticQCompQAnnealing}

An alternative to standard \textit{gate-based} quantum computing is \textit{adiabatic} quantum computing (AQC). In AQC, a controllable physical quantum system is initially prepared in the ground state of a simple Hamiltonian. Computation is then performed by changing the system Hamiltonian \textit{adiabatically}, such that the system remains (with high probability) in its ground state throughout the evolution~\cite{Farhi2001}. The idea of AQC is that the ground state of the final, non-trivial, Hamiltonian encodes the solution of a given computational problem. AQC is theoretically equivalent to universal quantum computing~\cite{Aharonov2004}, and, by definition, more robust against noise (though the trade-off between speed and protection against decoherence is not trivial~\cite{Ashhab2006}). For these reasons, AQC has attracted considerable theoretical and experimental interest. However, there is as of yet no proof that AQC actually gives an exponential speed-up compared to classical computation. In this section, we discuss some implementations of AQC in SQCs. For a recent detailed review on the theory of AQC, see Ref.~\cite{Albash2016}.

The first experimental steps towards realizing AQC with SQCs were taken with flux-qubit circuits, in which the ground state of the system was systematically mapped out as a function of flux in a three-qubit system~\cite{VanderPloeg2007IEEE}. Following this experiment, it has been shown that it may be possible to speed up AQC with SQCs using feedback control~\cite{Wilson2012}. More recently, an SQC experiment using up to nine qubits combined gate-based and adiabatic quantum computing to realize digitized AQC~\cite{Barends2016}. This experiment used up to $10^3$ quantum gates to solve instances of the 1D Ising problem and some other Hamiltonians with more complex interaction terms.  

Another approach to quantum computing, closely related to AQC, is \textit{quantum annealing} (QA)~\cite{Das2008}. Quantum annealing is similar to the classical method of \textit{simulated annealing}, where thermal fluctuations help to overcome energy barriers in the search for the global minimum in an optimization problem. In QA, a known initial configuration at a non-zero temperature evolves into the ground state of a Hamiltonian encoding an optimization problem. If the energy barriers in the configuration landscape are tall and narrow, QA can outperform simulated annealing thanks to quantum tunneling. QA differs from AQC in that, while both are methods to find the ground state of a problem Hamiltonian, the transitions in QA can be non-adiabatic and the initial configuration for QA need not be a ground state.

There have been several experimental implementations of QA in SQCs. An early demonstration used an array of eight flux qubits with programmable spin-spin couplings to find the ground state of the Ising model~\cite{Johnson2011a}. In this experiment, the temperature dependence of the evolution time at which the system dynamics freezes revealed a clear signature of QA, distinguishable from classical thermal annealing. Later, an experiment with 16 flux qubits demonstrated robustness of QA against noise~\cite{Dickson2013}. To further prove the viability of QA as a technology for large-scale systems of superconducting qubits, more experiments studied entanglement in a system with eight flux qubits~\cite{Lanting2014} and correlations in a setup with more than one hundred such qubits~\cite{Boixo2014}.

Ideally, a quantum annealer should be \textit{programmable}, i.e., the physical realization of QA should allow for changing parameters to implement various problem Hamiltonians. Experiments with SQCs have demonstrated programmable QA, scaling up from eight qubits to more than one hundred qubits~\cite{Boixo2013}. Quantum error correction has also been developed and demonstrated for such programmable QA in a system with antiferromagnetic chains formed by up to 344 flux qubits~\cite{Pudenz2014}. This method can improve the success probability of QA, but full-fledged fault-tolerant QA remains an open problem.

Recently, even larger circuits, containing 512 qubits (D-Wave~2), 1152 qubits (D-Wave~2X), and 2048 qubits (D-Wave~2000Q), have been fabricated and tested. It has been claimed that these large circuits can speed up certain algorithms, but it remains a much debated, open question whether this is a \textit{quantum speedup}~\cite{Ronnow2014, Zagoskin2014, Heim2015, Amin2015, Isakov2015, Boixo2016, Denchev2016}.

\subsubsection{Quantum error correction}
\label{sec:QEC}

Errors are inevitable in QIP. This makes \textit{quantum error correction} (QEC) a fundamental requirement for \textit{fault-tolerant quantum computing}~\cite{Nielsen2000, Devitt2013, Terhal2015}. One of the simplest examples of QEC is how bit- or phase-flip errors in single qubits can be corrected in a simple three-qubit code, which uses an entangled three-qubit state to encode a single-qubit state~\cite{Shor1995, Cory1998, Nielsen2000}. This error-correction code was first demonstrated in SQCs using three superconducting qubits coupled to a single-mode microwave field~\cite{Reed2012}. The three-qubit error correction code has also been used in SQCs to reduce the impact of intrinsic dissipation and improve the storage time of a quantum state~\cite{Zhong2014}. State preservation for qubits has also been extended to a setup with a 1D chain of nine superconducting qubits~\cite{Kelly2015}. In this (and other) setups, the occurrence of errors was tracked by repeatedly performing projective quantum non-demolition (QND) \textit{parity measurements} on pairs of qubits. Due to the importance of multi-qubit parity measurements for error correction codes, schemes for implementing such measurements in SQCs have been studied in several works~\cite{Kerckhoff2009, Lalumiere2010, Tornberg2010, Kockum2012, Nigg2013, DiVincenzo2013, Tornberg2014, Govia2015, Govenius2015}. In the experiment of Ref.~\cite{Kelly2015}, the outcomes of the repeated error detection were used to optimize gate parameters, showing that both the drift of a single qubit and even independent drifts in all qubits can be removed. This method provides a path towards controlling a large number of qubits for fault-tolerant quantum computing with SQCs~\cite{Kelly2016}.

Fault-tolerant quantum computing has been proposed to be implemented with various error-correction codes. In particular, \textit{surface codes}~\cite{Kitaev1998, Fowler2012} are considered a promising candidate for scalable quantum computing with SQCs due to their 2D layout with only nearest-neighbor qubit-qubit couplings and high error thresholds for fault tolerance. In a surface code, qubits are arranged in a square lattice. Half of the qubits in the lattice function as code qubits, while the other half are syndrome qubits, used to measure the stabilizer operators of surrounding code qubits.

Several significant steps towards large-scale QEC with SQCs have been taken in the last few years. For example, autonomous stabilization of a Bell state for two transmon qubits has been demonstrated using a combination of continuous drives and dissipation engineering~\cite{Shankar2013} and high-fidelity detection of the parity of two code qubits via the measurement of a third syndrome qubit has been achieved using three transmon qubits connected by two coplanar-waveguide resonators~\cite{Chow2014}. Furthermore, arbitrary quantum errors have been detected through two-qubit stabilizer measurements in a four-qubit sublattice of the surface code~\cite{Corcoles2015}, a four-qubit parity measurement has been implemented on a surface-code plaquette~\cite{Takita2016}, and a five-qubit setup (three code qubits, two syndrome qubits) has demonstrated protection of a logical qubit from bit-flip errors~\cite{Riste2015}. Considering these achievements together with the fact that a recent experiment with a five-qubit superconducting quantum processor demonstrated average fidelities of 99.92\% for single-qubit gates and 99.4\% for two-qubit gates~\cite{Barends2014}, it seems that SQCs are at the threshold of fault-tolerant quantum computing.

Although superconducting qubits are starting to reach quite long coherence times, the coherence times of microwave photons in resonators can be much longer. It is therefore natural to contemplate implementing quantum computation with \textit{microwave-photonic qubits} instead. Recently, some research in circuit QED is shifting towards this direction. When it comes to error correction for quantum computation with microwave photons, it has long been known that so-called \textit{bosonic codes} can protect quantum information stored in bosonic fields against amplitude damping~\cite{Chuang1996, Gottesman2001}. Furthermore, Schr\"odinger cat states, superpositions of two coherent states, can be used to encode logical qubits in schemes for correcting errors caused by spontaneous emission~\cite{Cochrane1999}, but such cat-state encodings do not protect against dephasing~\cite{Liu2004a}. However, in the last few years, renewed interest in the field has led to several new proposals for encoding and manipulating quantum information in bosonic modes~\cite{Leghtas2013a, Mirrahimi2014, Heeres2015, Krastanov2015, Mirrahimi2016, Terhal2016, Albert2016, Bergmann2016, Li2016}. One such proposal describes a new class of bosonic quantum codes that can correct errors stemming from both amplitude damping and dephasing~\cite{Michael2016}. For a recent detailed overview of various bosonic codes, see Ref.~\cite{Albert2017}.

With the recent theoretical work setting the stage for microwave-photonic quantum computing to reach new heights, experiments with SQCs have followed. After a few experiments demonstrated methods for generation and manipulation of cat states in a single resonator~\cite{Vlastakis2013, Sun2014, Leghtas2015}, a recent experiment demonstrated entanglement of coherent states of two microwave field in two separate resonators coupled to a superconducting qutrit~\cite{Wang2016}. In this experiment, the amplitudes of the entangled coherent states reached up to 80 photons. In another impressive achievement, a full QEC code was implemented for cat states in a 3D circuit-QED system, using real-time feedback to encode, decode, monitor, and correct errors from amplitude damping~\cite{Ofek2016}. This was the first experiment ever to reach ``break-even'' for QEC, meaning that the lifetime of the protected logical qubit \textit{exceeded} the lifetimes of \textit{all} parts of the system.

\subsection{Metamaterials}
\label{sec:Metamaterials}

Metamaterials are artificial composite structures that can go beyond the limitations of natural materials when it comes to manipulating electromagnetic waves~\cite{Bliokh2008, Zagoskin2011, Zheludev2012, Lapine2014}. Since superconducting devices have low losses, compact structures, and large nonlinearities, they can be utilized to make novel, nearly ideal metamaterial structures. For example, a tunable superconducting metamaterial has been implemented by embedding 480 Josephson junctions in a resonator, and shown to function as a parametric amplifier with high gain and large squeezing~\cite{Castellanos-Beltran2008}. Also, SQUID-based metamaterials~\cite{Savinov2012PRL, Jung2013} can exhibit strong nonlinearities and wide-band tunability along with unusual magnetic properties due to macroscopic quantum effects~\cite{Tsironis2014}. For extensive reviews on superconducting metamaterials, see Refs.~\cite{Anlage2011, Jung2014}. In this section, we limit ourselves to briefly describing \textit{quantum metamaterials}~\cite{Rakhmanov2008, Zagoskin2011, Zagoskin2016} constructed with superconducting qubits.

Superconducting qubits can couple coherently and strongly, even ultrastrongly, to microwave fields in resonators and transmission lines, as discussed in Secs.~\ref{sec:CircuitQED} and \ref{sec:WaveguideQED}. This offers unique opportunities for designing artificial quantum structures. For example, theoretical studies have shown that, in a quantum metamaterial made from an infinite chain of identical charge qubits inside a cavity, it is possible to implement a breathing photonic crystal with an oscillating band gap~\cite{Rakhmanov2008}. In such a metamaterial, the optical properties of the photonic crystal can be controlled by putting the qubits in various superposition states~\cite{Rakhmanov2008}. Other studies of quantum metamaterials with qubits have proposed setups without need for local control~\cite{Shvetsov2013}, explored the effect of dissipation in 1D setups~\cite{Wilson2013}, and investigated the effects of collective oscillations among many qubits in a resonator~\cite{Volkov2014}. Further theoretical proposals include making a low-loss left-handed quantum metamaterial with negative magnetic permeability using SQUID-based qutrits~\cite{Du2006}, generating nonclassical photonic states from the interaction between microwaves and a quantum metamaterial~\cite{Mukhin2013}, and combining a metamaterial with a regular transmission line to form a resonator that can achieve ultrastrong multimode coupling with a superconducting qubit~\cite{Egger2013}.

There have only been a few experiments on superconducting quantum metamaterials so far~\cite{Savinov2012, Macha2014, Kakuyanagi2016}. In the experiment of Ref.~\cite{Macha2014}, a prototype quantum metamaterial, consisting of 20 superconducting aluminum flux qubits embedded in a niobium microwave resonator, was demonstrated. In the more recent experiment of Ref.~\cite{Kakuyanagi2016}, 4300 flux qubits were coupled to an $LC$ resonator. A major challenge in fabricating this kind of superconducting quantum metamaterial is to make the qubits as identical as possible, since qubit-energy disorder degrades the metamaterial properties. Some recent theoretical works discuss the properties of many-qubit systems with disorder and how such disorder can be mitigated~\cite{Ian2012, Shapiro2015, Lambert2016}. However, to find real practical advantages of quantum metamaterials over classical metamaterials, many fundamental problems in this emerging field need to be answered both experimentally and theoretically in the near future.

\section{Summary and perspectives}
\label{sec:Summary}

In summary, both experimental and theoretical research on Josephson-junction-based superconducting quantum devices has made dramatic progress in the past $20$ years. Many phenomena from atomic physics and quantum optics have now been demonstrated in experiments in the microwave domain, often with unprecedented clarity, as shown in~\tabref{tab:Experiments}. Moreover, these superconducting quantum devices, functioning as macroscopic quantum machines, also provide a platform for exploring new phenomena and regimes that are hard to reach in traditional quantum optics and atomic physics, e.g., ultra-strong coupling and deep-strong coupling at the level of a single microwave photon. These new experimental capabilities open up new research directions like quantum nonlinear optics at the single-photon level, and will also help to improve understanding of the counterintuitive properties of quantum mechanics. Indeed, much new and interesting physics is emerging as circuit QED continues to develop. We note here that the study of microwave photonics using SQCs has benefitted greatly from the use of QuTiP (Quantum Toolbox in Python), which is a free open-source software (available on all major platforms) that is now being used at many institutions around the world for simulating the dynamics of open quantum systems~\cite{Johansson2012, Johansson2013a}.

Due to rapid development of device design, as well as progress in materials and fabrication methods, the coherence time of superconducting qubits has been improved by five orders of magnitude, going from nanoseconds to the threshold for error correction and fault-tolerant quantum computation. The coherence can be improved even further, e.g., by applying active feedback control methods~\cite{Zhang2014a}. Along with the improvements in coherence time, basic quantum-computation operations have been experimentally demonstrated with reasonable levels of performance using superconducting qubits. There does not seem to be any fundamental obstacle for realizing superconducting quantum computation on a larger scale. These developments are stimulating both academic laboratories and commercial companies, e.g., Google, Intel, and IBM, to announce aggressive strategies for designing and building practical quantum computers using superconducting qubits. For example, IBM is building a cloud-based, world-first commercial universal quantum computing service. Such a cloud quantum computer can be used by researchers around the world.

In order to successfully commercialize superconducting quantum computing, there are still many scientific and technological challenges to overcome; examples include error correction, quantum-information storage, and unwanted crosstalk between different circuit elements. The next crucial step for superconducting quantum computation is to implement error correction well enough that a complete algorithm for quantum digital logic can be run within the effective coherence time of a large scalable circuit. Since current error-correction approaches require large numbers of physical qubits to encode a single logical qubit, such an implementation would increase the complexity of circuit control, measurements, and fabrication beyond our current technological ability. While these challenges make building a large-scale universal quantum computer a long-term goal, efforts are being made to enable superconducting quantum-computing circuits to accomplish certain tasks, such as quantum simulation and quantum-assisted optimization, in the near future even without error correction. For example, D-Wave Systems has a quantum computing service which may be used for optimization. There is also work towards realizing quantum sampling and quantum machine learning using superconducting circuits. Researchers expect that quantum sampling with a $7\times7$ lattice of superconducting qubits can become the first example of quantum supremacy (a quantum computation surpassing a conventional computer), perhaps as early as 2017.

Superconducting quantum circuits can be coupled to many other solid-state quantum components to form hybrid devices, as has been demonstrated in several experiments~\cite{Xiang2013}. The main objective for such work is to harness the strengths of the different systems, such as the fast control possible in superconducting circuits, the long coherence times of nuclear spins, and the long-distance transfer possible with optical light, to better implement quantum information processing. A hybrid quantum computer might thus consist of a spin system acting as a quantum memory and a part with superconducting circuitry used as a quantum processor. Other hybrid devices being explored include superconducting circuits combined with semiconductors that can host Majorana fermions. If such an approach is successful, then topological quantum computing based on braiding of Majorana zero-modes might become possible.

Research on superconducting quantum devices and quantum computation is still developing at a rapid pace. New research topics continue to emerge in this field and the collaboration between industry and academia looks set to push the research on superconducting quantum computation significantly forward. We expect that these developments will have a profound impact both on fundamental physics and on applications of quantum microwave devices.


\section*{Acknowledgements}
We thank Ze-Liang Xiang and Sergey Shevchenko for helpful discussions. X.G. and Y.X.L. are supported by the National Basic Research Program of China (973 Program) under Grant No.~2014CB921401, the National Natural Science Foundation of China under Grant No.~91321208, the Tsinghua University Initiative Scientific Research Program, and the Tsinghua National Laboratory for Information Science and Technology (TNList) Cross-discipline Foundation. X.G. was also supported by the RIKEN IPA program. A.F.K. and A.M. gratefully acknowledge long-term fellowships from the Japan Society for the Promotion of Science (JSPS). A.M. is supported by the Grant No.~DEC-2011/03/B/ST2/01903 of the Polish National Science Centre. F.N. acknowledges support from the RIKEN iTHES Project, the MURI Center for Dynamic Magneto-Optics via the AFOSR award number FA9550-14-1-0040, the Japan Society for the Promotion of Science (KAKENHI), the IMPACT program of JST, JSPS-RFBR grant No.~17-52-50023, CREST grant No.~JPMJCR1676, the RIKEN-AIST ``Challenge Research'' program, and the John Templeton Foundation.

\setcounter{figure}{0}
\appendix
\label{sec:Appendix}

\section{Circuit Quantization}
\label{sec:circuit-quantization}

In several chapters of this review, we have written down Hamiltonians for artificial atoms and resonators formed by SQCs. In many cases, these Hamiltonians are the same as those used in quantum optics with natural atoms and optical cavities. To understand how SQCs are used to implement quantum optics with microwave photons, and also how SQCs allow us to design the artificial atoms and their interactions with those photons, it is instructive to see how these Hamiltonians are derived from circuit models. Following early work in the 1970s~\cite{Widom1979}, the process for quantizing circuits based on canonical quantization has been well described in Refs.~\cite{Yurke1984, Devoret1995, Burkard2003, Wendin2005, Nigg2012, QuantumMachines2014, Vool2016a}. To make this review more self-contained, we cover the main points in these references below, including some pertinent examples from the treatments in Refs.~\cite{TornbergPhD, JRJohanssonPhD, Anton2014}.

\subsection{Quantization procedure}

To arrive at a quantum description given an electrical circuit, the first step is to write down the classical \textit{Lagrangian} $\mathfrak{L}$~\cite{Goldstein} of the circuit. It turns out to be convenient to work with \textit{node fluxes}
\be
\Phi_n(t) = \int_{-\infty}^t V_n(t') \id t',
\ee
where $V_n$ denotes node voltage at node $n$, and \textit{node charges}
\be
Q_n(t) = \int_{-\infty}^t I_n(t') \id t',
\ee
where $I_n$ denotes node current. Note that Kirchhoff's laws apply and may reduce the number of degrees of freedom in the circuit. For example, the total voltage drop around a loop $l$ should be zero, and therefore
\be
\sum_{b \: \text{around} \: l} \Phi_b = \Phi_{\rm ext},
\ee
where $\Phi_{\rm ext}$ is an external magnetic flux threading the superconducting loop and the $\Phi_b$ are the \textit{branch fluxes} (not to be confused with the node fluxes) around the loop. The external flux is subject to the flux quantization condition $\Phi_{\rm ext} = n \Phi_0$, where $n$ is an integer and $\Phi_0 = h / 2e$, where $e$ is the elementary charge, is the flux quantum, which also enters in expressions for the Josephson junction below.

With the node fluxes as our \textit{generalized coordinates}, the Hamiltonian $H$ follows from the \textit{Legendre transformation}~\cite{Goldstein}
\be
H = \sum_n \frac{\partial \mathfrak{L}}{\partial \dot{\Phi}_n} \dot{\Phi}_n - \mathfrak{L}. \label{EqLegendreTransf}
\ee
The \textit{generalized momenta} ${\partial \mathfrak{L}}/{\partial \dot{\Phi}_n}$ will often, but not always, be the node charges $Q_n$.

Up to this point, everything (bar the quantization condition for the external flux) has been completely classical. To proceed to quantum mechanics, we promote the generalized coordinates and momenta to operators with the canonical commutation relation
\be
\comm{\Phi_n}{\frac{\partial \mathfrak{L}}{\partial \dot{\Phi}_m}} = i\hbar \delta_{nm},
\label{eq:CanonicalCommutationRelations}
\ee
where $\delta_{nm}$ is the Kronecker delta ($\delta_{nm} = 1$ if $n=m$, $\delta_{nm} = 0$ if $n \neq m$).

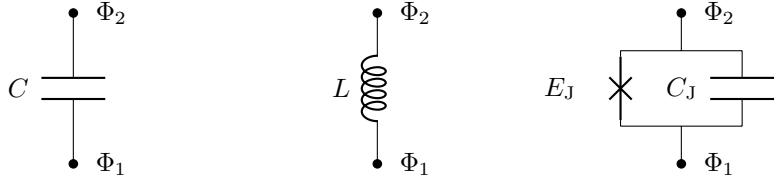
\begin{figure}\centering
	\begin{circuitikz} \draw
		(1,0) to[C, l=$C$,*-*] (1,2) 
		
		(5,0) to[L, l=$L$,*-*] (5,2)
		
		(9,0) to [short,*-] (9,0.5)
		
		(8.2,0.5) to (9.8,0.5)
		(8.2,0.5) to[barrier, l=$E_\text{J}$,] (8.2,1.5)
		(9.8,0.5)  to[C, l=$C_\text{J}$,] (9.8,1.5)
		(8.2,1.5) to (9.8,1.5)
		(9,1.5) to [short,-*] (9,2)
		
		(1.5,0) node[]{$\Phi_1$}
		(1.5,2) node[]{$\Phi_2$}
		(5.5,0) node[]{$\Phi_1$}
		(5.5,2) node[]{$\Phi_2$}
		(9.5,0) node[]{$\Phi_1$}
		(9.5,2) node[]{$\Phi_2$}
		;
	\end{circuitikz}
	\caption[The three basic circuit elements for superconducting circuits]{The three basic circuit elements used in quantum optics for superconducting circuits. From left to right: capacitance $C$, inductance $L$, and a Josephson junction with capacitance $C_\text{J}$ and Josephson energy $E_\text{J}$.\label{FigThreeCircuitElements}}
\end{figure}

For quantum optics with superconducting circuits, three basic elements are needed: capacitors, inductors, and Josephson junctions, illustrated in \figref{FigThreeCircuitElements}. A Josephson junction consists of a thin insulating barrier between two superconducting leads, and it can be modeled as a capacitor in parallel with a nonlinear inductor characterized by the Josephson energy $E_\text{J}$.

The Lagrangians for capacitors and inductors are straightforward. The energy of a capacitor with capacitance $C$ is
\be
\frac{C V^2}{2} = \frac{C\left(\dot{\Phi}_1-\dot{\Phi}_2\right)^2}{2},
\ee
where $V$ is the voltage across the capacitor, and for an inductor with inductance $L$ it is 
\be
\frac{LI^2}{2} = \left\{ V = L\dot{I} \right\} = \frac{\left(\Phi_1 - \Phi_2 \right)^2}{2L},
\ee
where $I$ is the current through the inductor. In the Lagrangian, capacitive terms (terms with $\dot{\Phi}$) represent kinetic energy and give a positive contribution, while inductive terms (terms with $\Phi$) represent potential energy and give a negative contribution. We thus have
\bea
\mathfrak{L}_C &=& \frac{C\left(\dot{\Phi}_1-\dot{\Phi}_2\right)^2}{2}, \label{EqLagrangianC} \\
\mathfrak{L}_L &=& -\frac{\left(\Phi_1 - \Phi_2 \right)^2}{2L}.  \label{EqLagrangianL}
\eea

For the Josephson junction, the contribution from the capacitive part with $C_\text{J}$ follows immediately from previous discussion. To get the contribution from the nonlinear inductor, we use the Josephson equations~\cite{Josephson1962, Waldram}
\bea
I_\text{J} &=& I_\text{c} \sin\phi, \label{EqJosephson1}\\
\dot{\phi} &=& \frac{2e}{\hbar} V(t), \label{EqJosephson2}
\eea
where $I_\text{J}$ is the supercurrent through the junction, $I_\text{c}$ is the critical current (the maximum value of $I_\text{J}$), $V(t)$ is the voltage across the junction, and $\phi = 2e \left(\Phi_1 - \Phi_2\right)/\hbar = 2\pi \left(\Phi_1 - \Phi_2\right)/\Phi_0$ is a phase difference across the junction. These equations give
\be
\int_{-\infty}^t I(t')V(t') \id t' = E_\text{J} \left(1 - \cos\phi \right),
\ee
where we have identified the Josephson energy $E_\text{J} = \hbar I_\text{c}/2e$. The Lagrangian for the Josephson junction is thus
\be
\mathfrak{L}_\text{JJ} = \frac{C_\text{J}\left(\dot{\Phi}_1-\dot{\Phi}_2\right)^2}{2} - E_\text{J} \left(1 - \cos\phi \right). \label{EqLagrangianJJ}
\ee
Note that the inductive term is a cosine function rather than the quadratic function for a normal inductor; this is why the Josephson junction can be seen as nonlinear inductance. This \textit{nonlinearity} is essential for making artificial atoms with various level structures. From normal capacitors and inductors we can only make harmonic $LC$ oscillators.

Once the device is fabricated, the Josephson energy is fixed. However, two Josephson junctions in a superconducting ring, i.e., a DC superconducting quantum interference device (SQUID) structure, effectively function as a single junction where the Josephson energy can be tuned by an external flux threading the superconducting ring.

\subsection{Hamiltonian for a Cooper-pair box}

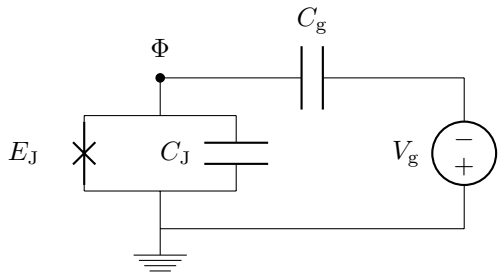
\begin{figure}\centering
	\begin{circuitikz}[american] \draw
		(0,0) node[ground] {}
		(0,2) to[C, l=$C_\text{g}$,*-] (4,2)
		(4,0) to [V, l=$V_\text{g}$] (4,2)
		
		(-1,0.5) to[barrier, l=$E_\text{J}$,] (-1,1.5)
		(1,0.5)  to[C, l=$C_\text{J}$,] (1,1.5)
		
		(-1,1.5) -- (1,1.5)
		(-1,0.5) -- (1,0.5)
		(0,0) -- (0,0.5)
		(0,1.5) -- (0,2)
		(4,0) -- (0,0)
		
		(0,2.4) node[]{$\Phi$}
		;
	\end{circuitikz}
	\caption[Circuit diagram for a Cooper-pair box]{Circuit diagram for a Cooper-pair box. The Josephson junction is modeled by the capacitance $C_\text{J}$ in parallel with a nonlinear inductor having Josephson energy $E_\text{J}$. The node between the gate capacitance $C_\text{g}$ and the Josephson junction is called the ``island''. \label{FigCPBCircuit}}
\end{figure}

The prototypical superconducting artificial atom is the \textit{Cooper-pair box} (CPB), discussed in~\secref{sec:SuperconductingChargeQubits}. The circuit diagram of a CPB is shown in \figref{FigCPBCircuit}. The CPB consists of a small superconducting island connected to a superconducting reservoir via a Josephson junction, which allows Cooper pairs to tunnel on and off the island. The model also includes an external voltage source $V_\text{g}$ coupled to the island via a gate capacitance $C_\text{g}$, to determine the background charge $n_\text{g} = C_\text{g}V_\text{g}/2e$ (measured in units of Cooper pairs) that the environment induces on the island.

Using Eqs.~(\ref{EqLagrangianC}) and (\ref{EqLagrangianJJ}) gives the CPB Lagrangian
\be
\mathfrak{L}_{\text{CPB}} = \frac{C_\text{g}\left(\dot{\Phi} - V_\text{g}\right)^2}{2} + \frac{C_\text{J}\dot{\Phi}^2}{2} - E_\text{J} (1 - \cos \frac{2e\Phi}{\hbar})
\ee
Applying the Legendre transformation, identifying the conjugate momentum (the node charge) $Q = (C_\text{J} + C_\text{g})\dot{\Phi} - C_\text{g}V_\text{g}$, and removing constant terms that do not contribute to the dynamics, we arrive at the Hamiltonian
\be
H_{\text{CPB}} = 4E_\text{C}(n-n_\text{g})^2 - E_\text{J} \cos\phi,
\ee
where $E_\text{C} = e^2/2(C_\text{g} + C_\text{J})$ is the electron charging energy, $n=-Q/2e$ is the number of Cooper pairs on the island, and $\phi = 2e\Phi/\hbar$.

We now promote $\Phi$ and $Q$ to operators as described around \eqref{eq:CanonicalCommutationRelations}. This translates into a commutation relation for $n$ and $\phi$, which since the Hamiltonian is periodic in $\phi$ should be expressed as~\cite{BishopPhD, GerryKnight}
\be
\comm{e^{i\phi}}{n} = e^{i\phi}.
\ee
From this follows that $e^{\pm i\phi} \ket{n} = \ket{n \mp 1}$, where $\ket{n}$ is the charge basis counting the number of Cooper pairs. Using the resolution of unity~\cite{Sakurai} and $\cos\phi = (e^{i\phi}+e^{-i\phi})/2$ we can then write the Hamiltonian in the charge basis as 
\be
H_{\text{CPB}} = \sum_n\left[ 4E_\text{C}(n-n_\text{g})^2 \ketbra{n}{n} - \frac{1}{2}E_\text{J} \left(\ketbra{n+1}{n} + \ketbra{n-1}{n}\right) \right],
\ee
which is \eqref{eq:HamiltonianCPBChargeBasis}.

\subsection{Quantizing a 1D infinite transmission line}
\label{app:Quantizing1DTransmissionLine}

\begin{figure}\centering
\begin{circuitikz} \draw
 (0,0) node[ground] {}
   to[C, l=$C_0 \Delta x$] (0,2)
   to[L, l=$L_0 \Delta x$] (3,2)
   to[C, l=$C_0 \Delta x$] (3,0)
  (3,2)  to[L, l=$L_0 \Delta x$] (6,2)
  to[C, l=$C_0 \Delta x$] (6,0)
  
  (6,2) -- (7,2)
  (7.3,2) -- (7.1,2)
  (7.6,2) -- (7.4,2)
  
  (-1,0) -- (7,0)
  (-1.3,0) -- (-1.1,0)
  (-1.6,0) -- (-1.4,0)
  (7.3,0) -- (7.1,0)
  (7.6,0) -- (7.4,0)
  
  (-1,2) -- (0,2)
  (-1.3,2) -- (-1.1,2)
  (-1.6,2) -- (-1.4,2)
  
  (0,2.3) node[]{$\Phi_{n-1}$}
  (3,2.3) node[]{$\Phi_{n}$}
  (6,2.3) node[]{$\Phi_{n+1}$}
  
  (4.5, -0.3) node[]{$\underbrace{\qquad\qquad\qquad\qquad}_{\Delta x}$}
;
\end{circuitikz}
\caption[Circuit diagram for a transmission line]{Circuit diagram for a transmission line. $C_0$ and $L_0$ denote capacitance per unit length and inductance per unit length, respectively, and $\Delta x$ is a small distance which will go to zero in the continuum limit. \label{FigTransmissionLineCircuit}}
\end{figure}
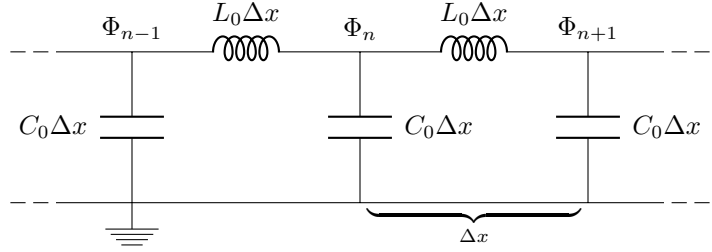

An open transmission line, the basis for the physics reviewed in~\secref{sec:WaveguideQED}, can be modeled by the circuit depicted in \figref{FigTransmissionLineCircuit}~\cite{Pozar2011, Collin2001}. From this circuit model, using Eqs.~(\ref{EqLagrangianC}) and (\ref{EqLagrangianL}), we immediately obtain the Lagrangian
\be 
\mathfrak{L}_{\text{TL}} = \sum_n \left[\frac{C_0 \Delta x}{2}\left( \dot{\Phi}_n(t)\right)^2 - \frac{1}{2L_0 \Delta x}\left(\Phi_{n+1}(t) - \Phi_n(t) \right)^2 \right], 
\ee
from which we can identify the conjugate momenta
\be
\frac{\partial \mathfrak{L}_{\text{TL}}}{\partial \dot{\Phi}_n} = C_0 \Delta x \dot{\Phi}_n(t), 
\ee
which are the node charges $Q_n(t)$. Applying the Legendre transformation [\eqref{EqLegendreTransf}] to $\mathfrak{L}_{\text{TL}}$, inserting the definition of the node charges, and taking the limit $\Delta x \rightarrow 0$ (or rather $\Delta x \rightarrow \rd x$) gives the Hamiltonian
\be
H_\text{TL} = \frac{1}{2}\int_{-\infty}^\infty \id x \left[ \frac{Q(x,t)^2}{C_0} + \frac{1}{L_0}\left(\frac{\partial \Phi(x,t)}{\partial x}\right)^2 \right],
\label{EqHTL}
\ee
where $Q(x,t)$ and $\Phi(x,t)$ denote charge density and flux density, respectively. The Lagrangian can also be written in a continuum form
\bea
\mathfrak{L}_{\text{TL}} &=& \int_{-\infty}^\infty   \id x \; \mathcal{L} = \int_{-\infty}^\infty   \id x \; \left[ \frac{C_0}{2}
\left(\dot{\Phi}(x,t)\right)^2 - \frac{1}{2L_0}\left(\frac{\partial \Phi(x,t)}{\partial x}\right)^2 \right],
\eea
and applying the Euler--Lagrange equations~\cite{PeskinSchroeder}
\be
\frac{\partial}{\partial \mu}\left(\frac{\partial \mathcal{L}}{\partial \left(\frac{\partial \Phi}{\partial \mu}\right)}\right) - \frac{\partial \mathcal{L}}{\partial \Phi} = 0, \quad \mu = x,t
\ee
to the Lagrangian density $\mathcal{L}$ gives the wave equation
\be
\frac{\partial^2\Phi(x,t)}{\partial t^2} - \frac{1}{L_0C_0} \frac{\partial^2\Phi(x,t)}{\partial x^2} = 0.
\ee
This tells us that there will be left- and right-moving flux waves,
\be
\Phi(x,t) = \Phi_\text{L}(kx + \omega t) + \Phi_\text{R}(- kx + \omega t),
\ee
moving in the transmission line with velocity $v = 1/\sqrt{L_0C_0}$ and wavenumber $k = \omega/ v$.

So far, all calculations have been classical. To quantize the field in the transmission line, we promote the generalized coordinates and momenta to operators with the commutation relation
\be
\comm{\Phi(x)}{Q(x')} = i\hbar \delta(x-x'),
\ee
where the delta function, rather than the Kronecker delta, appears since we are working with a \textit{continuum} model. From the form of the Hamiltonian in \eqref{EqHTL} it can be seen that we have a collection of harmonic oscillators. We can thus rewrite the generalized coordinates and momenta in terms of annihilation and creation operators. The left- and right-moving fluxes become~\cite{Yurke1984, Peropadre2013a}
\be
\Phi_\text{L/R}(x,t) = \sqrt{\frac{\hbar Z_0}{4\pi}} \int_0^\infty \frac{\id \omega}{\sqrt{\omega}} \left(a_{\text{L/R},\omega} e^{-i\left(\pm kx + \omega t\right)} + \text{H.c.} \right),
\label{EqPhiOpenTLOperatorExpansion}
\ee
where the annihilation and creation operators obey the commutation relations
\be
\comm{a_{\text{X},\omega}}{a_{\text{X'},\omega'}^\dag} = \delta(\omega-\omega')\delta_{\rm XX'},
\ee
and $Z_0 = \sqrt{L_0/C_0}$ is the \textit{characteristic impedance} of the transmission line.

It is illustrative to calculate the \textit{spectral density} of the voltage fluctuations in our open transmission line using \eqref{EqPhiOpenTLOperatorExpansion}. Using $V=\partial_t \Phi$, we have~\cite{Clerk2010}
\bea
S_{VV}[\omega] &=& \int_{-\infty}^\infty \id t e^{i\omega t}\expec{V(t)V(0)} = \int_{-\infty}^\infty \id t e^{i\omega t} \frac{\hbar Z_0}{4\pi} \int_0^\infty \frac{\id \omega'}{\sqrt{\omega'}} \int_0^\infty \frac{\id \omega''}{\sqrt{\omega''}}(-i\omega')(-i\omega'') \nn\\
&&\times \bigg\langle \left(a_{\text{L},\omega'} e^{-i\left(kx + \omega' t\right)} + a_{\text{R},\omega'} e^{-i\left(-kx + \omega' t\right)} - \text{H.c.} \right) \nn\\
&&\quad\times \left( a_{\text{L},\omega''} e^{-ikx} + a_{\text{R},\omega''} e^{ikx} - \text{H.c.} \right) \bigg\rangle \nn\\
&=& \frac{\hbar Z_0}{4\pi} \int_{-\infty}^\infty \id t e^{i\omega t} \int_0^\infty \id \omega' \sqrt{\omega'} \int_0^\infty \id \omega'' \sqrt{\omega''} 2 e^{-i\omega't} \delta(\omega'-\omega'') \nn\\
&=& \frac{\hbar Z_0}{2\pi}  \int_0^\infty \id \omega' \omega' 2\pi \delta(\omega-\omega') = Z_0 \hbar \omega,
\eea
where we assumed negligible temperature such that the only contribution from the expectation value is terms on the form $\expec{aa^\dag} = 1$. The result shows that the left- and right-traveling modes each contribute $\hbar\omega/2$ to the power spectral density $S_{VV}[\omega]/Z_0$, which agrees well with our expectations for the quantum vacuum fluctuations.

We can now proceed to introduce \textit{boundary conditions} in the open transmission line. Grounding one end at $x=0$ gives the boundary condition $\Phi(0,t) = 0$; it is equivalent to inserting a \textit{mirror} in open space. It is also possible to connect one end of the transmission line to ground via a capacitance or via a SQUID (the latter gives a tunable boundary condition, a \textit{moving} mirror, as discussed in~\secref{sec:Resonators}).

A semi-infinite transmission line still has a continuum of modes, but the boundary condition gives rise to a mode structure. This can be seen as an interference effect between waves approaching the mirror and waves that have been reflected off the mirror. As discussed in~\secref{sec:WaveguideQEDExpSCAtoms}, this mode structure, which is imposed also on the vacuum fluctuations, was explored in Ref.~\cite{Hoi2015} by placing an artificial atom close to a mirror and varying its resonance frequency. The relaxation rate of the atom is proportional to the spectral density of the voltage fluctuations at the atom transition frequency, which given the boundary condition $\Phi(0,t) = 0$ becomes $S_{VV}[\omega]/Z_0 = 2\hbar \omega \sin^2(kx)$.

If we introduce boundary conditions at two points $x=0$ and $x=d$ in an open transmission line, we create a \textit{resonator} (see \secref{sec:Resonators}). Using \eqref{EqPhiOpenTLOperatorExpansion} to satisfy $\Phi(0,t) = 0 = \Phi(d,t)$, we see that these boundary conditions enforce $a_{\text{L},\omega} = - a_{\text{R},\omega}$ and $\sin(kx) = 0$. Thus, only modes with frequencies
\be
\omega_n = \frac{n\pi v}{d} = \frac{n\pi}{d\sqrt{L_0C_0}},
\ee
where $n$ is an integer, remain. We now have a discrete, yet still infinite, collection of harmonic oscillators. In most applications only the fundamental mode $\omega_1 \equiv \omega_\text{r}$ is used, leading to the well-known harmonic-oscillator Hamiltonian
\be
H = \hbar\omega_\text{r} \left(a^\dag a + \frac{1}{2}\right),
\ee
where $a$ now is the annihilation operator for this localized mode. 

\section{Unitary transformations and the Jaynes--Cummings model}
\label{sec:UnitaryTransfJC}

The paradigmatic model for light-matter interaction is the Jaynes--Cummings Hamiltonian, as discussed in detail in~\secref{sec:CircuitQED}. As a complement to that section, we here give a few examples of how various unitary transformations and approximations are used to derive the Jaynes--Cummings Hamiltonian and related versions of it. The treatment here is mostly based on Ref.~\cite{Anton2014}. Throughout this section, we set $\hbar = 1$.

\subsection{From Rabi to Jaynes--Cummings with the rotating-wave approximation}
\label{app:RWA}

As a simple first example, we consider transforming the quantum Rabi Hamiltonian, \eqref{eq:Rabi},
\be 
H_{\text{Rabi}} = \omega_\text{r} a^\dag a + \frac{\omega_\text{q}}{2}\sigma_z + g\sigma_x\left(a+a^\dag\right),
\label{EqRabiHApp}
\ee
to a rotating frame by applying the unitary transformation
\be
U_{\text{rot}} = \exp\left(i\omega_\text{r} t a^\dag a + i \frac{\omega_\text{q}}{2}t\sz \right).
\label{EqRotFrameTransf}
\ee
This will clarify the time dependence of the coupling terms and show when the RWA is valid.

The transformed Hamiltonian $\tilde{H}$ is given by
\be
\tilde{H} = UHU^\dag + i\dot{U}U^\dag.
\label{EqTransfH}
\ee
Since the transformation in \eqref{EqRotFrameTransf} is time-dependent, we must include the second term of \eqref{EqTransfH}, which becomes
\be
i \dot{U}_\text{rot}U_\text{rot}^\dag = i \left(i\omega_\text{r} a^\dag a + i \frac{\omega_\text{q}}{2}\sz \right) U_\text{rot}U_\text{rot}^\dag = - \omega_\text{r} a^\dag a - \frac{\omega_\text{q}}{2}\sz.
\ee
Clearly, the first two terms in \eqref{EqRabiHApp} commute with $U_\text{rot}$, so we only need to find the transformations for the third term. Using the Baker--Hausdorff lemma and the commutation relation $\comm{\sm}{\sz} = 2\sm$, we obtain
\bea
U_\text{rot}\sm U_\text{rot}^\dag &=& \sm + i\frac{\omega_\text{q}}{2}t\comm{\sz}{\sm} + \frac{1}{2!}\left(i\frac{\omega_\text{q}}{2}t\right)^2 \comm{\sz}{\comm{\sz}{\sm}} + \dots \nn\\
&=& \sm\left(1 - i\omega_\text{q}t + \frac{1}{2!} \left(- i\omega_\text{q}t \right)^2 + \dots \right) = \sm e^{-i \omega_\text{q}t},
\eea
which also leads to
\bea
U_\text{rot}\sp U_\text{rot}^\dag = \left(U_\text{rot} \sm U_\text{rot}^\dag \right)^\dag = \sp e^{i\omega_\text{q}t}.
\eea
In the same way, remembering that $\comm{a}{a^\dag a} = a$, we obtain
\bea
U_\text{rot} a U_\text{rot}^\dag &=& a e^{- i\omega_\text{r}t}, \\
U_\text{rot} a^\dag U_\text{rot}^\dag &=& a^\dag e^{i\omega_\text{r}t}.
\eea

Combining all these results, we arrive at the transformed Hamiltonian
\bea
H_{\text{rot}} &=& U_{\text{rot}} H_{\text{Rabi}} U_{\text{rot}}^\dag  + i\dot{U}_{\text{rot}}U_{\text{rot}}^\dag = g \left(\sm e^{-i\omega_\text{q}t} + \sp e^{i\omega_\text{q}t} \right) \left(a e^{-i\omega_\text{r}t}+a^\dag e^{i\omega_\text{r}t} \right) \nn\\
&=& g \bigg(a\sp e^{i(\omega_\text{q} - \omega_\text{r})t} + a^\dag\sm e^{i(\omega_\text{r} - \omega_\text{q})t} + a\sm e^{-i(\omega_\text{q} + \omega_\text{r})t} + a^\dag\sp e^{i(\omega_\text{q}+\omega_\text{r})t} \bigg).\quad\quad
\label{EqRabiHAppRot}
\eea
The last two terms will always oscillate rapidly and can be discarded in the RWA provided that $g$ is small compared to
$\omega_\text{q} + \omega_\text{r}$. This leads to the Jaynes--Cummings Hamiltonian, \eqref{eq:JC}.

\subsection{The driven Jaynes--Cummings model}

The resonator mode in the Jaynes--Cummings model can be driven by a classical drive (pump) through the beam-splitter interaction described by $\left(a+a^\dag \right) \left[\epsilon^*(t) e^{i\omega_{\rm d} t} + \cc \right]$, where $\cc$ denotes the complex conjugate term, $\omega_\text{d}$ is the driving field frequency and $\epsilon(t)$ is the coupling strength between the resonator and drive. Under the RWA, this interaction simplifies to
\be
H_{\rm drive} = a \epsilon^*(t) e^{i \omega_\text{d} t} + a^\dag \epsilon(t) e^{-i \omega_\text{d} t}.
\label{eq:HDrive}
\ee
We can transform the driven Jaynes--Cummings Hamiltonian, Eqs.~(\ref{eq:JC}) plus (\ref{eq:HDrive}), to the rotating frame at the drive frequency $\omega_\text{d}$ using \eqref{EqTransfH} with the unitary operation $U = \exp \left[ i \omega_\text{d} t \left(a^\dag a + \sz \right) \right]$. This gives
\be
\tilde{H} = \left(\omega_{\rm r} - \omega_{\rm d} \right) a^\dag a + \frac{\left(\omega_\text{q} - \omega_\text{d} \right)}{2} \sz + \frac{\Omega_0}{2} \left(a \sp + a^\dag \sm \right) + a \epsilon^*(t) + a^\dag \epsilon(t).
\ee

The drive acting on the cavity field can be switched directly to the two-level atom with the displacement transformation $D=\exp \left[ \alpha(t) a^\dag - \alpha^*(t) a \right]$~\cite{Blais2007}, provided that $\alpha(t)$ satisfies the equation $-i \dot{\alpha}(t) + \left(\omega_\text{r} - \omega_\text{d} \right) \alpha(t) + \epsilon(t) = 0$. Thus, we have
\be
\tilde{H} = \left(\omega_{\rm r} - \omega_\text{d} \right) a^\dag a + \frac{\left(\omega_{\rm q} - \omega_\text{d} \right)}{2} \sz + \frac{\Omega_0}{2} \left(a \sp + a^\dag \sm \right) + \frac{\Omega_0}{2} \left[\sm \alpha^*(t) + \sp \alpha(t) \right].
\ee
When the drive frequency is detuned from the resonator frequency, we have effectively a single-qubit gate operation $H = \sm \alpha^*(t) + \sp \alpha(t)$~\cite{Blais2007}, i.e., $\sigma_x {\rm Re}\,\alpha(t) + \sigma_y {\rm Im}\,\alpha(t)$ can be interpreted as a rotation on the Bloch sphere.

\subsection{The dispersive regime of the Jaynes--Cummings model}
\label{app:JCDispDerivation}

As discussed in~\secref{sec:JaynesCummingsDispersive}, the Jaynes--Cummings Hamiltonian \eqref{eq:JC} can be approximated by \eqref{eq:dispersive} in the dispersive regime, where $g \ll \abs{\Delta} = \abs{\omega_{\rm q} - \omega_{\rm r}}$. Below, we show how \eqref{eq:dispersive} can be derived.

We begin from the Jaynes--Cummings Hamiltonian in \eqref{eq:JC}, dividing it into noninteracting and interacting parts:
\bea
H &=& H_0 + H_\text{I}, \\
H_0 &=&  \omega_\text{r} a^\dag a + \frac{\omega_\text{q}}{2}\sigma_z, \\
H_\text{I} &=& g\left(a\sigma_+ + a^\dag\sigma_- \right).
\eea
We then follow the main idea of the Schrieffer--Wolff transformation~\cite{SchriefferWolff1966}, i.e., we choose a unitary transformation $U = \exp(S)$ such that $\comm{S}{H_0} = - H_\text{I}$. This choice leads to a cancellation of terms in the expansion with the Baker--Hausdorff lemma:
\bea
\exp\left(S \right) H \exp\left(- S \right)
&=& H + \comm{S}{H} + \frac{1}{2!}\comm{S}{\comm{S}{H}} + \dots \nn\\
&=& H_0 + H_\text{I} + \comm{S}{H_\text{I}} - H_\text{I} + \frac{1}{2}\comm{S}{- H_\text{I} } + \dots \nn\\
&=& H_0 + \frac{1}{2}\comm{S}{H_\text{I}} + \dots,
\label{EqFrohlichNakajima}
\eea
This transformation is widely used in circuit QED (and other areas of physics), e.g., in Refs.~\cite{Grajcar2008, Hauss2008a}, where $\comm{S}{H_\text{I}}$ contributes to important physical phenomena. 

Noting that
\bea
\comm{a \sp - a^\dag \sm}{a^\dag a} &=& a \sp + a^\dag \sm,\\
\comm{a \sp - a^\dag \sm}{\sz} &=& - 2 \left( a \sp + a^\dag \sm \right), \\
\comm{a \sp - a^\dag \sm}{a \sp + a^\dag \sm} &=& 2 N \sz, \\
N &=& a^\dag a + \frac{1}{2} + \frac{\sz}{2},
\eea
where we have used $\comm{a}{a^\dag} = 1$, $\comm{\sp}{\sm} = \sz$, and $\comm{\sz}{\pm \sigma_\pm} = \pm 2 \sigma_\pm$, we can see that $S = \lambda \left(a \sp - a^\dag \sm \right)$, where $\lambda$ is to be determined, is a suitable choice for our Schrieffer--Wolff transformation. Actually, the Jaynes--Cummings Hamiltonian can be diagonalized exactly by choosing 
$\lambda = - \theta/ 2\sqrt{N}$, where $\tan\theta = 2 g \sqrt{N}/\Delta$~\cite{Carbonaro1979, Boissonneault2009}. The result is
\be 
e^S H e^{-S} = \omega_{\rm r} a^\dag a + \left(\omega_\text{r} - \sqrt{\Delta^2 + 4 g^2 N} \right) \frac{\sz}{2}.
\ee
To satisfy the condition $\comm{S}{H_0} = - H_\text{I}$, we instead use $\lambda = g/\Delta$. In the dispersive regime, $\abs{\lambda} = \abs{g/\Delta} \ll 1$ is the relevant small parameter in the perturbative expansion of the dispersive transformation $U_{\text{disp}} = \exp\left[\lambda\left(a\sp - a^\dag \sm \right)\right]$. Below, we show how individual operators transform under $U_{\text{disp}}$, which can be useful in some contexts~\cite{Boissonneault2009}.

The Baker--Hausdorff lemma gives
\bea
U_{\text{disp}} a U_{\text{disp}}^\dag &=& a + \lambda\comm{a\sp - a^\dag \sm}{a} + \frac{\lambda^2}{2} \comm{a\sp - a^\dag \sm}{\comm{a\sp - a^\dag \sm}{a}} + \mathcal{O}\left(\lambda^3\right) \nn\\
&=& a + \lambda \sm +  \frac{\lambda^2}{2}a\sz + \mathcal{O}\left(\lambda^3\right).
\eea
%
In a similar fashion, we calculate
\bea
U_{\text{disp}} \sz U_{\text{disp}}^\dag &=& \sz + \lambda\comm{a\sp - a^\dag \sm}{\sz} \nn\\
&&+ \frac{\lambda^2}{2} \comm{a\sp - a^\dag \sm}{\comm{a\sp - a^\dag \sm}{\sz}} + \mathcal{O}\left(\lambda^3\right) \nn\\
&=& \sz - 2\lambda\left(a\sp + a^\dag \sm\right) - \lambda^2 \comm{a\sp - a^\dag \sm}{a\sp + a^\dag \sm} + \mathcal{O}\left(\lambda^3\right) \nn\\
&=& \sz - 2\lambda\left(a\sp + a^\dag \sm\right) - \lambda^2\sz\left(1+2a^\dag a\right) + \mathcal{O}\left(\lambda^3\right),
\eea
where we used $\sp\sm = (1+\sz)/2$ and discarded constant terms since they do not affect the dynamics if they are included in the Hamiltonian. Finally, we also calculate
\be
U_{\text{disp}} \sm U_{\text{disp}}^\dag = \sm + \lambda a \sz + \mathcal{O}\left(\lambda^2\right),
\ee
where we only need to include the first-order terms since $\sm$ only appears in the weak interaction term of the Hamiltonian.

Applying our results to the transformation of the full Hamiltonian yields
\bea
U_{\text{disp}} H_\text{JC} U_{\text{disp}}^\dag &=& \omega_\text{r} a^\dag a + \lambda \omega_\text{r} \left(a\sp + a^\dag \sm\right) + \lambda^2 \omega_\text{r} \sz \left(a^\dag a + \frac{1}{2}\right) + \frac{\omega_\text{q}}{2}\sz \nn\\
&&- \lambda\omega_\text{q}\left(a\sp + a^\dag \sm\right) - \lambda^2 \omega_\text{r} \sz \left(a^\dag a + \frac{1}{2}\right) + g\left(a\sp + a^\dag \sm\right) \nn\\
&&+ \lambda g \sz \left(2a^\dag a + 1\right) + g\mathcal{O}\left(\lambda^2\right) \nn\\
&=& \omega_\text{r} a^\dag a + \frac{\omega_\text{q}}{2}\sz + \lambda g \sz \left(a^\dag a + \frac{1}{2}\right) + \mathcal{O}\left(\lambda^2\right).
\eea
Introducing the notation $\chi = g^2/\Delta$, the transformed Hamiltonian can, to first order in $\lambda$, be written
\be
H_\text{disp} = \left(\omega_\text{r} + \chi\sigma_z\right)a^\dag a + \frac{\omega_\text{q}+ \chi}{2} \sigma_z,
\label{EqHDispApp}
\ee
which is \eqref{eq:dispersive}.

\section{List of acronyms}
\label{sec:Acronyms}

\noindent
1D, 2D, 3D --- one-dimensional, two-dimensional, three-dimensional\\
2DEG --- 2D electron gas\\
AQC --- adiabatic quantum computing\\
ATS --- Autler--Townes splitting\\
CBJJ --- current-biased Josephson junction\\
CNOT --- controlled-NOT gate\\
CPB --- Cooper-pair box\\
CPHASE --- controlled-PHASE gate\\
CPT --- coherent population trapping\\
CPW --- coplanar waveguide\\
DCE --- dynamical Casimir effect\\
DQD --- double quantum dot\\
DSC --- deep-strong coupling\\
EIT --- electromagnetically induced transparency\\
ETA --- electromagnetically induced absorption\\
HOM --- Hong--Ou--Mandel\\
HBT --- Hanbury-Brown--Twiss (e.g., experiment)\\
HEMT --- high-electron-mobility transistor\\
JC --- Jaynes--Cummings (e.g., model)\\
JDA --- Josephson directional amplifier\\
JPC --- Josephson parametric converter\\
JPM --- Josephson photomultiplier\\
JRM --- Josephson ring modulator\\
LO --- local oscillator\\
LWI --- lasing without (population) inversion\\
LZSM --- Landau--Zener--St\"{u}ckelberg--Majorana (e.g., transitions)\\
MEMS --- microelectromechanical system(s)\\
NV --- nitrogen vacancy\\
QA --- quantum annealing\\
QAD --- quantum acoustodynamics\\
QEC --- quantum error correction\\
QED --- quantum electrodynamics\\
QIP --- quantum information processing\\
QND --- quantum nondemolition (e.g., measurement)\\
QST --- quantum-state tomography\\
RWA --- rotating-wave approximation\\
SAW --- surface acoustic wave\\
SME --- stochastic master equation\\
SNAP --- selective number-dependent arbitrary phase (gate)\\
SPDC --- spontaneous parametric down-conversion\\
SQ --- superconducting qubit\\
SQC --- superconducting quantum circuit\\
SQUID --- superconducting quantum interference device\\
TLS --- two-level system\\
USC --- ultrastrong coupling\\
VNA --- vector network analyzer\\
vSTIRAP --- vacuum-stimulated Raman adiabatic passage


\nocite{apsrev41Control}
\bibliography{ControlForBibstyle,library}

\end{document}